%% file: main.tex
\documentclass[twoside, openright, 12pt]{book}
\usepackage[a4paper,top=2.5cm,bottom=2.5cm,left=3.5cm,right=3.5cm]{geometry}

\input{Packages}
\usepackage[sorting=none]{biblatex}
\usepackage{csquotes}
\begin{document}

\frontmatter
\pagenumbering{gobble}      
\pagestyle{empty}
\input{Title}
\thispagestyle{empty}

\tableofcontents

\input{Abstract} 
\input{Notation}

\cleardoublepage
\mainmatter
\pagenumbering{arabic}      
\pagestyle{fancy} 
\input{Introduction}

\input{Math}

\input{Amplitudes}

\input{C4}
\input{Conclusions}

\printbibliography
\end{document}

%% file: Packages.tex
\PassOptionsToPackage{svgnames}{xcolor}
\usepackage{hyperref}
\usepackage[T1]{fontenc}
\usepackage[utf8]{inputenc}
\usepackage{amsfonts}
\usepackage{amssymb}
\usepackage[italian, english ]{babel}
\usepackage{graphicx}
\usepackage{colortbl}                
\usepackage{latexsym}
\usepackage{mathtools}
\usepackage{amsmath}
\usepackage{amsthm}
\usepackage{empheq}
\usepackage{slashed}
\usepackage{cases}
\usepackage{subfigure}
\usepackage{titlesec}
\usepackage{bbold}
\usepackage{float}
\usepackage[affil-it]{authblk}
\numberwithin{equation}{chapter}
\usepackage[parfill]{parskip}
\usepackage{lipsum}
\usepackage{mathdots}

\usepackage{circuitikz}
\usetikzlibrary{circuits.ee.IEC, circuits.logic.US, circuits.logic.IEC}
\usetikzlibrary{patterns}
\usepackage{graphicx}
\usepackage{enumerate}
\usepackage{enumitem}
\usepackage[toc,page]{appendix}
\usepackage{caption}
\usepackage{empheq}
\usepackage{multirow}
\usepackage[version=3]{mhchem}
\usepackage{braket}
\usepackage{centernot}
\usepackage{relsize}
\usepackage{amsmath}
\usepackage{geometry}
\usepackage{lmodern}
\usepackage[tone]{tipa}
\usepackage{hyperref}
\usepackage{array}
\usepackage{bm}
\usepackage{slashed}
\usepackage{multicol}
\usepackage{amscd}
\usepackage{xcolor}

\newtheorem{Ese}{Example}[section]

\newcommand{\n}{\nu}

\newcommand{\CF}{\mathcal{F}}
\usepackage{tcolorbox}
\usepackage{lipsum}
\tcbuselibrary{skins,breakable}
\usetikzlibrary{shadings,shadows}
\usepackage{tikz-cd}
\usepackage{tipa}
\usepackage{upgreek}
\usepackage{sectsty}
\usepackage[bottom]{footmisc}

    {\endtcolorbox}

    {\endtcolorbox}

    {\endtcolorbox}

	\titleformat{\part}
	[display]
	{\centering\normalfont\Huge\bfseries}
	{\vspace{3pt}\titlerule[2pt]\vspace{3pt}\MakeUppercase{\partname} \thepart}
	{0pt}
	{\titlerule[2pt]\vspace{50pt}\huge\MakeUppercase}

	\titlespacing*{\part}{0pt}{50pt}{20pt}

\catcode`,\active

\catcode`\,12

\newcommand{\nn}{\noindent}
\newcommand{\de}{\partial}
\newcommand{\V}{\mathcal{V}}
\newcommand{\B}{\mathcal{B}}
\newcommand{\A}{\mathcal{A}}
\newcommand{\Ha}{\mathcal{H}}
\newcommand{\La}{\mathcal{L}}

\newcommand{\R}{\mathbb{R}}

\newcommand{\Ol}{\mathcal{O}}
\newcommand{\K}{\mathbb{K}}
\newcommand{\Kan}{\mathcal{K}}
\newcommand{\Q}{\mathbb{Q}}
\newcommand{\Z}{\mathbb{Z}}
\newcommand{\N}{\mathbb{N}}

\newcommand{\C}{\mathbb{C}}
\newcommand{\F}{\mathcal{F}}

\newcommand{\M}{\mathcal{M}}

\newcommand{\Sh}{\mathcal{S}}
\newcommand{\Pro}{\mathbb{P}}
\newcommand{\z}{\mathbf{z}}
\newcommand{\U}{\mathcal{U}}

\newtheorem{definition}{Definition}[chapter]
\newcommand{\defn}[2][]{%
  \begin{definition}\textbf{(#1)}\\#2\end{definition}%
}

\newtheorem{Theorem}{Theorem}[chapter]
\newcommand{\theo}[2][]{%
  \begin{Theorem}\textbf{(#1)}\\#2\end{Theorem}%
}

\newtheorem{Lemma}{Lemma}[chapter]
\newcommand{\lem}[2][]{%
  \begin{Lemma}\textbf{(#1)}\\#2\end{Lemma}%
}

\newtheorem{Proposizione}{Proposition}[chapter]
\newcommand{\prop}[2][]{%
  \begin{Proposizione}\textbf{(#1)}\\#2\end{Proposizione}%
}

\definecolor{mycol}{RGB}{6,104,88}

\newcommand{\chapnumfontsize}{120}          
\newcommand{\chapboxinnersep}{10pt}        
\newcommand{\chapboxrounded}{4pt}          
\newcommand{\chaptitlesize}{\Huge}         
\newcommand{\chapnumvsep}{1pt}            
\newcommand{\chaprulelength}{0.65\linewidth}

\titleformat{\part}[display]
  {\normalfont\bfseries}                   
  { 
    \makebox[\linewidth][r]{%
      \vbox{%
        \hbox{%
          \tikz[baseline=(n.base)]{%
            \node[draw=none, fill=mycol, rounded corners=\chapboxrounded, inner sep=\chapboxinnersep] (n)
              {\color{white}\fontsize{\chapnumfontsize}{\chapnumfontsize}\selectfont\bfseries\thepart};
          }%
        }%
        \nointerlineskip
        \vskip -1pt
        \hbox{\tikz[baseline]{\draw[mycol,line width=3pt](-\chaprulelength,0)--(5,0);}}%
      }
    }
  }
  {\chapnumvsep}                             
  {\chaptitlesize\raggedleft}               
\titlespacing*{\part}{0pt}{30pt}{100pt}   

\titleformat{\chapter}[display]
  {\normalfont\bfseries}                   
  { 
    \makebox[\linewidth][r]{%
      \vbox{%
        \hbox{%
          \tikz[baseline=(n.base)]{%
            \node[draw=none, fill=mycol, rounded corners=\chapboxrounded, inner sep=\chapboxinnersep] (n)
              {\color{white}\fontsize{\chapnumfontsize}{\chapnumfontsize}\selectfont\bfseries\thechapter};
          }%
        }%
        \nointerlineskip
        \vskip -1pt
        \hbox{\tikz[baseline]{\draw[mycol,line width=3pt](-\chaprulelength,0)--(5,0);}}%
      }
    }
  }
  {\chapnumvsep}                             
  {\chaptitlesize\raggedleft}               
\titlespacing*{\chapter}{0pt}{30pt}{100pt}   


\usepackage{fancyhdr}
\renewcommand{\chaptermark}[1]{\markboth{\thechapter\ \ #1}{}}

\makeatletter
\let\ps@plain\ps@fancy
\makeatother

\titleformat{name=\chapter,numberless}[display]
  {\normalfont\huge\bfseries\centering}  
  {}                                     
  {0pt}                                  
  {\LARGE}                               


\hypersetup{
  colorlinks=true,        
  linkcolor=mycol, 
  citecolor=mycol,  
  urlcolor=mycol,    
  filecolor=mycol,      
  anchorcolor=mycol,
  breaklinks=true,        
  pdfborder={0 0 0},      
  unicode=true            
}

\makeatletter
\renewcommand\footnoterule{%
  \kern-3\p@ 
  \noindent\color{mycol}\rule{\textwidth}{1pt}
  \kern2.6\p@ 
}
\makeatother

\newcommand{\Important}[1]{\begin{center}
\fcolorbox{mycol}{gray!20}{\parbox{0.8\linewidth}{#1}}\end{center}}
  \newcommand{\amplitudeone}{                        
\draw (1,0.6) ellipse (0.1cm and 0.3cm);
\draw (1,-0.6) ellipse (0.1cm and 0.3cm);
\draw (3,0.6) ellipse (0.1cm and 0.3cm);
\draw (3,-0.6) ellipse (0.1cm and 0.3cm);
	\draw  (1.03,0.9) to[out=-45,in=-135] (2.97,0.9);
	\draw  (1.03,-0.9) to[out=45,in=135] (2.97,-0.9);
	
	\draw  (1.03,0.3) to[out=-25,in=90] (1.5,0) to[out=-90,in=25] (1.03,-0.3);
	\draw  (2.97,0.3) to[out=205,in=90] (2.5,0) to[out=-90,in=155] (2.97,-0.3);

	\draw (4,0.6) ellipse (0.1cm and 0.3cm);
\draw (4,-0.6) ellipse (0.1cm and 0.3cm);
\draw (6,0.6) ellipse (0.1cm and 0.3cm);
\draw (6,-0.6) ellipse (0.1cm and 0.3cm);
	\draw  (4.03,0.9) to[out=-45,in=-135] (5.97,0.9);
	\draw  (4.03,-0.9) to[out=45,in=135] (5.97,-0.9);
		\draw  (4.03,0.3) to[out=-25,in=90] (4.5,0) to[out=-90,in=25] (4.03,-0.3);
	\draw  (5.97,0.3) to[out=205,in=90] (5.5,0) to[out=-90,in=155] (5.97,-0.3);	
	\draw (5,0) ellipse (0.3cm and 0.3cm);

		\draw (7,0.6) ellipse (0.1cm and 0.3cm);
\draw (7,-0.6) ellipse (0.1cm and 0.3cm);
\draw (9,0.6) ellipse (0.1cm and 0.3cm);
\draw (9,-0.6) ellipse (0.1cm and 0.3cm);
	\draw  (7.03,0.9) to[out=-45,in=-135] (8.97,0.9);
	\draw  (7.03,-0.9) to[out=45,in=135] (8.97,-0.9);
	\draw  (7.03,0.3) to[out=-25,in=90] (7.5,0) to[out=-90,in=25] (7.03,-0.3);
	\draw  (8.97,0.3) to[out=205,in=90] (8.5,0) to[out=-90,in=155] (8.97,-0.3);	
	\draw (7.8,0) ellipse (0.18cm and 0.18cm);
	\draw (8.2,0) ellipse (0.18cm and 0.18cm);

      \node at(0,0) {$\mathcal{A}_4^{closed}=$};
           \node at(3.5,0) {$+$};
            \node at(6.5,0) {$+$};
             \node at(9.5,0) {$+......$};

                   }

%% file: Title.tex
\begin{titlepage}
\begin{figure}[!htb]
    \centering
    \includegraphics[width=5cm]{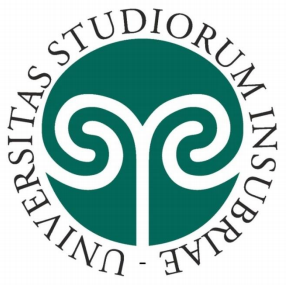}
\end{figure}

\begin{center}
    \Large{\textbf{UNIVERSITÀ DEGLI STUDI DELL'INSUBRIA}}
    \vspace{3mm}
    \\ \normalsize{DIPARTIMENTO DI SCIENZA E ALTA TECNOLOGIA}
    \vspace{6mm}
    \\ \normalsize{DOTTORATO}
    \\  \normalsize{\textbf{FISICA E ASTROFISICA}}
    \vspace{13mm}
\end{center}

\vspace{10mm}
\begin{center}
    \LARGE{\textbf{Exponential Periods for Integrals in Physics}}
\end{center}

\vspace*{\fill}

\begin{minipage}[t]{1\textwidth}
    \begin{multicols}{2}
    	{\normalsize{\textbf{Supervisor}}{\normalsize\vspace{1mm}
        \\ \normalsize{Prof. Sergio Luigi Cacciatori }}\\\\\normalsize{\textbf{Co-supervisor}}{\normalsize\vspace{1mm}
        \\ \normalsize{Prof. Pierpaolo Mastrolia }}} \\ 
        
          \columnbreak

         \begin{flushright}
            {\normalsize{\textbf{PhD Candidate}}{\normalsize\vspace{1mm}
            \\ \normalsize{Anthony Massidda}}} \\
         \end{flushright}
    \end{multicols}
\end{minipage}
\vspace{1cm}
\begin{center}
    {\normalsize{\textbf{Academic year}}{\normalsize\vspace{1mm}
    \\ \normalsize{2024/2025}}}  
\end{center}

\end{titlepage}

%% file: Abstract.tex
\clearpage
\chapter*{Abstract}
\thispagestyle{empty}                     

\addtocontents{toc}{\protect\contentsline{chapter}{\textcolor{mycol}{Abstract}}{}{}} 
\begin{center}

\begin{minipage}{4.5in}
The study of Feynman integrals through the lens of intersection theory offers a unifying framework for their analysis, capturing both the linear and quadratic relations that arise among integrals. In doing so, it provides a powerful method for systematically reducing them to the so called master integrals, a necessary strategy for multi-loop contributions, whose huge number make direct calculation unfeasible. The Twisted de Rham cohomology offers a powerful tool for describing integrals with multivalued integrands, arising in dimensional regularization. However, it fails whenever the underlying geometry shows richer structures, as singularities and intricate monodromies.
In this thesis we propose a systematic approach to identify and construct the appropriate homology and cohomology that allows to interpret Feynman integrals in parameter representation as exponential periods. This reformulation, together with the analytic continuation of the dimensional regularizator, provides a perfect framework to properly analyze the wall crossing structure and to correctly take into account Stokes’ phenomena for a sharp counting of the number of Master integrals.\\
This framework allows to embed within the same formalism not only perturbative integrals, coming both from quantum field theories and string theory, but also wide class of physically relevant integrals, from Fourier calculus to statistical mechanics partition functions, from quantum mechanics expectation values to conformal field theory correlators.
\end{minipage}
\end{center}
\clearpage

%% file: Notation.tex
\clearpage
\chapter*{Notation}
\thispagestyle{empty}                     

\addtocontents{toc}{\protect\contentsline{chapter}{\textcolor{mycol}{Notation}}{}{}} 
\textbf{FIs} Feynman Integrals\\
 \textbf{IBP} Integration by Parts \\
\textbf{MIs} Master Integrals \\
 \textbf{QFT} Quantum Field Theory \\
 \textbf{ST} String Theory\\
 \textbf{WCS} Wall Crossing Structure \\
 \textbf{(co)homology} homology and/or cohomology\\\\
 $\equiv$ Equality by definition\\
 $\Ol$ sheaf of holomorphic functions\\
 $\Omega^k$ sheaf of holomorphic $k-$forms\\
  $H_\bullet$ homology ring \\
  $H^\bullet$ cohomology ring \\
  $\Ha$ sheaf cohomology\\
  $\mathbb{H}$ Hypercohomology\\
  
\clearpage

%% file: Introduction.tex
\chapter{Introduction}
\thispagestyle{empty}                     

\chaptermark{Introduction}
Computing integrals is one of the main activities of physicists: evaluate fluxes of electric or magnetic fields, determine expectation values for processes in quantum mechanics, calculate averages in statistical physics, or infinite dimensional integrals in defining quantum field theories, and so on and so forth. More broadly, many physical problems reduce to solving systems of nonlinear partial differential equations \cite{Kotikov1991123,KOTIKOV1991158,Lunev_1994,Remiddi:1997ny,Argeri07} (see \cite{Grimm:2024tbg,Grimm:2025zhv} for recent developments in reduction methods). In very few cases, this can be done exactly, but generically it must be done perturbatively with various techniques. This is, in general, the case for scattering amplitudes. High-energy collisions are our most powerful microscope for probing the building blocks of matter and the forces that govern them: increasing the collision energy opens new production channels, reveals new degrees of freedom, and exposes the limits of our existing theories. The Standard Model remains extraordinarily successful, yet persistent experimental tensions and cosmological puzzles motivate searches for physics beyond it. Identifying tiny deviations from Standard Model predictions therefore requires theoretical control of scattering amplitudes at unprecedented precision, which in turn forces us to confront the combinatorial and analytic complexity of more and more difficult integrals.\\
Furthermore, many of these integrals exhibit cancellations and simplifications a posteriori, pointing to deep redundancies and the presence of hidden structures. Anticipating these structures in advance is crucial for reducing the otherwise overwhelming number of integrals, thereby making multiloop amplitude calculations tractable.
The importance of understanding these patterns goes far beyond providing a decisive computational advantage; it also allows us to gain deep insights into the structure of the theory itself, revealing properties that would otherwise remain hidden. By uncovering the underlying geometric and algebraic organization of amplitudes, we can identify symmetries, dualities, and consistency relations that are not manifest in the traditional Feynman diagrammatic formulation. In this sense, the study of such patterns does not merely simplify calculations: it reshapes our conceptual understanding of quantum field theory and string theory, highlighting the unity between their combinatorial, geometric, and physical aspects.
We just mention, that a very successful geometrical approach developed in recent years relies on the so called positive geometries \cite{Arkani-Hamed:2012zlh}, based on the observation that certain on-shell amplitudes can be identified with the unique canonical form \cite{Arkani17} of a geometrical object whose boundaries encode physical factorization channels. Prototypical examples are the amplituhedron \cite{Arkanijha} and momentum amplituhedron \cite{Arkanizlh,Britto:2004ap,Damgaard:2019ztj} in planar $N=4$ Super Yang-Mills; although recent development extended this formalism to cosmological wavefunctions and correlators \cite{Arkani-Hamed:2024jbp,Capuano:2025ehm}.\\
What we learn from Stokes' theorem is that when we consider fluxes of closeld forms over closed cycles, we can deform the cycle and add exact forms to the integrands without changing the final result. Here is where homology and cohomology start playing a central role.\\
In the simplest cases, mathematically, these integrals correspond to periods of elements of a de Rham cohomology over closed cycles of a dual homology (closed manifolds up to boundaries). However, more general cases can be considered by suitably identifying the correct cohomology theory entering the game. An important example is the case of hypergeometric type integrals and their generalizations, like GKZ equation systems and Euler integrals, where it has been realized that they can be again understood in terms of co-homology after replacing standard de Rham cohomology with its twisted version, and the dual homology with one realized in terms of open twisted cycles \cite{Aomoto11,yoshida2013hypergeometric,kita1,kita2,matsumoto1994,cho1995,Aomoto77_structure,matsumoto1998,adolphson_sperber_1997,Mimachi2003,Ohara98,Mimachi2004,OST2003,Goto:2013Laur,Yoshiaki_GOTO2015203,goto2015,matsubaraheo2019algorithm,Goto2022homology,Matsubara-Heo:2020lzo,Matsubara-Heo:2021dtm}.\\
In \cite{Mastrolia:2018uzb}, observing a similar underlying structure in Feynman integrals in QFT on Minkowski spacetime, it was proposed that the same strategies be applied to the computation of scattering amplitudes.  The key advantage lies in the fact that both homology and cohomology groups are finitely generated and endowed with a non-degenerate internal product, the topological intersection product for homology and its dual in cohomology, defined between closed forms. In this way, a given Feynman integral can be identified as an element of a finite dimensional vector space endowed with a double structure suitable for projecting any vector to a given basis, given precisely by the intersection products of (co)homology. Moreover, when a Hodge structure is available, important quadratic relations, generalizing the Riemann quadratic relations, can be determined, see \cite{Cacciatori:2021nli,Cattani:2008ec}. The identification of Feynman integrals with periods of a suitable cohomology \cite{BLOCH2015328,Lairez:2023nih,Bonisch:2021bfk,Bonisch:2022bdf,Klemm:2020knr,Duhr:2023dkn,pichon2025,Bogner:2007mn,Doran:2023yzu} thus allows to replace the integration by part identities (IBP) \cite{Chetyrkin:1981qh,Laporta:2001dd} with intersection projection, that provide a more systematic procedure, both for determining a basis of master integrals (MI) and to decompose the vector space accordingly, as well as for finding differential equations and quadratic identities satisfied by the MI, see \cite{Frellesvig:2019kgj,Mizera:2019vvs,Frellesvig:2019uqt,Frellesvig:2020qot,Mizera:2016jhj,Mizera:2017rqa,Kaderli:2019dny,Kalyanapuram:2020vil,Ma:2021cxg,Weinzierl:2020nhw,Gasparotto:2022mmp,Chen:2022lzr,Giroux:2022wav,Ahmed:2023htp,Duhr:2024bzt,Crisanti:2024onv} for several successful applications to physics. The efficiency of this strategy depends on identifying the appropriate cohomology and efficient ways to calculate the intersection product \cite{Brunello:2023rpq,Fontana_2023,Caron-Huot:2021xqj,Caron-Huot:2021iev,Chestnov:2022alh,Chestnov:2022xsy,Chestnov:2024mnw,brunello2024}. Recently, \cite{Cacciatori:2022mbi}, it has been shown that intersection theory plays a role in more general situations, beyond the realm of Feynman integrals, for integrals involving generalized orthogonal polynomials and computations of matrix elements in quantum mechanics, Green's functions in field theories, higher-order moments of probability distributions, suggesting a deep intertwinement between physics, geometry, and statistics.\footnote{It is not trivially expected an arbitrary generalization of the applicability of these methods to any situation, given that ``being a period'' is a special situation, see \cite{Kontsevich:2001kza} for an introduction to this idea.}\\
Driven by emerging perspectives that reveal integration theory as a unifying language in mathematical physics, in the present thesis, based on \cite{Angiusnostro}, we propose a quite general approach that can be applied to several questions of physical interest, like Feynman integrals, Fourier integrals \cite{Brunello2}, scattering in curved spacetime \cite{Cacciatori:2024zrv,Cacciatori:2024zbe}, string scattering \cite{Mizera:2019gea,Mazloumi:2024wys,Bhardwaj:2023vvm,Pokraka:2025zlh}(see also \cite{Massidda}) and the examples investigated in \cite{Cacciatori:2022mbi}: the exponential integrals.  Our analysis is based on the recent mathematical progresses presented in  \cite{Kontsevich:2024mks}, where exponential integrals, and the related wall-crossing structures, are analyzed in light of a series of isomorphisms between the twisted de Rham and Betti cohomologies in their local and global versions.\\
Our starting point are the exponential integral defined from a holomorphic function on a complex manifold $X$, $f:X\rightarrow \mathbb C$, so that
\begin{align*}
 I_\Gamma(f,\gamma)=\int_{\Gamma} e^{-\gamma f} \mu, 
\end{align*}
where we assume $f$ to have a finite number of isolated critical points, $\gamma$ is a non-vanishing complex parameter, $\mu$ is a holomorphic volume form over $X$, and $\Gamma$ is an open integration chain. This kind of integrals are met in several physical applications, where one evaluates them asymptotically for large values of $\gamma$, by using the saddle point approximations. The associated thimbles indeed represent selected basis of integration cycles \cite{Witten:2010cx}, relative to some subset $D_0\subset X$ of positive codimension, and allow to understand the cohomology structures underlying the integral. The twisted de Rham cohomology is determined by the exact form $df$, the twisting being given by the covariant differential $\nabla_f=d+df\wedge$. To the triple $(X,D_0,f)$, one can construct four Local Systems, given by the local and global twisted de Rham and Betti cohomologies, all deeply related. In these terms, exponential integrals can be interpreted as periods of such cohomologies. In proving this, a key role is played by the Wall Crossing Structures associated to the integrals, which appear when the parameter $\gamma$ meets Stokes lines in the complex plane. The proof of these facts, given in \cite{Kontsevich:2024mks}, relates on a generalization of the Riemann-Hilbert problem along the Stokes rays.\\
We will review these facts in Section \ref{WCSsection}, adapting the notation and formalism to the application to physics we have in mind. In a sequence of recent papers, it has been shown that the Skyrme model admits exact analytic solutions describing nuclear matter in different pasta states \cite{Alvarez:2020zui,Cacciatori:2021neu,Cacciatori:2022kag}. In particular, for the gauged Skyrme model one finds solutions representing baryonic layers at finite baryon density in the presence of a constant magnetic field \cite{Cacciatori:2024ccm}. The grand-canonical partition function of such system is expressed in terms of a  Pearcey integral. Its importance relies on the fact that it provides low energy non-perturbative effective description of chromodynamics and gives the occasion to replace cumbersome numerical analyses with analytical studies. The phase space of this system strongly depends on the Stokes lines of the partition function, which also determine critical curves in the $\mu_B-B_{ext}$ plane, $\mu_B$ being the (complexified) chemical potential of the external magnetic field $B_{ext}$. In Section \ref{PI} we will interpret the Pearcey integral as an exponential integral and apply the general theory to it in order to analyze the Stokes phenomenon.
This analysis is then extended to include exponential integrals involving multivalued functions, we will briefly review how the constructions related to the triple $\left( X, D_0, f \right)$ can be generalized to the triple $\left( X, D_0, \alpha \right)$, where the $1$-form $\alpha$, representing the twisting of the covariant differential, is now a generic closed holomorphic form rather than necessarily an exact one, as developed in \cite{Kontsevich:2024mks}. This setup, specialized to the case were $\alpha$ induces an exponential pairing involving the logarithmic of a polynomial function,$f=\log \mathcal{B}$, allows us to elaborate on a general strategy, independent on the specific underlying geometries, for dealing with Feynman integrals in dimensional regularization, without any special assumptions on the spacetime dimensions.\\
Any $D$ dimensional, $L$-loops Feynman integral with $E$ external legs, can indeed be rewritten in the so called Baikov representation \cite{Baikov96-1,Baikov96-2,Baikov05}

\begin{equation}
    I(\{\nu_i\})=\int_\Gamma \mathcal{B}^{-\gamma} \prod_{i=1}^n \frac{dz_i}{z^{\nu_i}},
    \label{Baikov}
\end{equation}
\nn
where the Baikov polynomial $\mathcal{B}$, is a polynomial on the integration variables which coefficients depends on the masses of the internal particles and the external momenta, and $\gamma=-(D-E-L-1)/2$. If $\gamma \in \mathbb{Q}$, the integral is still multivalued but its underlying geometry, i.e. the geometric space where the integrand is single-valued, is a computable Riemann foliation obtained by gluing a finite number of sheets. In such cases, the integral \eqref{Baikov} can be interpreted as a standard pairing between a twisted de Rham cohomology and the singular homology associated to this geometry; provided such space is smooth. Indeed, a first issue arises when singularities 
However, in the context of dimensional regularization, $\gamma\notin \Q$, the geometric intuition behind the integral is lost, and the previous interpretation of \eqref{Baikov} as a period no longer applies. Reformulating the integrand as an exponential
\begin{equation}
    I(\left\lbrace \nu_i \right\rbrace) = \int_{\Gamma} e^{-\gamma \log \mathcal{B}} \prod_{i=1}^n \frac{dz_i}{z^{\nu_i}}
    \label{Baikov:exp}
\end{equation}
solves the issue on $\gamma$ since it allows to interpret the integral as an exponential period for any $\gamma \in \mathbb{C}^{\ast}$ at the cost of dealing with a non-holomorphic function in the argument of the exponential.\\
Once the Feynman integral is expressed in such a exponential form, it can be interpreted as an exponential pairing between the a twisted de Rham cohomology and a dual Betti homology.
For the latter, we will describe a geometric construction of a basis in terms of thimbles $th_{i, \gamma}$, and we will define on it an internal product that can be computed purely from topological data. We will show that this construction remains valid even when the function $f$ is multivalued. This allows us to express the integration contour $\Gamma$ in \eqref{Baikov:exp} as a linear combination of this thimble basis:
\begin{equation}
    \Gamma = \sum_{i=1}^{dimH^{B}_n} \langle \Gamma, th_{i, \theta + \pi} \rangle \, \, th_{i, \theta}.
\end{equation}
Thus the integral \eqref{Baikov:exp} admits a decomposition with respect to the coefficients defined by the internal product in homology:
\begin{equation}
    I = \sum_{i=1}^{dim H_n^{B}} \langle \Gamma, th_{i, \theta + \pi} \rangle \int_{th_{i , \theta}} e^{- \gamma \log \mathcal{B}} \prod_{j=1}^n \frac{dz_j}{z_j^{\nu_j}}.
\end{equation}
From this perspective, the integrals evaluated over the thimbles that appear in the sum play the role of Master Integrals.\\\\
This thesis consists in three main chapters.\\
In Chapter \ref{C3} we provide a compound of the main algebraic-geometrical and topological objects and tools one needs handle to face the topic. A exhaustive detailed explanation of the subject would cover the material of different advanced course in mathematics and hundreds of pages; thereby, we selected a set of fundamental notions and concepts whose understanding is essential, emphasizing the explanation of the more basic tools that is mandatory to have an intuition of, to appreciate the perspective of our results. We will remand to the literature for concepts will unfortunately remain uncovered. After a short overview of intersection theory and sheaf theory, we will discuss dualities among homology and cohomology, representing the core of intersection pairings. After a brief round up on complex varieties and their proprieties, we will expose the Picard-Fuchs theory, providing a topological tool for concretely studying monodromies of complex analytic manifolds.\\
In chapter \ref{C2}, after reviewing how scattering amplitudes appear in Quantum Field Theory and String Theory, we will explore, in a possibly original fashion, where intersection theory enters the game. We will show that linear and quadratic relations among integrals, emerging naturally from the vector space structure induced by equivalence relations, can be elegantly described via (co)homological pairings. We will introduce the prototypical family of Feynman integrals, the Banana integrals, showing they are related to periods of families of Calabi-Yau manifolds.\\
We we will manly focus in emphasizing the possible sources of criticalities they may arise in the standard identification of FIs as twisted pairings, issues that we argue they are correctly treated, solved or avoid within the framework we propose in chapter \ref{C4}.\\
There, after a reformulation of exponential period methods and wall crossing structure analysis in a language more accessible to physicists, we will firstly apply this framework to exponential integrals of a holomorphic function, studying the Pearcy integral. Subsequently, we will elaborate on exponential periods involving the logarithm of a polynomial, studying the structure of the associated divisors, determining  their Betti cohomology and its variation with respect to $\gamma$. We will finally conclude applying this approach to the Legendre family of elliptic curves, that, being closely related to a two loops banana diagram,  it provides a excellent test bench for testing our setup.

%% file: Math.tex
\chapter{Mathematical Background}\label{C3}
\thispagestyle{empty} 
The goal of this chapter is to provide a concise and thorough explanation of the algebraic, geometric, and topological ideas and methods that we will require and employ in chapter \ref{C4} for exposing and conclude the main result of this thesis. The secondary objective is to offer a compendious introduction to selected concepts of advanced mathematics in the best tangible way possible for the author. Being unfeasible to completely cover the entire subject and to enter into the vary details, we will try to exhaustively and didactically explain the basic concepts we consider essential, while limiting ourself to mention and give the idea of some others, remanding the reader to the literature for the rest \cite{Arnold88,Bredon,Dundas,Ebeling1,hartshorne66,Lipman,Milnor:1963jmi,Porteous_1971,Nicolaescu2007,shafarevich1,shafarevich2,Voisin_2002}.\\
\section{Intersection theory}\label{co-ho}
One of the most powerful and recurring themes in physics is recognising when apparently different phenomena are really the same thing seen in different guises: showing that distinct objects or behaviours are equivalent in a deeper formalism lets us describe them as a single entity. This insight probably began with Newton’s realization that the falling apple and the Moon’s orbit are governed by the same gravitational law, continued with Maxwell’s unification of electricity and magnetism, and today appears in efforts to bring the fundamental forces under one roof, in the seek of a unified theory describing quantum gravity. Finding such equivalences, precisely formulating what it means for two things to be “the same” and building the mathematical structures that capture that “sameness” has repeatedly simplified our descriptions of nature and opened paths to new predictions and deeper understanding.\\
The notion of category was introduced by Eilenberg and Mac Lane  with the very precise intent to formulate a \textit{general theory of natural equivalences}\cite{Eilen45}. \\

\defn[Category]{
A category $\mathcal{C}$ consists of a class of \textbf{objects} $\mathrm{Obj}(\mathcal{C})$ and class of \textbf{morphisms} $\mathrm{mor}(\mathcal{C})$, whose elements $\mathrm{hom}_\mathcal{C}(X,Y)$ are all maps among objects $f:X\rightarrow Y$, together with a associative composition low $\circ: \mathrm{hom}_\mathcal{C}(X,Y)\times \mathrm{hom}_\mathcal{C}(Y,Z)\rightarrow \mathrm{hom}_\mathcal{C}(X,Z)$ such that for any object $X$ there exists a unique identity morphism $I_X$ such that for any other morpshim $f:X\rightarrow Y$: $f=I_Y\circ f=f\circ I_X$.\\\\

}
A category such that $\mathrm{hom}_\mathcal{C}(X,Y)$ is an abelian group for every $X$ and $Y$, and such that the composition low is distributive over the group operation, is called an \textbf{Abelian category}.\\ 
The simplest example of a category is $Set$, whose objects are sets, morphisms are functions, and the composition low is given by the ordinary composition of functions.\\
We call $\mathrm{Obj}(\mathcal{C})\ni I$ a \textit{initial object}, if there exists only one morphism $I\rightarrow X$, for every $X\in \mathrm{Obj}(\mathcal{C})$. Similarly, we call $\mathrm{Obj}(\mathcal{C})\ni T$ a \textit{terminal object} if there exists only one morphism $X\rightarrow T$, for every $X\in \mathrm{Obj}(\mathcal{C})$. An object that is both initial and terminal is called \textbf{zero object} and denoted by $0$. An abelian category has a unique zero object.\\
We want now to introduce maps among categories that preserve their structure, such a map is called \textbf{functor}.\\

\defn[Functor]{
Given two categories $\mathcal{A}$ and $\mathcal{B}$ a functor $F:\mathcal{A}\rightarrow \mathcal{B}$ is a map $F_o:\mathrm{Obj}(\mathcal{A})\rightarrow \mathrm{Obj}(\mathcal{B})$ such that $F(I_X)=I_{F(x)}$, for every $X \in\mathrm{Obj}(\mathcal{A})$, together with a map $F_m:\mathrm{mor}(\mathcal{A})\rightarrow \mathrm{mor}(\mathcal{B})$ that preserves the composition low $\circ$: we call it

\begin{itemize}
    \item \textbf{Covariant functor}, if  $F_m:\mathrm{hom}_\mathcal{A}(X,Y)\rightarrow \mathrm{hom}_\mathcal{B}(F_o(X),F_o(Y))$ is such that, for every for every $f,g \in\mathrm{hom}_\mathcal{A}(X,Y)$ the composition satisfies:
    \begin{equation}
        F(f\circ g)=F(f)\circ F(g);
    \end{equation}
      \item \textbf{Contravariant functor}, if  $F_m:\mathrm{hom}_\mathcal{A}(X,Y)\rightarrow \mathrm{hom}_\mathcal{B}(F_o(Y),F_o(X))$ is such that, for every for every $f,g \in\mathrm{hom}_\mathcal{A}(X,Y)$, the composition satisfies:
    \begin{equation}
        F(f\circ g)=F(g)\circ F(f).
    \end{equation}\\\\
\end{itemize}
}

\begin{figure}[h!]
    \centering    
    \includegraphics[scale=0.5]{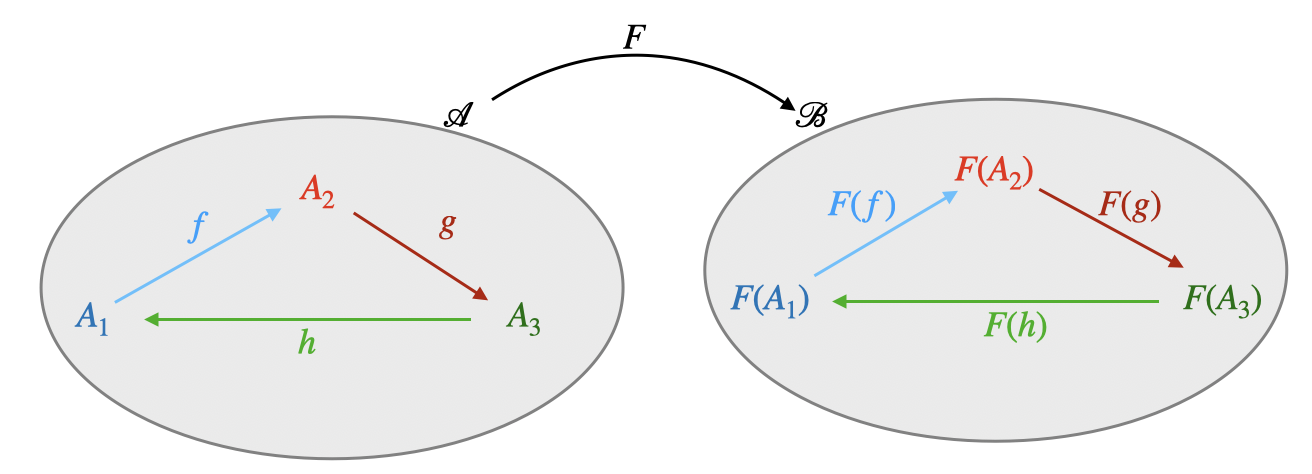}
    \caption{\small{Illustrative representation of a covariant functor.}}
    \label{figfunctor}
\end{figure}

A chain complex in a category $\mathcal{C}$, denoted by $C_\bullet\equiv (A_\bullet,\de_\bullet)$, is a sequence of objects $\{A_1,A_2,A_3,\dots\}$ and connecting morphisms, called \textbf{boundary operators} $\de_i: A_i \rightarrow A_{i-1}$, represented as 

\begin{equation}
    C_\bullet :\quad  \dots \xlongrightarrow[]{\de_{i+2}}A_{i+1} \xlongrightarrow[]{\de_{i+1}}A_{i}\xlongrightarrow[]{\de_{i}}A_{i-1}\xlongrightarrow[]{\de_{i-1}}\dots\, ,
    \label{ChainComplex}
\end{equation}

such that the image of each morphism is contained in the kernel of the next:
\begin{equation}
    \mathrm{Im}(\de_i)\subseteq \mathrm{ker}(\de_{i-1}),
    \label{nihil1}
\end{equation}
equivalent to the nilpotency of the differential: $\de^2=0$.\\
\begin{Ese} The spaces of singular $k-$chains $C_k(\M)$ on a smooth manifold $\M$, together with the boundary operator $\de:C_k (\M)\rightarrow C_{k-1}(\M)$, define a chain complex, called \textbf{singular chain complex} and denoted by $S_\bullet(\M)$.\\\end{Ese}

Analogously, we introduce a dual sequence $C^\bullet \equiv (A^\bullet,d^\bullet)$, called \textbf{cochain complex}, with $d^i$ nilpotent morphisms called \textbf{differentials}, obtained from \eqref{ChainComplex} by reversing all the arrows: 

\begin{equation}
C^\bullet :\quad  \dots \xlongrightarrow[]{d^{i-1}}A^{i} \xlongrightarrow[]{d^{i}}A^{i+1}\xlongrightarrow[]{d^{i+1}}A^{i-2}\xlongrightarrow[]{d^{i+1}}\dots\, .
    \label{coChainComplex}
\end{equation}

\begin{Ese} The spaces of differential $k-$forms $\Omega^k(\M)$ on a smooth manifold $\M$, together with the ordinary external derivative $d:\Omega^k(\M)\rightarrow\Omega^{k+1}(\M)$ define a cochain complex called \textbf{De Rham complex} and denoted by $\Omega^\bullet_{dR}(\M)$.\end{Ese} 

If the condition \eqref{nihil1} is saturated, i.e. equality holds, for any differential of the sequence, the latter is said to be \textbf{exact}. Consider the case of a finite exact sequence, starting and ending with the zero object. An exact sequence of only one object

\begin{equation}
    0 \rightarrow A \rightarrow 0,
\end{equation}

trivially implies $A\cong 0$, and for a two objects exact sequence 

\begin{equation}
    0 \rightarrow A \rightarrow B \rightarrow 0,
\end{equation}
immediately follows $A\cong B$. Therefore, the shortest, thus simplest, exact sequences containing non-trivial information, called \textbf{short exact sequences}, account for three objects

\begin{equation}
    0 \rightarrow A \xlongrightarrow[]{\de_{2}} B\xlongrightarrow[]{\de_{1}} C \rightarrow 0.
    \label{SES}
\end{equation}
For such a sequence, exactness implies

\begin{equation}
    C\cong B/\mathrm{Im}(\de_2)\cong B/\mathrm{ker}(\de_1).\\
\end{equation}

\lem[Poincaré]{
The de Rham complex $\Omega^\bullet_{dR}(U)$ over a contractible $U\in \R^n$ is an exact sequence. \\ 
}

Sequences longer then a short exact sequence, are called \textbf{long exact sequences}. \\
The position each objects occupies in a chain complex is called \textbf{degree}; so for instance in \eqref{SES} we naturally say $A$ is in degree $0$, $B$ in degree $1$ and $C$ in degree $2$.
Sometimes it is useful to shift this natural graduation and to assign the objects a different degree. Given a chain complex $C^\bullet$ we denote by $C^\bullet[k]$ the shifted complex whose objects are re-graduated according to the convention:

\begin{equation}
    (C[k])^i\equiv C^{i+k}.
    \label{shift}
\end{equation}
Thus, for instance, calling $C^\bullet$ the sequence \eqref{SES}, the shifted complex $C^\bullet[-1]$ as $A$ in degree $1$, $B$ in degree $2$ and $C$ in degree $3$.\\ 
Suppose now a chain complex of the form \eqref{ChainComplex} fails to be exact in some degrees, say $i$, meaning in the subsequence

\begin{equation}
    \dots\xlongrightarrow[]{}A_{i+1}\xlongrightarrow[]{\de_i} A_i\xlongrightarrow[]{\de_{i-1}} A_{i-1}\xlongrightarrow[]{}\dots
    \label{subseq}
\end{equation}
the kernel of $\de_{i-1}$ is larger then the image of $\de_{i}$. There is then some element $a_{(i)}\in A_i$ such that 
\begin{itemize}
    \item is \textbf{closed}: $\de_{i-1}a_{(i)}=0$,
    \item is \textbf{not exact}: $ \nexists \quad a_{(i+1)}\in A_{i+1}|\,\de_{i}a_{(i+1)}=a_{(i)} $. 
\end{itemize}
Intuitively, larger is the number of elements in $A_i$ that are closed but not exact, farer is the \eqref{subseq} from exactness. However, we must be careful and notice that if $a_{(i)}$ is closed but not exact, also $\tilde{a}_{(i)}= a_{(i)}+\hat{a}$, for any $\hat{a}\in \rm{im}[\de_{i+1}]$, is closed but not exact. But this is trivially implied by the nilpotencty of $\de$ and it adds no information on exactness; thus the whole equivalence class generated by $a_{(i)}$ via the above equivalence relation is the object we have to take into account. The correct measure of the “distance” of a chain complex from exactness is thus the space of closed elements modulo the exact ones; and we arrive to the following definition. \\

\defn[Homology]{
Given the chain complex \eqref{ChainComplex} we call the $i-$th \textbf{homology} group\footnote{If $\mathcal{A}$ is an additive/abelian category they are indeed groups.} the quotient 

\begin{equation}
    H_i(A_\bullet,\de)\equiv\frac{\rm{ker}(\de_i)}{\rm{im}(\de_{i+1})}.
\end{equation}
}
\begin{figure}[h!]
    \centering    
    \includegraphics[scale=0.35]{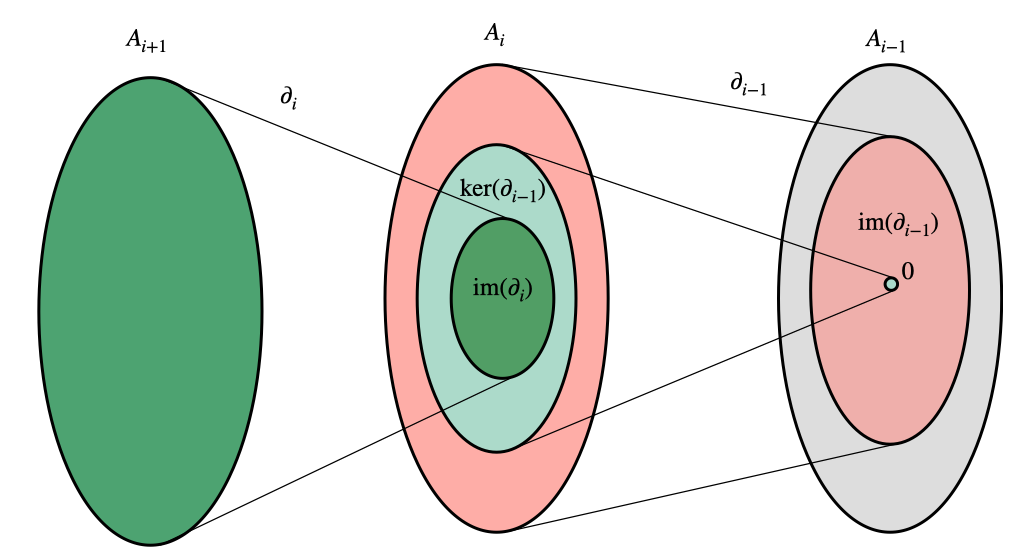}
    \caption{\small{Illustrative representation of a chain complex.}}
    \label{figcoho}
\end{figure}
The dual notion is defined in the same fashion:\\

\defn[Co-homology]{
Given the cochain complex \eqref{coChainComplex} we call $i-$th \textbf{cohomology} group the quotient 

\begin{equation}
    H^i(A^\bullet,d)\equiv\frac{\rm{ker}(d_{i+1})}{\rm{im}(d_{i})}.
\end{equation}
}

\subsection{Sheaves}
Physical experiments test local phenomena, and theoretical models attempt to infer universal laws from those observations. This immediately suggests a mathematical question: how can locally defined data be glued together to produce a well-defined global object? The theory of sheaves provides the precise formalism for this local-to-global passage. In the following section we introduce the basic definitions and illustrative examples; for proofs and a broader treatment the reader is referred to \cite{shafarevich1,shafarevich2,hartshorne1977algebraic}.\\
A presheaf $\F$ on a topological space $X$ assigns to every open set $U\subset X$ a set (or group, ring, module, \dots) $\F(U)$ whose elements are called \textbf{sections} over $U$, together with restriction maps $\rho_{U,V}:\F(U)\to\F(V)$ for each inclusion $V\subset U$. These restriction maps satisfy the compatibilities $\rho_{U,U}=\mathrm{id}_{\CF(U)}$ and $\rho_{V,W}\circ\rho_{U,V}=\rho_{U,W}$ for $W\subset V\subset U$.\\\\
\begin{figure}[h!]
    \centering    
    \includegraphics[scale=0.5]{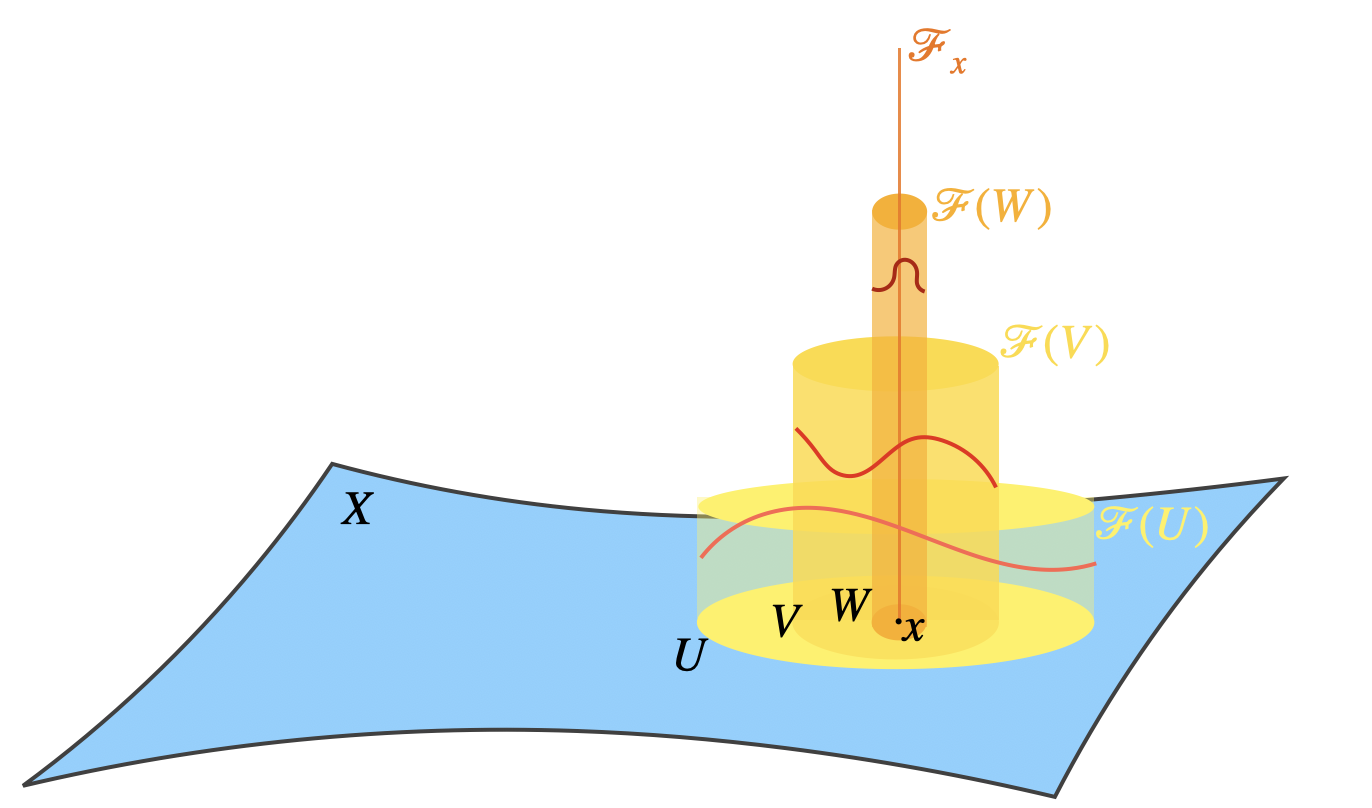}
    \caption{\small{Illustrative representation of a sheaf $\F$ on a space $X$, and its stalks $\F_x$.}}
    \label{figSheaf}
\end{figure}
\defn[Sheaf]{
A sheaf on a topological space $X$ is a presheaf $\mathcal{F}$ whose restriction maps satisfy the further conditions:
\begin{itemize}
\item[(i)] Locality: If two sections $s,t \in \CF (U)$ are equal when restricted to every $V_i$ in an open cover $\left\lbrace V_i \right\rbrace$ of $U$, then $s=t$.
\item[(ii)] Gluing: If we have a collection of sections $s_i \in \CF (V_i)$ over an open cover $\left\lbrace V_i \right\rbrace$ of $U$ that agree on the overlaps 
$V_i \cap V_j$, then there exists a unique section $s \in \CF (U)$ that restricts to each $s_i$.\\\\
\end{itemize}
}

These conditions express the proprieties that sections are determined locally and that locally compatible data paste together uniquely.\\
A standard example of a sheaf is the assignment $\mathcal C$ defined by $\mathcal C(U)=C^0(U)$, the set (or ring) of continuous real-valued functions on $U\subset\R$: restriction is ordinary restriction of functions, locality follows because equality of continuous functions is a local property, and two continuous functions that agree on overlaps paste to a continuous function on the union.\\ 
Suppose now to restrict the open $U$ with a limit procedure around a point $x\in \R$. As $U$ shrinks, more and more functions satisfy continuity on $U$ and enter into $C^0(U)$. At the same time, more and more functions coincide on $U$ and thus identify the same element of $C^0(U)$; precisely, two functions $f$ and $g$ will overlap in the limit if there exists a neighborhood $W$ of $x$ such that $f|_W=g|_W$. What we get at the end is then the space of germs of continuous functions at $x$.\\
For a generic sheaf $\F$, this procedure, called direct limit

\begin{equation}
    \F_x \,=\, \varinjlim_{x\in U}\F(U),
\end{equation}
captures the behaviour of sections arbitrarily near $x$ and defines the \textbf{stalk} $\F_x$ at $x$.\\
A sheaf $\mathcal{F}$ on $X$ whose stalk $F_x\equiv S$ is the same at any point $x\in X$, in the sense that the restriction maps are the identity, is called \textbf{constant sheaf} and denoted by $\underline{S}$. Concretely, its sections are locally constant maps $f:U\rightarrow S$.
On the other extreme, one can define the so called \textbf{skyscraper sheaf} with support only at $x\in X$, as the sheaf whose stalk is zero everywhere except at $x$.
Let $\F$ and $\mathcal{G}$ be two sheaves on a topological space $X$. A morphism $\phi:\F\rightarrow\mathcal{G}$ is a family of maps $\phi_U: \F(U)\rightarrow\mathcal{G}(U)$, for every $U\in X$ such that for every $V\subset U$, the restriction maps compose as

\begin{equation}
    \rho^\mathcal{G}_{U,V}\circ \phi_V=\phi_U \circ  \rho^\mathcal{F}_{U,V}.
\end{equation}
Composition of morphisms is defined componentwise on each open set, and the identity morphism on a sheaf is the family of identity maps.
The class of sheaves on a topological space $X$ together with the morphisms of sheaves form a category, denoted by $Sh(X)$. Let us now introduce three fundamental functors acting on $Sh(X)$.
The central problem we care about is reconstructing global data from local ones. For a constant sheaf on a connected manifold every locally defined section extends trivially to a global one; in general, however, that situation is the exception rather than the rule: topological features of the underlying space \footnote{Nontrivial fundamental group, nontrivial coverings, nonzero cohomology classes, etc.)} create obstructions that prevent locally defined sections from gluing into global sections. Algebrically this problem is encoded by the global section functor, assigning to each sheaf the set (group, ring,\dots) of global sections.\\\\  

\defn[Global section functor]{
Let $Sh(X)$ be the category of sheaves on a topological space $X$ taking values in a category $\mathcal{C}$. The \textbf{global section functor} $\Gamma: Sh(X)\rightarrow \mathcal{C}$ is the covariant functor defined by 

\begin{itemize}
    \item $\Gamma(X,\F)=\F(X)$, for every $\F \in \mathrm{Obj}\left(Sh(X)\right )$;
    \item $\Gamma(\phi): \F(X)\rightarrow \mathcal{G}(X)$, for every $\mathrm{mor}\left(Sh(X)\right )\ni \phi: \F\rightarrow\mathcal{G}$.\\\\
\end{itemize}
}

\subsection{Coherent sheaves}
A important class of sheaves showing very specific behavior and playing central role in physics is given by the so called \textbf{Coherent sheaves} (see \cite{Decataldo09,Decataldo02,fulton2013riemann} for further details).\\
Let $X$ be a topological space and $\Ol_X$ a sheaf of rings on $X$. The pair $(X,\Ol_x)$ is called a ringed space and $\Ol_X$ its structure sheaf.\\\\
\defn[Coherent sheaf]{
A sheaf $\mathcal{F}$ of $\Ol_X-$modules on $X$ is said to be \textbf{coherent} if there is an open neighborhood $U\subset X$ for every point $x \in X$, such that the morphism 

\begin{equation}
     \Ol^n|_U \rightarrow \mathcal{F}|_U
\end{equation}
\begin{itemize}
    \item[(i)] is surjective for some $n\in \N$;
    \item[(ii)] has a finitely generated kernel for any $n\in \N$.\\\\
\end{itemize}
}
A coherent sheaf is a generalization of a vector bundle, where fibers can be modules with different dimensions; in fact, if $\Ol_X$ is a field\footnote{A module over a field is a vector space.} and condition $(i)$ above is satisfied for the same $d$ for any $x$, then $\mathcal{F}$ is a vector bundle.
These are precisely the objects one needs for studying the (co)homology of singular spaces.\\
A coherent sheaf on $X$ that is locally free of rank $r$, namely for any $x\in X$ there is a neighborhood $U\subset X$ such that 

\begin{equation}
    \F|_U\cong \Ol^r|_U,
\end{equation}
is called \textbf{locally constant sheaf} or \textbf{local system}.\\
As we will see later on, the choice and the construction of a local system encoding the relevant data is the key to correctly interpret integrals in a cohomological language.\\
Roughly speaking, a local system looks like a trivial vector bundle on small neighborhoods but can twist globally, via monodromy.
In facts, local systems are equivalent to representations of the fundamental group, and one can construct a local system from a such a representation.\\
Let $p:\widetilde X \to X$ be the universal cover of a connected,
locally path-connected space $X$, and let
\[
\rho:\pi_1(X,x_0)\longrightarrow \mathrm{GL}(V)
\]
be a representation of its fundamental group on a finite-dimensional vector space $V$. We can associate to $\rho$, a local system $\La_\rho$ defined as the sheaf
whose sections over an open set $U \subset X$ are given by
\begin{equation}
\mathcal L_\rho(U)=
\left\{
f : p^{-1}(U) \to V
\big |\text{$f$ is l.c. and}\,
 f(\gamma \tilde x) = \rho(\gamma)\, f(\tilde x)
\,\forall\, \gamma \in \pi_1(X)
\right\},
\end{equation}
where a locally constant (l.c.) section is a section that is constant when restricted to a sufficient small neighborhood of any point.\\
Equivalently, one may form the quotient bundle
\begin{equation}
Q = (\widetilde X \times V)/\pi_1(X),
\end{equation}
where $\pi_1(X)$ acts diagonally by
\begin{equation}
\gamma \cdot (\tilde x, v)
=
(\gamma \tilde x,\; \rho(\gamma)v).
\end{equation}

Then $Q \to X$ is a flat vector bundle, and $\mathcal L_\rho$
is the sheaf of locally constant (equivalently, flat) sections of $Q$.
The stalk of $\mathcal L_\rho$ at any point is naturally
isomorphic to $V$, and the monodromy action of a loop
$\gamma$ on the stalk is given by $\rho(\gamma)$.\\\\

\begin{Ese}
    Let $X=S^1$, $\R$ its universal cover and 
\begin{equation}
    \rho: \pi_1(S^1)\cong \Z \rightarrow {\rm{GL}}(1,\C)\cong \C^\ast,
\end{equation}
a one dimensional representation of the fundamental group. The quotient bundle is 
\begin{equation}
    Q=(\R\times \C^\ast)/\Z,
\end{equation}
where $n\in \Z$ acts by $n\cdot(t,z)=(t+2\pi n,\lambda^n z)$, with $\lambda \in\C^\ast$.
Thus, the sheaf of locally constant sections of  $Q$

\begin{equation}
\mathcal L_\rho(U)=
\left\{
f : p^{-1}(U)\subset \R \to \C
\big |
\text{$f$ is l.c. and}\,
 f(t+2\pi) = \lambda f(t)
\right\}.
\end{equation}
defines a local system on $S^1$.
\end{Ese}

\paragraph{Cotangent sheaf}\mbox{}\\
Let $(X,\Ol_X)$ be a ringed space. Consider the diagonal map 

\begin{equation}
    \Delta: X\rightarrow X\times X,
\end{equation}
\nn
sending any point $x$ to $(x,x)$. The ideal sheaf $\mathcal{I}_\Delta$ is defined to be the kernel of the map 

\begin{equation}
    \Ol_{X\times X}\rightarrow \Delta_*\Ol_X,
\end{equation}
\nn
that is, the sheaf whose stalk $\mathcal{I}(U)$ on $U\subset X\times X$ consists of elements vanishing on the diagonal $\Delta(U\cap X)$.\\
Then one defines the \textbf{conormal sheaf} as the quotient $\mathcal{I}_\Delta/\mathcal{I}_\Delta^2$, consisting of elements vanishing “at first order” on the diagonal.\\ The cotangent sheaf on $X$, also called sheaf of K\"ahler differentials, and denoted by $\Omega_X^1$, is defined to be the pullback on $X$ of the conormal sheaf:

\begin{equation}
    \Omega_X^1 \equiv \Delta^*(\mathcal{I}_\Delta/\mathcal{I}_\Delta^2).
\end{equation}
\nn
Suppose $X=\rm{Spec}(\mathcal{R})$, is an affine scheme, with $\mathcal{R}=\mathcal{A}/(f)$ and $\mathcal{A}$ a $k-$algebra, one can prove the sheaf of K\"ahler differentials is isomorphic to

\begin{equation}
    \Omega^1_X \cong \frac{\Omega^1_\mathcal{A}\otimes_\mathcal{A}\mathcal{R}}{\mathcal{R}\cdot df}.
\end{equation}
Roughly speaking, differentials on $X$ are given by the one in the ambient space modulo the relation $df=0$. As well as providing a useful computational prescription, this relation make evident the mismatch, and thus the possible source of troubles, between differential and algebraic forms in singular cases. \\\\

\begin{Ese}
Consider the affine nodal curve $X=Spec (\mathcal{R})$, with coordinate ring $\mathcal{R}=\C[x,y]/(f)$, with
\begin{equation}
    f(x,y)=y^2-x^2(x+1).
\end{equation}
\nn
We have 
\begin{equation}
    X\times X= Spec(\mathcal{R}\otimes_\C\mathcal{R})=Spec\left (\frac{\C[x,y,x',y']}{(f(x,y),f(x',y'))} \right ),
\end{equation}
The image of the diagonal map is given by 
\begin{equation}
    \Delta(X)=\{(x,y,x',y')\subset X\times X |x=x',y=y'\},
\end{equation}
thus the ideal sheaf is $\mathcal{I}_\Delta=(x-x',y-y')\subset \mathcal{R}\otimes_\C\mathcal{R}$. Quotienting by $\mathcal{I}^2_\Delta=\{i=\sum_k a_kb_k |a_k,b_k \in I\}$ we obtain the conormal sheaf is generated by two elements, the equivalent class $\langle x-x'\rangle$ and $\langle y-y'\rangle$; after pulling them back to $X$ we can identifying them we with $dx$ and $dy$. However $dx$ and $dy$ do not freely generated an $\Ol_X-module$, because there are constrained by the relation $df=0$. Thus we find:

\begin{equation}
    \Omega^1_X= \frac{\Omega^1_\C \otimes_\C \mathcal{R}}{df}
\end{equation}

\end{Ese}
If X is an affine variety and $\mathcal{R}=\C[x_1,...x_n]/(f)$ is its coordinate ring, we can define the sheaf $\mathcal{E}^1_X$ of sections of the cotangent bundle of $X$, whose stalk at $x\in X$ is given the space freely generated by differential $df$:

\begin{equation}
    \mathcal{E}^1_X \ni \varepsilon = \sum g_i df ,\quad  \mbox{with}\quad g_i \in \C^\infty(X). 
\end{equation}

For a complex manifold $\mathcal{E}^1_X\cong \Omega^1_X$, simply because we can take the Taylor expansion of $df$ in local coordinates. On the contrary, if $X$ is singular, this isomorphism does not longer hold true: indeed, on one side the tangent space at the singular locus turns to be ill-defined, on the other, $\Omega^1_X$ is not longer free, because the annihilation of the Jacobian introduces a relation $\de_i f=0$ among $dx_i$, reducing the dimension by one and constraining them.\\
Notice, the torsion turns to be isomorphic to the Tjurina module \cite{Tjurina}:
\begin{equation}
    Tor(\Omega^1_X)\cong \frac{\mathcal{R}}{(\de_if)}=T(f).
\end{equation}

\paragraph{Canonical Sheaf}\mbox{}\\
By canonical sheaf of a space $X$, we denote the sheaf whose stalk at $x\in X$, is the space of germs of sections of the canonical bundle of $X$ at $x$. To keep notation as easy as possible, we will always denote the canonical bundle and the canonical sheaf by the same symbol $\Kan_X$, being clear from the context to which one it is referred.\\
Let denote by $\Ol_X(d)$, for $d\in \Z$, the locally trivial line bundle on $X$ whose transition functions are given by $d-$ degree homogeneous polynomials, for $d$ positive, or rational functions with poles of order $d$, for $d$ negative. Any line bundle on the projective space $\C\Pro^n$ is isomorphic to $\Ol_{\C\Pro^n}(d)$, for some $d$; in particular the canonical bundle of $\C\Pro^n$ can be proved to be $\Kan_{\C\Pro^n}\simeq \Ol(-n-1)$. \\
If $X$ is a smooth subvariety of a smooth algebraic variety or complex manifold $Y$, their canonical bundles are related by the so called \textbf{adjunction formula}

\begin{equation}
   \Kan_X \simeq i^*\Kan_{Y} \otimes det(\mathcal{I}/\mathcal{I}^2)^\vee,
   \label{adjunction formula}
\end{equation}

\nn
where $\mathcal{I}$ is the ideal sheaf of $X$ in $Y$, and $i:X\rightarrow Y$ the inclusion map.\\
In case $X$ is a complete intersection in $\C\Pro^n$ defined by $r$ polinomials $f_i$ of degrees $d_i$, the adjunction formula \eqref{adjunction formula} reduces to: 

\begin{equation}
    \Kan_X \simeq \Ol(-n-1+\sum d_i)|_X.
\end{equation}

\subsection{Sheaf cohomology}
Starting from the same local-to-global question that motivates the previous section, assuming locally compatible data on overlapping neighborhoods form a sheaf, we can ask: when do they actually glue to a single global object, and if they do not, how can we measure the obstruction? Sheaf cohomology is the canonical tool for answering these questions.\\  
Let $\mathcal{F}:\mathcal{A}\rightarrow\mathcal{B}$ be a covariant functor. One can proceed analogously, mutatis mutandis, for $\mathcal{F}$ a contravariant functor.\\
Applying $\mathcal{F}$ to a short exact sequence in $\mathcal{A}$

\begin{equation}
     0 \rightarrow A_3 \xlongrightarrow[]{\de_2} A_2\xlongrightarrow[]{\de_1} A_1 \rightarrow 0,
     \label{SES2}
\end{equation}
one does not generally obtain a short exact sequence in $\mathcal{B}$, but rather a short sequence that fails to be exact. One can thus seek for a long exact sequence of the form

\begin{equation}
     \dots \rightarrow \F(A_3) \xlongrightarrow[]{\F(\de_2)} \F(A_2)\xlongrightarrow[]{\F(\de_1)} \F(A_1) \rightarrow \dots.
     \label{s}
\end{equation}
It actually turns there is a unique canonical way for filling the dots, in such a way the resulting long sequence is exact, consisting in introducing the so called \textbf{left} ($L^i\F$) and \textbf{right} ($R^i\F$) \textbf{derived functors}:

\begin{equation}
\begin{split} 
\dots & \longrightarrow L^2\F(A_1) \longrightarrow L^1\F(A_3) \longrightarrow L^1\F(A_2)\longrightarrow L^1\F(A_1)\longrightarrow\\ & \longrightarrow \F(A_3) \xlongrightarrow{\F(\de_2)} \F(A_2)\xlongrightarrow{\F(\de_1)} \F(A_1) \longrightarrow\\&\longrightarrow R^1\F(A_3) \longrightarrow R^1\F(A_2) \longrightarrow R^1\F(A_1)\longrightarrow R^2\F(A_3)\longrightarrow \dots.
\end{split}
\label{LES}
\end{equation}
If $L^i\F=0$ for all $i$, $\F$ is said to be left-exact, and similarly if $R^i\F=0$ for all $i$, $\F$ is said to be right-exact. If $\F$ is both left and right exact, and then it maps exact sequences in $\mathcal{A}$ to exact sequences in $\mathcal{B}$, it is called exact.\\
Consider a left-exact functor.\\
An object $A\in \mathcal{A}$ such that $R^i\F(A)=0$ for all $i>0$ is called \textbf{acyclic}. Suppose $A_2$ and $A_1$ in \eqref{SES2} are acyclic, then \eqref{LES} splits in
\begin{equation}
    0\longrightarrow \F(A_3)\xlongrightarrow[]{\F(\de_2)}\F(A_2)\xlongrightarrow[]{\F(\de_1)}\F(A_1)\xlongrightarrow[]{\phi} R^1\F(A_3)\longrightarrow 0.
    \label{LES1}
\end{equation}
and $R^i\F(A_3)=0$ for $i>1$.\\
Exactness condition applied on the last object of \eqref{LES1} implies $ R^1\F(A_3)\cong \rm{ker}[\phi]/\rm{im}[\F(\de_1)]$, that is, by definition, the first right derived functor of $\F$ on $A_3$ must coincide with the first degree cohomology group of the complex $\F(A^\bullet):0\rightarrow\F(A_2)\rightarrow\F(A_1)\rightarrow 0$:

\begin{equation}
    R^1\F(A_3)\cong \Ha^1(F(A^\bullet)).
\end{equation}
So if $A_3$ fits into a short exact sequence together with two object acyclic with respect to $\F$,  
Consider now a longer exact sequence of the form

\begin{equation}
     0 \rightarrow A \xlongrightarrow[]{\de_0} I_0\xlongrightarrow[]{\de_1} I_1 \xlongrightarrow[]{\de_2} I_2 \xlongrightarrow[]{} 0,
     \label{SES3}
\end{equation}
with $I_i$ acyclic object with respect to $\F$. We can introduce an auxiliary object $C\equiv \rm{ker}[\de_2]=\rm{im}[\de_1]$ and split \eqref{SES3} as

\begin{equation}
    \begin{split}
     &0 \rightarrow A \xlongrightarrow[]{\de_0} I_0\xlongrightarrow[]{\de_1} C \xlongrightarrow[]{} 0,\\&   0 \rightarrow C \xlongrightarrow[]{\de_1} I_1\xlongrightarrow[]{\de_2} I_2 \xlongrightarrow[]{} 0.
    \end{split}
\end{equation}
The long exact sequence \eqref{LES1} associated to the first of these short exact sequences splits, and gives us

\begin{equation}
    \begin{split}
        &0\longrightarrow \F(A)\xlongrightarrow[]{}\F(I_0)\xlongrightarrow[]{}\F(C)\xlongrightarrow[]{} R^1\F(A)\longrightarrow 0\\& 0\longrightarrow R^i\F(C)\longrightarrow R^{i+1}\F(A)\longrightarrow 0,
    \end{split}
    \label{LES2}
\end{equation}
while the one associated to the second gives
\begin{equation}
        0\longrightarrow \F(C)\xlongrightarrow[]{}\F(I_1)\xlongrightarrow[]{}\F(I_2)\xlongrightarrow[]{} R^1\F(C)\longrightarrow 0.
        \label{LES3}
\end{equation}
Comparing \eqref{LES2} and \eqref{LES3}, and proceeding as before we deduce:

\begin{equation}
\begin{split}
    &R^1\F(A)\cong H^1(\F(I^\bullet)),\\
    &R^2\F(A)\cong H^2(\F(I^\bullet)),\\
    &R^i\F(A)\cong 0 \quad \mbox{for} \, i>2.
\end{split}
\end{equation}
The idea is clear! Longer is the starting exact sequence, higher is the degree of the (possibly) non-vanishing right derived functors. 
In general, $A\in\mathcal{A}$ does not fit into a short or almost short exact sequence of acyclic objects, however, if $\mathcal{A}$ satisfies the so called requirement of having \textbf{enough injectives}, it is always possible to find a bounded-below exact sequence, called \textbf{resolution}, of acyclic objects

\begin{equation}
    0 \longrightarrow A \xlongrightarrow[]{} I_0\xlongrightarrow[]{} I_1 \xlongrightarrow[]{} \dots,
\end{equation}
and use it to compute right derived functors of $\F$ on $A$ as: 
\begin{equation}
    R^i\F(A)\cong H^i(\F(I^\bullet)).
\end{equation}
We are now ready for introducing the following definition\\\\

\defn[Sheaf Cohomology]{
Let $\mathcal{S}$ be an abelian sheaf on a topological space $X$, and $I^\bullet$ a resolution of acyclic objects of $\mathcal{S}$. The sheaf cohomology groups $\Ha^i(X,\mathcal{S})$ are defined as the right-derived functors of the global section functor of $\mathcal{S}$:
\begin{equation}
    \Ha^i(X,\mathcal{S})\equiv R^i\Gamma(X,\mathcal{S})\cong H^i\left (\Gamma(X,I^\bullet)\right ).
    \label{SheafCoho}
\end{equation}
}
Notice, we can always find a acyclic resolution of $\mathcal{S}$, because $Sh(X)$ can be proven to have enough injectives (see part $III$ section $2$ of \cite{hartshorne1977algebraic}).\\
The failure of $\Gamma$ to be right exact corresponds to the existence of topological obstructions.\\\\
\begin{Ese}
Suppose we have a global magnetic field $\mathbf{B}$ on a space $X$ with $\nabla \cdot \mathbf{B} = 0$, can we always find a single global vector potential $\mathbf{A}$ such that $\nabla \times \mathbf{A} = \mathbf{B}$?\\
On any open set $U\subset X$, the set of all smooth vector potentials $\mathbf{A}$ is $\Omega^1(U)$, the set of all smooth magnetic fields $\mathbf{B}$ that satisfy $\nabla \cdot \mathbf{B} = 0$ is the set of closed two forms $Z^2(U)$ and the set of all vector potentials $\mathbf{A}$ that are “curl-free” ($\nabla \times \mathbf{A} = 0$) is the set of closed $1-$forms $Z^1(U)$.
The exterior derivative $d$ (curl) defines a sheaf morphism

\[ d: \Omega^1 \longrightarrow Z^2, \]

 that takes a local potential $\mathbf{A} \in \Omega^1(U)$ and gives its corresponding local magnetic field $\mathbf{B} = \nabla \times \mathbf{A}$. The resulting $\mathbf{B}$ automatically satisfies $\nabla \cdot \mathbf{B} = 0$, so it lands in $Z^2(U)$.
By Poincaré Lemma, $d: \Omega^1 \to Z^2$ is surjective at the sheaf level, that is locally any field $\mathbf{B}$ with $\nabla \cdot \mathbf{B} = 0$ can be written as the curl of some potential $\mathbf{A}$.
This gives us a short exact sequence of sheaves on $X$:
    \[ 0 \to Z^1 \stackrel{i}{\longrightarrow} \Omega^1 \stackrel{d}{\longrightarrow} Z^2 \to 0 ,\]
that perfectly encodes the local physics.\\
For extracting the global data, let us now apply the global section functor $\Gamma(X, -)$ to this sequence:
\[
0 \to \Gamma(X, Z^1) \stackrel{\Gamma(i)}{\longrightarrow} \Gamma(X, \Omega^1) \stackrel{\Gamma(d)}{\longrightarrow} \Gamma(X, Z^2) \rightarrow \Ha^1(X,Z^1) \rightarrow \dots 
\]
relating global curl-free potentials $\Gamma(X, Z^1)$, global vector potentials $\Gamma(X, \Omega^1)$ and global magnetic fields $\Gamma(X, Z^2)$. 
If $\Gamma(X,-)$ is right exact, the map $\Gamma(d): \Gamma(X, \Omega^1) \to \Gamma(X, Z^2)$ is surjective and any global magnetic field is indeed image of a global smooth potential; however if $\Ha^1(X,Z^1)$ does not vanish there exist $\mathbf{B}\in\Gamma(X, Z^2)$ that is not exact.\\
Suppose we are interested in the physics outside a magnetic monopole sitting at the origin $\{0\}$ of a three dimensional space, thus $X = \mathbb{R}^3 \setminus \{0\}$. Consider the magnetic field
\[ \mathbf{B} = \frac{g}{r^3} \mathbf{r}, \]

it is defined, smooth and closed everywhere on $X$. thus it is a valid global section of $ Z^2$.
The magnetic flux through a $S^2$ sphere centered at the origin is given by 

\[\Phi= \int_S^2 \mathbf{B} d\mathbf{S}=4\pi g.\]

However, if a global $\mathbf{A}$ such that $\nabla \times \mathbf{A} = \mathbf{B}$ existed, we could use Stokes' theorem, obtaining a contradiction, unless $g=0$:
\[ \Phi = \iint_{S^2} (\nabla \times \mathbf{A}) \cdot d\mathbf{S} = \oint_{\partial S^2} \mathbf{A} \cdot d\mathbf{l}=0. \]
Thus we found a global section $\mathbf{B} \in \Gamma(X, Z^2)$ that is not in the image of the global map $\Gamma(d)$; the obstruction to finding a global potential $\mathbf{A}$ is precisely the non-zero magnetic charge $g$, which is a topological invariant of the field on the space $X$, measure by the non vanishing of $H^1(X, Z^1)$, due to the puncture at the origin. \\\\ \end{Ese}

\paragraph{De Rham cohomology again.}\mbox{}\\
We are now ready for reviewing de Rham cohomology introduced in \ref{co-ho} via sheaf cohomology.
Consider the constant sheaf $\underline{\R}(X)$ on a topological space $X$. Let $\Omega^k(X)$ be the sheaf of smooth $k-$ forms on $X$ and $\Omega^\bullet(X)$ the \textbf{De Rham complex}:
\begin{equation}
    \Omega^\bullet(X):   0 \longrightarrow \Omega^0(X) \xlongrightarrow[]{d} \Omega^1(X)\xlongrightarrow[]{d} \Omega^2(X) \xlongrightarrow[]{} \dots.
    \label{deRhamcomplex}
\end{equation}
Then the following theorem \cite{DeRham} holds.\\

\theo[de Rham Theorem]{
The de Rham complex $\Omega^\bullet(X)$ is an acyclic resolution of $\underline{\R}(X)$.
\label{DeRhamTheo}
}
It immediately follows by \eqref{SheafCoho} that 

\begin{equation}
    H_{dR}(X)\cong \Ha^i(X,\underline{\R}).
    \label{DeRhamcoho}
\end{equation}


\paragraph{Image Functors.}\mbox{}\\
Let $f:X\rightarrow Y$ be a continous map of topological spaces. Just like in differential geometry, one introduces the pushforward and pullback of $f$ to transport the differential structure from one space to the other and viceversa, in this context one introduces the \textbf{direct image} $f_\ast$ and \textbf{inverse image} $f^\ast$ functors to transport sheaves from one space to the other:

\begin{equation}
    f_\ast: Sh(X)\rightarrow Sh(Y) \quad \quad \mbox{and}\quad \quad  f^\ast: Sh(Y)\rightarrow Sh(X).
\end{equation}
The direct image sheaf $f_\ast\F$ on $Y$ of a sheaf $\F$ on $X$ is defined the sheaf whose sections are such that 
\begin{equation}
    f_\ast \F(U)\equiv \F(f^{-1}(U)),
    \label{DIF}
\end{equation}
for open $U\subset Y$.

The direct image functor is left exact, but usually not exact. Its right derived functors are called \textbf{higher direct images} and one can prove they turn to be sheaves associated to the presheaves 

\begin{equation}
    R^qf_\ast\F(U)=H^q(f^{-1}(U),\F).
    \label{HigherDirectImages}
\end{equation}

Let us finally just mention, because we will use it later, the \textbf{direct image with compact support} functor $f_!$, defined as: 

\begin{equation}
f_!(\F)(U)\equiv \{\sigma \in \F(f^{-1}(U))| f|_{supp(s)}: supp(s)\rightarrow U \quad \mbox{is proper}\}.
\end{equation}

Where proper here, refers to the property that every compact subset of $U$ has a compact preimage inside the support of $\sigma$. 
\subsection{Hypercohomology}
In the previous section we introduced sheaf cohomology by starting from a short exact sequence of sheaves \eqref{SES2}. Hypercohomology provides a generalization to an arbitrary complex. \\
Starting from a complex of sheaves $\Sh^\bullet$ on $X$ and applying the global section functor, we get the long exact sequence

\begin{equation}
\begin{split} 
&0\longrightarrow \Gamma(X,\Sh_1) \longrightarrow \Gamma(X,\Sh_2)\longrightarrow\Gamma(X,\Sh_3)\longrightarrow\dots\\&\dots \longrightarrow R^1\Gamma(X,\Sh_1) \longrightarrow R^1\Gamma(X,\Sh_2) \longrightarrow R^1\Gamma(X,\Sh_3)\longrightarrow\dots\\& \dots \longrightarrow R^2\Gamma(X,\Sh_1)\longrightarrow \dots.
\end{split}
\label{LES4}
\end{equation}
That we could rewrite compactly as 

\begin{equation}
\begin{split} 
&0\longrightarrow \Gamma(X,\Sh^\bullet) \longrightarrow R^1\Gamma(X,\Sh^\bullet) \longrightarrow R^2\Gamma(X,\Sh^\bullet)\longrightarrow \dots.
\end{split}
\label{LES5}
\end{equation}
This object is not a complex, because there is no canonical differential such as $R^i\Gamma(X,\Sh^\bullet)\rightarrow R^{i+1}\Gamma(X,\Sh^\bullet)$. However, we can write the double complex: 

\[\begin{tikzcd}
	0\hspace{0.5cm} & \Gamma(X,\Sh_1)  & \Gamma(X,\Sh_2) & \hspace{0.5cm} \,\,\cdots \\
	0\hspace{0.5cm}& R^1\Gamma(X,\Sh_1) &  R^1\Gamma(X,\Sh_2) & \hspace{0.5cm} \,\,\cdots \\
	0\hspace{0.5cm} & R^2\Gamma(X,\Sh_1) &  R^2\Gamma(X,\Sh_2) & \hspace{0.5cm} \,\,\cdots \\
   \hspace{0.5cm}&\vdots&\vdots&\\
	\arrow[from=1-2, to=2-2]
	\arrow[from=1-3, to=2-3]
	\arrow[from=2-2, to=3-2]
	\arrow[from=2-3, to=3-3]
    \arrow[from=1-1, to=1-2]
    \arrow[from=1-2, to=1-3]
    \arrow[from=2-1, to=2-2]
    \arrow[from=2-2, to=2-3]
    \arrow[from=3-1, to=3-2]
    \arrow[from=3-2, to=3-3]
    \arrow[from=1-3, to=1-4]
    \arrow[from=2-3, to=2-4]
    \arrow[from=3-3, to=3-4]
    \arrow[from=3-2, to=4-2]
	\arrow[from=3-3, to=4-3]
\end{tikzcd}\]
where the horizontal differentials $d_h$ are the original ones and the vertical differentials $d_v$ in the $i-th$ column comes from a injective resolution of the sheaf $\Sh_i$.\\
We can then define the complex $R\Gamma(X,\Sh^\bullet)$ with objects
\begin{equation}
    R\Gamma(X,\Sh^\bullet)^k\equiv \bigoplus_{p+q=k}R^p\Gamma(X,\Sh_q),
\end{equation}
where now the morphism $D:R\Gamma(X,\Sh^\bullet)^k\rightarrow R\Gamma(X,\Sh^\bullet)^{k+1}$ can be shown to be the well defined differential 

\begin{equation}
    D= d_h+(-1)^kd_v.\\\\
\end{equation}

\defn[Hypercohomology]{
Let $X$ be a topological space and $\Sh^\bullet$ a complex of sheaves on $X$. The \textbf{Hypercohomology} of $\Sh^\bullet$ is defined as: \\\\
\begin{equation}
    \mathbb{H}^k(X,\Sh^\bullet)\equiv H^n(R\Gamma(X,\Sh^\bullet)).\\\\
\end{equation}
}
In light of this, we can reformualate De Rham theorem \ref{DeRhamTheo} as\\\\

\theo[de Rham Theorem revisited]{
Let \eqref{deRhamcomplex} be the de Rham complex on a manifold $X$, then: 
\begin{equation}
     H_{dR}(X)\cong \mathbb{H}^k(X,\Omega^\bullet).\\\\
     \label{HyperdeRham}
\end{equation}
\label{DeRhamTheo2}
}
This isomorphism together with the machinery of spectral sequences we will see in the next section, offers a practical prescription for computing de Rham cohomology.\\
Finally, notice that any sheaf $\F$ can be though of as a complex with only one element, that is a complex that has zero elements everywhere except in some position $n$:

\begin{equation}
     \F^\bullet:\, \dots\,\rightarrow 0\rightarrow \F\rightarrow 0\rightarrow\, \dots.
\end{equation}
We say $\F^\bullet$ is concentrated in degree $n$ and according to the notation \eqref{shift} we write $\F[-n]$. \\
The sheaf cohomology of a sheaf $\F$ can then been seen as the hypercohomology of the complex $\F^\bullet$ concentrated in degree $0$:

\begin{equation}
    \mathbb{H}^k(X,\F^\bullet)\cong \Ha^k(X,\F).
\end{equation}

\subsection{Spectral sequences}
Spectral sequences are generalization of exact sequences providing a powerful tool used in homological algebra for computing (co)homology groups by an iterative process of successive approximations. Formally, a spectral sequence is a collection $\{E_r,d_r\}_r$ of bigraded objects $E_r^{p,q}$ and nilpotent differentials

\begin{equation}
    d_r^{p,q}: E_r^{p,q}\longrightarrow E_r^{p+r,q-r+1},
\end{equation}
organized as follow: the collection of objects and differential for fixed $r$ form a double complex, called \textbf{page}; the next page is obtain from the previous one by taking (co)homology with respect to $d_r$:

\begin{equation}
    E_{r+1}^{p,q}\equiv \frac{{\rm{ker}}(d_r^{p,q})}{{\rm{im}}(d_r^{p-r,q+r-1})}.
\end{equation}

In many standard situations, this “turning page” process \textbf{converges}, in the sense that for large $r$ the groups stop changing, and they stabilize to a page, called $E_\infty^{p,q}$. It may happen that the spectral sequence stabilized for finite $r=n$, corresponding to the vanishing all subsequent differentials ($d_r=0$ for $r>n$); in that case we say the spectral sequence \textbf{degenerates} at the $n-$page, and one has $E_n^{p,q} = E_\infty^{p,q}$.\\
This stable page is related to the target (co)homology group by 

\begin{equation}
    H^{k}\cong \bigoplus_{p+q=k}  E_\infty^{p,q}.
\end{equation}

In practice, when we have to compute some complicated (co)homology groups, we first hope there is a spectral sequence express in terms of easily computable objects that converges to it and then we turn pages hoping for its soon degeneracy.\\
Let us see some simple example of  spectral sequences. \\
Consider a fibration $f:X\rightarrow B$, and suppose one want to compute the cohomology of the total space $X$, starting from the knowledge of the cohomologies of the base space $B$ and of the fiber $F$. Serre \cite{Serre51} showed that the spectral sequence with second page

\begin{equation}
    E_2^{p,q}=H^p(B,H^q(F)),
    \label{SerreSQ}
\end{equation}

converges to $H^{p+q}(X)$. This is the so called \textbf{Serre spectral sequence}, and we will use it in section \ref{TDC}.
As a trivial application, useful for having a taste of how it works, consider the torus $T_2=S^1\times S^1$ as a trivial fiber bundle with $B=F=S^1$. We have 

\begin{equation}
    H^0(S^1,\Z)\cong H^1(S^1,\Z)\cong \Z \quad \mbox{and} \quad H^i(S^1,\Z)\cong 0 \quad \mbox{for}\, i>1,
\end{equation}
thus the only non-zero entries of the second page defined by \eqref{SerreSQ} are

\begin{equation}
\begin{split}
    &E_2^{0,0}=H^0(S^1,\Z)\cong \Z \quad \quad E_2^{1,0}=H^1(S^1,\Z)\cong \Z\\
     &E_2^{0,1}=H^0(S^1,\Z)\cong \Z \quad \quad E_2^{1,1}=H^1(S^1,\Z)\cong \Z\\
    \end{split}
\end{equation}
and the differentials  

\begin{equation}
\begin{split}
    &d_2^{0,0}:E_2^{0,0}\longrightarrow E_2^{2,-1}\cong 0 \quad \quad d_2^{1,0}:E_2^{1,0}\longrightarrow E_2^{3,-1}\cong 0  \\
     &d_2^{0,1}:E_2^{0,1}\longrightarrow E_2^{2,2}\cong 0  \quad \quad d_2^{1,1}:E_2^{1,1}\longrightarrow E_2^{3,2}\cong 0  \\
         \end{split}
\end{equation}
all vanishes. Thus, the spectral sequence degenerates at $E_2$ page, and one finds:

\begin{equation}
\begin{split}
    &H^0(T_2,\Z)\cong E_2^{0,0}\cong \Z,\\
    &H^1(T_2,\Z)\cong E_2^{0,1}\oplus E_2^{1,0}\cong \Z\oplus\Z,\\
    &H^2(T_2,\Z)\cong E_2^{1,1}\cong \Z;
    \end{split}
\end{equation}
obtaining the expected answer for the integral cohomology of the Torus. \\
Another example of relevant spectral sequence is:\\

\prop[Leray spectral sequence]{
Let $f:X\rightarrow Y$ be a continous map of topological spaces,$f_\ast:Sh_{Ab}(X)\rightarrow Sh_{Ab}(Y)$ its direct image on abelian sheaves and $\F \in \mathrm{Obj}(Sh_{Ab}(X))$.\\
The spectral sequence with second page
\begin{equation}
    E_2^{p,q}=H^p(Y,R^qf_\ast(\F)))
    \label{LareySQ}
\end{equation}
computes to sheaf cohomology $H^{p+q}(X,\F)$.
}
Comparing \eqref{SerreSQ} and \eqref{LareySQ}, having in mind \eqref{HigherDirectImages}, it's clear Leray spectral sequence reduces to Serre spectral sequence whenever $\F$ is a constant sheaf and the presheaf $H^q(f^{-1}(U),\F)$ glues with trivial transition functions: $\F$ \textit{ has trivial monodromy}. \\
\subsection{Intersection Pairings}\label{IT}
The central objective of intersection theory is to construct and study well-defined pairings between homology and cohomology groups, enabling the precise measurement of how subvarieties or cycles intersect within a given space. Through these pairings, geometric intersections are translated into algebraic invariants that capture profound topological and algebraic information about the underlying space. The effectiveness of such pairings hinges on the identification of dual objects, which allows for the formation of non-degenerate bilinear forms essential to this translation process.
A fundamental result in algebraic geometry is the so called \textbf{universal coefficient theorem}, that establishes a relation between (co)homology groups with different coefficients. Let $X$ be a topological space. \\\\

\theo[Universal coefficient theorem(homology)]{
The short sequence
\begin{equation}
0 \longrightarrow H_i(X,\Z)\otimes A \xlongrightarrow[]{} H_i(X,A)\xlongrightarrow[]{} \mathrm{Tor}_1(H_{i-1}(X,\Z),A) \xlongrightarrow[]{} 0,
\label{UCT-homology}
\end{equation}
is exact.
}
This implies that the homology with integer coefficient completely determines the homology with coefficient in any abelian group $A$. The $\mathrm{Tor}$ functor measures the presence of torsion elements in the $(i-1)$th homology group, essentially $Z_k$ pieces, that can produce new elements in the $i$th homology group with coefficient in $A$. Thus, if $H_i(X,\Z)$ is flat, one gets the isomorphism 

\begin{equation}
    H_i(X,A)\cong H_i(X,\Z)\otimes A ,
\end{equation}
otherwise, in the general case, one has 

\begin{equation}
    H_i(X,A)\cong \left ( H_i(X,\Z)\otimes A\right )\oplus \mathrm{Tor}_1(H_{i-1}(X,\Z),A).
\end{equation}

There is also a version of \eqref{UCT-homology} for cohomology.
For every ring $R$ and $R-$module $M$.
\theo[Universal coefficient theorem(cohomology)]{
The short sequence
\begin{equation}
0 \longrightarrow \mathrm{Ext}^1_R (H_{i-1}(X,R),M) \xlongrightarrow[]{} H^i(X,M)\xlongrightarrow[]{} \mathrm{Hom}_R(H_i((X,R),M)) \xlongrightarrow[]{} 0,
\label{UCT-cohomology}
\end{equation}
is exact.
}
Also, this sequence always splits, although not naturally, giving 

\begin{equation}
    H^i(X,M)\cong \mathrm{Hom}(H_i(X,R),M)\oplus \mathrm{Ext}(H_{i-1}(X,R),M).
\end{equation}
Beyond being a practical tool for computations, this result reveals the presence of a duality between cohomology and homology groups; in particular for any field $\K$, because of $\mathrm{Ext}(-,\K)=0$, one gets the isomorphism: 

\begin{equation}
    H^i(X,\K)\cong \mathrm{Hom}(H_i(X,\K),\K).
\end{equation}
Within algebraic geometry, Grothendieck provided a unified and powerful framework to understand how cohomology groups of coherent sheaves relate to one another through natural dualities \cite{grothendieck62,hartshorne66}.\\\\

\defn[Dualizing sheaf]{
Let\footnote{The actual statement is more general and abstract of the one reported here, that is a simplified version avoiding further technicalities but accomplishing our claim.}  $X$ be an $n-$dimensional projective variety over a field $\K$ and $\mathcal{F}$ any coherent sheaf on $X$. There exists a coherent sheaf $\omega_X$, called \textbf{dualizing sheaf} such that 

\begin{equation}
    \mathcal{H}om_X(\mathcal{F},\omega_X)\simeq \mathcal{H}^n(X,\mathcal{F})^\vee.
    \label{Coherentduality}
\end{equation}}

The dualizing sheaf together with the linear functional $\mathcal{T}_X: \Ha^n(X,\omega_X)\rightarrow \K$, called trace map, induce a non degenerate pairing, called \textbf{coherent duality}:

\begin{equation}
    \mathcal{H}om_X(\mathcal{F},\omega_X)\times \mathcal{H}^n(X,\mathcal{F})\rightarrow \Ha^n(X,\omega_X)\rightarrow \K.
\end{equation}

There is topological analogue of Grothendiek duality provided by Verdier. \\\\

\theo[Verdier Duality \cite{verdier69}]{
Let $f:X\rightarrow Y$ be a map between locally compact Hausdorff spaces; and let $\F^\bullet$ be a complex of sheaves of abelian groups. There exist two complexes $\omega^\bullet_X$ and $\omega^\bullet_Y$, called \textbf{dualizing complexes} such that
\begin{equation}
    Rf_! \left (R\mathrm{Hom}_X\left (\F^\bullet,\omega^\bullet_X \right )\right)\cong R\mathrm{Hom}_Y\left (Rf_\ast \F^\bullet,\omega^\bullet_Y \right ).\\\\
    \label{VD}
\end{equation}
}
We stress that \eqref{Coherentduality} and \eqref{VD} are the fundamental tools one should use to properly define intersection product when dealing, for instance, with singular spaces. Unfortunately the general explicit determination of the dualizing sheaf/complex is not known; in most cases they can only be described abstractly through their defining universal properties, and despite they provide powerful theoretic tools, their explicit formulas exist only for special classes of spaces.\\
Indeed, if $X$ is “nice enough”, its dualizing sheaf can be proved to be described in term of the canonical sheaf $\Kan_{\C\Pro^n}$ of the embedding space; in particular, if $X$ is normal, the dualizing sheaf turns to be isomorphic to its canonical sheaf:

\begin{equation}
    \omega_X\simeq\Kan_X;
\end{equation}
thus specifically, for a smooth variety, where the canonical sheaf is isomoprhic to the top exterior power of the cotangent bundle, one recovers:\\

\theo[Serre duality \cite{Serre55}]{
If $X$ is an $n-$dimensional non singular projective variety over an algebraically illustrad field, the dualizing sheaf coincides with the sheaf $\Omega^n_X$ of top forms:

\begin{equation}
    \Ha^i(X,\mathcal{F})\simeq \Ha^{n-i}(X,\mathrm{Hom}(\mathcal{F},\Omega^n_X))^\vee\simeq \Ha^{n-i}(X,\F^\vee\otimes \Omega^n_X)^\vee,
    \label{Serreduality}
\end{equation}}
with $\F^\vee= \mathrm{Hom}(\mathcal{F},\Ol_x)$.\\
Let us apply \eqref{VD} to the case of $X$ a locally compact, oriented $n-$dimensional manifold, $Y=p$ a single point and $\F^\bullet$ a single sheaf (placed in degree 0). The direct image gives the global sections and the direct image with compact support gives the global sections with compact support. For an oriented manifold one can prove the dualizing complex is the orientation sheaf in degree $-n$
\begin{equation}
    \omega_X\cong or_X[n]\cong \underline{\C}[n],
\end{equation}

thus \eqref{VD} becomes 

\begin{equation}
    R\Gamma_c\left (R\mathrm{Hom}\left (\F,\C[n] \right )\right )\cong \left (R\Gamma\left (\F\right ) \right)^\vee,
    \label{THIS}
\end{equation}
where $\Gamma_c$ is the \textit{global section functor with compact support}.\\
Specializing it for the case $\F=\underline{\C}$ is the constant sheaf, we have

\begin{equation}
    R\mathrm{Hom}\left (\F^\bullet,\underline{\C}[n] \right )\cong R\mathrm{Hom}\left (\underline{\C},\underline{\C}[n] \right )\cong \C[n],
\end{equation}
since the $\textrm{Hom}$ of any constant sheaf to any sheaf is equal to the latter. Thus \eqref{THIS} reduces to

\begin{equation}
    R\Gamma_c\left (\underline{\C}[n]\right )\cong R\Gamma\left (\underline{\C}\right ).
\end{equation}
Thus, finally, taking the cohomologies of this complexes we get 

\begin{equation}
    \Ha^{i+n}\left (X,\underline{\C} \right )\cong \Ha^{-i}\left (X,\underline{\C} \right ),
\end{equation}

and after reindexing by setting $k=-i$, we obtain :\\\\

\theo[Poincaré duality]{
In a locally compact oriented $n-$dimensional manifold $X$, there is an isomorphism, called \textbf{Poincaré duality}:
\begin{equation}
    H_k(X,\C)\cong H_c^{n-k}(X,\C).\\\\
    \label{PoincarèDuality}
\end{equation}
}
Suppose now $\F=\La$ is a locally constant sheaf. Because of universal coefficient theorem, the computation is pretty much the same and it gives a so called \textbf{twisted version} of \eqref{PoincarèDuality}:

\begin{equation}
    H_k(X,\La)\cong H_c^{n-k}(X,\La^\vee).\\\\
    \label{TwistedPoincarèDuality}
\end{equation}


\section{Complex varieties}

In this section we present a brief review of the main results  
regarding complex varieties. 
We refer the reader to \cite{scheidemann}.\\
Let us start thinking about this obvious fact: in order to say that the real plane and the complex line are “the same thing”, $\R^2\cong \C$, we need to introduce the imaginary unit $i$ and perform the identification $\R^2\ni (x,y)\leftrightarrow z=x+iy$. The imaginary unit is nothing but linear operator $i:\R^2\rightarrow \R^2$ acting on $(x,y)$ as $i(x,y)=(-y,x)$ and such that $i^2=-1$. This idea can be generalized to a generic even dimensional real manifold $M$: take its tangent space $TM$ and define a endomorphism $J:TM\rightarrow TM$ that squares to minus the identity; provided $J$ fits nicely with a chart subdivision of $M$ \footnote{This is a integrability requirement, given by the vanishing of Nijenhuis tensor $N_J[X,Y]\equiv [JX,JY]-[X,Y]-J([JX,Y]-[X,JY])$ for any two vector fields $X$ and $Y$ on $M$.},  it defines a so called \textit{complex structure} on $M$; what we get is a \textbf{complex manifold}.\\

\subsection{Dolbeault cohomology and Hodge structure}

Let $X$ be a complex manifold of complex dimension $n$. The complex structure on $X$ induces a decomposition of the complexified cotangent bundle into its holomorphic and antiholomorphic parts, inducing a decomposition of the space of smooth differential forms into types. Any smooth $k$-form may be uniquely written as a finite sum of $(p,q)$-forms with $p+q=k$, where a local expression for a $(p,q)$-form is
\[
\Omega \,=\, \sum f_{I,J}(z,\overline z)\,dz^{i_1}\wedge\cdots\wedge dz^{i_p}\wedge d\overline z^{j_1}\wedge\cdots\wedge d\overline z^{j_q}.
\]
Let us denote by $\Omega^{p,q}$ the sheaf of such forms.\\ 
The exterior derivative splits as $d=\partial+\overline{\partial}$ with $\partial: \Omega^{p,q}\to \Omega^{p+1,q}$ and $\overline{\partial}: \Omega^{p,q}\to \Omega^{p,q+1}$.\\
The operator $\overline{\partial}$ satisfies $\overline{\partial}^2=0$, so for each fixed $p$ we obtain a complex of sheaves, called \textbf{Dolbeault complex}:

\begin{equation}
    \Omega^{p,\bullet}:0\to\Omega^{p,0}\xrightarrow{\ \overline{\partial}_{p,0}\ }\Omega^{p,1}\xrightarrow{\ \overline{\partial}_{p,1}\ }\cdots,
    \label{DolbeaultComplex1}
\end{equation}

\defn[Dolbeault cohomology]{
The Dolbeault cohomology groups are defined by
\begin{equation}
    H^{p,q}_{\overline{\partial}}(X)=\frac{\ker\big(\overline{\partial}_{p,q}\big)}
{\mathrm{im}\big(\overline{\partial}_{p,q-1}\big)}.
\end{equation}}

A fundamental result due to Dolbeault states that this complex provides a fine resolution of the sheaf of holomorphic $p$-forms $\Omega^p$, thus taking global sections computes sheaf cohomology:\\\\

\theo[Dolbeault]{
For $X$ a complex manifold, there is a canonical isomorphism:
\begin{equation}
    \Ha^p(X,\Omega^p_X)\cong H^{p,q}_{\overline{\de}}(X).
\end{equation}
\label{DolbeaultTheo}
}
Now, also the operator $\de$ satisfies $\de^2=0$, thus, analogously to \eqref{DolbeaultComplex1}, we can write a complex of sheaves $\Omega^{\bullet,q}$ for each fixed $q$. Combining the two, we obtain a double complex $\Omega^{\bullet,\bullet}$. Consider the spectral sequence with first page $E_1^{p,q}\equiv \Omega^{p,q}$. This sequence can be shown to converge to 

\begin{equation}
    \mathbb{H}^k(X,\Omega^\bullet) \cong \bigoplus_{p+q=k}E_{\infty}^{p,q}, 
    \label{DolbeaultSpecrtral}
\end{equation}
and thus to de Rham cohomology, due to De Rham theorem \ref{DeRhamTheo2}.\\
We introduce now the concept of \textbf{K\"ahler manifold}.\\
Given a complex manifold $X$ and a complex structure $J$, a metric $g$ is said to be hermitian with respect to $J$ if $g(Ju,Jw)=g(u,w).$ To such a metric we can associate a $2-$form $\kappa$, called {K\"ahler} form, that in local coordinates reads:
\begin{equation}
    \kappa=\frac{i}{2\pi}g_{ij}dz^i\wedge d\overline{z}^j.
\end{equation}
A complex manifold $X$ equipped with a hermitian metric with closed {K\"ahler} form is called a \textbf{K\"ahler manifold}.\\
Refering the reader to \cite{Fre}, we skip the details and we just highlith that for a K\"ahler manifold one can prove that any Dolbeault class admits a harmonic representative and this implies all the turning page differentials of the spectral sequence $E_r^{p,q}$ with first page $E_1^{p,q}=\Omega^{p,q}$ vanish. Thus, the following statement holds.\\\\

\theo[Hodge theorem]{
For $X$ a compact K\"ahler manifold the spectral sequence defined above degenerate at the first page. Thus \eqref{DolbeaultSpecrtral}, combined with De Rham theorem \ref{DeRhamTheo2} and Dolbeault theorem \ref{DolbeaultTheo}, gives the \textbf{Hodge decomposition}:

\begin{equation}
    H_{dR}^k(X)\cong \bigoplus_{p+q=k}H^{p,q}_{\overline{\de}}.
\end{equation}

}
The dimensions $h^{p,q}=\mathrm{dim} H^{p,q}_{\overline{\de}}$ are called \textbf{Hodge numbers}.\\
It is important to observe that the decomposition above is not merely an isomorphism of vector spaces but interacts with additional structures. Complex conjugation exchanges $H^{p,q}$ and $H^{q,p}$, yielding the so called Hodge symmetry $h^{p,q}=h^{q,p}$. Hodge numbers are pictorially organized is the so called \textbf{Hodge diamond} (figure \ref{HodgeDiamond}).
\begin{figure}
\begin{equation}
    \begin{array}{ccccccccc}
 & & & & h^{0,0} & & & & \\[2pt]
 & & & h^{1,0}&  &h^{0,1} & & & \\[2pt]
 & & h^{2,0}& & h^{1,1} & &h^{0,2} & & \\[2pt]
& \iddots& & & \vdots & & &\ddots & \\[2pt]
h^{n,0}& &h^{n-1,1} &\cdots &  &\cdots & h^{1,n-1}& & h^{0,n}\\[2pt]
& \ddots& & & \vdots & & &\iddots & \\[2pt]
  & & h^{n,n-2}& & h^{n-1,n-1} & &h^{n-2,n} & & \\[2pt]
 & & & h^{n,n-1}&  &h^{n-1,n} & & & \\[2pt]
 & & & & h^{n,n} & & & & \\[2pt]
\end{array}
\end{equation}
    \caption{\small{Hodge diamond for a generic $n-$dimensional K\"ahler manifold.}}
    \label{HodgeDiamond}
\end{figure}
Cup product and the existence of a Kähler class furnish hard Lefschetz and Hodge--Riemann bilinear relations, which place strong algebraic and metric constraints on the decomposition. These constraints are encoded abstractly by the notion of a pure Hodge structure.

\paragraph{Pure Hodge structure}\mbox{}\\
Let $\mathbb{V}$ be a finite dimensional vector space.\\
A \textbf{(decreasing) filtration} is a family of nested subspaces $F^p$ of \(V\) that organizes the vector space into layers:

\begin{equation}
    \mathbb{V}= F^0\subset F^1\subset \dots \subset \{0\}.
\end{equation}
Analogously, we define a \textbf{increasing filtration} $W^p$ as
\begin{equation}
    \mathbb{V}= W^n\subset W^{n-1}\subset \dots \subset \{0\}.
\end{equation}
Once a filtration (say increasing) $F^p$ for $\mathbb{V}$ is given, one can associate a graduation $\mathbb{V}^p\equiv F^p/F^{p+1}$, such that 
\begin{equation}
    \mathbb{V}=\bigoplus_p \mathbb{V}^p
\end{equation}
\defn[Pure Hodge Structure]{
Let $\mathbb{V}$ be a finite dimensional vector space over $\K$ and $\mathbb{V}_\C\equiv \mathbb{V}\otimes_\K\C$ its complexification. A pure Hodge structure on $\mathbb{V}$ is a bigraduation $\mathbb{V}^{p,q}$ preserving the graduation 

\begin{equation}
    \mathbb{V}_\C =\bigoplus_k\left ( \bigoplus_{p+q=k}\mathbb{V}^{p,q}\right ),
\end{equation}
 such that $\mathbb{V}^{p,q}=\overline{\mathbb{V}^{q,p}}$.\\\\
}

\theo[Hodge]{
Let $X$ be a compact K\"ahler manifold and $H(X,\C)$ its de Rham cohomology. Define the vector spaces $H^k(X)\equiv H^k(X,\C) $ and $H(X)=\bigoplus_k H^k(X)$. The vector space $H(X)$ admits a pure Hodge structure given by 

\begin{equation}
    H^k(X)\otimes \C = \bigoplus_{p+q=k} H^{p,q}(X),
\end{equation}
where $H^{p,q}(X)$ are the vector spaces associated to dolbeault cohomology groups.\\\\
}

Let $\Omega^\bullet_X$ be the de Rham complex on a $n-$dimensional complex analytic space $X$. We can introduce a trivial filtered complex $F^p\Omega^\bullet$, by replacing with zero all objects in degree $k<p$:

\begin{equation}
    F^p\Omega_X^\bullet: 0\rightarrow\dots\rightarrow 0\rightarrow\Omega^p\rightarrow \Omega^{p+1}\rightarrow\dots \rightarrow\Omega^n.
\end{equation}
Let now 
\begin{equation}
    \Phi^{k}_p: \mathbb{H}^k(X,F^p\Omega_X^\bullet)\longrightarrow\mathbb{H}^k(X,\Omega_X^\bullet)
\end{equation}
be the morphism, induced by the inclusion of complexes, among the $k-$ hypercohomologies of the filtered and full de Rham complexes, and consider the spaces

\begin{equation}
    S^k_p\equiv \mathrm{im}\left (\Phi^k_p\right ).
\end{equation}

Because $F^0\Omega_X^\bullet = \Omega_X^\bullet$, $\Phi^k_0$ is an isomoprhism for any $k$ an thus  $S^k_0 = \mathbb{H}^k(X,\Omega_X^\bullet)$; on the other extreme, $F^n\Omega_X^\bullet=0$, thus $S^k_n =\{0\}$. Moreover, one can also see $S^k_p\subset S^k_{p-1}$, so $S^k_p$ is a well defined decreasing filtration of $\mathbb{H}^k(X,\Omega_X^\bullet)$:

\begin{equation}
    S^k_p \equiv F^p\mathbb{H}^k(X,\Omega_X^\bullet).
\end{equation}
In particular, one shows

\begin{equation}
    F^pH^k(X,\C) = \bigoplus_{r\geq p}H^{r,k-r}(X), 
\end{equation}

and it follows that this filtration induces the Hodge decomposition

\begin{equation}
    H^k(X,\C) \cong \bigoplus_{p} F^pH^k(X,\C).
\end{equation}
It is worthwhile to stress, that if singularities appear and K\"ahler structure crushes, the isomorphism above fails to hold. However, one menage to still find a decomposition by introducing another increasing filtration, leading to \textbf{Hodge-Deligne splitting}.\\
Since the K\"ahler form $\kappa$ is a closed $2-$ form, it induces a well defined cohomology homomorphism $L: H^{k}(X,\C)\rightarrow H^{k+2}(X,\C)$ given by

\begin{equation}
    L(\omega)= \omega \wedge \kappa,
\end{equation}
for any $\omega \in H^{k}(X,\C)$.\\\\

\theo[Hard Lefschetz]{
For a $n-$dimensional compact K\"ahler manifold $X$, the map 
\begin{equation}
    L^k: H^{n-k}(X,\C)\rightarrow H^{n+k}(X,\C)
\end{equation}
is an isomorphism.\\\\
}
This theorem reveals a deep symmetry in the cohomology of K\"ahler manifolds: cohomology groups in complementary degrees (relative to the middle dimension) are isomorphic.\\\\ 

\defn[Primitive cohomology]{
For each degree $k$, define the \textbf{primitive cohomology} as

\begin{equation}
    \mathcal{P}H^k(X,\C) \equiv \mathrm{ker}\left (L^{n-k+1} \right ).\\\\
\end{equation}
}
A primitive form $\pi\in \mathcal{P}H^k$ satisfies $\kappa^{n-k+1}\wedge \pi=0$. Intuitively, primitive cohomology captures the “genuine new” part of a cohomology group which does not arise by wedging lower degree classes with $\kappa$. Conversely, once all primitive cohomologies are known, the entire cohomology can be reconstructed via the Lefschetz decomposition \eqref{LefDeco}:

\begin{equation}
    H^k(X,\C)\cong \bigoplus_{r\geq 0}L^r\mathcal{P}H^{k-2r}.
    \label{LefDeco}
\end{equation}
Since the K\"ahler form is of type $(1,1)$, the Lefschetz operator preserves the Hodge decomposition. Hence, also the primitive cohomology admits a Hodge decomposition:

\begin{equation}
    \mathcal{P}H^k = \bigoplus_{p+q=k}\mathcal{P}H^{p,q},
\end{equation}
where $\mathcal{P}H^{p,q}$ consists of classes whose harmonic representatives are primitive $(p,q)-$forms.\\
For smooth projective varieties, Griffiths' theorem \cite{griffiths1} provides a powerful algebraic description of primitive cohomology:\\\\

\theo[Griffiths]{
Let $\mathcal{V}\equiv\mathcal{V}[f]\subset \C\Pro^n$ be a complex projective variety defined by an (irreducible) homogenous polynomial $f$ of degree $d$. Let $J$ be the Jacobian ring of $f$ and $J_i$ its graded part in degree $i$, with the natural graduation inherited from $\C[x_0,...,x_n]$. For $\mathcal{V}$ smooth, the $(p,q)$-part of the primitive cohomology is given by

\begin{equation}
    \mathcal{P}H^{p,q}\cong J_{qd+d-n-1}.\\\\
    \label{GriffithsTheo}
\end{equation}}
Griffiths's theorem, and its generalization to singular varieties \cite{dimca2008}, translates geometric questions into algebraic ones, enabling to compute cohomology groups via commutative algebra.

\subsection{Calabi-Yau manifolds}\label{CY}

The mismatch between the $D=10$ critical dimension of super string theory and the observed four dimensional space-time,  can be fixed by assuming the extra six dimensions are compactified on a small internal manifold \cite{CANDELAS198546}. 
Supersymmetry generators are spinors on the full spacetime and decompose into products of external and internal spinors; compactification preserves a supercharge only if the internal manifold admits a globally defined internal spinor (equivalently a spin structure and, for unbroken supersymmetry, a covariantly constant spinor). If no such internal spinor exists, the corresponding supercharge is lost and supersymmetry is broken in the lower-dimensional theory. Preserving some supersymmetry in the effective theory is desirable because it improves the theory’s behaviour: it protects hierarchies, controls quantum corrections (non-renormalization theorems), and aids vacuum stability.
In order for having a globally defined spinor bundle on the internal manifold, one requires the second Stiefel-Whitney class to vanish. Moreover if a global nowhere vanishing section of such bundle is required to exist, the holonomy group must be a subgroup of $SU(n)$. 
In \cite{Calabi1957}, Calabi conjectures that on a compact K\"aler manifold, each K\"aler class contains a unique K\"ahler metric whose Ricci form equals any prescribed real $(1,1)-$ form representing the first Chern class; in \cite{Yau1978},  Yau was able to prove Calabi conjecture and in particular that the vanishing of the first Chern class implies the existence of a Ricci-flat K\"ahler metric, which makes the nowhere-vanishing holomorphic volume form parallel and hence enforces the holonomy group to be a subgroup of $SU(n)$. \\
This stringy motivation was the first reason for complex K\"aler manifold with vanishing first Chern class, called \textbf{Calabi-Yau manifolds}, to receive the attention of physicists and to be largely studied along the last decades. As we will see in chapter \ref{C2}, recent developments have shown that they also play a crucial role in scattering amplitudes computation.\\
The vanishing of the first Chern class of a $n-$dimensional compact K\"ahler manifold (Calabi-Yau $n-$fold) can be proven to be equivalent to the uniqueness of a non-vanishing holomorphic $n-$form $\Omega$, that is $h^{n,0}=1$.\\
Morevoer, Hodge numbers turn to be constrained by the relation $h^{p,0}=h^{0,n-p}$ and, if the homolomy group is exactly $SU(n)$, one has

\begin{equation}
    h^{p,0}=0.
\end{equation}
Therofe, the general form of the Hodge diamond for a Calabi-Yau $n-$fold with $SU(n)$ holonomy group is

\begin{equation}
    \begin{array}{ccccccccc}
 & & & & 1 & & & & \\[2pt]
 & & & 0&  &0 & & & \\[2pt]
 & & 0& & h^{1,1} & &0 & & \\[2pt]
& \iddots& & & \vdots & & &\ddots & \\[2pt]
1& &h^{n-1,1} &\cdots &  &\cdots & h^{1,n-1}& & 1\\[2pt]
& \ddots& & & \vdots & & &\iddots & \\[2pt]
  & & 0& & h^{n-1,n-1} & &0 & & \\[2pt]
 & & & 0&  &0 & & & \\[2pt]
 & & & & 1 & & & & \\[2pt]
\end{array}
\end{equation}

A simple and widely used class of CY manifolds arises as smooth projective hypersurfaces or complete intersections in complex projective space. Because the propriety to be K\"ahler is automatically induced by projectivity, the Calabi-Yau condition reduces to the vanishing of the first Chern class.\\
The complete intersection of $r$ homogeneous polynomials of degrees $\alpha_1,\dots,\alpha_r$ in $\mathbb{C}\mathbb{P}^n$ has first Chern class (see \cite{Fre})

\begin{equation}
    c_1(X)=\left (n+1-\sum_{i=1}^r \alpha_i \right ) H_X,
\end{equation}
with $H_X$ the hyperplane class.\\
The vanishing first Chern class is then subjected to the relation

\begin{equation}
    n+1-\sum_{i=1}^r \alpha_i =0.
    \label{InterChern}
\end{equation}
For a smooth hypersurface $X_d\subset\mathbb{C}\mathbb{P}^n$ of degree $d$, \eqref{InterChern} reduces to $d=n+1$, thus one has that:\textit{
The projective algebraic variety associate to a polynomial of degree $n+1$ in $\C\Pro^{n}$ is a Calabi-Yau $(n-1)$-fold.}\\
The simplest example of a hypersurface satisfying this condition is a cubic in $\C\Pro^{2}$, that, rather then $1-$fold Calabi-Yau, is better known as \textbf{Elliptic curve}

\subsection{Periods}\label{PeM}
Consider the Legendre family of elliptic curves
\begin{equation}
    y^2=\B(x,\lambda)=x(x-1)(x-\lambda)
    \label{LegendreFamily}
\end{equation}
and the standard elliptic basis of the De Rham cohomology

\begin{equation}
    \{\omega_1,\omega_2\} \equiv \left \{\frac{dx}{y},\frac{x\,dx}{y}\right \}.
    \label{EllipticBasis}
\end{equation}
For any locally constant $1-$cycle $\gamma(\lambda)$ on the fiber, we call \textbf{period} the function of $\lambda$:

\begin{equation}
    \Pi(\lambda)=\int_{\gamma(\lambda)} \omega_1(\lambda).
\end{equation}

Differentiating with respect to $\lambda$ we get

\begin{equation}
    \de_\lambda\omega_1=-\frac{1}{2} \B^{-3/2}(\de_\lambda\B) dx = \frac{1}{2} \frac{x(x-1)}{y^3}dx,
    \label{PeM-1}
\end{equation}
and after noticing that the numerator can be written as

\begin{equation}
    x(x-1)=\frac{1}{3}\B'(x,\lambda)+\frac{2\lambda-1}{3}x-\frac{\lambda}{3},
    \label{PeM-2}
\end{equation}
we can rewrite \eqref{PeM-1} as

\begin{equation}
    \de_\lambda\omega_1 = \left [ \B'(x,\lambda)\frac{dx}{6y^3} \right ]+\frac{1}{6}\frac{(2\lambda-1)x-\lambda}{\B(x,\lambda)}\omega_1.
    \label{PeM-3}
\end{equation}
The term in square brackets is an exact differential\footnote{$\B'(x,\lambda)\frac{dx}{6y^3}=-\frac{1}{3}d\left (\frac{1}{y}\right )$}
thus, in cohomology, we can drop it from \eqref{PeM-3}, obtaining that $\de_\lambda\omega_1$ is related to $\omega_1$ by a rational coefficient function:

\begin{equation}
    \de_\lambda\omega_1 = \frac{1}{6}\frac{(2\lambda-1)x-\lambda}{\B(x,\lambda)}\omega_1.
    \label{PeM-4}
\end{equation}

Because $\de_\lambda\omega_1$ belongs to vector space spanned by \eqref{EllipticBasis}, there exist rational functions $a(\lambda)$ and $b(\lambda)$, independent on $x$, such that

\begin{equation}
     \de_\lambda\omega_1 = a(\lambda)\omega_1 +b(\lambda)\omega_2, 
     \label{LinearComb}
\end{equation}
Comparing \eqref{PeM-1} and \eqref{LinearComb}, we get  

\begin{equation}
    \frac{1}{2}x(x-1)=(a+bx)\B + \delta,
\end{equation}
with $\delta$ a closed $1-$form.
Assuming the ansatz $\delta=d(P/y)=(\B P'-1/2 P\B')$, with $P(x)$ a three degree polynomial, and solving term by term for the unknown coefficient, one finds:

\begin{equation}
    a(\lambda)=-\frac{1}{2(\lambda-1)} \quad\quad \mbox{and}\quad\quad b(\lambda)=\frac{1}{2\lambda(\lambda-1)}.
\end{equation}
Applying the same reduction procedure to express $\de_\lambda\omega_2$ in the above basis we obtain that the $\lambda$ derivation acts on the vector basis $(\omega_1,\omega_2)^T$ by a $2\times 2$ matrix with rational entries:

\begin{equation}
    \de_\lambda \begin{pmatrix}\omega_1\\\omega_2\end{pmatrix}=\begin{pmatrix}-\frac{1}{2(\lambda-1)}& \frac{1}{2\lambda(\lambda-1)}\\-\frac{1}{6(\lambda-1)}& \frac{2-\lambda}{6\lambda(\lambda-1)}\end{pmatrix}\begin{pmatrix}\omega_1\\\omega_2\end{pmatrix}.
    \label{GM1}
\end{equation}
Commuting with integration in \eqref{PeM-1}, we can rewrite this as a linear differential equation for the period vector $ \Pi_i \equiv \int_{\gamma(\lambda)}\omega_i$ as 

\begin{equation}
    \de_\lambda \Pi_i = M_{ij}\Pi_j.
\end{equation}

Solving for one of the periods, say $\Pi_2$, and substituting back, we get a second order differential equation for the other period, called \textbf{Picard-Fuchs equation}: 

\begin{equation}
    \lambda(1-\lambda)\Pi''+(1-2\lambda)\Pi'-\frac{1}{4}\Pi=0.
    \label{PFeq1}
\end{equation}

This is not an accident occurring for elliptic curves, but a consequence of the structure of the cohomological bundle.\\
Let $f:X\rightarrow \M$ be a one parameter ($\lambda\in \M$) family of smooth $n-$dimensional complex manifolds $X_\lambda=f^{-1}(\lambda)$.
The relative de Rham cohomology sheaf

\begin{equation}
    \Ha_{dR}^n(f)\equiv R^nf_\ast(\Omega^\bullet_{X/\M}),
\end{equation}
is a vector bundle on $\M$ equipped with a flat connection 

\begin{equation}
    \nabla_{GM}: \Ha_{dR}^n(f)\rightarrow \Omega_S^1\otimes \Ha_{dR}^n(f)
\end{equation}
called \textbf{Gauss-Manin connection}.\\
The periods of relative closed forms are the flat sections of the dual local system: 

\begin{equation}
    \frac{d}{d\lambda}\int_{\gamma(\lambda)}\omega(\lambda) =\int_{\gamma(\lambda)}\nabla_{GM}\omega(\lambda).
\end{equation}
The steps we applied from \eqref{PeM-1} to \eqref{GM1} are actually part of a general precedure, called \textbf{Griffiths-Dwork reduction}, whose closure is granted by \textbf{Griffiths transversality} of the Gauss-Manin connection:

\begin{equation}
    \nabla_{GM}(F^p\Ha^n_{dR}(f))\subset \Omega^1_\M \otimes F^{p-1}\Ha^n_{dR}(f).
\end{equation}

\paragraph{The Brieskorn lattice}\mbox{}\\
Let $f\in\mathcal{R}\equiv\C[x_0,..x_n]$ a polynomial in $n+1$ variables and $\mathcal{K}\equiv \mathcal{K}(\de f, \mathcal{R})$ the Koszul complex \cite{Choudary1994KoszulCA,dimca2024} associated to the sequence of partial derivatives

\begin{equation}
    \mathcal{K}: 0\rightarrow \bigwedge^{n+1}\mathcal{R}^{n+1} \xrightarrow{d_{n+1}}...\xrightarrow{d_2} \mathcal{R}^{n+1}\xrightarrow{d_1} \mathcal{R} \rightarrow 0,
\end{equation}

\nn
where $\mathcal{R}^{i}\equiv\otimes_i \mathcal{R}$ and the differential acts as:

\begin{equation}
    d_i(e_1\wedge ...\wedge e_k )= \sum_{j=1}^i (-1)^{j+1}(\de_j f)e_1\wedge...\wedge \hat{e}_j \wedge ...\wedge e_k.
\end{equation}
\nn
For a variety $\mathcal{V}[f]$ with singular locus $\Sigma=\{\sigma_i\}_i$ the homology of the associated Koszul complex satisfies

\begin{equation}
    \Ha_i(\mathcal{K})\neq 0\quad \mbox{for}\quad i\leq dim(\Sigma)+1;
\end{equation}
\nn
i.e. it is concentrated in degree zero for $\mathcal{V}[f]$ smooth, and for isolated singularities the only non trivial groups are:

\begin{equation}
    \Ha_0(\mathcal{K})=J_f \quad ,\quad \Ha_1(\mathcal{K})=\frac{\{\vec{a}\in \mathcal{R}^{n+1}|\sum a_i\de_if=0\}}{\{\vec{b}\in \mathcal{R}^{n+1}| b_k=\sum_{i<j}r_{ij}(\de_if e_j-\de_jfe_i)\}}.
\end{equation}
\nn
Remind that for any graded ring $A=\bigoplus A_j$, the associated Poincaré series $\mathcal{P}[A](t)$ is defined to be:

\begin{equation}
    \mathcal{P}[A](t)=\sum_{j\geq 0} (dim A_j)t^j.
\end{equation}

\nn
For isolated singularities, the “smooth” part of the Jacobian is encoded in the difference 
\begin{equation}
    \mathcal{P}[\Ha_0(\mathcal{K})](t)- \mathcal{P}[\Ha_1(\mathcal{K})](t)=\frac{(1-t^{d-1})^{n+1}}{(1-t)^{n+1}}.
    \label{Poncaréseries}
\end{equation}

\nn
The form of the full Poincaré series of the Jacobian is not known in general. However, if the singular locus is a 0-dimensional complete intersection of type $(a_1,...a_n)$, it can be expressed in a closed rational form: 

\begin{equation}
    \mathcal{P}[\Ha_0(\mathcal{K})](t)=\frac{(1-t^{d-1})^{n+1}}{(1-t)^{n+1}}+t^{(n+1)(d-1)-\sum a_i}\left [\frac{\prod_i (1-t^{a_i})}{(1-t)^{n+1}}\right ].
    \label{Poincarèseries2}
\end{equation}

\nn
The global Jacobian ring $J$ is not finitely generated unless the variety is smooth. However, if only isolated singularities arise, the dimensions of the graded pieces stabilize for $q>(n+1)(d-2)$ to a constant dimension $\tau(\mathcal{V})$, called Tjurina number of the hypersurface. We will write:

\begin{equation}
    J= dim(J_0) \oplus dim(J_1) \oplus...\oplus dim(J_{(n+1)(d-2)-1})\oplus \overline{\tau(\mathcal{V})},
\end{equation}
\nn
using the overline for denoted all the next graded pieces have the same dimension.     
The Tjurina number $\tau(\mathcal{V})$ of the hypersurface is the sum of the Tjurina numbers\footnote{For homogenous polynomial Milnor and Tjurina numbers are equal, because of Euler relation.} of the singularities $\tau(\mathcal{V},\sigma_j)$, corresponding, by definition, to the dimension, as a $\C$ vector space, of the localization of the Jacobian ring at the singular locus: 

\begin{equation}
    \tau(\mathcal{V}) =\sum\tau(\mathcal{V},\sigma_j)= \sum dim_{\C} (J)_{\sigma_i},
\end{equation}

\nn
with $(J)_{\sigma_i}$, for $\sigma_i\in\Sigma$, is 

\begin{equation}
    (J)_{\sigma_i}=\left \{\frac{f}{g}| f,g \in J, g \notin \sigma_i\right \}=\frac{\C\{x_0,...,x_n\}}{(\de_i f)},
\end{equation}
\nn
where $\C\{x_0,...,x_n\}$ is the ring of formal power series. 
There is an exact sequence
\begin{equation}
    0 \rightarrow \frac{df \wedge \Omega^{n}}{df\wedge d\Omega^{n-1}}\rightarrow \mathcal{B}_f \rightarrow J_f \rightarrow 0,
\end{equation}
\nn
where $\B_f$ is called \textbf{Brieskorn lattice}.
In the case of isolated singularities the torsion is just the jacobian ring again, who is injectively mapped onto the Brieskorn lattice, thus the sequence splits 

\begin{equation}
    \mathcal{B}_f = J_f \oplus J_f.
\end{equation}




\section{Picard-Lefshetz theory}\label{PLT}
The \textbf{Picard-Lefschetz theory} \cite{HuseinSade} is the complex analogous of the Morse theory that studies the topology of level sets of complex analytic functions and provides a concrete tool for computing monodromies.\\
Let us begin by considering the following simple example of a two variable function:
\begin{equation}
    f(z,w)=z^2+w^2.
    \label{function_ex1}
\end{equation}

The function $f(z,w)$ has a unique critical point:

\begin{equation}
    \begin{cases} \partial_z f(z,w)=0 \\ \partial_w f(z,w)=0 \end{cases} \quad \quad \longrightarrow \quad \quad (z,w) =(0,0).
\end{equation}

We refer to the value of the function $f(z,w)$ at a critical point as a \textbf{critical value}, in the present case $f(0,0)=0$. The \textbf{critical set} is the set of points in $\mathbb{C}^2$ where the function $f$ takes the critical value:
\begin{equation}
    V_0= \left\lbrace (z,w) \vert z^2+w^2=0 \right\rbrace. 
\end{equation}

For any other value $f(z,w)=t$, different from the critical one, we call \textbf{level sets} the loci
\begin{equation}
    V_{t} = \left\lbrace (z,w) \vert z^2 + w^2 = t \right\rbrace.
\end{equation}

In order to figure out the topology of these level sets we can consider the Riemann surfaces associated with the function \eqref{function_ex1}
\begin{equation}
    w = \sqrt{(t - z^2)}.
\end{equation}

\begin{figure}[h!]
    \centering    
    \includegraphics[scale=0.15]{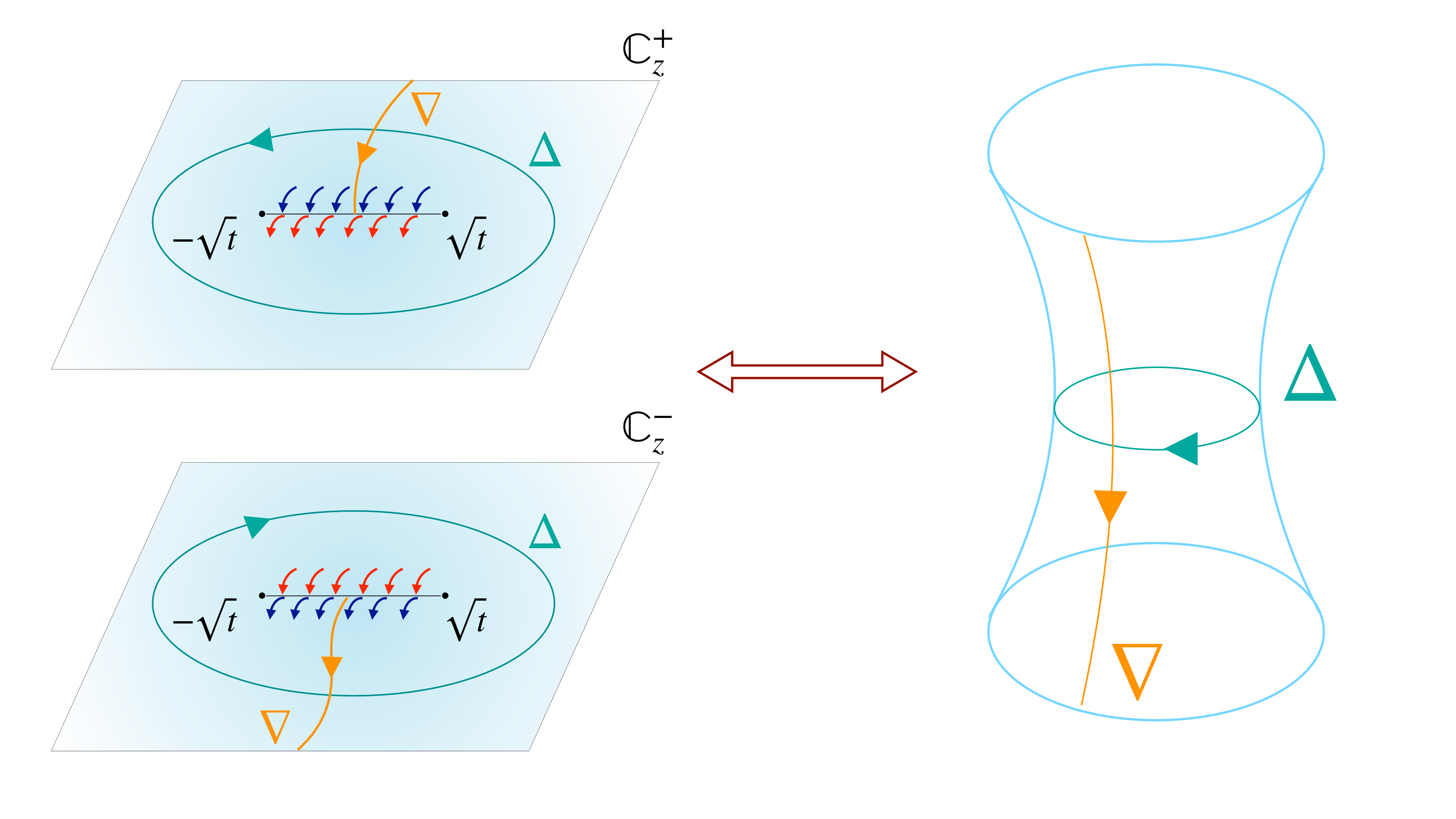}
    \caption{\small{Gluing of the two Riemann sheets along the two edges of the cut from $- \sqrt{t}$ to $\sqrt{t}$. The resulting surface is topologically equivalent to a cylinder.}}
    \label{fig1}
\end{figure}

These surfaces can be obtained gluing together two copies of the complex plane $z$ with a cut along the segment $(- \sqrt{t}, \sqrt{t})$, as showed in Figure \ref{fig1}, resulting in a surface topologically equivalent to a cylinder. When $t=0$, the corresponding critical level set consists of two lines intersecting at the point $0$.

Consider now the fibration $f: \mathbb{C}^2 \longrightarrow \mathbb{C}^\ast_{t}$ over the space $\mathbb{C}_{t}^{\ast} = \mathbb{C} \setminus \left\lbrace 0 \right\rbrace $, which fibers are Riemann surfaces representing the non-critical level sets $V_{t}$. Note that we removed from the base the point $t=0$, which correspond to the singular fiber $V_0$. Let us now consider the circular path around $t=0$

\begin{equation}
    t (\tau)= e^{2 \pi i \tau} \alpha \quad \quad 0 \leq \tau \leq 1 , \quad \alpha >0.
    \label{closed_curve}
\end{equation}

\begin{figure}[h!]
    \centering    
    \includegraphics[scale=0.2]{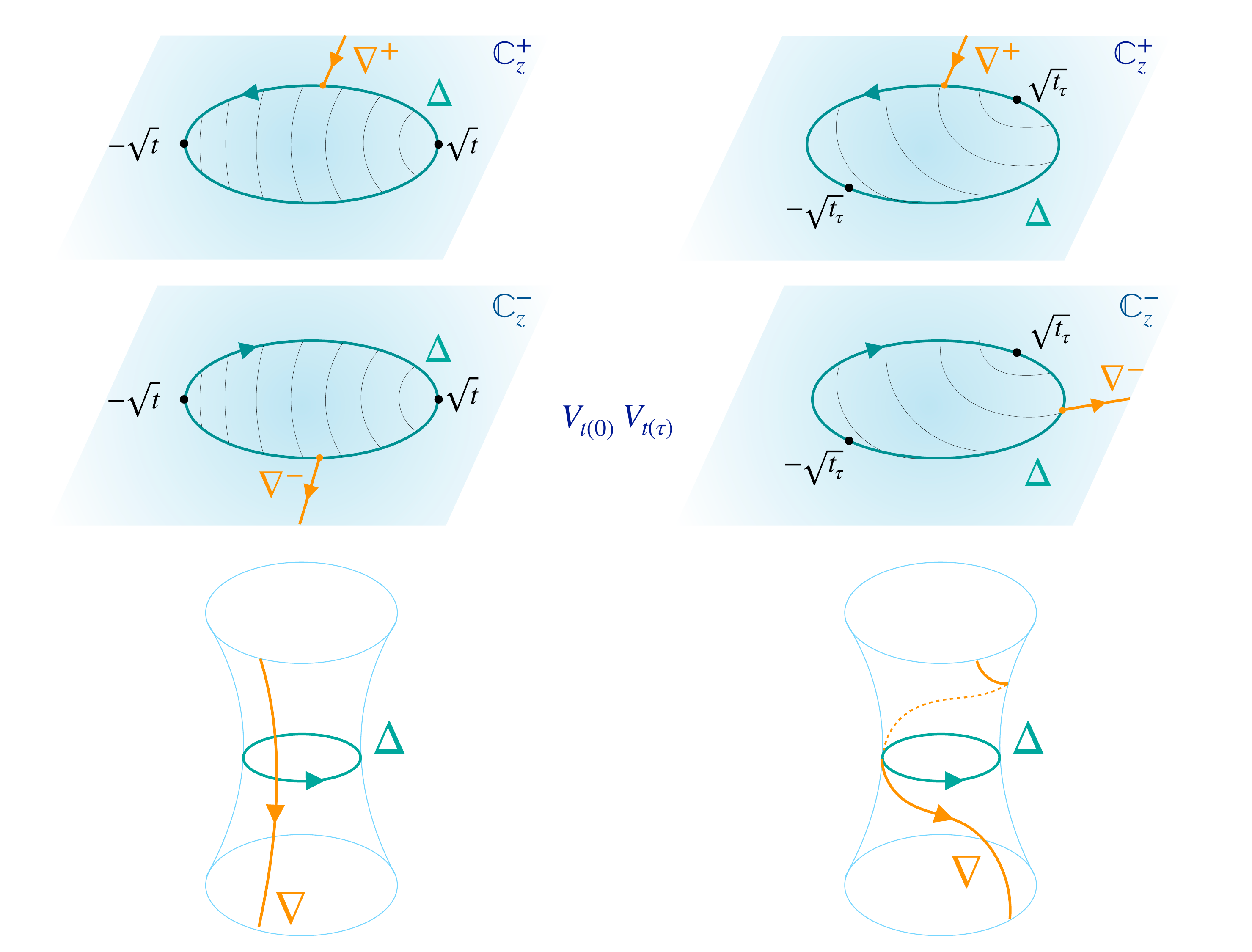}
    \caption{\small{Action of the monodromy on the vanishing and covanishing cycles. }}
    \label{fig2}
\end{figure}

\begin{figure}[h!]
    \centering    
    \includegraphics[scale=0.2]{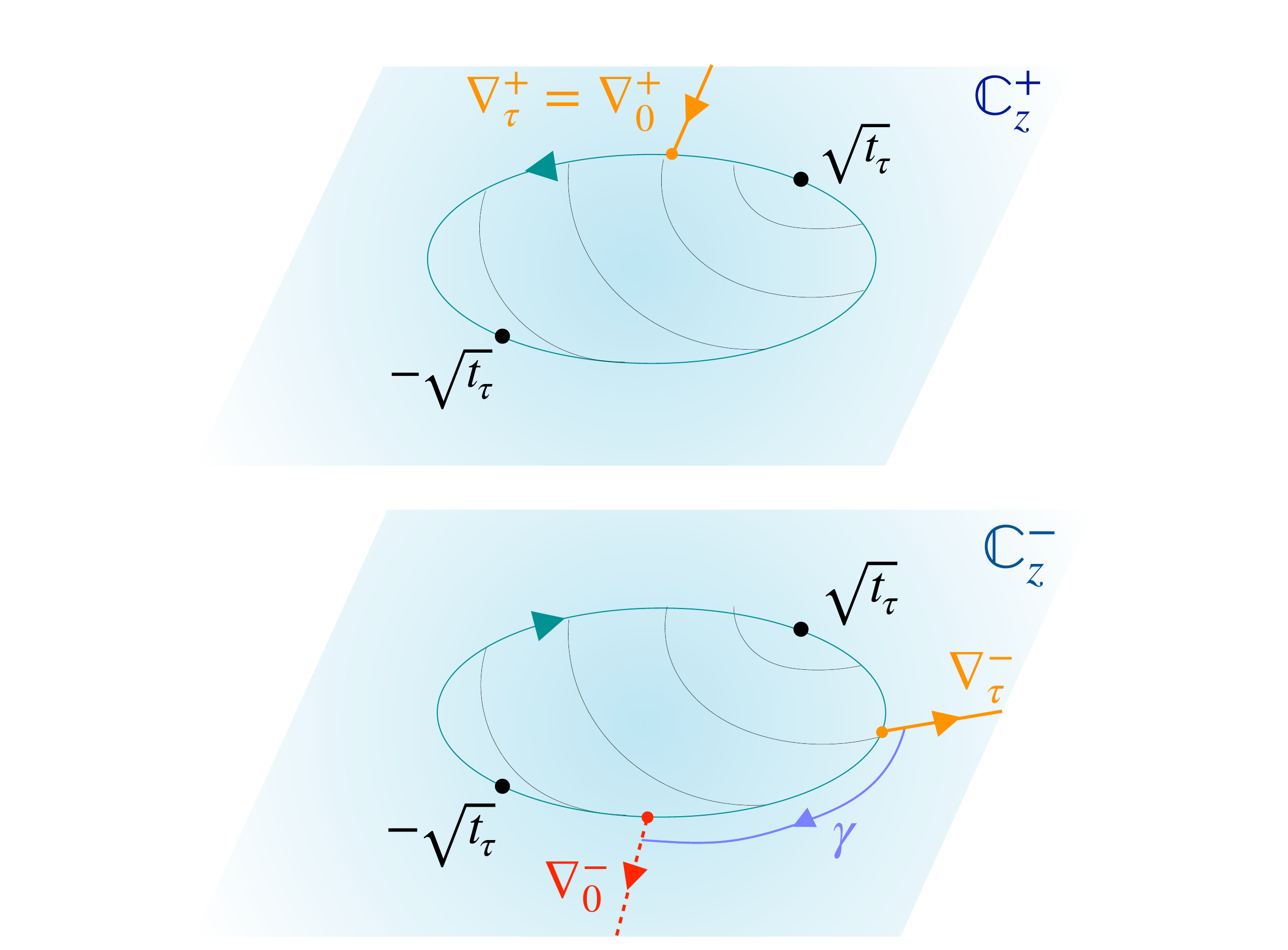}
    \caption{\small{Homotopically deformation of $\nabla^-_\tau$ over $\nabla^-_0$. }}
    \label{fig3}
\end{figure}

We aim to study how the fiber $V_{t}$ varies along this path  by tracking the motion of the branch points $z= \pm \sqrt{t(\tau)}$ in the complex planes $\mathbb{C}_z^\pm$ as the parameter $\tau$ evolves. We observe that these branch points rotate counterclockwise around $z=0$ undergoing a half-turn (rotation by $\pi$) at $\tau=1$ (see Figure \ref{fig2}). 
Considering the fibration $V\longrightarrow I$, with $I\equiv [0,1]$, defined by $V_{t(\tau)}\longrightarrow \tau$, we can associate to the closed curve $t(\tau)$ in \eqref{closed_curve} a continuous map
\begin{align}
    \Gamma: V\longrightarrow V_{t(0)},
\end{align}
such that for any $\tau$, the map $\Gamma_\tau: V_{t(\tau)}\longrightarrow V_{t(0)}$ defined by $\Gamma_\tau(\cdot)=\Gamma(\tau,\cdot)$ is a diffeomorphism, and $\Gamma_0=id$. Since $t(1)=t(0)$, the corresponding map $h=\Gamma_1: V_{t(0)}\longrightarrow V_{t(1)}$ is called \textbf{monodromy map}.\\
We are interested to know how this map acts on the first homology group of the fiber $V_{t}$, which is generated by the 1-cycle $\Delta$ represented in Figure \ref{fig1}. This $1$-cycle $\Delta$ is called  \textbf{Picard-Lefschetz vanishing cycle}, due to the fact it shrinks to a point when $t \rightarrow 0$. Its transversely intersecting cycle $\nabla$ is called \textbf{covanishing cycle} and it generates the first homology group $H^{BM}_1 \left( V_{t}, \mathbb{Z} \right) \simeq \mathbb{Z}$. Here, $H^{BM}_1 (V_{t})$ is the first Borel-Moore homology of the non-compact space $V_{t}$. This homology admits chains that may be infinite in extent but are restricted to be finite in any compact region. A vertical line in the cilinder $V_{t}$ is locally finite because, in any compact sub-interval of the vertical direction the line is finite. Even though the line can extend indefinitely along the cylinder, within any small, bounded region it is just a finite segment. Since $H^{BM}_1 \left( V_{t}, \mathbb{Z} \right) \simeq H_1 \left( C, \partial C ; \mathbb{Z} \right)$, where $C$ is the compact cylinder, it is easy to prove that $H^{BM}_1 \left( V_{t}, \mathbb{Z} \right) \simeq \mathbb{Z}$ with generator $\nabla$.  \\
In Figure \ref{fig2} we show how the diffeomorphism $\Gamma_\tau$ acts on these two cycles. In the $\mathbb{C_z^-}$ foil, after the action of $\Gamma_\tau$, $\nabla^-$ is rigidly transported along the counterclockwise direction of the rotation. As depicted in Figure \ref{fig3} we can deform homotopically the support of $\nabla^-_{\tau}$ to the support of $\nabla^-_0$ through a connected path $\gamma$ in $\mathbb{C}_z^-$ such that:
\begin{equation}
    \nabla_\tau = \nabla^+_\tau + \nabla^-_\tau  \thickapprox \nabla^+_0 + \tilde{\nabla}^- = \nabla^+_0 + \gamma + \nabla_0^-.
\end{equation}
In terms of the homology we have:
\begin{equation}
    \left[ \nabla_\tau \right] \sim \left[ \nabla_0 \right] + \left[ \gamma \right] \sim \left[ \nabla_0 \right],
\end{equation}
if $\gamma$ in contractible. If $\tau=1$ the path $\gamma$ closes around the hole and 
\begin{equation}
    \left[ \nabla_1 \right] = \nabla_0 + \Delta.
\end{equation}

\nn In particular, one gets that, up to homotopies, the monodromy map $h$ acts as the identity outside a compact set around $z=0$ and non-trivially inside this set. More precisely, it maps the vanishing cycle $\Delta$ into itself and it acts as the (homotopically equivalent) identity  in the part of $\nabla$ extending outside the compact set and with the following transformation inside the set:
\begin{equation}
    h \nabla = \nabla - \Delta.
\end{equation}
It is worth to mention that while the map $h$ defined on the fibration $V\to I$ depends on the choice of the diffeomorphisms $\Gamma$, when induced to the (co)homology it becomes independent on such a choice.
This action allows us to define a function, called \textbf{variation map}, mapping a cycle with closed support to a cycle with compact support:
\begin{equation}
    \begin{split}
         \rm{Var} \quad : \quad H_1^{BM} \left( V_{t}, \mathbb{Z} \right) \quad & \longrightarrow \quad H_1 \left( V_{t}, \mathbb{Z} \right), \\
         \nabla  \quad & \longrightarrow \quad (\Delta \circ \nabla) \Delta, \\
    \end{split}
\end{equation}
where $(\Delta \circ \nabla)$ denotes the intersection pairing between the cycles $\Delta$ and $\nabla$, defined as the number of topological intersections counted with a sign depending by the relative orientation of the two cycles.\\
The main objects we defined so far are the \textbf{vanishing cycles}, the \textbf{monodromy} and the \textbf{variation map} for a two-variables function. Defining these objects for arbitrary functions of several variables is a challenging problem that remains unsolved in general. Picard-Lefschetz theory offers a powerful method to address this problem by using \textbf{deformation theory} techniques. 

\subsection{Monodromy, variation operators and vanishing cycles}
In this paragraph, we extend the above discussion to the case of holomorphic functions in several complex variables and provide formal definitions of the concepts introduced in the previous example.\\
Let
\begin{equation}
    f : \mathcal{M}^n \quad \longrightarrow \quad \mathbb{C}
\end{equation}
be a holomorphic function on a $n-$dimensional complex manifold $\mathcal{M}^n$.  Let $U$ be a contractible compact region in the target space with smooth boundary $\partial U$, and let us assume $f$ has a finite number of critical points $\Sigma= \left\lbrace \sigma_i \right\rbrace_{i=1}^{\mu}$ with critical values $t_i = f \left( \sigma_i \right)$ on $\overline{U}$. Let us indicate with $F_t$ the level set of the function $f$ at $t \in U$:
\begin{equation}
    F_t = \left\lbrace \mathbf{z}\in \C^n \vert f(\mathbf{z})=t \right\rbrace.
\end{equation}
If $t \in U$ is not a critical value, then $F_t$ is a $(n-1)-$dimensional complex manifold with smooth boundary. Let $t_0$ be a non-critical value in the boundary $\partial U$ and let us construct for each class of loops $\left[  \gamma \right] \in \pi_1 \left( U \setminus \left\lbrace t_1,\dots ,t_\mu \right\rbrace , t_0 \right)$ a continuous family of mappings $\Gamma_\tau : F_{t(\tau)} \longrightarrow F_{t_0}$,  for which $\Gamma_0= id$. Then, $h_{\gamma} \equiv \Gamma_1$, transforming the non-singular level $F_{t_0}$ into itself, defines the \textit{monodromy map}  along the loop $\gamma$. Note that the map $h_{\gamma}$ depends on the specific path $\gamma$ we are considering.\\\\

\defn[Monodromy operator]{
We call monodromy operator of the loop $\gamma$ the action $h_{\gamma \ast} = h_{\left[ \gamma \right]}$ of the transformation $h_{\gamma}$ on the homology of the non-singular level set $H_{\ast} \left( F_{t_0} \right)$.}

The transformation $h_{\gamma}$ also induces an automorphism $h_{\left[ \gamma \right]}^{(r)}$ in the relative homology group $H_{\bullet} \left( F_{t_0}, \partial F_{t_0} \right)$ of the non-singular level set $F_{t_0}$ modulo its boundary. This homology is isomorphic to the homology of cycles with closed support:

\begin{equation}
    H_{\bullet} \left( F_{t_0}, \partial F_{t_0} \right) \simeq H^{BM}_{\bullet} \left( F_{t_0} \setminus \partial F_{t_0} \right).
\end{equation}

Since the action $h_{\gamma}$ is trivial on the boundary $\partial F_t$, then the difference between $h_{ \left[ \gamma \right]}^{(r)} \delta$ and $\delta \in H_{\bullet} \left( F_{t_0}, \partial F_{t_0} \right)$ is a cycle in $H_{\bullet} \left( F_{t_0} \right)$.\\\\

\defn[Variation operator]{
The homomorphism 
\begin{equation}
    \rm{Var}_{\gamma} \quad : \quad H_{\bullet} \left( F_{t_0}, \partial F_{t_0} \right) \quad \longrightarrow \quad H_{\bullet} \left( F_{t_0}\right)\\
\end{equation}
is called the variation operator over the loop $\gamma$.}
Using the natural homomorphism
\begin{equation}
    i_{\ast} \quad : \quad H_{\bullet} \left( F_{t_0}\right) \quad \longrightarrow \quad H_{\bullet} \left( F_{t_0}, \partial F_{t_0} \right)
    \label{map1}
\end{equation}

induced by the inclusion $F_{t_0} \subset \left( F_{t_0}, \partial F_{t_0} \right)$, we can write the following relations connecting the automorphisms $h_{\left[ \gamma \right]}$ and $h_{\left[ \gamma \right]}^{(r)}$:

\begin{equation}
  \begin{split}
      & h_{\left[ \gamma \right]} = id + \rm{Var}_{\gamma} \cdot i_{\ast} \,, \\
      & h_{\left[ \gamma \right] }^{(r)} = id + i_{\ast} \cdot \rm{Var}_{\gamma}\, .
  \end{split}  
\end{equation}

If the class $\left[ \gamma \right] \in \pi_1 \left( U \setminus \left\lbrace t_1,\dots,t_\mu \right\rbrace , z_0 \right)$ is given by $\left[ \gamma \right] = \left[ \gamma_1 \right] \cdot \left[ \gamma_2 \right]$,\footnote{Note that in the composition of homology classes, we follow the convention of right multiplication.} then 
\begin{equation}
\begin{split}
    & \rm{Var}_{\gamma} = \rm{Var}_{\gamma_1} + \rm{Var}_{\gamma_2} + \rm{Var}_{\gamma_2} \cdot i_{\ast} \cdot \rm{Var}_{\gamma_1}, \\
    & h_{\left[ \gamma \right]} = h_{\left[ \gamma_2 \right]} \cdot h_{\left[ \gamma_1 \right]},\\
    & h_{\left[ \gamma \right]}^{(r)} = h_{\left[ \gamma_2 \right] }^{(r)} \cdot h_{\left[ \gamma_1 \right] }^{(r)}.\\
\end{split}
\end{equation}
\nn
Let us suppose all the critical points are non-degenerate, i.e. the Hessian of $f$ at any critical point is non-degenerate, and suppose the corresponding critical values are different:\footnote{This second requirement is not strictly necessary to define a Morse function.} such a function is said to be \textbf{Morse}.\\
\defn[Monodromy group]{
The map
\begin{equation}
\begin{split}
    \pi_1 \left( U \setminus \left\lbrace t_1,\dots,t_\mu \right\rbrace , z_0 \right) \quad &\longrightarrow \quad Aut \left( H_{\bullet} (F_{t_0})\right),\\
    [\gamma] \quad & \longrightarrow h_{\gamma \ast}\\\\
\end{split}
\end{equation}

is called monodromy representation of $\pi_1 \left( U \setminus \left\lbrace t_1,\dots,t_\mu \right\rbrace , t_0 \right)$. The imagine of this map defines what we call the Monodromy group of the Morse function $f$.}
Now, we construct a path $u: \left[0,1 \right] \longrightarrow U$ joining the non-critical value $t_0=u(0) \in \partial U$  to some critical value $t_i = u(1) \in U$ without crossing any other critical value. The Morse lemma tells us that, given a holomorphic Morse function $f$, it always exists a local set of coordinates in a neighbourhood of the non-degenerate critical point $p_i$ such that the function takes the form

\begin{equation}
    f(z_1 ,\dots, z_n) = t_i + \sum_{j=1}^n  z_j^2.
\end{equation}

Then, for each path $u$, we can define a family of $(n-1)-$dimensional spheres in the level manifolds $F_{u(\tau)}$. For each point of the path $u(\tau)$ the level set $F_{u(\tau)}$ is a hyperboloid equivalent to a trivial fibration with base a $(n-1)-$dimensional sphere of radius$\sqrt{|u(\tau)-t_i|}$:
\begin{equation}
    S(\tau)= \sqrt{u(\tau)-t_i} S^{n-1}.
\end{equation}
In particular, we have that the sphere $S(1)$ reduces to the critical point $p_i$. \\

\defn[Picard-Lefschetz vanishing cycle]{
The homology class $\Delta \in H_{n-1} \left( F_{t_0} \right)$ represented by the $(n-1)-$dimensional sphere $S(0)$ in $F_{t_0}$ is called vanishing cycle of Picard-Lefschetz along the path $u$.} 

Note that the homotopy class of $u \in U$ uniquely defines the homology class of the vanishing cycle $\Delta$ modulo orientation. \\[0.5cm]
\defn[Distinguished Basis]{ 
The set of cycles $\Delta_1, \dots, \Delta_{\mu} \in H_{n-1} \left(F_{t_0} \right)$, with $t_0$ non-singular, is called distinguished if:
\begin{itemize}
    \item[(i)] The cycles $\Delta_i$ are vanishing along non-self-intersecting paths $u_i$ reaching the critical values $t_i$;
    \item[(ii)] The unique common point of $u_i$ and $u_j$ for $i \neq j$ is $u_i(0)=u_j(0)=t_0$; 
    \item[(iii)] The paths $u_i, \dots, u_{\mu}$ are numbered in the order in which they enter to the point $t_0$ counting clockwise starting from the boundary $\partial U$ of $U$.
\end{itemize}}

\noindent
\begin{Ese}\label{Esempio1} $f(z)=z^3-3 \lambda z$.\\
Let us consider the Morse function $f(x)=z^3-3 \lambda z$, with $\lambda \in \mathbb{R}_+$, and let us construct a distinguished basis of vanishing cycles. This function is a deformation of the function $f(z)=z^3$ and it has two critical points in the real line 
\begin{equation}
    \mathcal{C}: \quad \bar z_1= \sqrt{\lambda}, \quad \quad \bar z_2 = -\sqrt{\lambda},
\end{equation}
with corresponding critical values are
\begin{equation}
    t_1= - 2 \lambda \sqrt{\lambda}, \quad \quad t_2 = 2 \lambda \sqrt{\lambda}.
\end{equation}
Let us choose as non-critical reference point $t_0=0$ and let us construct the paths $u_1$ and $u_2$ connecting the critical values with $t_0$. The level manifold at $t_0=0$ consists of three points:
\begin{equation}
    F_{t_0}: \quad z^3-3 \lambda z =0 \quad \rightarrow z_1=- \sqrt{3 \lambda}, \quad z_2=0, \quad z_3= \sqrt{3 \lambda}.
\end{equation}
In this example the level manifold for a generic regular point $t$ is given by the condition $f(z)=t$ which admits three point solutions. The vanishing cycles, when we approach the critical values $t_1$ and $t_2$, are the differences
\begin{equation}
    \Delta_1 = \left\lbrace z_3 \right\rbrace - \left\lbrace z_2 \right\rbrace, \quad \quad \quad \Delta_2 = \left\lbrace z_2 \right\rbrace - \left\lbrace z_1 \right\rbrace 
\end{equation}
between the zeroth homology classes represented by the points.\\
Choosing the set $U$ as depicted in Figure \ref{fig4} the cycles $\Delta_1$ and $\Delta_2$ form a distingushed basis for $H_{1} \left(F_{t_0} \right)$.\\[0.5cm]

\begin{figure}[h!]
    \centering    
    \includegraphics[scale=0.15]{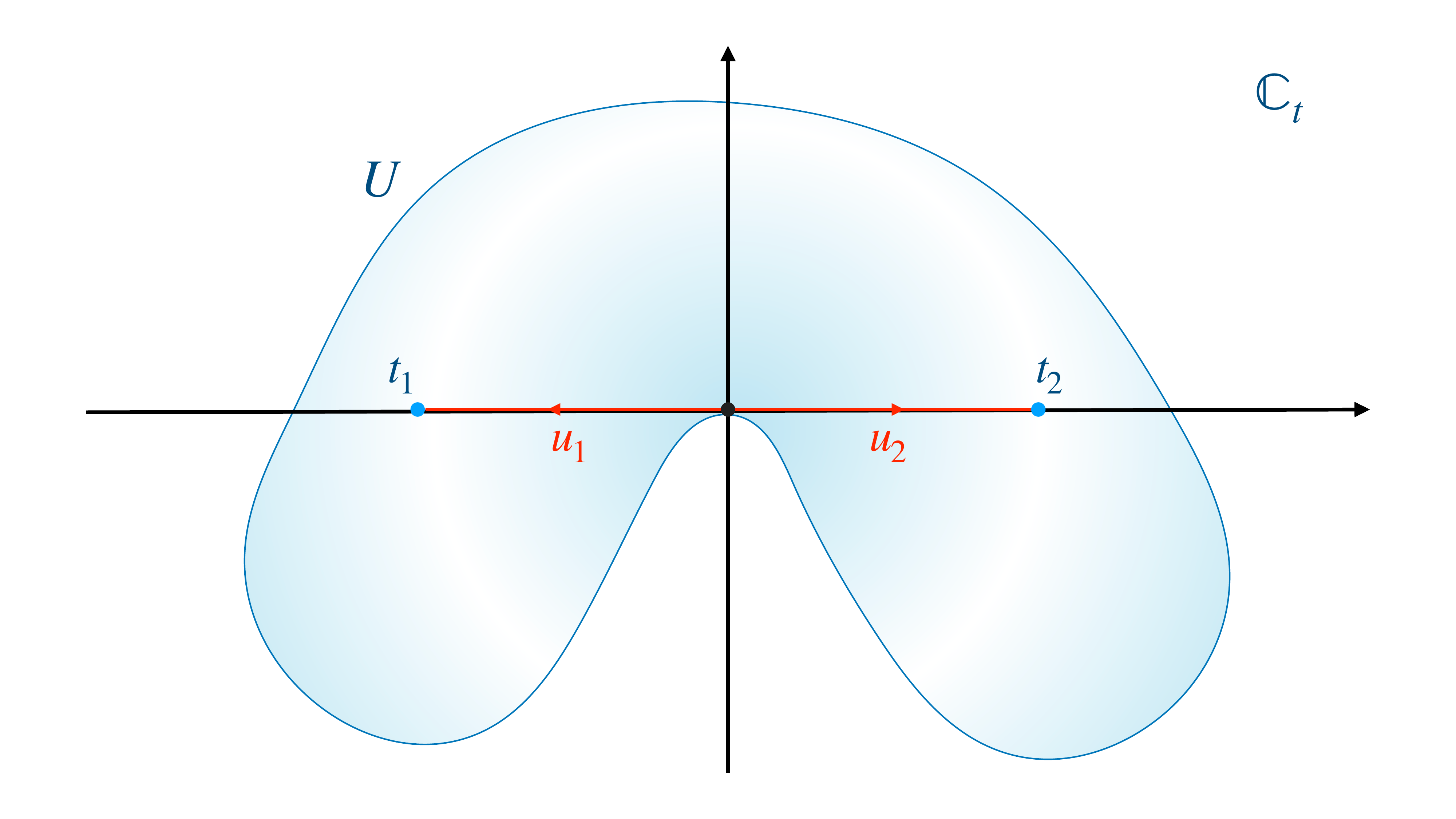}
    \caption{\small{Choice of the set U and the paths $u_1$ and $u_2$ in the codomain of the function $f(z)=z^3-3 \lambda z$.}}
    \label{fig4}
\end{figure}
\end{Ese}

\noindent
\begin{Ese}\label{Esempio2} $f(x,y)=x^3-3 \lambda x + y^2$.\\
In this second example we consider the function of two variables $f(x,y)=x^3-3 \lambda x + y^2$, which is a deformation through the small real parameter $\lambda$ of the function $f(x,y)=x^3+y^2$. The set of critical points in $\mathbb{C}^2$ with their corresponding critical values is
\begin{equation}
    \mathcal{C}: \quad \begin{cases}  \, P_1 \, : \, \, \left( x, y\right) = \left( \sqrt{\lambda} , 0 \right)  \quad &\rightarrow \quad t_1= -2 \lambda \sqrt{\lambda}\\ 
     \, P_2 \, : \, \, \left( x, y\right) = \left( - \sqrt{\lambda} , 0 \right)  \quad & \rightarrow \quad t_2= 2 \lambda \sqrt{\lambda}.\\
    \end{cases}
\end{equation}

 As in the previous example we can consider the paths $u_1$ and $u_2$ joining the two critical values with the non-critical value $t_0=0$. 
The level manifold in this regular point is the graph of the two-valued function $y= \pm \sqrt{x^3-3 \lambda x}$, namely the double-covering of the $x$ complex plane branched between the points $x_1= - \sqrt{3 \lambda}$ and $x_2=0$ and $x_3= \sqrt{3 \lambda}$ and infinity. \\
As we move the value of $t$ from $0$ to one of the two critical values, the level manifold $f(x,y)=t$ is deformed and becomes singular at $t=t_1$ and $t=t_2$. In particular, when we approach $t_1$ we have that the branch point $x_2$ moves until it overlaps $x_3$, while, when we approach $t_2$ the point $x_2$ moves towards the point $x_1$. From this construction, we can draw the vanishing cycles corresponding to the paths $u_1$ and $u_2$: we obtain $\Delta_1$ encircling the points $x_2$ and $x_3$, and $\Delta_2$ encircling the points $x_1$ and $x_2$ (see Figure \ref{fig5}).

\begin{figure}[h!]
    \centering    
    \includegraphics[scale=0.15]{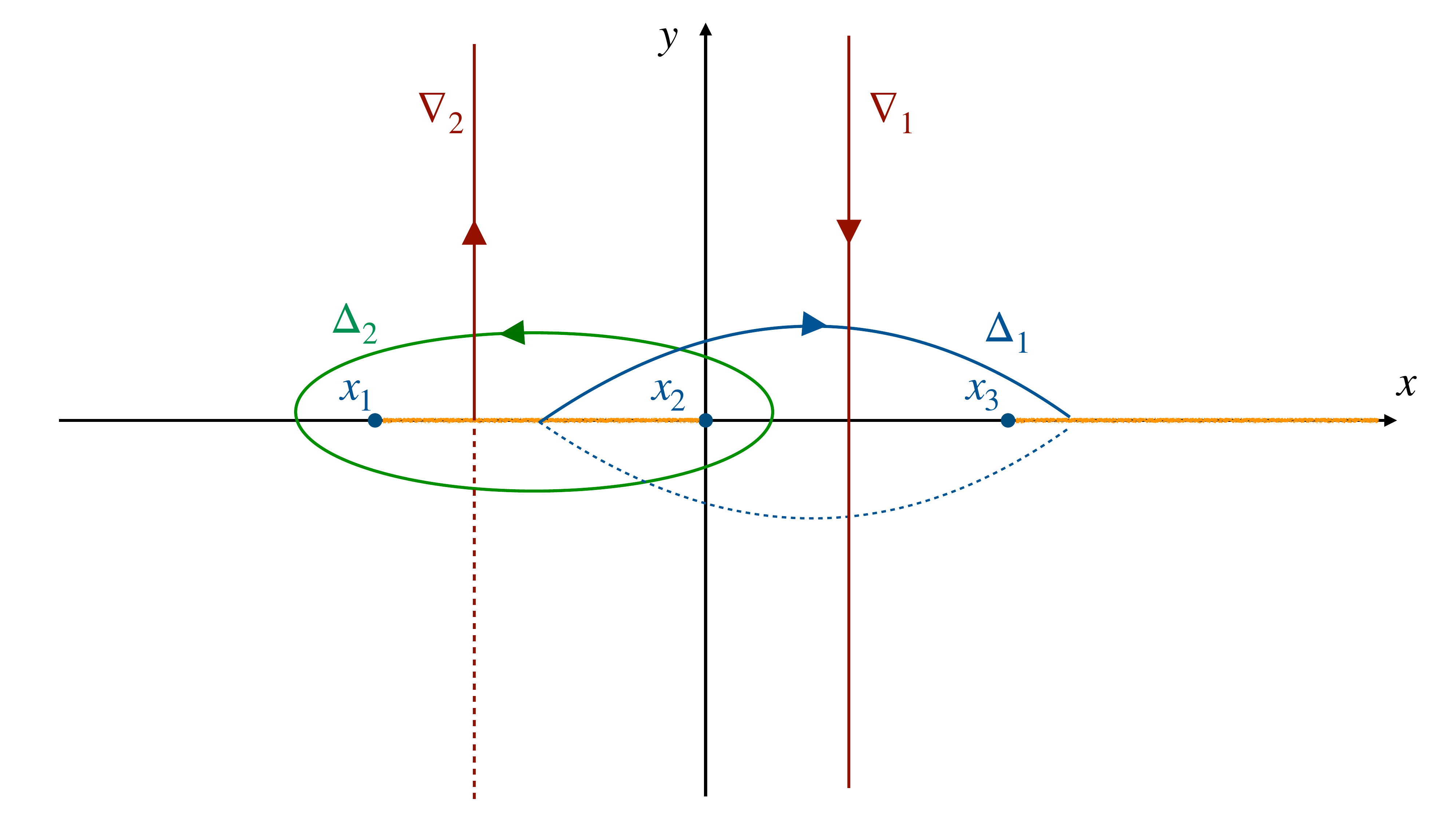}
    \caption{\small{Vanishing and Covanishing cycles.}}
    \label{fig5}
\end{figure}
\end{Ese}
\defn[Simple loops]{A simple loop is an element $\tau_i$ of $\pi_1 \left( U \setminus \left\lbrace t_1,\dots,t_\mu \right\rbrace, t_0 \right)$ represented by the loop going along the path $u_i$ from $t_0$ to $t_i$, then encircling $t_i$ with a anticlockwise path and returning along $u_i$ to $t_0$.}
The region $\left( U \setminus \left\lbrace t_1,\dots,t_\mu \right\rbrace, t_0 \right)$ is homotopically equivalent to a bouquet of $\mu$ circles. Then, the fundamental group $\pi_1 \left( U \setminus \left\lbrace t_1,\dots,t_\mu \right\rbrace, t_0 \right)$ is a free group with $\mu$ generators $\tau_1, \tau_2,\dots,\tau_{\mu}$. \\[0.5cm]
\defn[Weakly distinguished]{The set of vanishing cycles $\Delta_1,\dots,\Delta_{\mu}$ defined by the paths $u_1,\dots,u_{\mu}$ is called weakly distinguished if $\pi_1 \left( U \setminus \left\lbrace t_1,\dots,t_\mu \right\rbrace, t_0 \right)$ is the free group generated by the simple loops $\tau_1,\dots,\tau_{\mu}$ associated to the paths $u_1,\dots,u_{\mu}$.}

We have that if the paths $\left\lbrace u_i \vert i=1,\dots, \mu \right\rbrace$ define a weakly distinguished set of vanishing cycles $\Delta_i$ in the $(n-1)-$homology group of the non-singular level manifold, then, the monodromy group of the function $f$ is generated by the monodromy operators $h_{\tau_i \ast} = h_{\left[ \tau_i \right]}$. Hence, the monodromy group of $f$ is always a group generated by $\mu$ generators.\\
\defn[Picard-Lefschetz operator]{The monodromy operator
\begin{equation}
    h_i =h_{\tau_i \ast} \, : \, \, H_{\bullet} \left( F_{t_0} \right) \, \longrightarrow \, H_{\bullet} \left( F_{t_0} \right) 
    \label{monodromy_ops_simple_loops}
\end{equation}
of the simple loop $\tau_i$ is called the $i^{\rm th}$ Picard-Lefschetz operator.}
In the example \ref{Esempio1}, we can take trace of the change of the position of the three points $z_i$ when we move $t$ along the paths $\tau_1$ and $\tau_2$. We observe that along the path $\tau_1$ the point $z_2$ approaches the point $z_3$, then they make a half-turn around a common centre and move again away one from the other. The point $z_1$ stays fixed. Then, we deduce the following monodromy action on the vanishing cycles:

\begin{equation}
    h_1 \Delta_1 = - \Delta_1, \quad \quad \quad \quad \quad \quad h_1 \Delta_2 = \Delta_1 + \Delta_2.
\end{equation}

In the same way we can deduce

\begin{equation}
    h_2 \Delta_1 = \Delta_1 + \Delta_2, \quad \quad \quad \quad \quad \quad h_2 \Delta_2 = - \Delta_2.    
\end{equation}

The cycles $\Delta_i$ are in the homology group $H_{n-1} \left( F_t \right)$ of the non singular level manifold $F_t$. Moreover, we are interested also in the homology group $H_{n-1} \left( F_t , \partial F_t \right)$, which is dual to the group $H_{n-1} \left( F_t \right)$. In the present case it is generated by two cycles $\nabla_i$ such that
\begin{equation}
    \left( \nabla_i \circ \Delta_j \right) = \delta_{ij}.
\end{equation}
We can choose
\begin{equation}
    \nabla_1 =  \left\lbrace z_3 \right\rbrace \quad , \quad \nabla_2 = - \left\lbrace z_1 \right\rbrace,
\end{equation}

for which we have the following variations:

\begin{equation}
    \begin{split}
         &\rm{Var}_{\tau_1} \nabla_1 = \left\lbrace z_2 \right\rbrace - \left\lbrace z_3 \right\rbrace = - \Delta_1,\\ &\rm{Var}_{\tau_1} \nabla_2 = 0, \\
         &\rm{Var}_{\tau_2} \nabla_1 = 0,\\ &\rm{Var}_{\tau_2} \nabla_2 = - \left\lbrace z_2 \right\rbrace + \left\lbrace z_1 \right\rbrace = - \Delta_2. \\        
    \end{split}
\end{equation}
We can now consider the loop $\tau = \tau_2 \tau_1$ that turns around the point $t_0=0$ encircling the two critical values $t_1$ and $t_2$ in a positive counterclockwise direction. The monodromy transformation associated to this loop permutes the points $z_3 \longrightarrow z_2 \longrightarrow z_1 \longrightarrow z_3$, then,
\begin{equation}
    \begin{split}
        & h_{\tau} \Delta_1 = \left\lbrace z_2 \right\rbrace -\left\lbrace z_1 \right\rbrace = \Delta_2, \\
        & h_{\tau} \Delta_2 = \left\lbrace z_1 \right\rbrace -\left\lbrace z_3 \right\rbrace =- \Delta_1 - \Delta_2, \\
    \end{split}
\end{equation}
and
\begin{equation}
    \begin{split}
        & \rm{Var}_{\tau} \nabla_1 = \left\lbrace z_2 \right\rbrace -\left\lbrace z_3 \right\rbrace = - \Delta_1,\\
        & \rm{Var}_{\tau} \nabla_2 = - \left\lbrace z_3 \right\rbrace +\left\lbrace z_1 \right\rbrace = - \Delta_1 - \Delta_2.
    \end{split}
\end{equation}

The monodromy group of the Morse function $f$ is generated by the Picard-Lefschetz operators $h_{\tau_1}$ and $h_{\tau_2}$. All the elements of this group preserve the intersection product of the group $H_0 \left( F_t \right)$, for $t$ non-critical, generated by the vanishing cycles $\Delta_i$. The monodromy group is the group $S_3$ of permutations of three elements.\\
Now, let us construct the monodromy group for the example \ref{Esempio2}. Drawing the analogous of the Figure \ref{fig3}, we can deduce the action of the Picard-Lefschetz operator on the vanishing cycles to be
\begin{equation}
    \begin{split}
       &h_1 \Delta_1 = \Delta_1,\\ &h_1 \Delta_2 = \Delta_1 + \Delta_2, \\
       &h_2 \Delta_1 = \Delta_1 - \Delta_2 ,\\ &h_2 \Delta_2 = \Delta_2,
    \end{split}
\end{equation}
and the following variation on the dual cycles:
\begin{equation}
    \begin{split}
        &\rm{Var}_{\tau_1} \nabla_1 = - \Delta_1 ,\\  &\rm{Var}_{\tau_1} \nabla_2 = 0, \\
        &\rm{Var}_{\tau_2} \nabla_1 =0,\\
        &\rm{Var}_{\tau_2} \nabla_2 =- \Delta_2.
    \end{split}
\end{equation}
The monodromy group of the Morse function $f(x, y)$ is isomorphic to the group of non-singular $2 \times 2$ integer matrices with determinant $1$. The group is generated by the action of the Picard-Lefschetz operators on the vanishing cycles, given by
\begin{equation}
    M_1 = \left( \begin{matrix} 1 & 0 \\ 1 & 1 \end{matrix} \right)
    \quad \quad \mbox{and}\quad \quad M_2 = \left( \begin{matrix} 1 & -1 \\ 0 & 1 \end{matrix} \right).
\end{equation}

These methods, explicitly shown in one or two complex dimensions, can in principle be extended to higher dimensions to determine vanishing cycles, their duals, and the action of the monodromy group on them. 

\subsection{Picard-Lefschetz Theorem}
The Picard-Lefschetz theorem establishes a relation between the variation of (co)-vanishing cycles due to the action of the monodromy operator with their intersection product in $H_{n-1} \times H_{n-1}^{\vee} \longrightarrow \mathbb{Z}$.\\
Let us start considering the simple loop $\tau_i$ associated with the path $u_i$ connecting  the non-critical reference point $t_0 \in \mathbb{C}_t$ to the critical value $t_i \in \mathbb{C}_t$. Let us assume that the critical value is $t_i=0$, so that in some local coordinates around the critical point $P_i \in \mathbb{C}^{n}$, we can write the function $f$ in the form
\begin{equation}
    f(z_1, \dots, z_n) = \sum_{j=1}^n z_j^2.
\end{equation}
If we intersect $f^{-1} (t_0)$ with the ball $\sum_j \vert z_j \vert^2 \leq 4 \epsilon^2$, the non-critical value $t_0$ is sufficiently close to the critical value $0$, say $\vert t_0 \vert = \epsilon^2$. We can suppose that all other critical values of $f$ are outside the disk of radius $4 \epsilon^2$ in $\mathbb{C}_t$, so that our simple loop encircles just one singularity.\\
Let us define the ball $\overline{B}_{2\epsilon}$ of radius $2\epsilon$ in the space $\mathbb{C}^n$,
\begin{equation}
    \overline{B}_{2\epsilon} = \left\lbrace \left( z_1 , \dots , z_n \right) \vert r \leq 2\epsilon \right\rbrace,
\end{equation}
and let us call $\tilde{F}_t$ the intersection of the level set $F_t$ with this ball. \\
\lem[Transversality]{For $\vert t \vert < 4\epsilon^2$, the level set $F_t$ is transverse to the $(2n-1)$-dimensional sphere $\partial \overline{B}_{2\epsilon}$.}

From this lemma it follows that for $0 < \vert t \vert <4\epsilon^2$ the sets $\tilde{F}_t = F_t \cap \overline{B}_{2\epsilon}$ are diffeomorphic manifolds with boundary, while $\tilde{F}_0$ is a cone with vertex in zero.\\

\lem[Bouquet of Spheres]{For $0 < \vert t \vert < 4 \epsilon^2$, the manifold $\tilde{F}_t$ is diffeomorphic to the disk sub-bundle of the tangent bundle of the standard $(n-1)$ dimensional sphere $S^{n-1}$.}
From this second lemma follows the following result:\\

\lem[Self-intersection]{The self-intersection number of vanishing cycle $\Delta$ in the complex manifold $\tilde{F}_{\epsilon^2}$ is equal to}
\begin{equation}
\left( \Delta \circ \Delta \right) = (-1)^{(n-1)(n-2)/2} \left( 1 + (-1)^{n-1}\right) \, = \, \begin{cases}
    0 \quad & \text{for} \quad n=0 \, \mathrm{mod} \, 2, \\
    +2 \quad & \text{for} \quad n=1 \, \mathrm{mod} \, 4, \\
    -2 \quad & \text{for} \quad n=3 \, \mathrm{mod} \, 4. \\
\end{cases}
\label{SelfIntersection1}
\end{equation}

Poincar\'e duality \ref{PoincarèDuality} for a compact manifold $X$ of dimension $n$ states that $H^k(X)\sim H_{n-k}(X)$. If $X$ is noncompact, while for cohomology it is not a problem, for homology one has to introduce Borel-Moore homology for which one has $H^k(X)\sim H^{BM}_{n-k}(X)$, see \cite{Bredon}. Hence, in our case, we get $H_{k} \left(  \tilde{F}_{\epsilon^2}, \partial \tilde{F}_{\epsilon^2} \right)\sim H^{BM}_{n-k} \left(  \tilde{F}_{\epsilon^2} \right)$, and $H_{k} \left(  \tilde{F}_{\epsilon^2}, \partial \tilde{F}_{\epsilon^2} \right)\sim H^{k} \left(  \tilde{F}_{\epsilon^2} \right)\sim H^k(S^{n-1})$.
Therefore, the relative homology group $H_k  (\tilde{F}_{\epsilon^2} , \partial \tilde{F}_{\epsilon^2})$ is zero for $k \neq n-1$, while $H_{n-1}(\tilde{F}_{\epsilon^2} , \partial \tilde{F}_{\epsilon^2})$ is isomorphic to the $\mathbb{Z}$. Moreover the latter is generated by the relative cycle $\nabla$ dual to $\Delta$ such that $\Delta \circ \nabla =1$. \\
In general, a relative cycle $\delta \in H_k\left( F_{\epsilon^2}, \partial F_{\epsilon^2}\right)$ can be represented in the form
\begin{equation}
    \delta = \delta_1 + \delta_2
\end{equation}
\nn
where $\delta_1 \in H_k \left( \tilde{F}_{\epsilon^2}, \partial \tilde{F}_{\epsilon^2} \right)$ and $\delta_2$ is a chain in $F_{\epsilon^2} \setminus B_{2\epsilon}$. The transformation $h_{\tau}=\Gamma_1$ is the identity in $F_{\epsilon^2} \setminus B_{2\epsilon}$, hence, it acts non-trivially only on the cycle $\delta_1$. Therfore, $\rm{Var}_{\tau} (\delta) = \rm{Var}_{\tau} (\delta_1)$. \\
Since $H_k \left( \tilde{F}_{\epsilon^2} , \partial \tilde{F}_{\epsilon^2} \right)= \langle \nabla \rangle \simeq \mathbb{Z} $, then $\delta_1 = m \cdot \nabla$, with $m \in \mathbb{Z}$ and $m= \delta \circ \Delta $, and, in order to compute the action of the variation operator on $H_k\left( \tilde{F}_{\epsilon^2} , \partial \tilde{F}_{\epsilon^2} \right) $, it is sufficient to calculate its action on $\nabla$.\\[0.5cm]
\theo[Picard-Lefschetz] {Under the above hypotheses
\begin{equation}
    \rm{Var}_{\tau} \left( \nabla \right) = (-1)^{n(n+1)/2} \Delta.
\end{equation}}
\nn
It follows that for $a \in H \left( F_{t_0}, \partial F_{t_0}\right)$:
\begin{align}
    \rm{Var}_{\tau} (a) &= (-1)^{n(n+1)/2} \left( a \circ \Delta \right) \Delta,\\
    h^{(r)}_{\tau} (a) &= a+(-1)^{n(n+1)/2} \left( a \circ \Delta \right) i_{\ast}\Delta;
\end{align}
and for $a \in H_{n-1} \left( F_{t_0}\right)$
\begin{equation}
    h_{\tau} (a) = a+(-1)^{n(n+1)/2} \left( a \circ \Delta \right) \Delta,
    \label{PicardLefschetzFormula}
\end{equation}
\textit{where $i_{\ast}$ is the homomorphism \eqref{map1}.}\\\\

\begin{Ese}\label{A3}
In this example we compute the monodromy matrices and intersection numbers, needed in section \ref{PearceyIntegrals} for the analysis of the Pearcey integral.\\
Let $\Delta_1=\{z_3\}-\{z_4\}$ be the vanishing cycle associated to the singular point $\sigma_1$, meaning the two points $\{z_3\}$ and $\{z_4\}$ coalesce when approaching to $\sigma_1$. The self intersection number

\begin{equation}
    \Delta_1 \circ \Delta_1 = 2, 
\end{equation}
\nn
can be easily computed by setting $n=1$ in equation \eqref{SelfIntersection1}.\\
For the sake of clarity, being careful not to make confusion among vanishing and co-vanishing cycles, is useful here to re-wright \eqref{PicardLefschetzFormula} with adapted notation:

\begin{equation}
     h_{j} (\Delta_i) = \Delta_i+(-1)^{n(n+1)/2} \left( \Delta_i \circ \Delta_j \right) \Delta_j,
     \label{PicardLefschetzFormula2}
\end{equation}
\nn
We can now apply \eqref{PicardLefschetzFormula2}, for evaluating the monodromy action on $\Delta_1$ when going around the singular point $\sigma_1$:

\begin{equation}
    h_1(\Delta_1)=\Delta_1-(\Delta_1\circ \Delta_1)\Delta_1=- \Delta_1.
    \label{Delta1}
\end{equation}
\nn
The transformed vanishing cycle

\begin{equation}
     \Delta_1^{(1)}\equiv h_1(\Delta_1)=\{z_4\}-\{z_3\}.
\end{equation}
\nn
reveals the flipping $\{z_3\}\leftrightarrow\{z_4\}$. We can now use this information to determine the action on the co-vanishing cycle $\Delta_2=\{z_1\}-\{z_4\}$:

\begin{equation}
    \Delta_2^{(1)}=\Delta_2(\{z_3\}\leftrightarrow\{z_4\})=\{z_1\}-\{z_3\}.
\end{equation}
In order to express $\Delta_2^{(1)}$ in terms of $\Delta_1$ and $ \Delta_2$, we use the trick to insert $\{z_4\}-\{z_4\}=0$ in the previous equation, getting:

\begin{equation}
    \Delta_2^{(1)}=\{z_1\}+(\{z_4\}-\{z_4\})-\{z_3\} = (\{z_1\}-(\{z_4\})+(\{z_4\})-\{z_3\})=\Delta_2-\Delta_1.
    \label{Delta2}
\end{equation}
\nn
Proceeding analogously for $\Delta_3$, we get

\begin{equation}
    \Delta_3^{(1)}=\Delta_3-\Delta_1.
    \label{Delta3}
\end{equation}
\nn
Packaging relations \eqref{Delta1},\eqref{Delta2} and \eqref{Delta3} in the representation of the monodromy, around $\sigma_1$, acting on the basis of vanishing cycles, i.e.

\begin{equation}
    \Delta_i^{(1)}=(M_1)_{ij}\Delta_j,
\end{equation}
\nn
we get the first of the matrices reported in \eqref{MonodromyMatricesPositiveDet}:

\begin{equation}
    M_1=\begin{pmatrix}-1&0&0\\-1&1&0\\-1&0&1\end{pmatrix}.
\end{equation}
\nn
In order to obtain $M_2$ and $M_3$ we simply repeat the very same procedure considering the singular points $\sigma_2$ and $\sigma_3$ cases, respectively.\\
Finally, we can use monodromy matrices to extract the intersection numbers among vanishing cycle. Inverting \eqref{PicardLefschetzFormula2}, we get 

\begin{equation}
    \Delta_i\circ \Delta_j= [\Delta_i-h_j(\Delta_i)]\Delta_j^{-1}.
\end{equation}
\nn
By contruction, we have $h_j(\Delta_i)={(M_j)}_{ik}\Delta_k$, so when can rewrite Picard-Lefschetz formula with an explicit dependence on the monodromy matrices as

\begin{equation}
    \Delta_i\circ \Delta_j= [\Delta_i-{(M_j)}_{ik}\Delta_k]\Delta_j^{-1}.
    \label{ExplicitPicardLef}
\end{equation}
\nn
Let us apply \eqref{ExplicitPicardLef} to compute, for instance, the intersections 
\begin{equation}
    \begin{split}\Delta_1\circ \Delta_2&=[\Delta_1-(M_2)_{1k}\Delta_k]\Delta_2^{-1}\\&=[\Delta_1-\Delta_1+\Delta_2]\Delta_2^{-1}=1, \end{split}
\end{equation}
\nn
and 
\begin{equation}
    \begin{split}\Delta_2\circ \Delta_1&=[\Delta_2-(M_1)_{2k}\Delta_k]\Delta_1^{-1}\\&=[\Delta_2+\Delta_1-\Delta_2]\Delta_1^{-1}=1, \end{split}
\end{equation}
\nn
revealing the intersection product is symmetric, as expected for even $n-1$.\end{Ese}

%% file: Amplitudes.tex
\chapter{Perturbative theory via Intersection theory}\label{C2}
\thispagestyle{empty} 
Given a classical model described by fields $\bm{\phi}$ with action $S[\bm{\phi}]$, the behavior of the corresponding quantum model is encoded in the partition function
\begin{equation}
    Z=\int \mathcal{D}\bm{\phi}\, e^{i S[\bm{\phi}]}.
    \label{partitionfunction}
\end{equation}
Because of Haag's theorem, the only well-defined physical states of a quantum theory are the free (non-interacting) states. The best we can do, for a interacting theory, is thus to define the states in the asympotic regions where the coupling tends to vanish and to study the interaction by observing the evolution of a state from an asymptotic region to another, that is via a scattering process. In practice, what one does is to fix the initial (incoming) and final (outgoing) states, and to compute the probability that such evolution occurs, encoded in the \textbf{Scattering matrix}. For a scattering process involving $n$ states, the scattering matrix can be shown to be related to the so called $n-$ points Green function:

\begin{equation}
    G^{(n)}\equiv Z^{-1}\int  \mathcal{D}\bm{\phi}\,\phi_1\dots \phi_n \,e^{i S[\bm{\phi}]}.
    \label{npointsgreen}
\end{equation}

If we knew a general closed-form solution for path integral computation, this chapter could end here. Unfortunately, we don't; even worse, we can compute it only in one case: when $S$ is at most quadratic in $\phi$ and the path integral is gaussian. Except for the free theory, only in some very special case, with extra structure, this is enough to fully solve the theory; examples include certain topological field theories (e.g. Chern-Simons), integral QFTs and statitistical models in low dimension (e.g. sine-Gordon, Liouville),  theories amenable to supersymmetric localization (e.g. $N=2$ SUSY on $S^4$). In generic quantum field theories, statistical systems and string theory, exact path-integral evaluation is not available; one must therefore resort to non-perturbative techniques, as instanton calculus, Lattice field theory and mean-field methods, or approximation techniques, expanding  either around the classical solution (Saddle-point) in powers of $\hbar$, or around the free theory in powers of the coupling constant: this is \textbf{perturbative theory}.\\
In order to perturbatively expand the $n-$ points Green's functions, considering only one scalar field for the sake of simplicity, it is convenient to slightly modify the partition function \eqref{partitionfunction} by adding an external auxiliary source $J$, defining the partition functional\footnote{By $\langle -\rangle$ we denote space-time integration.}

\begin{equation}
    Z[J]=\int \mathcal{D}\phi\, e^{i \left ( S[\phi]+\langle J\phi\rangle\right )},
    \label{partitionfunction2}
\end{equation}
with respect to which, the $n-$ points Green's functions can be written as 
\begin{equation}
    G^{(n)}= (-i)^n\frac{\delta^n}{\delta J_1\dots\delta J_n} Z[J]\Big |_{J=0}.
    \label{npointsgreen2}
\end{equation}
Let us now split the action as $S=S_0+S_I$, with $S_0$ the free quadratic part and $S_I$ the interacting one. Denoting by $Z_0[J]$ the partition functional of the free theory, it is quite easy to show that one can rewrite the full partition functional as 

\begin{equation}
\begin{split}
    Z[J]&=e^{iS_I\left [-i\frac{\delta}{\delta J}\right ]}Z_0[J]\\&=\sum_{k=0}^\infty \frac{1}{k!}(i)^k \left ( S_I\left [-i\frac{\delta}{\delta J}\right ]\right )^k Z_0[J],
    \end{split}
\end{equation}
obtaining an expansion in terms of a gaussian partition function, whose standard solution, encoding Wick's theorem, is written in terms of the inverse of the quadratic operator, called \textbf{Feynman propagator} $\Delta_F$, as the sum over all possible pairings $P$ (contraction):

\begin{equation}
    G^n(x_1,\dots,x_n)_0=\sum_{P} \prod_{\{i,j\}\in P} \Delta_F(x_i-x_j).
\end{equation}

Remarkably, all order of the expansion can be systematically written by the two-point propagator, the interaction vertex and suitable prescriptions (\textbf{Feynman rules}); this leads to \textbf{Feynman diagrams}. \\
The story is a little more involving in the case of string theory, because interaction cannot be switch off and a free theory cannot be isolated: the interaction coupling constant is an internal variable arising from dynamic, precisely the vacuum expectation value of a particular string state, the dilaton; as a consequence string theory is intrinsically interacting and free string theory is meaningless. Moreover, conformal invariance constrains strings to be on-shell, then off-shell amplitudes are not well-defined.\footnote{See \cite{Ashoke} for a review}.\\
However, the advantage of conformal invariance allows to perturbatively treat the partition function associated to the Polyakov action and to express the bosonic $n-$string amplitude as a topological expansion in the worldsheet genus $g$ as:

\begin{equation}
\A_n= \sum_g \int  DX Dh e^{-S_p[X,h,\lambda]} \prod_{i=1}^n V_i =\sum_g e^{-\lambda \chi(g)}\int  DX Dh e^{-S_p[X,h]} \prod_{i=1}^n V_i,
\label{stringAmplitude}
\end{equation}
where the $V_i$, called vertex operators, encode the Lorentz degree of freedom of the external strings and depend on their energy state. After quantization, each term of the above expansion, in the closed string case, can be written as an integral over the $(3g-3 +n)-$dimensional moduli space $\M_{g,n}$ of genus $g$ Riemann surfaces with $n$ punctures as

\begin{equation}
\A_n^{closed}\sim\int_{\M_{g,n}} d\lambda_i\int_{\Sigma^n}\prod_{i=1}^nd^2z_i \prod_{i<j=1}^n |z_{ij}|^{2\alpha '\, k_ik_j} |F(z_{ij},k)|^2,
\label{stringAmplitudeCLOSED}
\end{equation}
where $F$ is a rational function and $z_{ij}\equiv z_i-z_j$.\\
A similar expression holds for a partial open amplitude, one finds:

\begin{equation}
\A_n^{open}(\sigma)\sim\int_{\M^o_{g,n}} d\lambda_i\int_{D(\sigma)}\prod_{i=1}^n\,dx_i \prod_{i<j=1}^n |x_{ij}|^{\alpha '\, k_ik_j} F(x_{ij},k),
\label{stringAmplitudeOPEN}
\end{equation}
with $D(\sigma)$ the oriented punctured boundary of the open worldsheet and $M^o_{g,n}$ the corresponding moduli space. 

 \begin{figure}[h!]
\centering
\begin{tikzpicture}
\amplitudeone{(0,0)}
\end{tikzpicture}
\caption{Diagrammatic expansion of four closed string scattering.}
\label{amplitudepicture1}
 \end{figure}
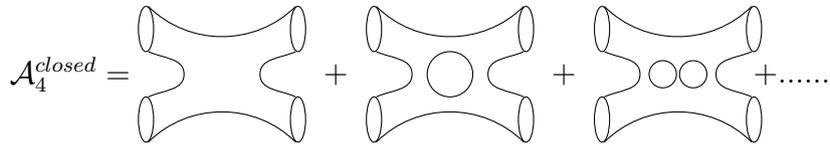

\section{Feynman Integrals}
Despite manifest locality and causality, the configuration space quantities arising from \eqref{npointsgreen2}, are very complicated convolutions of Bessel functions, whose evaluation is not known in the general case. In curved-space quantum field theories \cite{Cacciatori:banana}, and more generally whenever translation invariance is broken, global Fourier transform is not available and configuration-space methods become indispensable; however, when this is not the case, it is convenient to work in momentum space, where the Feynman propagator is a rational function. We define the momentum space $n-$\textbf{scattering amplitude} as the transform $n-$point Green function with all external states information cut out:\footnote{By "amplitude" we will also refer to the full quantity, with external states included.}

\begin{equation}
    i\mathcal{A}\equiv \tilde{G}_2(p_1)^{-1}\dots \tilde{G}_2(p_n)^{-1}\tilde{G}_n^c(p_1\dots p_n).
\end{equation}

So far, our discussion has been limited to scalar fields; however, realistic models require amplitudes to transform non-trivially under the symmetries of the theory, in order to capture the full set of degrees of freedom (i.e., spin, colour). Thus, the amplitude carries a multi-index in the appropriate representation (to be contracted with external states). Remarkably, all the extra structure can always be isolated by means of suitable projection techniques (tensor reduction) and all the topology information can be packaged into some scalar quantities $F_i$, called \textbf{form factors}:

\begin{equation}
    \mathcal{A}^{\alpha}= \mathcal{M}_i^{\alpha}F_i.
\end{equation}
Each form factor turns to be a linear combination of scalar multi-loop integral, called \textbf{Feynman integrals}(FIs).
These integrals are often ill-defined because of divergences arising for large (ultraviolet) or small (infrared) momentum. The latter are controlled by Kinoshita-Lee-Neuemberg theorem, ensuring they cancel when summed over all degenerate physical states while; on the other hand, in order to govern ultraviolet divergences, one identifies the analytic dependence of the divergence in term of auxiliary parameters (regularization) and then redefine the fields and the coupling constant (order by order) in such a way to absorb the poles (renormalization). The most standard regularization used in perturbative calculus is dimensional regularization, consisting in slightly shifting the space-time dimension away from the integer value $\lfloor D\rfloor$ by setting $D=\lfloor D\rfloor+k\varepsilon$.\\
A generic $l-$loop Feynman integral in dimensional regularization, associate to a diagram with $N$ external legs, has momentum space representation\footnote{Here: $\gamma_E$ is Euler-Mascheroni constant, $k=-2$ and $\mu$ is mass scale parameter}: 

\begin{equation}
    \mathcal{I}_{\nu_1\dots\nu_n} = e^{l\varepsilon\gamma_E}\left( \mu^2 \right)^{\nu-lD/2} \int \prod_{i=1}^l\frac{d^Dk}{i\pi^{D/2}}\frac{1}{D_1^{\nu_1}\dots D_n^{\nu_n}}, 
    \label{Qrep}
\end{equation}

with $\nu_i\in \Z$, $n=l(l+1)/2+lN$ and where

\begin{equation}
    D_j = q_j^2-m_j^2,
\end{equation}
with $q_j=q_j(k_i,p_k)$ a linear function of loop momenta $k_i$ and external momenta $p_k$. \\
Depending on the proprieties one wants manifestly to appear and the analysis one wants to carry on, the momentum space representation is often not the most suitable. Via some manipulations applied on \eqref{Qrep} one can recover Feynman integrals expressed in different representations. 
The detailed of the various derivations can be found in any text on the topic (see \cite{Weinzierl:2022eaz} for example), we just list here the essential information.
By using the following identity 

\begin{equation}
    \label{Schwinger_trick}
 \frac{1}{A^\nu} = \frac{1}{\Gamma\left(\nu\right)} \int\limits_0^\infty d\alpha \, \alpha^{\nu-1} \, e^{-\alpha A},
\end{equation}
called Schwinger trick, one menages to rewrite \eqref{Qrep} in the so called \textbf{Schwinger representation}:

\begin{equation}
 \mathcal{I}_{\nu_1\dots\nu_n} =
 \frac{e^{l \varepsilon \gamma_E}}{\prod\limits_{j=1}^{n}\Gamma(\nu_j)}
 \int\limits_{\alpha_j \ge 0}  d^{n}\alpha
 \left( \prod\limits_{j=1}^{n} \alpha_j^{\nu_j-1} \right)
 \left[ {\mathcal U}\left(\alpha\right) \right]^{-\frac{D}{2}}
 \exp\left( - \frac{{\mathcal F}\left(\alpha\right)}{{\mathcal U}\left(\alpha\right)}\right), 
 \label{SchwingerRep}
\end{equation}
with $\U$ and $\F$, called first and second \textbf{Symanzik polynomials} (or graph polynomials) degree $l$ and $l+1$ (respectively) homogeneous polynomials, given by 

\begin{equation}
    \U=\det (Q) \quad \quad \mbox{and}\quad\quad \F=\mu^{-2}\det (Q)(J+\boldsymbol{\nu}^TQ^{-1}\boldsymbol{\nu}),
\end{equation}
where $Q$ and $J$ are defined by

\begin{equation}
    \sum_{j=1}^n\alpha_jD_j=\boldsymbol{k}^TQ\boldsymbol{k}+2\boldsymbol{k}\boldsymbol{\nu}+J.
\end{equation}
Expression \eqref{SchwingerRep} can be further manipulated: by inserting the identity 

\begin{equation}
    \int_0^\infty dt\delta\left(t-\sum \alpha_j\right )=1,
\end{equation}
and making use of graph polynomials homogeneity, one can rewrite it as a projective integral in the homogeneous coordinates $x_i$ of $\Pro^{n-1}$, obtaining the so called \textbf{Symanzik representation}, also named (Projective) Feynman parameter representation:

\begin{equation}
    \mathcal{I}_{\nu_1\dots\nu_n}=\frac{e^{l \varepsilon \gamma_E}\Gamma\left(\nu-\frac{l D}{2}\right)}{\prod\limits_{j=1}^{\n}\Gamma(\nu_j)} \int_\Delta \prod_ix_i^{\nu_i-1}\frac{\mathcal{U}(x_i)^{\nu-(l+1)\frac{D}{2}}}{\mathcal{F}(x_i)^{\nu-l\frac{D}{2}}}\mu,
    \label{SimanzikRep}
\end{equation}
where $\mu$ is the measure of $\Pro^{n-1}$
\begin{equation}
    \mu = \sum_{i=1}^n (-1)^{i-1} x_i dx_1\wedge...\wedge \hat{dx_i}\wedge ... \wedge dx_n
\end{equation}
and the integration domain is the simplex

\begin{equation}
    \mathbb{P}^{n-1} \supset \Delta = \{[x_1:x_2:...:x_{n}]\in \C\mathbb{P}^{n} |x_i\geq 0\}.
\end{equation}

Consider now a diagram where the number of internal legs is related to the number of loops by

\begin{equation}
    n=\frac{1}{2}l(l+1)+ el,
\end{equation}
with $e=\mathrm{dim}(\mathrm{span}\{p_1,\dots,p_N\})$.\footnote{Notice $e=n-1$ for generic momenta and $D\geq N-1$.} \\
In the momentum representation integral of such a diagram, the number of integration variables equals the number of irreducible scalar products $s_j$, namely all possible independent Lorentz invariants built up using loop and external momenta. 
Furthermore, suppose any internal inverse propagator can be written as a linear combination of such invariants; that is, it exist an invertible matrix $C$ such that $D_i=C_{ij}s_j+c_i$.\\
In this case, it is possible to use the denominators of \eqref{Qrep} as new integration variables, setting $z_i=D_i$, and to express the integral in the so called \textbf{standard Baikov representation} \cite{Baikov96-1,Baikov96-2,Baikov05}(see also \cite{Grozin:2011mt} for a review):

\begin{equation}
    \mathcal{I}_{\nu_1\dots\nu_n}= \frac{e^{l \varepsilon \gamma_E} \left(\mu^2\right)^{\nu-\frac{l D}{2}} \left[ \det G\left(p_1,...,p_{e}\right)\right]^{\frac{-D+{e}+1}{2}}}{\pi^{\frac{1}{2}\left(n-l\right)} \left( \det C \right) \prod\limits_{j=1}^{l} \Gamma\left(\frac{D-e+1-j}{2}\right)}
 \int\limits_{\mathcal C} d^{n}z \,
 \left[{\mathcal B}\left(\z\right)\right]^{\frac{D-l-{e}-1}{2}}\prod\limits_{i=1}^{n} z_s^{-\nu_i};
 \label{BaikovRep-1}
\end{equation}
where $G(p_1,\dots,p_n)=\det(p_i\cdot p_j)$ denotes the Gram determinant, the \textbf{Baikov polynomial} $\B(\z)$ is given by 

\begin{equation}
    \B(\z)=G(k_1,\dots,k_l,p_1,\dots,p_e),
\end{equation}
and the integration domain is $\mathcal{C}\equiv \cap \mathcal{C}_j$ with

\begin{equation}
    \mathcal{C}_j\equiv \left\{\z \in \R^n \big | \frac{det G(k_j,\dots,k_l,p_1,\dots,p_e)}{det G(k_{j+1},\dots,k_l,p_1,\dots,p_e)}\geq 0\right\}.
\end{equation}
From now on, we will rewrite \eqref{BaikovRep-1}, in the compact form:

\begin{equation}
 \mathcal{I}_{\nu_1\dots\nu_n}\equiv K\int \B^{\gamma}\frac{d^nz}{z^{\nu_i}}.
 \label{BaikovRep}
\end{equation}
A few comments are in order. 
Despite the very strict conditions required for the Baikov change of variables (and its representation's initial appearance as applicable only to a specialized subset of Feynman diagrams), it can be shown that any non-conforming graph is a subgraph of one that does satisfy the conditions, allowing it to inherit a Baikov representation from the larger graph.
While in most parametric representations of Feynman integrals, the number of integration variables equals the number of propagators, in the (standard) Baikov representation the number of variables grows quadratically with the number of loops, making it widely inefficient for high loop calculations. The attempt to reduce the variable count led to a modified version called loop-by-loop Baikov representation \cite{Frellesvig:2017aai,Frellesvig:2021vem}.\\
A \texttt{Mathematica} package, named \texttt{BaikovPackage}, able to generate both standard and Loop-by-loop Baikov representations has been presented by Frellesvig in \cite{Frellesvig:2024ymq}.

\paragraph{Unitariry cuts.}\mbox{}\\ 
The unitarity of the $S-$matrix, $S^\dagger S=1$, constrains the analytic structure of scattering amplitudes, tying together their imaginary part and the sums over intermediate on-shell states. The simplest and best known implication, concerning a scattering with one initial particle, is the \textbf{optical theorem}, forcing the imaginary part of a forward elastic scattering amplitude to equate the total total transition rate (total cross section) into all possible final states. In the complex energy plane this statement is equivalent to relating the discontinuity of the amplitude across its physical branch cut to the sum over on-shell intermediate contributions.
Cutkosky \cite{Cutkosky60} proved a general diagrammatic prescription (\textbf{Cutkosky rules}) to compute the discontinuity of a perturbative Feynman amplitude: given a Feynman diagram, consider every possible cut that separates the diagram into two parts, called \textbf{unitarity cuts}, and places a subset of internal lines on shell. Each cut line with momentum $p$ and mass $m$ is replaced in the integrand by 

\begin{equation}
    \frac{1}{q^2-m^2+i0}\longrightarrow -2\pi i\delta_+(q^2-m^2),
\end{equation}
enforcing the positive-energy on-shell condition. One then recovers the imaginary part by multiplying the left and right sub-amplitudes, integrating over the on-shell phase space of the cut lines, and summing over all allowed cuts.\\
This idea can been further extended beyond the Cutkosky sense, allowing for cuts not corresponding to physical discontinuities, including the one that do not separate the diagram. They are implemented as algebraic constrains at the integrand level by the imposition of on-shell conditions on a arbitrary subsets of propagators. These methods are the base of \textbf{generalized unitary}
\cite{Bern:1994cg,Bern96-1,Bern96-2,Bern00-3,Bern04-4,Britto05,Britto08,Anastasiou06} (see \cite{Bern:2011qt} for a review).
A special role is played by the so called \textbf{maximal cut} \cite{Bosma:2017ens}, where the largest possible number (compatibly with the topology) of internal propagators
are imposed on-shell. Maximal cuts compute the leading singularities of the integrand, ie. the residues at the global poles of the loop integrand in complexified momentum space. These residues are highly constrained objects that serve as robust building blocks for systematically reconstructing the integrand. As we will see, FIs turn be solutions of differential equations in the kinematic variables; the corresponding maximal cut has the propriety to solve the associated homogenized differential equation.\\
Unitarity cuts are very naturally implemented in Baikov representation: a $k-cut$ on a Feynman integral expressed as in \eqref{BaikovRep-1} is obtained by taking the multiple residue at the simultaneous pole $z_i=\dots=z_{i+k}=0$:

\begin{equation}
     \mathcal{I}_{\nu_1\dots\nu_n}|_{k-cut}\sim 
 \int\limits_{\mathcal{C}'} d^{n-k}z \,\oint d^kz
 \left[{\mathcal B}\left(\z\right)\right]^{\frac{D-l-{e}-1}{2}}\prod\limits_{i=1}^{n} z_s^{-\nu_i}.
\end{equation}

In \cite{vanhove18}, Vanhove shows, through toric-geometric arguments, that the maximal cut also admits a nice Symanzik parametrization obtained by simply replacing the simplex $\Delta$ by the $l-$torus $T^l=\{z_i\in \Pro^{n+1}|z_i=1\}$:

\begin{equation}
    I= \int_{T^l} \prod_ix_i^{\nu_i-1}\frac{\mathcal{U}(x_i)^{\nu-(l+1)\frac{D}{2}}}{\mathcal{F}(x_i)^{\nu-l\frac{D}{2}}}\mu\,.
    \label{SimanzikMaxCut}
\end{equation}

\section{Master Integrals and periods}
High order perturbative integrals are extremely tough to compute analytically and their numerical evaluation requires huge computational effort. Moreover, the number of integrals contributing to the amplitude at a given order, grows at least exponentially with the latter. These two facts make direct high order corrections computation impossible. 
The crucial simplification allowing to bypass this obstacle, firstly observed \cite{Chetyrkin:1981qh}, is to realise that the Feynman integrals in a given family are not all independent, but satisfy linear relations coming from Stokes’ theorem, known as integration by parts identities, IBP's. Implemented algorithmically by Laporta in \cite{Laporta:2001dd}, IBP reduction turns the problem of handling an enormous set of integrals into a finite linear-algebra task: one generates many IBP relations, orders integrals by complexity and eliminates systematically to express every integral in the family as a linear combination of a small, finite set of irreducible objects, called \textbf{master integrals}. Once the masters are identified, their dependence on kinematics can be obtained by differential equations or other means and the entire family is reconstructed from their solutions. More conceptually, it was observed by Mastrolia and Mizera in \cite{Mastrolia:2018uzb}, and later refined in \cite{Frellesvig:2019kgj,Frellesvig:2019uqt}, that Feynman integrals naturally span a finite-dimensional vector space  which can be understood in cohomological terms: integrals are pairings of (twisted) cycles and cocycles, IBP relations correspond to exact forms or boundaries, and intersection theory provides a nondegenerate pairing that lets one project any integrand onto a chosen basis of masters. This geometric viewpoint not only explains why the number of masters is finite but also suggests computational alternatives to brute force elimination by computing intersection numbers and exploiting algebraic geometry one can often obtain reduction coefficients more efficiently and gain structural insight into when and why certain masters evaluate to polylogarithmic, elliptic, or more general period functions.\\
Let us see how this ideas comes in the most naturally way.\\
Consider a set of integrals

\begin{equation}
    I_{ij} = \int_{c_i} \omega_j(\z),
    \label{StartingFamily}
\end{equation}

where $\omega_j$, with $j\in \{1,\dots, N_\omega\}$, are multi-variable $n-$forms depending on $\z=(z_1,\dots,z_n)$ and $c_i$, with $i\in \{1,\dots, N_c\}$, are $n-$dimensional integration contours. \\
If we find a equivalence relation $\sim$ among integrands such that 
\begin{equation}
    \omega_j \sim \omega_k \quad \quad \Longrightarrow \quad \quad I_{ij}= I_{ik},
\end{equation}

we identify $N_1\leq N_\omega$ equivalence classes $\braket{\omega}_J$, reducing the number of independent integrals from $N_\omega N_c$ to $N_1N_c$.\\
There is still one source of interdependence: due to the linearity of integration, the equivalence classes we have defined are actually subjected to the further relation:

\begin{equation}
    \braket{\omega}_J \sim  \braket{\omega}_K + \braket{\omega}_L \quad \quad\Longrightarrow\quad \quad I_{iJ}=I_{iK}+I_{iL},
\end{equation}

reducing the number of independent equivalence classes to $\nu\leq N_1$.
Let us pick up one representative element $\ket{\phi_\beta}$ per equivalence class.\\
We can apply the very same reasoning to contours, identifying $\tilde{\nu}$ equivalence classes $[\gamma_{\tilde{\alpha}}|$. \\
In order for integration not to be degenerate there must be $\tilde{\nu}=\nu$.\\
Defining now the vector spaces $\mathbb{V}=\mathrm{span}\{ \ket{\phi_\beta} \}$ and $\mathbb{V}^\vee=\mathrm{span}\{[\gamma_\beta|\}$, we can express any integrand and any contour appearing in \eqref{StartingFamily} as a linear combinations of the respective basis elements as

\begin{equation}
[c_i| = a_i^{\,\alpha} [\gamma_\alpha| \quad\quad \mbox{and}\quad\quad \ket{\omega}_j = b_j^{\,\beta}\ket{\phi_\beta}.
\label{InterNumbers}
\end{equation}

Integration naturally induces a scalar product among these vector spaces, that actually turn to be dual.\\
We finally end up with a subset of independent $\nu^2$ integrals 

\begin{equation}
    I_{\alpha\beta}= [\gamma_\alpha\ket{\phi_\beta}
    \label{MI-1}
\end{equation}
called (a basis of) \textbf{Master integrals}, completely determining a whole family of integrals, even larger then the starting one, by 

\begin{equation}
    I_{ij} = a_i^{\,\alpha} b_j^{\,\beta} I_{\alpha\beta}.
\end{equation}

These vector spaces are not just ad-hoc linear reductions: they exactly are the (co)homology $n$-th groups of (co)chain complex, over some space $\M$, with (co)boundary operators the one inducing the equivalence relations above. \\
The master integral matrix is then nothing else that the period matrix we saw in section \ref{PeM}.\\\\

\Important{We call \textbf{geometry} associated to a set of integrals the space whose periods provide a master integral basis for it.}
\vspace{1cm}

Integrals $I_{ij}(\boldsymbol{\lambda})$ appearing in perturbative calculations actually depends on extra parameters $\boldsymbol{\lambda}$, i.e. the kinematic variables and regularization parameter,
thus also the basis of master integrals would depend on them, so following the above reasoning and section \ref{PeM}, that would be a family of complex manifold $M_{\boldsymbol{\lambda}}$ whose periods are such that:

 \begin{equation}
    \Pi_{\alpha\beta}(\boldsymbol{\lambda})=I_{\alpha\beta}(\boldsymbol{\lambda}).
\end{equation}

Summarizing, the challenges one has to face for successfully manage to apply intersection theory to handle this families of integrals are:
\begin{itemize}
    \item Identify which (co)homology groups capture the equivalence relations among the considered integrals;
    \item Compute such groups and in particular determining their dimension, corresponding to the number of MIs of the family;
    \item Evaluate the coefficients appearing in the decomposition \eqref{InterNumbers} by defining suitable \textbf{intersection pairings};
    \item Identify the underlying geometry, that is finding a family $M_{\boldsymbol{\Lambda}}$ of algebraic varieties and its moduli space whose periods computes the MIs.
    \item Determine the Picard-Fuchs equation governing the periods of $M_{\boldsymbol{\Lambda}}$.
\end{itemize}

\subsection{The sought (co)homology}
The first step is to recognize the appropriate (co)homology groups capturing the structure of the integral family we are considering.\\
Let $\M$ be a smooth $n-$dimensional manifold and $\La$ a local system on $\M$ with connection $1-$form $\omega$. A flat sections $u$ of $\La$, solves the differential equation $\nabla_\omega u\equiv (d+\omega)u(\z)=0$, in a coordinate system $z_i$, and can thus be written as

\begin{equation}
    u= c\,e^{\lambda},
\end{equation}
with $d\lambda=\omega$.\\
Global sections $\phi_R$ of the sheaf $\Omega^k(\M,\La)$ of $k-$forms taking value in $\La$, because of the isomorphism $\Omega^k(\M,\La)\cong \Omega^k(\M,\C)\otimes \La$, can be expressed as

\begin{equation}
   \phi_R(\z) = u(\z)\phi(\z),
\end{equation}
for $\phi(\z)$ a global section of $\Omega^k(\M,\C)$. Let us call $\phi_R$ a ($k-$)\textbf{right form}.\\
In order for a right form to be closed 

\begin{equation}
    d(\phi_R)=du\phi+u d\phi=-\omega u\phi +u d\phi =u\nabla_{-\omega}\phi,
\end{equation}
$\phi$ must be closed with respect to the covariant derivative $\nabla_{-\omega}\equiv d-\omega$, that is it must belong to the $k-$th cohomology group of the complex 

\begin{equation}
    \Omega^\bullet_{-\omega}: \dots\,\rightarrow \Omega^{k-1}(\M,\C)\xrightarrow[]{\nabla_{-\omega}}\Omega^{k}(\M,\C)\xrightarrow[]{\nabla_{-\omega}}\dots 
\end{equation}
Denote by $H^k(\M,\nabla_{-\omega})\equiv H^k(\Omega^\bullet_{-\omega})\otimes\La$ the group of closed right $k-$ forms $\ket{\phi_R}$, called \textbf{twisted cocycles}.\\
The homology groups dual to this are given by $H_k(\M,\nabla_{\omega})\equiv  H_k(\Omega^\bullet_{-\omega})^\vee\otimes \La^\vee$, where $H_k(\Omega^\bullet_{-\omega})^\vee$ is the cohomology of the dual complex and $\La^\vee$ is the local system with connection $\omega$. Let us call the elements of this group \textbf{twisted cycles} and denote them by $[C_L|$.\\
There is then a non degenerating pairing given by: 

\begin{equation}
    [C_L\ket{\phi_R} = \int_C \phi_R.
    \label{TwistedPairing}
\end{equation}
Now, Feynman Integrals in Baikov representation \eqref{BaikovRep}\footnote{After complexification.}, as well as string amplitudes \eqref{stringAmplitudeCLOSED} and \eqref{stringAmplitudeOPEN}, have the form 
\begin{equation}
    I = \int_C \B^\gamma(\z)\phi(\z), 
    \label{Twistedintegral}
\end{equation}
where $\phi(\z)$ is a rational $n-$form,
\begin{equation}
    \phi(\z)=\frac{dz_1\wedge\dots\wedge dz_n}{z_1^{\nu_1}\dots z_n^{\nu_n}}
\end{equation} 
and $\B^\gamma$ is a multivalued function, with $\B(\z)$ a polynomial and $\gamma \in \C\backslash\Z$.\\
Thus they can be recognized as a twisted pairing \eqref{TwistedPairing} up to the identification of the suitable space $\M$ and local system $\La$.\\
We need $\B^\gamma$ to be a flat section of $\La$, thus solving $\nabla_\omega\B^\gamma=0$ for $\omega$, and hence we get 

\begin{equation}
    \omega= dlog \B^\gamma=\gamma \left(\frac{\de_i\B}{\B} \right )dz_i\,.
\end{equation}
Moreover, we need $\omega$ to be globally defined on $\M$. Thus, if $\V[\B]\equiv\{\B=0\} $ is smooth, we have $\M=\C\Pro^n\setminus\{\B=0\} $. 
We stress here that a first problem may arise if $\V[\B]$ is singular.\\
Moreover, being $\V[\B]$ non compact, we need to control the behavior at infinity; the conclusion proved by Pham \cite{Pham:1967}, is that the right cohomology for describing an integral of the form \eqref{Twistedintegral}, up to the assumptions discussed above, is the cohomology group $H_n(\M,D_\infty)$ relative to a neighbordhood of infinity $D_\infty$.

\paragraph{Number of MIs.}\mbox{}\\

In practical calculations, one generally does not need to algebraically determine the (co)homology groups, but rather to just compute their dimension $\nu$, corresponding to the number of MIs. A lack of the algorithms based on IBPs relations is indeed the impossibility to know $\nu$ a priori, thus one has to keep trying generating independent elements until the procedure reach the closure. However, if the number of MIs is known, the generation of a basis is a relative easy task.\\
Under the assumptions given in \cite{Aomoto11} (see also \cite{Aomoto2,Aomoto75vanishing,Esnault1992}) one can prove the following vanishing theorem.\\\\
\theo[Vanishing theorem]{
Let $\M = \C\mathbb{P}^n \backslash \mathcal{P}$ be the $n-$dimensional complex manifold complementary to the projective variety $\mathcal{P}=\mathcal{P}[P]$ defined by $m$ polynomials $P_j: \C^n \rightarrow \C$, and $\nabla_\omega$ the covariant derivative, with connection $\omega$, of a rank-1 local system $\La$ on $\M$.\\
The only possibly non vanishing twisted cohomology group is the one in middle degree: 

\begin{equation}
H^k(\M,\nabla_\omega)=0 \quad \quad \mbox{for} \quad k\neq n.
\end{equation}
\label{vanishing} 
}

\nn
By Euler-Poincaré formula\footnote{$\chi(X)=\sum (-1)^k H_k(X,\C)$.} we then have 

\begin{equation}
\dim H^n(\M,\nabla_\omega)= (-1)^n\chi(\M).
\end{equation}

\nn
and thus:

\begin{equation}
\nu= (-1)^n (n+1-\chi(\mathcal{P})).  
\end{equation}

\nn
The computation of the Euler characteristic for the projective variety $\mathcal{P}$ is actually non-trivial.\\
However, we can compute $\chi(\M)$ by means of Morse theory \cite{Milnor:1963jmi}(see \cite{Massidda} for a simple introduction).
Let $f$ be a Morse function for $\M$, and $k_p$ the number of its critical points with Morse index $p$. Because of the vanishing theorem there are no critical points with index $k\neq n$ and thus all critical points must have the same, i.e. the highest, Morse index; $k_n$ is therefore the total number of critical points of $f$, and it must equal the $n-$th Betti number $b_n$:

\begin{equation}
\nu = b_n = k_n.
\end{equation}

\nn
In \cite{Lee:2013hzt}, Lee and Pomeransky show that $\log{\B^\gamma}$ is a Morse function for $\M$, thus the number of its critical points, the solutions $\omega=0$, gives the Euler characteristic of $\M$. Therefore, we have reduced the problem of determining the number of master integrals to the resolution of a system of equations, \textit{provided the vanishing theorem \ref{vanishing} holds.} We stress indeed, the latter is necessary for the Euler  characteristic of $\M$ to coincide, up to a possible sign, with $\nu$; if some other cohomology groups appear, we would eventually need to know their dimensions for evaluating $\nu$ via Euler-Poincaré formula.\\
Among the assumptions of the above theorem, it is required that the monodromy of $\La$ around each irreducible component of $\mathcal{P}$ and their intersections must \textbf{not} be trivial, i.e. equal to one. This is a necessary condition for killing cohomology group outside the middle one.

\subsection{Intersection numbers}\label{InterNSection}
Once a basis of Master integrals is provided, we need to be able to decompose any other integral of the family in terms of the basis, that is to compute the coefficients appearing in the \eqref{InterNumbers}.
We need to identify dual vector spaces $\mathbb{W}^\vee= \mathrm{span}\{\bra{\phi^\beta}\}$ and $\mathbb{W}= \mathrm{span}\{|\gamma_\alpha]\}$, and scalar products 

\begin{equation}
    \braket{\cdot|\cdot}: \mathbb{V}\times \mathbb{W}^\vee \longrightarrow \C \quad\quad \mbox{and}\quad\quad[\cdot|\cdot]: \mathbb{V}^\vee\times \mathbb{W} \longrightarrow \C,
\end{equation}

such that, provided the basis are orthonormally chosen

\begin{equation}
    [\gamma_\alpha| \gamma^\beta]=\braket{\phi^\beta|\phi_\alpha}=\delta^\beta_\alpha,
\end{equation}

we can write:

\begin{equation}
    a_i^\alpha = [c_i| \gamma^\alpha]\quad\quad \mbox{and}\quad\quad b_j^\beta =\braket{\phi^\beta|\omega_j}.
    \label{IntersectionNumebers}
\end{equation}

For a general reasonable topological space $\M$, we have seen in section \ref{IT} that such dual spaces are provided by Verdier duality \eqref{VD}; however, no general construction of such dual objects is known, and its concrete and explicit realization for practical purposes is an open problem. 
However, if we restrict to local systems on locally compact oriented manifolds, the twisted version of Poincaré duality \eqref{TwistedPoincarèDuality} gives us

\begin{equation}
\begin{split}
    &H^k(\M,\La)\times H^k_{BM}(\M,\La^\vee)\longrightarrow \C,\\
    & H_k(\M,\La)\times H_k^{lf}(\M,\La^\vee)\longrightarrow  \C,
    \end{split}
\end{equation}
revealing the sought dual spaces are the one of compactly supported twisted cocycles and locally finite twisted cycles.
The computation of homological intersection numbers is performed via diagrammatical methods for which we remand to \cite{kita1,kita2,Mimachi2003,Mizera:2016jhj,Mizera:2019gea} (see also \cite{Massidda}), while below we will report a brief review of cohomological intersection numbers evaluation methods.

\paragraph{Cohomological intersection numbers}\mbox{}\\
Let $\M$ be $n-$dimensional, $\phi_R \in H^k(\M,\nabla_{-\omega})$  and $\phi^c_L\in H^k_{BM}(\M,\nabla_{\omega})$, with $n=2k$. A realization for the pairing 
\begin{equation}
    H^k(\M,\nabla_\omega)\times H^k_{BM}(\M,\nabla_{-\omega})\longrightarrow \C,
\end{equation}
is simply provided by the integration

\begin{equation}
\braket{\phi^c_L|\phi_L}=\int_\M \phi^c_L \wedge \psi_R,
\label{Intercompact}
\end{equation}
induced by the cup product.\\ 
Now, because $H^k_{BM}(\M,\nabla_{\omega})\cong H^k(\M,\nabla_{\omega})$, we can extend the pairing to all twisted cocycles

\begin{equation}
    \braket{\phi_R|\phi_L}=\braket{\iota(\phi_R)|\phi_L},
\end{equation}

where $\iota:H^k(\M,\nabla_\omega)\rightarrow H^k_{BM}(\M,\nabla_{\omega}) $, is called \textbf{regularization map}. Roughly speaking, any cohomology class has a representative with compact support; the formula \eqref{Intercompact} provides a way to compute the intersection with that form, but due the to invariance of the integral map inside the same class, the intersection must be the same for any chosen representative.
The problem then reduces to find a realization of the regularization map.\\
First consider the case of a holomorphic one form.\\
Let $U_j \subset V_j \subset \M$ be two sets of neighborhoods of $z_j \in \M$, and let $h_j$ a partition of unity

\begin{equation}
h_j(z)=\begin{cases} &1 \quad\quad z \in U_j, \\ &0 \quad\quad z\notin V_j .\end{cases}
\end{equation}  

\nn
Let $\phi_L(z) \in \Omega^1(\M,\nabla_\omega)$. By Poincaré lemma, it exists $\psi_j(z) \in \Omega^0(V_j,\nabla_\omega)$ such that $\nabla_\omega \psi_j(z) = \phi_L(z)$, for $z\in V_j$. Introduce the one form
 
 \begin{equation}
 \phi'_L \equiv \phi_L - \nabla_\omega (h_j\psi_j) = \begin{cases} &0 \quad \mbox{if}\, z \in U_j \\& \phi_L  \quad \mbox{if}\, z \notin V_j,\end{cases}
 \label{regularizedmaprealization}
 \end{equation}

with support 

\begin{equation}
\mathrm{supp}[\phi'_L] \subset \left ( \bigcup_j U_j\right )^c,
\end{equation}  

\nn
where the superscript $c$ here stands for the complementary set in $\mathbb{P}^1$. Because $U_j$ are open, their complementary is closed, and thus $\phi'$ has compact support. Therefore, we get the regularization map:

\begin{equation}
\iota(\phi_L)= \phi_L - \nabla_\omega (h_j\psi_j).
\end{equation}

\nn

Plugging this result into \eqref{Intercompact}, after some algebra, and using residue theorem, one obtains

\begin{equation}
\langle \phi_L | \phi_R \rangle =\sum_j \mathrm{Res}_{z=z_j} (\psi_j \phi_R).
\label{intersectionnumber1}
\end{equation}

\nn
The computation of the intersection number for one forms then reduces to solve a differential equation around each pole and to compute residues at each of those poles. \\
Expression \eqref{intersectionnumber1} assumes a further simplified form whenever $\phi_L$ or $\phi_R$ have only simple poles:
\begin{equation}
    \langle \phi_L | \phi_R \rangle =\sum_j \frac{\mathrm{Res}_{z=z_j} \psi_j \mathrm{Res}_{z=z_j}\phi_R}{\alpha_j},
    \label{1LogInter}
\end{equation}
where $\alpha_j$ are the residues of $\omega$ at $z_j$. \\
In \cite{cho1995}, Cho and Matsumoto proved that \eqref{1LogInter} can be generalized to the case of logarithmic $n-$form. Moreover, in \cite{Mizera:2017rqa} Mizera showed it can be expressed in terms of the zeroes $z_{ij}^*$ of $\omega$, instead of its poles:

\begin{equation}
\langle \phi_L |\phi_R \rangle = \int \prod_{i=1}^n dz_i\delta[\omega_i]\phi_L\phi_R=\sum_{z^*_{ij}} |\de_i\omega_j|^{-1} \phi_L\phi_R \Big |_{z_i=z^*_{ij}},
\label{Mizerainter}
\end{equation}

which actually turns to be a big computational advantage, due to the greater ease to practically identify the zeros rather than the poles. \\
Deligne's theorem 
\begin{equation}
    H^k(U,\C)\cong H^k(X,\Omega^\bullet(\log D))
\end{equation} 
guarantees every cohomology class has a logarithmic form representative. Thus \eqref{Mizerainter} should suffice to evaluate any intersection number. However, except in the particular case of a hyperplane arrangement and the trivial case one variable case, there is no known realization of such map, in the sense of a general computationally implementable prescription to find a logarithmic equivalent to a general rational $n-$form.
To overcome this, a fibration-based method was introduced in \cite{Ohara98,Frellesvig:2019kgj,Frellesvig:2020qot} and recently refined in 
\cites{Caron-Huot:2021xqj,Caron-Huot:2021iev,Fontana_2023,Brunello:2023rpq}, consisting in an iterative procedure of degree reduction, through which $n-$forms intersection numbers are reduced to $1-$ forms intersections.

\subsection{Canonical differential equation}
Consider the period vector $\Pi_i(\boldsymbol{\lambda},\varepsilon)$, satisfying the system of Picard-Fuchs equations 

\begin{equation}
    \de^\nu\Pi_i(\boldsymbol{\lambda},\varepsilon) = \Omega_{ij}^{\nu}(\boldsymbol{\lambda},\varepsilon)\Pi_j(\boldsymbol{\lambda},\varepsilon),
    \label{PFcanonical}
\end{equation}
with $\de^\nu =\frac{\de}{\de \lambda_\nu}$. 
Because Gauss-Manin connection is flat (covariant derivatives commutes), the connection 1-forms $\Omega^\nu$ obey the integrability condition

\begin{equation}
    \de^\mu \Omega^\nu-\de^\nu \Omega^\mu +[\Omega^\nu,\Omega^\mu].
\end{equation}
Under a (holomorphic) change of basis $\widetilde{\Pi}=\Lambda\Pi$, the connection matrices transform by the usual guage rule

\begin{equation}
    \widetilde{\Omega}^{\mu}= \Lambda\Omega^\mu\Lambda - \Lambda^{-1}\de^\mu \Lambda.
\end{equation}
In \cite{Henn:2013pwa}, Henn prososed that there exists a gauge $\Lambda$ such that the dependence on the dimensional regulator  factors out:

\begin{equation}
    \widetilde{\Omega}^{\mu}(\boldsymbol{\lambda},\varepsilon)=\varepsilon \Omega^\mu(\boldsymbol{\lambda}),
\end{equation}

requiring the condition\footnote{One can easily see non commutativity of connection matrices is a obstraction to higher order in $\varepsilon$ to identically vanish.}
\begin{equation}
    [\widetilde{\Omega}^\mu,\widetilde{\Omega}^\nu]=0.
    \label{Commutativity}
\end{equation}

In this basis, the differential equation \eqref{PFcanonical} reduces to the so called $\varepsilon-$ factorized form 

\begin{equation}
    \de^\nu\widetilde{\Pi}_i(\boldsymbol{\lambda},\varepsilon) = \varepsilon \widetilde{\Omega}_{ij}^{\nu}(\boldsymbol{\lambda})\widetilde{\Pi}_j(\boldsymbol{\lambda},\varepsilon),
    \label{PFcanonical2}
\end{equation}

whose solution can be written in terms of path-ordered exponential

\begin{equation}
    \widetilde{\Pi}(\boldsymbol{\lambda},\varepsilon)=\left (\mathcal{P}e^{\varepsilon\int_C d\widetilde{\Omega}}\right )\Pi(\boldsymbol{\lambda},0).
\end{equation}
If this trivializing gauge transformation is found, then the task of solving $\varepsilon-$expanded master integrals is essentially solved.

The global existence of a such gauge transformation implies that the structure group of the cohomology bundle admits a reduction to an abelian subgroup, that is, that monodromy matrices commute. However, being interested in study MIs for $\varepsilon\rightarrow 0$, we just need this transformation to exist locally, in which case the above strong condition, although clearly sufficient, it is not necessary any more\cite{Levelt61,Turrittin55,Wasow65,Balser94}:\\\\

\Important{Provided $\varepsilon=0$ is a regular singular point for \eqref{PFcanonical} and \eqref{Commutativity} holds, local $\varepsilon-$factorized form exist.}

If $\Omega(\lambda,\varepsilon)$ has at most simple poles in $\varepsilon$, there exist local holomoprhic gauge transformation putting \eqref{PFcanonical} in the \textbf{Levelt form}:

\begin{equation}
    \frac{d \Pi}{d\boldsymbol{\Lambda}}=\frac{R}{\varepsilon}\Pi,
\end{equation}
with $R$ in Jordan normal form.\\
On the other hand, if $\varepsilon=0$ is an irregular point, Levelt-Turrittin theorem states that, possibily after a finite ramified cover $\tilde{\varepsilon}=\varepsilon^p$ ($p\in \Z$), and a formal\footnote{Not necessary converging Laurent power series.} gauge transform, the system decomposes into blocks whose formal solutions involves exponential factors $exp(P_j(\tilde{\varepsilon})^{-1})$. In this case, as we will discuss in the next chapter, Stoke's phenomena may arise, introducing a possible obstruction to $\varepsilon-$factorization.\\
Assuming $\varepsilon-$factorization exists, no general strategy for reaching it is known.  
In \cite{Frellesvig:2021hk}, Frellesvig conjectured the desired gauge transform is the one trivializing the period matrix, showing it is the case for elliptic FIs (see also \cite{chen25}). However, in order to impose this condition, one must be able to compute the periods in advance, and to known their $\varepsilon-$dependence for gauging it away from the differential equation. This make the algorithm tough to be applied in more complicated cases, and essentially inapplicable in those cases where the differential equation is sought to obtain the master integrals.
Imposing that the connection matrices be of $\mathrm{d}\log$ form in the kinematic variables \cite{Primo17,Adams18,Dlapa:2020cwj}, and, more recently, extending the notions of pure functions and uniform transcendental weight \cite{Gorges23,duhr2025-CY,Broedel:2018qkq}, lead to the introduction of the so-called \textbf{canonical form}.
Elaborating on this, different algorithms, consisting in the subsequently applications of ad hoc constructed rotation matrices, have been successfully proposed on parameter families of K3 surfaces and CY $3-$folds.\cite{Duhr:2025xyy,Pogel22,Pogel23,Pogel23-2,,bree25}

\section{Quadratic relations: Double Copies}
In \cite{Kawai:1985xq}, Kawai, Lewellen and Tye found that tree-level closed-string amplitudes could be expressed as a sum of products of open-string amplitudes. This open/closed string duality stems from the fact that a cylinder shaped worldsheet can be interpreted as either a closed string propagator or an open string vacuum diagram. When applied to first excited string states, KLT relation reveals duality between amplitudes involving spin two states (gravitons) and two copies of amplitudes involving spin 1 states (gluons). 
In \cite{BCJ}, Bern, Carrasco and Johansson proposed that the KLT relations were a manifestation of a more fundamental property of gauge theories themselves. They showed that gauge theory amplitudes can be arranged in a color-kinematics duality representation, such that, once color factors are replaced with a second copy of kinematic factors, gravity amplitudes are produced. This allows to compute complex, multi-loop graviton scattering amplitudes by "squaring" the much better understood amplitudes of Yang-Mills theory. This has been proven to work at tree level and higher orders, including at fourth post-Minkowskian order.\\
On the other hand, no full proof beyond three level has been found for KLT relations yet; however important developments have been recently done in this direction \cite{Britto:KLT, Stieberger:KLT, Mazloumi:2024wys, Bhardwaj:2023vvm,Pokraka:2025zlh}, mostly using the language of intersection theory.\\
Quadratic relations in intersection theory arise quite naturally by the vector space structure. \\
Consider two new basis $\{\ket{\psi_\alpha}\}$ and $\{\bra{\psi^\alpha}\}$ for $\mathbb{V}$ and $\mathbb{W}^\vee$ respectively, with 

\begin{equation}
\braket{\psi^\alpha|\psi_\beta}=\delta^\alpha_\beta.
\end{equation}
Expanding with respect to the old basis we have

\begin{equation}
\begin{split}
&\bra{\psi^\alpha}=\braket{\psi^\alpha|\phi_\beta}\bra{\phi^\beta}\equiv  (C)^\alpha_{\,\beta}\bra{\phi^\beta},\\&\ket{\psi_\beta}=\braket{\phi^\alpha|\psi_\beta}\ket{\phi_\alpha}\equiv \left (C^{\alpha}_{\,\beta}\right)^{-1}\ket{\phi_\alpha}.
\end{split}
\end{equation}
The matrix $C$ is called \textbf{(cohomology) intersection matrix}.\\
Analogously, introducing two new basis $\{[\Gamma^\alpha|\}$ and $\{|\Gamma_\alpha]\}$ for $\mathbb{V}^\vee$ and $\mathbb{W}$ respectively, one defines the \textbf{(homology) intersection matrix} as

\begin{equation}
    H^\alpha_{\,\beta}\equiv [\Gamma^\alpha|\gamma_\beta].
\end{equation}

Combining the expansions in the two different basis of each space, for any $\ket{\omega_i}\in \mathbb{V}$, $\bra{\omega_i}\in \mathbb{V}^\vee$, $[c_i|\in \mathbb{W}^\vee$ and $|c_i]\in \mathbb{W}$, we find the quadratic relations:

\begin{equation}
\braket{\omega^i|\omega_j}=\bra{\omega^i}\Gamma_\alpha]H^\alpha_{\beta}[\gamma^\beta \ket{\omega_j}\quad\mbox{and}\quad[c^i|c_j]=[c^i\ket{\psi_\alpha}C^\alpha_\beta\bra{\phi^\beta}c_j].
\label{Quadratic}
\end{equation}

Introducing two different basis for each vector space has a practical computational advantage: we are generically able to identify a basis $\{\ket{\phi_\beta}\}$ for $\mathbb{V}$ and a basis $\{\bra{\psi^\alpha}\}$ for $\mathbb{W}^\vee$, but we cannot explicitly find their orthonormal dual basis $\{\bra{\phi^\alpha}\}$ and $\{\ket{\psi_\beta}\}$ respectively; idem for contours. Thus, in practice one computes the intersection matrices $H$ and $C$, and use their inverses after rewriting \eqref{Quadratic} as 
\begin{equation}
\braket{\omega^i|\omega_j}=\bra{\omega^i}\gamma_\alpha]\left (H^\alpha_{\beta}\right )^{-1}[\Gamma^\beta \ket{\omega_j}
\end{equation}
and 
\begin{equation}
[c^i|c_j]=[c^i\ket{\phi_\alpha}\left (C^\alpha_\beta\right )^{-1}\bra{\psi^\beta}c_j].
\label{Quadratic2}
\end{equation}

\section{Banana integrals}\label{SecBanana}
The banana family, shown in figure \ref{banana-2}, is the prototypical class of multi-loop two-point Feynman integrals that already display a wide range of geometric phenomena encountered in contemporary amplitude computations.

\begin{figure}[h!]
    \centering
\begin{tikzpicture}
\draw[black, thick] (0,0) -- (1,0);
\draw[black, thick] (5,0) -- (6,0);
\draw[black] plot [smooth,tension=1.5] coordinates {(1,0) (3,1) (5,0)};
\draw[black] plot [smooth,tension=1.5] coordinates {(1,0) (3,-1) (5,0)};
\draw[black,dashed] plot [smooth,tension=1.5] coordinates {(1,0) (3,0.5) (5,0)};
\draw[black,dashed] plot [smooth,tension=1.5] coordinates {(1,0) (3,-0.5) (5,0)};
\draw node [above] at (3,1) {$m_1$};
\draw node [above] at (3,0.5) {$m_2$};
\draw node [below] at (3,-0.5) {$m_{l}$};
\draw node [below] at (3,-1) {$m_{l+1}$};
\draw node [above] at (0.5,0) {$p$};
\draw node [above] at (5.5,0) {$p$};
\node at (3,0.2) [circle,fill,inner sep=0.5pt]{};
\node at (3,0.0) [circle,fill,inner sep=0.5pt]{};
\node at (3,-0.2) [circle,fill,inner sep=0.5pt]{};
\end{tikzpicture}
    \caption{\small{l loops banana diagrams.}}
    \label{banana-2}
\end{figure}

The Symanzik representation \eqref{SimanzikRep} becomes  

\begin{equation}
    I_B^{(D)}= \int_\Delta \frac{\mathcal{U}(x_i)^{\frac{1}{2}(l+1)(2-D)}}{\mathcal{F}(x_i)^{\frac{1}{2}l(2-D)+1}}\mu,
    \label{Simanzikbanana}
\end{equation}

with Symanzik polynomials 
\begin{equation}
\begin{split}
    &\mathcal{U}(x_i)=\left (\prod_{i=1}^{l+1}x_i\right )\left (\sum_{i=1}^{l+1}\frac{1}{x_i}\right ),\\
    &\mathcal{F}(x_i)=-p^2\left (\prod_{i=1}^{l+1}x_i\right )+\left (\sum_{i=1}^{l+1}m_i^2x_i\right )\mathcal{U}(x_i).
    \end{split}
\end{equation}

The commonly adopted strategy is to study this family in $D=2-2\varepsilon$, being confident to obtain back four dimensional integrals by dimensional-shift relations 
\cite{Tarasov96,Lee:2009dh}.
The great advantage is that for $\varepsilon=0$, \eqref{Simanzikbanana} simplifies to 

\begin{equation}
    I\equiv I_B^{(2)}= \int_\Delta \frac{\mu}{\mathcal{F}(x_i)}.
\end{equation}
The next step is to analyze the integral's singularity structure by taking its maximal cut, as prescribed in \eqref{SimanzikMaxCut}. After introducing an external mass scale $\Lambda$, and the dimensionaless parameters $\xi=p^2/\Lambda$ and $\mu_i=m_i/\Lambda$, the integral becomes

\begin{equation}
    I(\xi, \mu_i)=\int_{T^l}\frac{\tilde{\mu}}{\left [\left (\sum_{i=1}^{l+1}\mu_i^2z_i\right )\left (\sum_{i=1}^{l+1}\frac{1}{z_i}\right )-\xi\right ]\left (\prod_{i=1}^{l+1}z_i\right )}, 
\end{equation}
where $\z$ are complex homogeneous coordinates and $\tilde{\mu}$ is the transformed integration measure.
The denominator above, a $l+1$ degree polynomial in $\Pro^l$ (see section \ref{CY}), defines a family of $(l-1)-$fold Calabi-Yau manifolds

\begin{equation}
    M_{\boldsymbol{\lambda}}=\{\z \in \C\Pro^l |\F(\z;\boldsymbol{\lambda})=0\},
\end{equation}
with $l^2$ complex structure moduli $\boldsymbol{\lambda}= \boldsymbol{\lambda}(\xi,\mu_1,\dots,\mu_{l+1})$. We stress that the relation between kinematic parameters and moduli is very complicated and, in general, unknown; even in the simplest case of elliptic curves it invokes modular functions. Thus one restricts the analysis to subfamilies by making simplification assumptions on the parameters, such as zero masses, equal masses, large momentum.
The integral can now be evaluated using the Poincaré residue theorem, and expressed as a period of the holomorphic $(l-1,0)-$form $\Omega$ on $M_{\boldsymbol{\lambda}}$:

\begin{equation}
    I(\xi, \mu_i)= \int_{T^{l-1}} \Omega|_{\F=0}.
\end{equation}

\paragraph{The sunrise diagram.}\mbox{}\\
Consider the two loops banana, also called \textbf{sunrise diagram} (or sunset, depending on the mood): 

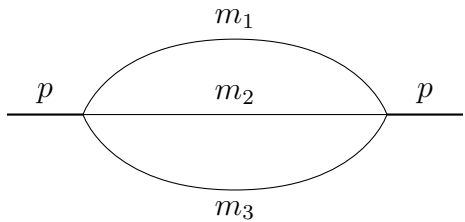
\begin{figure}[h!]
    \centering
\begin{tikzpicture}
\draw[black, thick] (0,0) -- (1,0);
\draw[black, thick] (5,0) -- (6,0);
\draw[black] plot [smooth,tension=1.5] coordinates {(1,0) (3,1) (5,0)};
\draw[black] plot [smooth,tension=1.5] coordinates {(1,0) (3,-1) (5,0)};
\draw[black](1,0) -- (5,0);
\draw node [above] at (3,1) {$m_1$};
\draw node [above] at (3,0) {$m_2$};
\draw node [below] at (3,-1) {$m_{3}$};
\draw node [above] at (0.5,0) {$p$};
\draw node [above] at (5.5,0) {$p$};
\end{tikzpicture}
    \caption{\small{2 loops banana diagrams.}}
    \label{Sunrise}
\end{figure}
The corresponding Feynman integral in Symanzik representation and $D=2$ takes the form

\begin{equation}
    I=\int_\Delta \frac{\mu}{(\mu_1x+\mu_2y+z)(xz+yz+xy)-\xi xyz},
    \label{2DSunriseSymanzik}
\end{equation}
where we set $\Lambda=m_3$.\\

The second Symanzik polynomial defines the family of elliptic curves

\begin{equation}
     \mathcal{E}(\xi;\mu_i)\equiv\{(x,y,z)\in \C\Pro^2 |(\mu_1x+\mu_2y+z)(xz+yz+xy)-\xi xyz=0\}.
\end{equation}
We can study the structure of singularities by considering the fibration \\$\mathcal{E}:\M\rightarrow\C_\xi$, identifying a smooth family over $\xi$, except for the six singular fibers located at
\begin{equation}
    \xi \in \Xi = \left \{0,\left(1\pm\sqrt{\mu _1}\pm\sqrt{\mu _2}\right)^2,\infty\right\}.
\end{equation}
\nn

The singular loci $\Sigma_i\equiv \Sigma(\xi_i)\subset \C\Pro^2$ of the critical fibers on $\xi\in\Xi$ are computed to be: 
\begin{equation}
\begin{split}
    &\Sigma_0 =\left\{\left [\frac{1+\mu_1-\mu_2\pm\Delta(\mu)}{2\sqrt{\mu _1}},\frac{1-\mu_1+\mu_2\mp\Delta(\mu)}{2\sqrt{\mu _2}},-1\right ]\right\}, \\ &\Sigma_1 =\left\{\left [\frac{1}{\sqrt{\mu _1}},\frac{1}{\sqrt{\mu _2}},1\right ]\right\},   \\&\Sigma_2 =\left\{\left [\frac{1}{\sqrt{\mu _1}},-\frac{1}{\sqrt{\mu _2}},1\right ]\right\},  \\&\Sigma_3 =\left\{\left [-\frac{1}{\sqrt{\mu _1}},\frac{1}{\sqrt{\mu _2}},1\right ]\right\},    \\& \Sigma_4 =\left\{\left [-\frac{1}{\sqrt{\mu _1}},-\frac{1}{\sqrt{\mu _2}},1\right ]\right\},\\& \Sigma_\infty =\left\{\left [1,0,0\right ],\left [0,1,0\right ],\left [0,0,1\right ]\right\},  
    \end{split}
\end{equation}
\nn
where we set
\begin{equation}
    \Delta(\mu)\equiv\sqrt{-4\mu_2+(1-\mu_1+\mu_2)^2}.
\end{equation}
\nn

Let $\mathcal{K}_\xi$ be the Koszul complex of fiber over $\xi$, the smooth part of Jacobian ring, coinciding with the full Jacobian ring for $\xi\notin \Xi$, is clearly independent on $\xi$ and given by \eqref{Poncaréseries}:
\begin{equation}
    \mathcal{P}[\Ha_0(\mathcal{K_\xi})](t)- \mathcal{P}[\Ha_1(\mathcal{K_\xi})](t)=(t+1)^3.
\end{equation}
\nn
On the other hand, for $\xi\in\Xi$, the Jacobian ring turns not to be finitely generated and it depends on the type of the singularity, thus indirectly on $\xi$.
We can make use of \eqref{Poincarèseries2} to compute the Poincaré series of the Jacobians for $\xi\in \Xi\backslash\{\infty\}$, in facts $\Sigma_0$ is a complete intersection of type $(2,1)$ and $\Sigma_i$ for $i=1,..,4$ are complete intersection of type $(1,1)$.
On the other hand, $\Sigma_\infty$ is not a complete intersection\footnote{$xy,xz,yz$ is not a regular sequence.} and no formula is known to the authors to compute the associated Poincaré series, we then proceeded by explicitly calculate the graduation of the Jacobian ring, or directly computing its Tjurina numbers, to obtain its Poincaré series a posteriori. For instance, for $\sigma_\infty^1=[1:0:0]$, we have

\begin{equation}
    \tau(\mathcal{V},[1:0:0])=dim\left(\frac{C\{1,y,z\}}{(yz,z,y)}\right)=dim(\C)=1,
\end{equation}
\nn
and proceeding similarly for all singular loci we find $\tau(\mathcal{V}_{\infty})=1+1+1=3$.\\
Finally, one finds: 

\begin{equation}
\begin{split}
    &\mathcal{P}[J(0)](t)= (1-t)^3+ \frac{t^3(1-t^2)}{(1-t)^2},\\ &\mathcal{P}[J(\xi_i)](t)= (1-t)^3+ \frac{t^4}{1-t},\\ &  \mathcal{P}[J(\infty)](t)= (1-t)^3+ \frac{t^3(t+2)}{(1-t)},
    \end{split}
\end{equation}

\nn
that is:

\begin{equation}
    \begin{split}
    &J(0)= 1\oplus 3\oplus 3\oplus \bar{2},\\
    &J(\xi_i)= 1\oplus 3\oplus 3\oplus \bar{1},\\
    &J(\infty)= 1\oplus 3\oplus 3\oplus \bar{3}.
    \label{graduation1}
    \end{split}
\end{equation}
\nn 
Therefore the corresponding Tjurina numbers are: 
\begin{equation}
    \tau(0)=2, \quad \tau(\xi_i)=1, \quad \tau(\infty)=3. 
\end{equation}

The singular fibers degenerate for specific values of the masses ratios $\mu_1$ and $\mu_2$, shown in figure \ref{muspace}.\\
\begin{figure}[h!]
    \centering
\begin{tikzpicture}
\draw[black, thick,->] (0,0) -- (5,0);
\draw[black, thick,->] (0,0) -- (0,5);
\draw[red, thick] (1,0) --(1,5);
\draw[red, thick] (0,1) -- (5,1);
\draw[red, thick] (0,0) -- (4,4);
\draw[domain=0:5, smooth, samples=100, variable=\x, orange] plot ({\x}, {1+\x-2*sqrt(\x)});
\draw[domain=0:1.5, smooth,samples=50, variable=\x, green] plot ({\x}, {1+\x+2*sqrt(\x)});
\draw node [below] at (4,-0.5) {$\mu_1$};
\draw node [left] at (-0.5,4) {$\mu_2$};

\node at (1,1) [circle,fill=blue,inner sep=1pt]{};
\end{tikzpicture}
\caption{\small{$\mu$ space.}}
\label{muspace}
\end{figure}

For $\mu_1=1$, $\mu_2=1$ and $\mu_1=\mu_2$, corresponding to two equal masses, two of the critical points coincides and the singular fibers reduce to five. A similar thing happens when $\mu_1$ and $\mu_2$ sits on the parabola

\begin{equation}
    \mathcal{S}(\mu_1,\mu_2,\mu_3)\equiv\mu_1^2+\mu_2^2-2\mu_1\mu_2-2\mu_1-2\mu_2+1=0, 
\end{equation}
\nn
that is, when masses are zeroes of the Symanzik polynomial:

\begin{equation}
   \mathcal{S}(m_1,m_2,m_3)= m_1^4+m_2^4+m_3^4-2m_1^2m_2^2-2m_1^2m_3^2-2m_2^2m_3^2,
\end{equation}
\nn
namely, where the triangular inequality is saturated:
\begin{equation}
    m_i^2=m_j^2+m_k^2,
\end{equation}
Finally, for $\mu_1=\mu_2=1$, i.e. equal masses case, critical points reduce the well studied four points: $\bar{\Xi}=\{0,1,9,\infty\}$, corresponding to singular fibers of Kodaira types, respectively $I_2,I_3,I_1,I_6$ \cite{Beauville}, with associated monodromy matrices \cite{barth2012}:

\begin{equation}
    M(I_k) = \begin{pmatrix}
        1 &k\\0&1
    \end{pmatrix}.
\end{equation}

%% file: C4.tex
\chapter{Exponential Periods}\label{C4}
\thispagestyle{empty}

Exponential integrals are ubiquitous in physics, particularly in path integral computations across any quantum field theory, including conformal field theory correlators and non-perturbative analyses in string theory. After providing an overview of the mathematical techniques developed to handle these type of integrals \cite{Kontsevich:2024mks,Kontsevich:2022ana}, we will show how Feynman integrals in Baikov representation can be interpreted as exponential pairings among suitable (co)homology classes.
We stress this interpretation could be given for any parameter representation of FIs and in general for any perturbative integral appearing in field theories and string theory, providing a very general result for any scattering amplitude evaluation.
As we saw in the previous chapter, the identification of FIs as pairings among locally finite holomology classes and compacted supported cohomology classes, fails to capture the full structure whenever the underlying geometry displays singularities or a non trivial behavior appear at the infinity. 
Furthermore, the evaluation of the number of master integrals via the Euler characteristic of $\M$ requires the vanishing theorem \ref{vanishing} to hold, and thus the cohomology to be concentrated in the middle degree. This fails to be true whenever trivial monodromies appear around the divisors or their intersection; and again, if the geometry is not smooth: in this case the Hodge-Deligne splitting may lead to non vanishing cohomology classes outside the middle degree.\\ 
Finally, the standard strategy to complexify the dimensional regularizator parameter must carefully take into account the possible emergence of non trivial phenomena inherited from the analytic continuation of the integrand.\\
We argue that the approach presented here provides the correct framework for properly deal with and overcome these issues. 

\section{Exponential Integrals}
Let $X$ be a smooth $n-$dimensional complex affine algebraic variety, $\Ol_X$ its structure sheaf and $\Omega^k_X$ the sheaf of differential $k-$forms on $X$. Given a holomorphic function $f\in \Ol(X)=\Gamma(X,\Ol_X)$, a Borel-Moore $n-$chain $\Gamma$ (locally compact) and an algebraic volume form $\mu\in\Gamma(X,\Omega^n_X)$, one defines the exponential integral of $f$ over $\Gamma$ with respect to $\mu$ as:

\begin{equation}
    I \equiv I_{\Gamma}(f)= \int_{\Gamma} e^{-f} \mu.
    \label{ExpIntegral1}
\end{equation}
\nn
Since we are working with smooth algebraic varieties, we can assume the support of $\Gamma$ to be an integer linear combination of closed oriented submanifolds. However, differently from ordinary homology, $\Gamma$ may have a nonempty boundary $\de \Gamma$. We will assume that the boundaries of the integration cycles are contained in a closed algebraic subset
$D_0\subset X$ of strictly positive codimension, ($Supp(\de\Gamma)\subset D_0$). Therefore, if the integration chain $\Gamma$ is such that the map
\begin{equation}
    \mathrm{Re}(f)|_{Supp(\Gamma)}:Supp(\Gamma)\rightarrow\R
\end{equation}
\nn
is proper\footnote{The pre-image of any compact is compact.} and bounded from below, the exponential integral \eqref{ExpIntegral1} is absolutely convergent.\\
Furthermore, we can even generalize the notion of exponential integral by rescaling the function $f \rightarrow \gamma f$ and studying how the structure of the resulting integral 

\begin{equation}
    I(\gamma) = I_{\Gamma} (f, \gamma) = \int_{\Gamma} e^{-\gamma f} \mu
    \label{ExpIntegralgamma}
\end{equation}
\nn
depends on the complex parameter $\gamma \in \C^*=\C\backslash \{0\}$\footnote{In order to emphasize the variable we will often use the notation $\C_\gamma$ to mean the copy of $\C$ where $\gamma$ takes values.}.\\
For generic $\gamma$, the integral $I(\gamma)$ can be expressed as a linear combination of exponential integrals over special integration cycles, called \textbf{thimbles}. These are real, non-compact cycles formed by the gradient flow lines of $\mathrm{Re}(\gamma f)$ with respect to an auxiliary Hermitian metric on $X$.   In general, these gradient flow lines, which originate from a critical point, do not cross any other critical point along their trajectory. However, as the argument $arg(\gamma)$ varies, there exist special values of $\gamma \in \mathbb{C}^{\ast}$ at which this condition fails, leading to a change in the number of independent gradient flow lines. When one of these special loci, known as \textbf{Stokes lines}, is crossed in $\mathbb{C}^{\ast}$, the linear combination of thimbles undergoes a discontinuous change (jump) described by a Stokes automorphism. \\
The collection of Stokes automorphisms along the plane $\mathbb{C}^{\ast}_{\gamma}$ forms the wall crossing structure associated to the integrals \eqref{ExpIntegralgamma}, which coincides with the one arising from the holomorphic version of Morse theory, see \cite{Nicolaescu2007} for an introduction to complex Morse theory. Exponential integrals can also be placed within the framework of exponential Hodge theory and interpreted as periods. In particular, they can be embedded into a generalized Riemann-Hilbert correspondence to study the relationship between de Rham and Betti cohomologies, both at the local and global levels.  In the global setting, the isomorphism between these two cohomologies associated with the triple $(X,D_0,f)$ is precisely realized through the exponential integral.
In the next subsections, we will study in detail the four cohomologies associated with this triple, each of which defines a vector bundle over $\mathbb{C}^{\ast}_{\gamma}$, and we will discuss their mutual relations. \\
Let us  define the \textbf{bifurcation set} $S^*\subset\mathbb{C}$ as the minimal finite set of points such that for any $t \in \mathbb{C} \setminus S^*$ there exists an open neighborhood $U$ of $t$ (in analytic topology\footnote{Algebraic geometry makes use of the Zariski topology. However, since we have invoked smoothness, we can always view $X$ as a complex manifold and use the corresponding topology. This is called analytification of the topology.}) and a homeomorphism $f^{-1}(U) \simeq U \times f^{-1}(t)$ which is compatible with the natural projections on both spaces to $U$ and such that it induces a homeomorphism:
\begin{equation}
    f^{-1}(U) \cap D_0 \, \simeq \, U \times \left( f^{-1}(t) \cap D_0\right).
\end{equation}
\nn
Smoothness implies $X$ is a complex manifold, i.e. it locally looks like $\C^n$. Since $f$ is continuous, and $U\subset \C$ is open, $f^{-1}(U)$ can be identified with an open set of $\C^n$. If $U$ does not contain bifurcation points, the above definition states that $f:\C^n \supset f^{-1}(U)\rightarrow U$, defines a local fibration on $\C$ (indeed a fibration on $U$), whose fiber is $f^{-1}(t)\cong Spec[\C[x_1,\dots,x_n]/\langle f\rangle]$,\footnote{With $\langle f \rangle$ we mean the ideal generated by $f$ in $\C[x_1,\dots,x_n]$, e.g. the elements of the form $gf$, $g\in \C[x_1,\dots,x_n]$. Therefore, $R:=\C[x_1,\dots,x_n]/\langle f\rangle$ is a ring. On the other hand, $f=0$ determines an affine subvariety $Y$ of $\C^n$. If $x\in Y$ and $h\in R$, we have an evaluation map $ev_x:R\rightarrow \C$ which is a homomorphism. The map $ev: Y\rightarrow R,\ x\rightarrow ev_x$, gives a bijection between $Y$ as a subset of $\C^n$, and the set of homomorphisms $R\rightarrow \C$. The latter is called $Spec[\C[x_1,\dots,x_n]/\langle f\rangle]$.} with $t\in U$, up to constant deformations (that are deformations depending trivially on $t$). Such fiber is smooth whenever the Milnor algebra\footnote{Also called local Jacobian ring} (Chiral ring)

\begin{equation}
    \mathcal{M}_f= \frac{\Ol_{\C^n,\mathbf{z}}}{\langle \de_{x_1} f, \ldots, \de_{x_n} f\rangle}
\end{equation}
\nn
is trivial, that is, if its dimension $\mu$, called (local) \textbf{Milnor number}, vanishes \cite{Porteous_1971}. Here $\Ol_{\C^n,\mathbf{z}}$ is the stalk at $\mathbf{z}$ of $\Ol(X)$, that is the ring of germs of power series converging in some neighborhood of $\mathbf{z}$, $f(\mathbf{z})=t$.
If $\mu=0$ for any $t'\in U(f(\mathbf{z}))$, Ehresmann's lemma \cite{Dundas} implies that $f$ is a locally trivial fibration over $U$. On the other hand, if non zero Milnor numbers arise, the transition functions are constrained by elements of the Jacobian and the fibration cannot be trivial: the bifurcation set contains at least the set $S\equiv \{t_i\equiv f(\sigma_i)\}_i$ of critical values of $f$.\\
In the latter case, one can still define a locally trivial fibration $M$ on $U^\odot_i\equiv U(t_i)\backslash\{t_i\}$, called \textbf{Milnor fibration} \cite{Ebeling,Ebeling1,ebeling2005monodromy,Gabrielov1979, Kulikov_1998}, whose fiber $M_{\mathbf{z}_i}$ is a CW complex homotopy equivalent to a bouquet of $\mu_{i}\equiv \mu|_{t_i}$ copies of $(n-1)-$spheres. Each of such spheres, or equivalently each element $\Delta_i$ of $H_{n-1}(M_{\mathbf{z}_i},\C)$\footnote{$H_k(S^{n-1},\C)=\C$, for $k=0,n-1$ and vanishes otherwise.}, is called (algebraic)\textbf{vanishing cycle}. Their denomination follows from the fact that they shrink to zero when approaching the critical point.  
The importance of the role they play here derives from Brieskorn and Malgrange's\cite{Brieskorn1970,Malgrange1994} proof of the isomorphism between the homology generated by the vanishing cycles and the hypercohomology of the De-Rham complex twisted by middle extended Gauss-Manin connection:

\begin{equation}
    H_{n-1}(M_z,\C)\cong \mathbb{H}(\Omega^\bullet_{X},\nabla^{mid}_{GM}), 
\end{equation}

\nn
which means that, in some sense, the homology of the full space is determined by the homology of the fiber. One can get an idea as follows. In the above local fibration all fibers are isomorphic, so they have equal (co)homology. This determines a vector bundle over the complement of the bifurcation points, with fibers the (co)homology groups, and whose transition functions are nontrivial only around the bifurcation points, so they are locally constant. In this sense, one can think of (co)homology classes as functions of the base point $t$ and the Gauss-Manin connection is the flat connection telling how the (co)homology classes change along the basis, i.e. how to take their covariant derivative in $t$.\\ As we saw in section \ref{PLT}, \textbf{Picard-Lefschetz theory} \cite{Lefschetz:1924,HuseinSade} provides a concrete tool for determining and studying these vanishing cycles.\\
As we will see explicitly in Section \ref{sec:the relative homology}, the set of $n-$dimensional(real) manifolds $\mathcal{T}_i \simeq\Delta_i\times \R^+\subset X$ corresponding to the traces of the vanishing cycles along the vanishing directions in the base space, called Picard-Lefschetz thimbles, provide a basis of thimbles for the global Betti cohomology associated with the triple $\left( X, D_0, f \right)$. \\
Suppose now $X$ can be compactified to a smooth projective variety $\overline{X}$ such that $f$ extends to a regular map (t.i. everywhere defined):
\begin{equation}
    \overline{f} \, \, : \, \, \overline{X} \, \longmapsto \, \mathbb{P}^1.
\end{equation}
\nn
We can decompose $\overline{X}-X = D_h \sqcup D_v$, where the \textbf{vertical divisor} $D_v = \overline{f}^{-1}(\infty)$ is the locus at infinity where $f$ diverges, and the \textbf{horizontal divisor} $D_h$ is the locus at infinity where $f$ has finite limit.
In the following we will assume the set $D_0 \cup D_v \cup D_h$ is a normal crossing divisor and that no critical points lie at infinity nor at $D_0$. With this we mean that the restriction of $f$ to $D_0$ or to the infinity locus \footnote{In general, the infinity locus is not a submanifold but rather a stratifold. Thus, one has to check that the restrictions of $f$ to each open stratum has no critical points.} has no critical points, which implies $S^*=S$. 
Finally, suppose that no degeneration of critical points occurs.

\section{Wall Crossing Structure}\label{WCSsection}
It is a very well known fact in complex analysis that analytic functions may show different asymptotic behavior in different regions of the complex plane. The prototypical example of such a propriety, called \textbf{Stokes' phenomenon}, is provided by the \textbf{Airy functions} $\mathrm{Ai}(z)$ and $\mathrm{Bi}(z)$, standard solutions of the differential equation 

\begin{equation}
    y''(z)-zy(z)=0.
\end{equation}
Two linear independent formal asymptotic solutions of this equation are given by

\begin{equation}
    \Phi_{\pm}=z^{-1/4} e^{\pm\left (\frac{2}{3}z^{3/2}\right )},
\end{equation}
alternatively showing a exponential grow or a exponential decay in a given direction. 
Indeed, the real part of the exponent changes sign as the argument of $z$ varies. In particular the directions $\mathrm{Re}(z^{3/2})=0$, namely

\begin{equation}
    \mathrm{arg}(z)\in \{-\frac{\pi}{3},\frac{\pi}{3},\pi\},
\end{equation}
correspond to lines on the complex plane where $\Phi_+$ and $\Phi_-$ interchange the dominant and subdominant regime, identifying three Stokes' lines. Choosing a basis of solutions that matches the formal series $\Phi_\pm$ in a given sector, the analytic continuation across the boundary, through an adjacent region, gives a jump of the form 

\begin{equation}
    \Phi_- \longrightarrow \Phi_- \pm i \Phi_+.
\end{equation}
This facts are known since the 19-th century, discovered by Stokes himself, and several analytic approaches have been developed to study them; however, only recently Kontsevich and Soibelman \cite{Kontsevich:2013rda} provided a algebraic-geometrical framework that packages these behaviors into the notion of a \textbf{Wall Crossing Structure}(WCS). 
We will review below what it consists of and how to define it, specializing to the case of our interest related to exponential integrals. 

\subsection{Global and Local Twisted de Rham Cohomologies}
We introduced above the triple $\left( X, D_0, f \right)$ and the exponential integral \eqref{ExpIntegral1}, where $\mu$ is a holomorphic top form on $X$. From a cohomological perspective $\mu$ is closed with respect to the differential
\begin{equation}
    \nabla_f \, = \, d \, \, - \, \, df \, \wedge,
\end{equation}
thus there is a well-defined de Rham complex (in Zariski topology):
\begin{equation}
    \Omega_X^{\bullet} \, = \, \Omega_X^0 \, \xrightarrow{\nabla_f} \, \Omega^1_X \, \xrightarrow{\nabla_f}\, \dots \, \xrightarrow{\nabla_f} \, \Omega^n_X.
\end{equation}
\nn
In order to incorporate the boundary divisor $D_0$, we restrict to the subcomplex $\Omega^{\bullet}_{X,D_0}$ of $\Omega^{\bullet}_X$ of forms with support on $X \setminus D_0$. 
Thus we have:\\\\

\defn[Global twisted de Rham]{
The global twisted de Rham cohomology is the graded abelian group
\begin{equation}
    H^{\bullet}_{GdR} \left( (X,D_0),f \right)\equiv \mathbb{H}^{\bullet} \left( X_{Zar}, (\Omega^{\bullet}_{X, D_0}), \nabla_f \right)\\\\
    \label{GlobalDerham}
\end{equation}

of equivalence classes of forms on $X \setminus D_0$ with respect to the differential $\nabla_f$.}
Notice, for instance, that any $1$-form $\alpha$ closed  with respect to the standard de Rham differential, yields a $\nabla_f$-closed $1$-form $e^{f} \alpha$:
\begin{align}
        \nabla_f (e^f \alpha) \, & = d (e^{f} \alpha) - df \wedge e^f \alpha
        = e^f d \alpha + e^f df\wedge \alpha - df \wedge e^f \alpha = 0 .
\end{align}
\noindent
If we now fix $\gamma \in \mathbb{C}^{\ast}_{\gamma}$, and replace $f \rightarrow \gamma f$, we obtain the graded $\mathbb{C}-$vector space
\begin{equation}
H^{\bullet}_{GdR,\gamma} \left( (X, D_0), f\right)=H^{\bullet}_{GdR} \left( (X, D_0), \gamma f \right).
\end{equation}

\noindent
In addition to this global version of cohomology, one may also study the cohomology localized near each critical point of $f$.\\\\

\defn[Local twisted de Rham]{
Let $\Sigma = \left\lbrace \sigma_i \in X \setminus D_0 \vert df(\sigma_i)=0 \right\rbrace$ be the set of critical loci of $f$ in $X \setminus D_0$. The local twisted de Rham cohomology associated to the triple $(X,D_0,f)$ is the $\mathbb{C} [[1/\gamma]]$-module \footnote{ Notation: $\mathbb{C}[[1/\gamma]]$ is the ring of formal power series of $\frac{1}{\gamma}$ with coefficients in $\mathbb{C}$. A $\mathbb{C}[[1/\gamma]]-$module is an abelian group equipped with an induced action of $\mathbb{C}[[1/\gamma]]$.}

\begin{equation}
    H^{\bullet}_{LdR} \left( (X, D_0), f \right) \equiv \bigoplus_{i \in \Sigma} \mathbb{H}^{\bullet} \left(  U_{form} (\sigma_i), (\Omega^{\bullet}_{X,D_0} \left[ \left[\gamma \right] \right], (1/\gamma) d - df \wedge )\right)
\end{equation}

where $U_{form}(\sigma_i)$ is the formal neighborhood of the critical locus $\sigma_i \in X$. Each summand is called local de Rham cohomology associated with $\sigma_i$ (or $t_i=f(\sigma_i) \in S$) and it is denoted with $H^{\bullet}_{LdR,\sigma_i} (X, D_0, f)$.\\\\ }
Assume that $f$ is proper, and set $\tau=1/\gamma$.\\
The coherent sheaf $\mathcal{H}^{\bullet}_{GdR}(X,f)$ on $\mathbb{C}$, defined as
    \begin{equation}
        \mathcal{H}^{\bullet}_{GdR} (X,f) \equiv \mathbb{H}^{\bullet}_{Zar} \left(  X \times \mathbb{C}_{\tau} , \left(pr^{\ast}_X ( \Omega^{\bullet}_X, \tau d_X - df \wedge ) \right)\right),
    \end{equation}
gives rise to a graded vector bundle over $\mathbb{C}_{\tau}$.\footnote{Notation: $pr^{\ast}_X$ is the map induced from the projection of $X \times \mathbb{C}_{t}$ onto $X$.} Its restriction to $\mathbb{C}^{\ast}_{\tau}$ carries a Gauss-Manin connection and a covariant derivative $\nabla_{\tau}$, encoding how cohomology varies with the parameter $\tau$. Such connection has a regular singularity at $\tau = \infty$ (i.e. $\gamma=0$) and a second order pole at $\tau=0$ (i.e. $\gamma = \infty$).\\
As we approach the point $\tau=0$, the global connection $\nabla_{\tau}$ splits into a direct sum of blocks, each of which is the tensor product of an exponential factor $e^{t_i \gamma} $ (rank 1 irregular D-module on $\mathbb{C}$) and a regular connection:
    \begin{equation}
        \left( \mathcal{H}^{\bullet}_{GdR}(X,f), \nabla_{\tau} \right) \simeq \bigoplus_{i \in S} e^{t_i \gamma} \otimes \left( E_i, \nabla_i \right).
    \end{equation}

In physical jargon this is the statement that, as $\tau \rightarrow 0 $ (i.e. $\gamma \rightarrow \infty$), the integral localizes around each critical point $\sigma_i$ giving an irregular contribution $e^{t_i \gamma}$ times a regular contribution solution of the system $\left( E_i, \nabla_i \right)$.\\
 The fiber of $\mathcal{H}^{\bullet}_{GdR} (X,f)$ at $\tau = 0$ (i.e. $\gamma=\infty$) is isomorphic to the sum
    \begin{equation}
        \bigoplus_{i \in S} \mathbb{H}^{\bullet} \left(  U_{form}(\sigma_i), \left( \Omega^{\bullet}, -df \wedge   \right)\right).
    \end{equation} 
Formally near $\tau=0$, the global twisted de Rham cohomology can be reconstructed using the local pieces around each critical point via the following global-to-local isomorphism:
    \begin{equation}
        \varphi_{dR} \, \, : \, H^{\bullet}_{GdR} ((X,D_0),f) \otimes_{\mathbb{C}[\gamma]}  \mathbb{C[[\gamma]]} \, \simeq \, H^{\bullet}_{LdR} ((X,D_0),f).
        \label{dR:global_to_local}
    \end{equation}
Moreover, for any $\gamma \in \mathbb{C}^{\ast}$ there is a non-degenerate pairing 
    \begin{equation}
        H^{\bullet}_{GdR,- \gamma} (X,f) \otimes H^{\bullet}_{GdR, \gamma} (X,f) \quad \longmapsto \quad \mathbb{C} \left[ -2\,dim_{\mathbb{C}} X\right]
    \end{equation}
    which extends to a non-degenerate pairing at $\gamma = \infty$ (i.e. $\tau=0$). This is the twisted Poincaré duality, shifting cohomological degree by $-2\,dim_{\mathbb{C}} X=-2n$.\\
We will show how to compute it concretely in section (\ref{PI}).

\subsection{Global and Local Betti (Co)Homologies}\label{sec:betti}
To complement the twisted de Rham picture, we now introduce the corresponding Betti (co)homology groups, which capture the topology of chains on $X$ relative to the level sets of $f$ at infinity. We begin by fixing a real constant $c >0$ and considering the singular relative homology
\begin{equation}
    H_{\bullet} \left( X, D_0 \cup f^{-1} (\mathrm{Re}(z) \geq c), \mathbb{Z} \right) \simeq H_{\bullet} \left(  X, D_0 \cup f^{-1} (c), \mathbb{Z}\right).
\end{equation}
Once $c > \mathrm{max}_{\sigma \in \Sigma} \mathrm{Re}(f(\sigma))$, the critical points do not lie on the boundary, and then the relative homology stabilizes (i.e. it is the same replacing $c$ with any $c'>c$).\\\\
\noindent
\defn[Global Betti (co)homology]{
The global Betti homology of $(X,D_0,f)$ is
\begin{equation}
    H_{\bullet}^{GB} \left( (X,D_0), f, \mathbb{Z} \right) := H_{\bullet} \left( (X,D_0), f^{-1} (\infty), \mathbb{Z}\right)
\end{equation}
and, similarly, the global Betti cohomology is
\begin{equation}
    H^{\bullet}_{GB}  \left( (X,D_0), f, \mathbb{Z} \right) \equiv H^{\bullet} \left( (X,D_0), f^{-1} (\infty), \mathbb{Z}\right),
\end{equation}
where the infinity means selecting the stabilized (co)homology.\\\\}

As we did before, we consider the rescaling of the function $f \rightarrow \gamma f$, and we extend the global Betti (co)homology to any point in the plane $\mathbb{C}^{\ast}_{\gamma}$.\\
Let $\gamma \in \mathbb{C}^{\ast}_{\gamma}$. For each fixed $\gamma$, we define the graded abelian group
\begin{equation}
    H^{\bullet}_{GB,\gamma} \left( (X, D_0),f , \mathbb{Z} \right) \equiv H^{\bullet} \left( (X, D_0), (\gamma f)^{-1} (\infty), \mathbb{Z} \right).
    \label{Def:Betti_global_gamma}
\end{equation}

On $\mathbb{Q}$, one can prove that the following Poincaré duality holds.\\\\

\prop[Poincaré duality \cite{Kontsevich:2024mks}]{
Let $X'=\overline{X}-D_v- \overline{D}_0$ and $D_0' = D_h - (D_h \cap D_v)$. We have the following isomorphism:
\begin{equation}
    H_{\bullet}^{GB} \left( (X,D_0),f \right) \simeq H^{\bullet}_{GB} \left( (X',D'_0),-f\right) \left[ 2\,dim_{\mathbb{C}} X \right].
    \label{Betti:Poincare_duality}
\end{equation}}
The family of abelian groups \eqref{Def:Betti_global_gamma} over the whole space $\mathbb{C}^{\ast}_{\gamma}$ defines a local system over $\mathbb{C}^{\ast}_{\gamma}$ denoted as $\mathcal{H}^{\bullet}_{GB} \left( (X,D_0), f\right)$.\\
Now, we want to relate these groups to local data. In order to do this, we look at the codomain $\C_t$ of the function $f$ as the real plane $\R^2$ and choose an open region $U$, whose closure $B=\bar U$ is a submanifold of $\R^2$ isomorphic to a unit disc. Assuming that the boundary $\partial B$ does not intersect the critical locus of $f$, we fix an arbitrary point $t_0 \in \partial B$. The idea is the following.\\ For each $k \in \mathbb{Z}_{\geq 0}$ and $k<n$, we associate to the pair $(B, t_0)$ the abelian group

\begin{equation}
    V^k(B,t_0) := H^{k} \left( f^{-1} (B), (D_0 \cap f^{-1} (B)) \cup f^{-1}(t_0), \mathbb{Z}\right)
\end{equation}

and we look at it as a vector space. We now assume that all finite, non-degenerate critical values of $f$ lie in the interior of $B$. Since they are isolated points, we can find a finite number of subsets $B_i=\bar U_i \subset B$, each containing exactly one critical value and such that they have vanishing intersection in $B$ and intersect $\partial B$ precisely in the same marked point $b$. By retracting $B$ to the bouquet of $B_i$, one gets the isomorphism 

\begin{equation}
    V(B, b) \simeq \bigoplus_j V(B_j,b).
\end{equation}

This allows us to explore each component separately and study cohomologies with a single critical point. This leads one to introduce the following definitions.
Let us assume $D_0 = \emptyset$ ($X$ is projective) to lighten notation.\\\\  

\defn[Local Betti cohomology]{
For each critical value $t_i$, a small positive $\epsilon$ and $\gamma \in \mathbb{C}^{\ast}$, we define the local Betti cohomology associated with the pair $(t_i,\gamma)$ as the graded abelian group
\begin{equation}
    H^{\bullet}_{LB,t_i, \gamma} (X,f) = V\left( D( \gamma t_i, \epsilon), t_{\theta_\gamma}\right) = H^{\bullet} \left( (\gamma f)^{-1}(D(\gamma t_i, \epsilon)) , f^{-1} (t_{\theta_\gamma}), \mathbb{Z}\right)
    \label{Betti_local:ti}
\end{equation}

where $D( \gamma t_i, \epsilon)$ is a closed disc in $\mathbb{C}$ of radius $\epsilon$ centered in $\gamma t_i$ and $t_{\theta_\gamma}$ is the point on the boundary of the disc  such that $t_{\theta_\gamma}= \gamma t_i+ \epsilon e^{i \theta_\gamma}$ with $\theta_\gamma = \pi-\arg (\gamma)$.\\
At a fixed $\gamma$, the direct sum of these cohomology groups for each $t_i \in S$ form the local Betti cohomology $H^{\bullet}_{LB, \gamma} (X,f)$:
\begin{equation}
    H^{\bullet}_{LB, \gamma} (X,f) = \bigoplus_{t_i \in S} H^{\bullet}_{LB, t_i, \gamma} (X, f).
\end{equation}}

\begin{figure}[h!]
    \centering
    \includegraphics[width=0.8\linewidth]{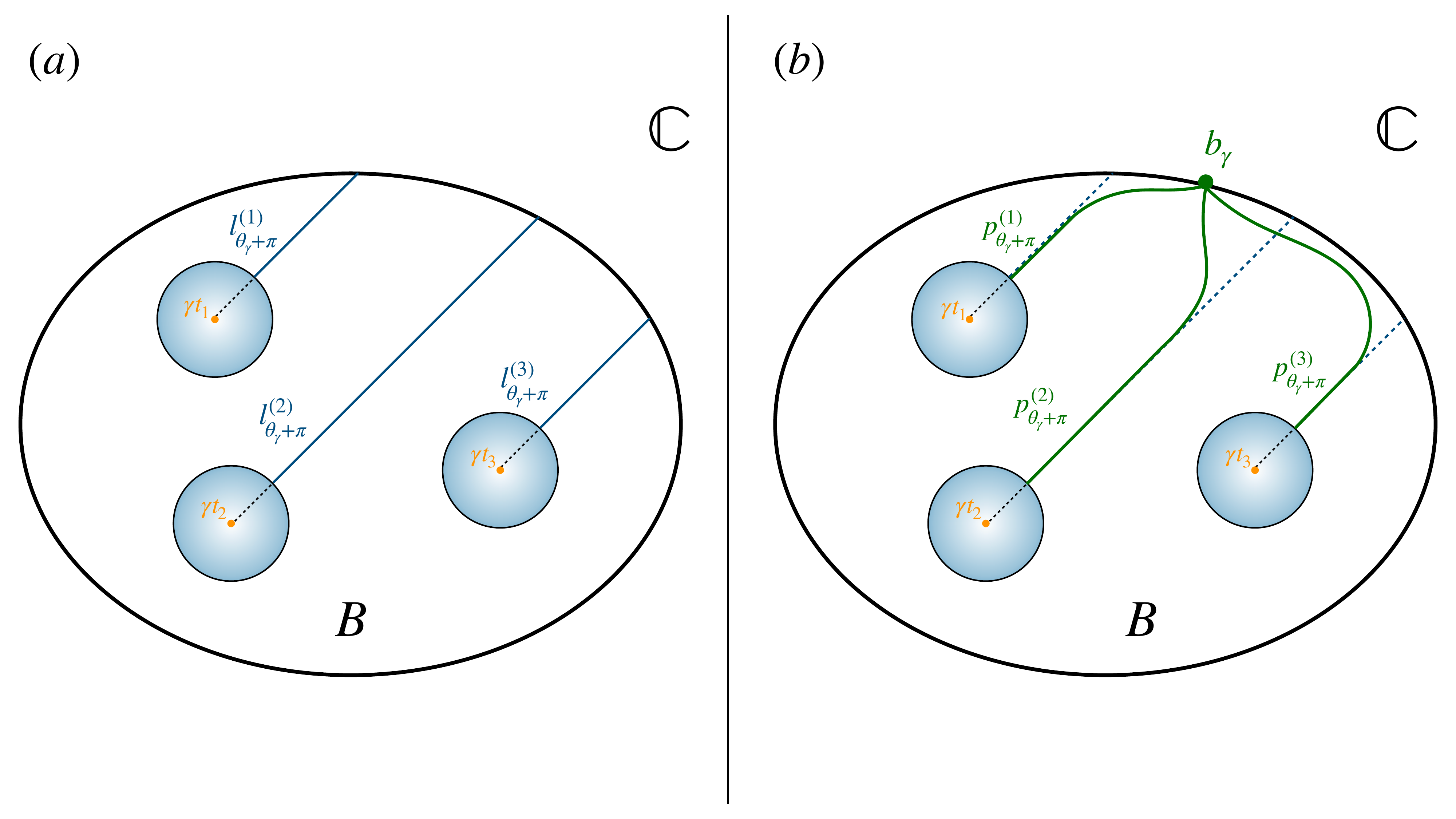}
    \caption{\small{Pictorial representation of the homotopy deformation implementing the boundary retraction that induces the isomorphism \eqref{Betti:global_to_local}.}}
    \label{fig:Betti_local_to_global}
\end{figure}

Similar to the global case, the family of local Betti cohomologies over the space $\mathbb{C}^{\ast}_{\gamma}$ forms a Local System denoted as $\mathcal{H}^{\bullet}_{LB}(X,f)$.\\
 Now we will make use of the description of the local and global cohomology groups as vector spaces $V(B_j,t_j)$ in order to relate them to each other. First, let us construct a sufficiently large disc $B \subset \mathbb{C}$ containing all the critical values of $S$, and for each critical value in $t_i \in S$ let us construct its proper disc $D(\gamma t_i, \epsilon)$ with marked point $t^{(i)}_{\theta_{\gamma}}$ on its boundary. From each of these points $t^{(i)}_{\theta_{\gamma}}$ let us construct a ray $l^{(i)}_{\theta_{\gamma}+\pi}$ in the direction $\theta_{\gamma}+\pi$. The resulting configuration consists of a set of parallel lines originating from the small discs and terminating at the boundary of the large disc, as depicted in figure \ref{fig:Betti_local_to_global}-(a). At this point, we can construct a homotopy of the large disc such that the deformed rays $l^{(i)}_{\theta_{\gamma}+\pi} \rightarrow p^{(i)}_{\theta_{\gamma}+\pi}$ intersect at a unique point $b_{\gamma}$ on the boundary of the large disc (figure \ref{fig:Betti_local_to_global}-(b)). \\
For all the $\gamma \in \mathbb{C}^{\ast}$ that do not belong to the Stokes rays, defined below, the retraction of the complement of the big disc with respect to the union of the small discs and the paths $p^{(i)}_{\theta_{\gamma}+\pi}$ gives rise to the Betti local to global isomorphism:
\begin{equation}
    \varphi_{Betti} \, \, : \, \,  \,  \mathcal{H}^{\bullet}_{LB} (X,f) \stackrel{\sim}{\longrightarrow} \mathcal{H}^{\bullet}_{GB} (X,f).
    \label{Betti:global_to_local}
\end{equation}

\noindent
We call $\varphi_{\theta}$ the restriction of this isomorphism along a specific ray $\theta$ in $\mathbb{C}^{\ast}_{\gamma}$. We can think of it as generated by the embedding of a neighborhood of the ray (intersected with $B$) in $B$.\\\\

\defn[Stokes ray]{
We call the ray $s_{\theta}=\left\lbrace \gamma \vert \arg(\gamma)= \pi-\theta_{ij} = \theta \right\rbrace= \mathbb{R}_{\geq 0} \cdot (t_i-t_j)^{-1} \subset \mathbb{C}_{\gamma}$ with $\theta_{ij}= \arg(t_j-t_i)$ a Stokes ray.\\
Rays with vertex at the origin that are not Stokes rays are called generic rays. }

Notice that there can be more copies $t_a,t_b\in \C_t$ of critical points such that $\arg(t_a-t_b)=\arg(t_i-t_j)$. All these copies give the same Stokes ray.
 Whenever $\gamma$ lies on the Stokes ray of slope $\pi-\arg(t_i-t_j)$ in the plane $\C_\gamma$, the corresponding line $l^{(i)}_{arg(t_i-t_j)}$ in the $\C_{\gamma t}$-plane, used to construct the Betti isomorphism $\varphi_{Betti}$, passes through both points $\gamma t_i$ and $\gamma t_j$ before reaching the boundary of the large disc. For all other points nothing special happens.  
Therefore, we see that for each Stokes ray $s_{\theta}$ (namely a Stokes ray with slope $\theta$), we have an isomorphism $T_{\theta}$ among the graded abelian groups $H^{\bullet}_{LB, \gamma}(X,f)$ with $\arg(\gamma)$ sufficiently close to $\theta$.  Concretely, choose a sector in the $\gamma$-plane with boundary rays at angles  $\theta_{\pm}= \theta \pm \epsilon$. These two rays lift to sectors in the $\mathbb{C}_{\gamma t}$-plane whose retraction paths avoid all but one critical value, so for each of those values the two deformations give homotopic maps and hence the same identification of local Betti groups.\\
However, when $\theta= \pi - \arg(t_i-t_j)$, the corresponding rays in $\mathbb{C}_{\gamma t}$ pass through both $\gamma t_i$ and $\gamma t_j$. In that case the edges cannot be deformed into one another without crossing the line $l^{(i)}_{\theta_{ij}}$.  The required isomorphism for the jump, given by $\varphi_{\theta^-}^{-1} \circ \varphi_{\theta^+}$, is implemented by the operator:
\begin{equation}
    T_{\theta} = \mathbb{1} + \sum_{\begin{array}{c}
       i \neq j   \\ \arg(t_i-t_j)= -\theta 
    \end{array}} T_{ij},\label{Tij}
\end{equation}
where
\begin{equation}
    T_{ij} \, \, : \,\ \, H^{\bullet} \left( D (t_i, \epsilon), t_{\theta_{ij}}; \mathbb{Z} \right) \quad \longmapsto \quad H^{\bullet} \left( D(t_j, \epsilon), t_{\theta_{ij}}; \mathbb{Z}\right).
\end{equation}
With this isomorphism we can glue the Local System $\mathcal{H}_{LB} (X,f)$ across the Stokes rays. Equipped with these Stokes automorphisms, the local system $\mathcal{H}_{LB} (X,f)$ over the circle $S^1 = \left\lbrace \vert \gamma \vert = \mathrm{const} \right\rbrace$ provides a concrete example of an analytic wall-crossing structure.
To this point we have introduced four cohomology theories, de Rham and Betti in their global and local versions, each pair related by its own isomorphism. In the statements that follow, we will establish the comparison isomorphisms between the de Rham and Betti frameworks.\\\\

\defn[Exponential period map]{
The integration over cycles defines a non-degenerate pairing

\begin{equation}
   \int \, \, : \,\,  H_{\bullet}^{GB} \left( (X,D_0), f \right) \, \otimes  \, H^{\bullet}_{GdR} \left( (X, D_0), f \right) \quad \longmapsto \quad \mathbb{C},
   \label{GlobalPairing}
\end{equation}
called Exponential Period Map.}

From this pairing, for each $\gamma \in \mathbb{C}^{\ast}_{\gamma}$, we can construct the following isomorphism
\begin{equation}
    \varphi_{\gamma} \, \, : \, \, H^{\bullet}_{GdR, \gamma} (X,f) \, \simeq \, H^{\bullet}_{GB, \gamma} (X,f) \otimes \mathbb{C}.
    \label{isomorphism:dR_Betti_global}
\end{equation}

Then we can refer to the integrals \eqref{ExpIntegralgamma} as exponential periods of de Rham cocycles over Betti cycles. \\
Finally, by promoting the Local Systems over $\mathbb{C}_{\gamma}^{\ast}$ to vector bundles with connection $\nabla_{\tau=1/\gamma}$, we have the following local version of the isomorphism \eqref{isomorphism:dR_Betti_global}

\begin{equation}
    RH^{-1}_{loc} \left( \mathcal{H}^{\bullet}_{LB} ((X,D_0),f) \otimes \mathbb{C} \right) \, \simeq \, \mathcal{H}^{\bullet}_{LdR} ((X,D_0), f),
\end{equation}

where $RH^{-1}_{loc}$ is the Riemann-Hilbert inverse functor from the category of Local Systems of complex vector spaces to the category of regular singular connections of vector spaces over $\C[[t]]$. 

\subsection{WCS for Exponential Integrals} \label{Sec:WCS}
\begin{figure}[h!]
    \centering
    \includegraphics[width=0.9\linewidth]{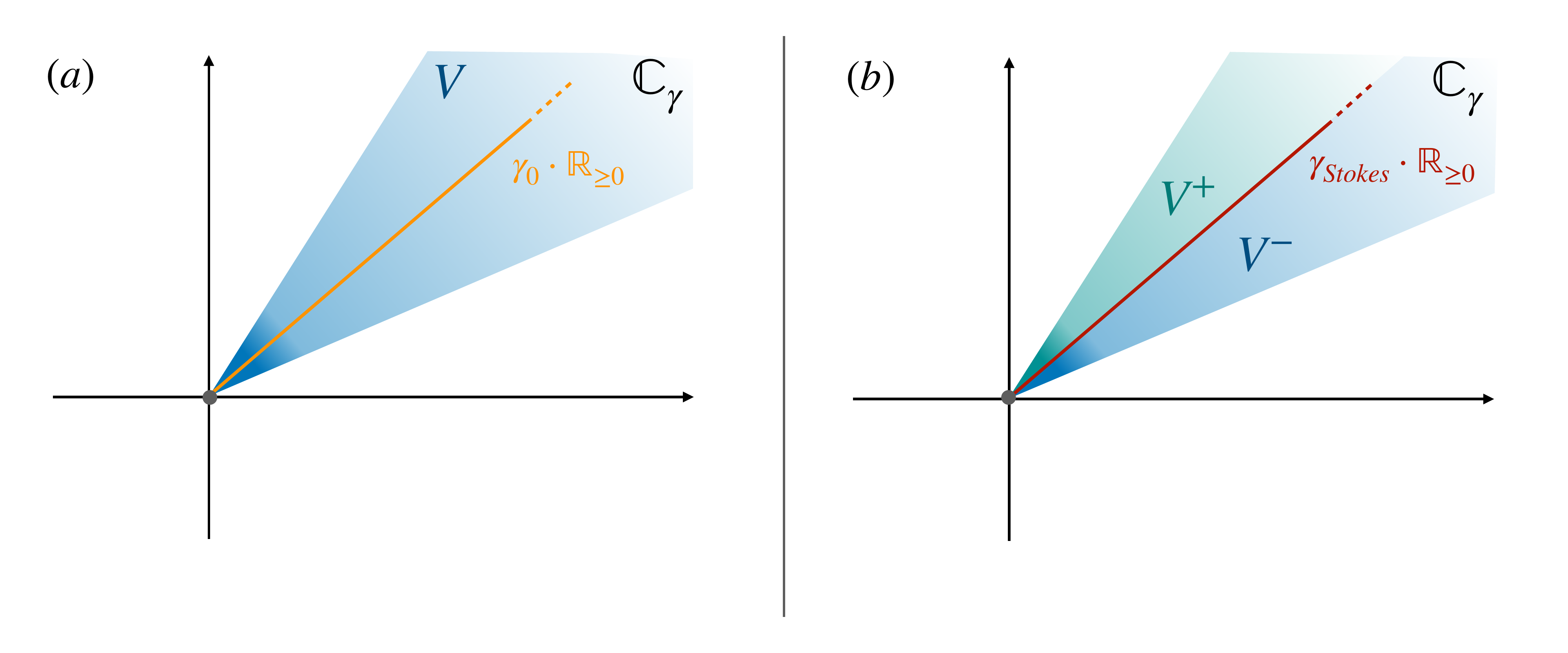}
    \caption{\small{(a) Sector $V$ in the complex plane $\mathbb{C}_{\gamma}$ containing the generic ray $\gamma_0 \cdot \mathbb{R}_{\geq 0}$. (b) Sectors $V^+$ and $V^-$ in the complex plane $\mathbb{C}_{\gamma}$ separeted by the Stokes ray $\gamma_{Stokes} \cdot \mathbb{R}_{\geq0}$}}
    \label{fig:sectors}
\end{figure}

One of the main consequences of the generalization of the exponential integral \eqref{ExpIntegral1} to the one-parameter family \eqref{ExpIntegralgamma}, achieved by rescaling the function $f \rightarrow \gamma f$, is the emergence of Stokes phenomena for specific values of the parameter $\gamma \in \mathbb{C}^{\ast}$. For these special values $\gamma^{\ast}$, the number of independent lines $l^{(i)}_{\theta_{\gamma^{\ast}+ \pi}}$ used to construct the Betti local to global isomorphism decreases, leading to discrete changes in the graded abelian groups $H^{\bullet}_{LB, \gamma} (X,f)$. These changes are controlled by wall crossing formulas. In this section we discuss the wall crossing structure for exponential integrals \cite{Kontsevich:2022ana, Kontsevich:2024mks}, which provides a generalization of the $2d$ version used by Cecotti and Vafa in \cite{Cecotti:1992ccv}.\\
Let us fix a region $\mathcal{R}\subset\mathbb{C}_{\gamma}$, such that for any $\gamma\in \mathcal{R} $, the exponential integral \eqref{ExpIntegralgamma} is an analytic function of $\gamma$, depending on $\Gamma$. Notice that if $\mathrm{Supp}(\Gamma)$ is compact, then the region is unrestricted. Otherwise, it is necessary to ensure that $\gamma f$ is bounded from below. In general, if $\gamma_0\in \mathcal R$, then $\R_{\geq 0}\gamma_0 \subset \mathcal R$. \\
If we do not fix the integration cycle but we keep the volume form fixed, we can interpret $I(\gamma)$ as a morphism of sheaves of abelian groups on $\mathbb{C}^{\ast}_{\gamma}$
\begin{equation}
    \begin{split}
        \mathcal{H}_{\bullet}^{GB} \left( (X,D_0),f\right) \, \, \, & \longmapsto \, \, \, \mathcal{O}^{an}_{\mathbb{C}_{\gamma}^{\ast}}\\
        \Gamma \, \, \, & \longmapsto \, \, \, \int_{\Gamma} \, e^{-\gamma f} \, \mu. \\
    \end{split}
    \label{morphism:1}
\end{equation}
\nn
If we choose $\gamma_0$ lying on a generic ray in $\mathbb{C}^{\ast}_{\gamma}$, then, for any $\gamma$ in a small sector $V\subset \mathcal R$ containing the ray $\mathbb{R}_{\geq 0} \cdot \gamma_0$ (see figure \ref{fig:sectors}-(a)), the canonical isomorphism between global and local Betti homologies induced by \eqref{Betti:global_to_local} is well defined and it gives rise to the following morphism among sheaves:
\begin{equation}
    \bigoplus_{t_i \in S} \mathcal{H}^{LB, \gamma, t_i} \left( f^{-1} (V), f \right) \, \, \, \longmapsto \, \, \, \mathcal{O}^{an}_{\mathbb{C}^{\ast}_{\gamma}}(V).
\end{equation}
\noindent
Let us now choose a Stokes ray $s_{\theta}$ and consider a new small sector $V= V^+ \cup V^-$ in the plane $\mathbb{C}^{\ast}_{\gamma}$ containing the ray (see figure \ref{fig:sectors}-(b)). We choose two bases $\left\lbrace \Gamma^+_{(i)} \right\rbrace$ and $\left\lbrace \Gamma^-_{(i)} \right\rbrace$  for the local Betti homology in the sectors $V^+$ and $V^-$, respectively, corresponding to the angles $\theta^+=\arg(\gamma^+ )= \theta + \epsilon$ and $\theta^-=\arg(\gamma^- )= \theta - \epsilon$. With these choices, we can define two vector valued analytic functions:
\begin{equation}
    \overline{I}^+ (\gamma) =\left( \begin{matrix} \int_{\Gamma_{(1)}^+} e^{-\gamma f} \mu \\ \int_{\Gamma_{(2)}^+} e^{-\gamma f} \mu \\ \dots \\ \dots \\ \int_{\Gamma_{(k)}^+} e^{-\gamma f} \mu \end{matrix} \right) \quad \quad \quad \quad \quad \overline{I}^- (\gamma) = \left( \begin{matrix} \int_{\Gamma_{(1)}^-} e^{-\gamma f} \mu \\ \int_{\Gamma_{(2)}^-} e^{-\gamma f} \mu \\ \dots \\ \dots \\ \int_{\Gamma_{(k)}^-} e^{-\gamma f} \mu \end{matrix} \right)
\end{equation}
related by the following wall crossing formulas
\begin{equation}
    \int_{\Gamma^-_{(i)}} e^{-\gamma f} \mu= \int_{\varphi^{\ast}_{\theta_+}(\Gamma^+_{(i)})} e^{-\gamma f} \mu + \sum_{\substack{j \neq i \\ \arg(t_i - t_j) = \theta}}\int_{(\varphi^{\ast}_{\theta^+} \circ T_{ij}) \Gamma^+_{(i)}} e^{-\gamma f} \mu,
    \label{wcf:general}
\end{equation}
where $\varphi^{\ast}$ denotes the dual isomorphism to the one in \eqref{Betti:global_to_local}. Roughly speaking, these formulas describe the analytic continuation of the function $\overline{I}^- (\gamma)$ from the sector $V^-$ to the adjacent sector $V^+$ across the Stokes ray.

In the special case in which $f$ is a Morse function with $k$ different critical points, there is a special basis for the local and global Betti homologies for each direction $\theta=\pi-\arg(\gamma)$ which is the one of Lefschetz thimbles $th_{i, \theta + \pi}$. By definition, $th_{i, \theta + \pi}$ is the union of gradient lines of the function $Re (e^{-i \theta} f)$ emerging from the critical point $\sigma_i$, while $f(th_{i, \theta + \pi})$ is the line with direction $\theta + \pi$ emerging from the critical value $f(\sigma_i)=t_i$. Using this basis of thimbles, we can define the following collection of integrals for any generic direction $\gamma \in \mathbb{C}^{\ast}$, with $\theta= \pi-\arg(\gamma)$, such that $\gamma$ does not lie on a Stokes ray:
\begin{equation}
    I_i (\gamma)= \int_{th_{i, \theta + \pi}} e^{-\gamma f} \mu.
    \label{expInt:thimble}
\end{equation}
\nn Let us suppose to have defined them along the direction $\theta_-=\theta-\epsilon$, where $\theta$ identifies now a Stokes ray. Then, when we move toward $\theta_+=\theta+\epsilon$ through the Stokes line, the integrals undergo a discontinuous jump according to \eqref{wcf:general}
\begin{align}
    I_i\rightarrow I_i+n_{ij}I_j,
\end{align}
where $n_{ij}$ are integers counting the number of gradient trajectories of $\mathrm{Re}(e^{i \theta_{ij}}f)$ joining the critical values $\sigma_i$ and $\sigma_j$. Equivalently, $n_{ij}$ is the intersection index of the opposite thimbles $th_{i,\theta_++\pi}$ and $th_{j, \theta_--\pi}$ emerging from the critical points $\sigma_i$ and $\sigma_j$.\\
As $\gamma \rightarrow \infty$, the integrals \eqref{expInt:thimble} admit a power expansion:
\begin{equation}
    I_i (\gamma) = e^{-\gamma t_i} \sum_{\lambda} c_{i,\lambda} \gamma^{- \lambda -1},
    \label{series:Exp_integral_gamma}
\end{equation}
for some $c_{i,\lambda} \neq 0$. \\
In order to analyze this series let us start isolating the exponential dependence at the critical point
\begin{equation}
   I_i(\gamma)= e^{-\gamma t_i} I_i^{mod} (\gamma)
   \label{Exp_integral:mod}
\end{equation}
and define the new variable
\begin{equation}
    s= f (\mathbf{z}) -t_i.
    \label{s:definition}
\end{equation}
Since the function ${\mathrm{Im}}(e^{-i \theta} f)$  remains constant along the cycle $th_{i, \theta + \pi}$, the variable $s$ ranges over the real interval from zero to infinity. Let us denote by $\Delta_i(s)$ the $(n-1)$-dimensional closed hypersurfaces defined by the level equations $f(\mathbf{z})=s=\mathrm{const}$. These level sets identify the vanishing cycles of the homology group $H_{n-1} \left( f^{-1}(s), \mathbb{Z} \right)$ \cite{Pham:1967} (see section \ref{PLT}). When $\gamma$ does not lie on a Stokes ray, the trace of these vanishing cycles along the variation of $s$ in the range $\left[ 0; + \infty \right)$ span the thimble
\begin{equation}
    th_{i, \theta + \pi} = \bigcup_{s \geq 0} \Delta_i (s).
    \label{family:vanish_cycles}
\end{equation}
Using the Gelfand-Leray form $\frac{\mu}{ds} \Big\vert_{\Delta_i (s)}$, the exponential integral \eqref{Exp_integral:mod} can be rewritten as
\begin{equation}
    I_i^{mod}(\gamma) = \int_{0}^{\infty} ds  e^{- \gamma s} vol_{\Delta_i} (s),
\end{equation}
where 
\begin{equation}
    vol_{\Delta_i} (s) = \int_{\Delta_i (s)} \frac{\mu}{ds} \Big\vert_{\Delta_i (s)}
    \label{volume:vanishing_cycle}
\end{equation}
denotes the volume of the $(n-1)-$dimensional vanishing cycles $\Delta_i(s)$ in the family \eqref{family:vanish_cycles}. Note that the modified integral $I^{mod}_i(\gamma)$ can be interpreted as the Laplace transform of $vol_{\Delta_i}(s)$. On the other hand, the function $vol_{\Delta_i}(s)$ can be read as the pairing between the holomorphic cohomology class $\left[ \frac{\mu}{ds}\right] \in H^{n-1} \left( f^{-1}(s), \mathbb{C}\right)$ and the homology class $\left[ \Delta_i \right] \in H_{n-1} \left( f^{-1}(s), \mathbb{Z} \right)$. According to the resolution of singularities theorem (see for example \cite{Arnold88}), this function admits an absolutely convergent power series expansion for $0 < s \ll \varepsilon$ of the form:
\begin{equation}
    vol_{\Delta_i}(s) = \sum_{\lambda } \sum_{ 0 \leq k \leq k_{max}} a_{\lambda,k} s^{\lambda} \log (s)^k.
     \label{expansion:vol_Delta}
\end{equation}
The numbers $\lambda$ correspond to the eigenvalues $e^{2 \pi i \lambda}$ of the monodromy operator $M_i$ acting on $H_{n-1}(f^{-1}(s), \mathbb{Z})$ when we turn around the singularity $s=0$, while the integer $k_{max}+1$ determines the size of the largest Jordan block associated with that eigenvalue. 
Taking the total differential of the definition \eqref{s:definition}, we obtain
\begin{equation}
    ds = \frac{\partial f}{\partial z^1} dz^1 + \frac{\partial f}{\partial z^2} dz^2 + \dots + \frac{\partial f}{\partial z^n} dz^n.
\end{equation}
This expression shows that the function $vol_{\Delta_i}(s)$ develops potential singularities in the complex $s$-plane whenever all partial derivatives of $f$ vanish simultaneously, that is, at the critical points of $f$. Consequently, a series expansion of $vol_{\Delta_i} (s)$ in powers of $s$ will have a radius of convergence determined by the distance to the nearest singularity on the same Riemann sheet. \\
Substituting the expansion \eqref{expansion:vol_Delta} into the exponential integral \eqref{Exp_integral:mod}, and using the following identity:
\begin{equation}
    \int_{0}^{\infty} e^{- \gamma s} s^{\lambda} (\log s)^k ds = \frac{d^k}{d \lambda^k} \left[ \gamma^{-(\lambda +1)} \Gamma \left( \lambda + 1\right) \right]
\end{equation}
we obtain
\begin{equation}
     I_{i} (\gamma) \,= e^{- \gamma t_i} \sum_{\lambda} \sum_k a_{\lambda, k} \int_{0}^{\infty} ds e^{- \gamma s} s^{\lambda} (\log s)^k = e^{- \gamma t_i} \sum_{\lambda} \sum_k a_{\lambda, k} \frac{d^k}{d \lambda^k} \left[ \gamma^{-(\lambda +1)} \Gamma \left( \lambda + 1\right) \right]. 
\end{equation}
Comparing this result with the power series in \eqref{series:Exp_integral_gamma}, we have
\begin{equation}
    c_{i, \lambda} = \frac{1}{\gamma^{-(\lambda+1)}}\sum_k a_{\lambda,k} \frac{d^k}{d \lambda^k} \left[ \gamma^{- (\lambda +1)} \Gamma \left( \lambda +1 \right) \right].
    \label{coefficients:c_series_MIs}
\end{equation}
One of the main advantages of constructing the $\gamma$-expansions of the same integral evaluated in the thimble basis is that it enables a direct comparison among the various integrals $I_i (\gamma)$(the Master Integrals) associated with different homology classes of integration contours within the same sector of the $\mathbb{C}_{\gamma}$ plane. In some sectors, certain thimbles may dominate over the others.


\section{Holomorphic Morse theory }\label{PI}

\begin{figure}[h!]
    \centering
    \includegraphics[width=1\linewidth]{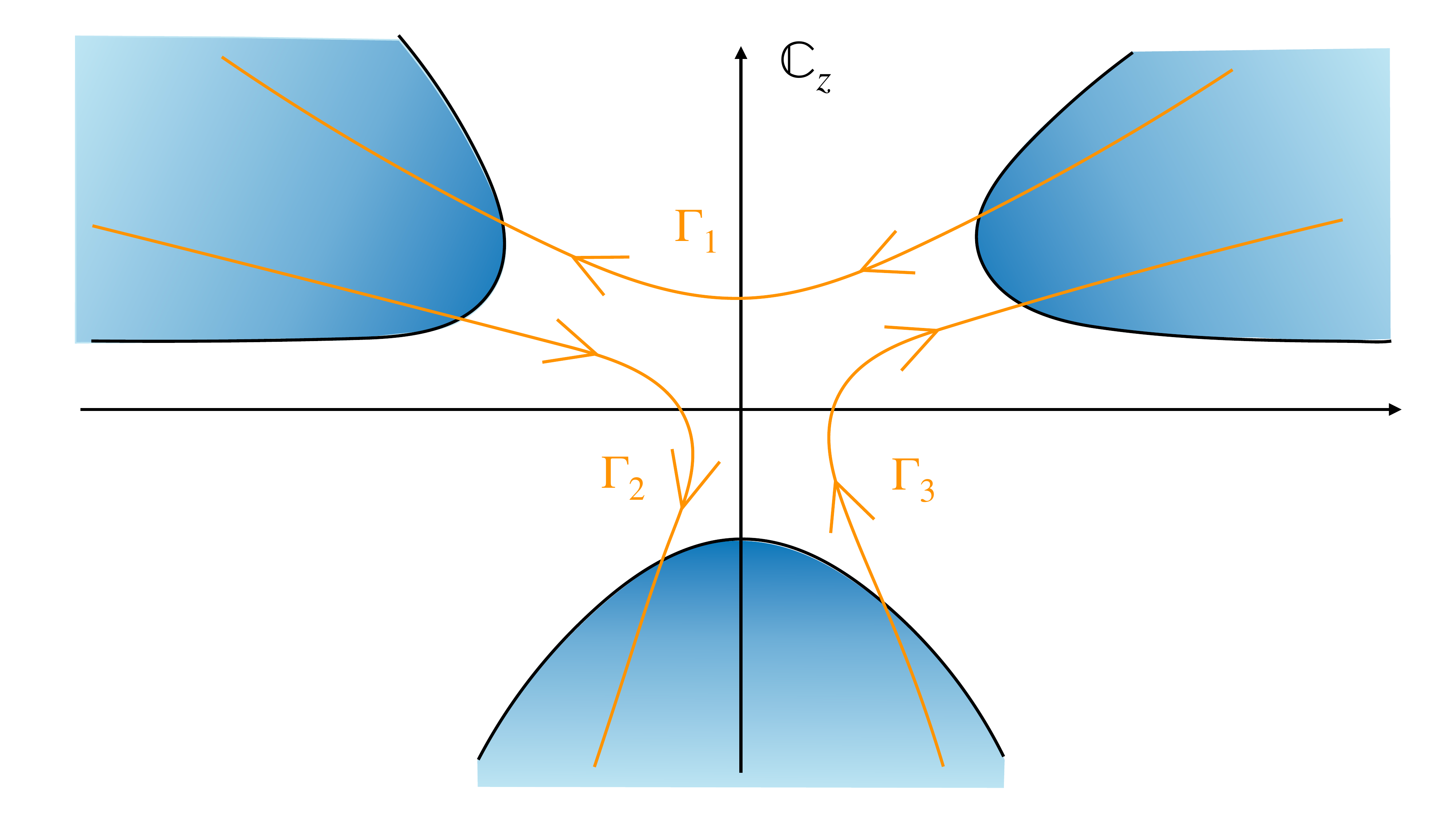}
    \caption{\small{The blue areas in the complex plane represent the regions $D_N$ for $f(\mathbf{z}=Ai(\mathbf{z}))$ and large $N$. Despite $X= \mathbb{C}$ is contractible, $D_N$ can be the union of different disjoint pieces. The $1-$cycles $\Gamma_i \in H_1 (X, D_N , \mathbb{Z})$ must connect distinct components of $D_N$.  }}
    \label{fig_th1}
\end{figure}

We want now to present a concrete application of the formalism developed above, accompanied by a discussion of its connection to holomorphic Morse theory. This connection not only provides deeper geometric insight into the structure of the theory but also clarifies the role of the abstract objects we introduced in the construction of Betti cohomology.\\
In order to build geometric intuition and develop familiarity with the setup, we now focus on the case $X\cong \C^n$ and consider exponential integrals of the form
\begin{equation}
    I(f,\gamma)=\int_\Gamma e^{-\gamma f(\mathbf{z})}g(\mathbf{z})d^n\mathbf{z},
    \label{ExpIntegral3}
\end{equation}
where $f(\mathbf{z})$ is a holomorphic function and $g (\mathbf{z}) d^n \mathbf{z}$ is a holomorphic $n$-form on $X$.
The aim of the procedure is the one to provide a basis for the integration contours, for any $\gamma \in \mathbb{C} \setminus \left\lbrace 0 \right\rbrace$, such that the integral \eqref{ExpIntegral3} converges. As $\gamma$ varies over its domain, the admissible integration contours $\Gamma$ must be deformed accordingly to ensure convergence.\\
Let us define the set $D_N$ in $X$ as

\begin{equation}
    D_N = \left\lbrace \mathbf{z} \in \mathbb{C}^n \vert \mathrm{Re}(\gamma f(\mathbf{z})) \geq N \right\rbrace,
\end{equation}
\nn
for $N \in \R$, with $ |N|>>1$. This subset of $X$ consists in general of different disconnected components (an illustrative example is given by the blue regions in Figure \ref{fig_th1}). Any reasonable cycle $\Gamma$ for \eqref{ExpIntegral3} should connect two distinct regions of this subset, namely it should be a non-compact $n-$cycle of $X$ with boundaries in $D_N$, i.e. an element of the relative homology $H_n \left( X, D_N , \mathbb{Z} \right)$.
\nn
The condition on the boundaries is just part of the requirements that our integration contours have to satisfy. To ensure that the integrals are well-behaved, we must also impose conditions on the portions of the cycles extending into the complementary region $X \setminus D_N$. In particular, the cycles must avoid regions of $X$ where $\mathrm{Re}(\gamma f (\mathbf{z}))\rightarrow - \infty$,  as such behavior would lead to divergence. Furthermore, to prevent oscillations, we must impose the condition that $\rm{Im}(\gamma f (\mathbf{z}))$ remains constant along $\Gamma$, ensuring that we can factor out the phase $e^{i c}$ and reduce the problem to a real-valued integral. \\
The techniques described in Section \ref{WCSsection} provide a systematic method for analyzing the cycles in this relative homology, constructing a basis for them, and defining a well-behaved intersection pairing. 

\subsection{Relative Homology}\label{sec:the relative homology}
Let $H_k \left( X, D_N, \mathbb{Z}\right)$ be the $k-$homology group of $X$, on $\Z$, relative to $D_N$. The elements of this group, called relative cycles, are equivalence classes of $k$-chains in $X$ whose boundaries lie in $D_N$, modulo those chains that are homologous to chains entirely contained in $D_N$. Notice that, in the limit $N \rightarrow + \infty$, this homology group corresponds, up to Poincaré duality as given in \eqref{Betti:Poincare_duality}, to the Betti homology group defined in \eqref{Def:Betti_global_gamma}, with $D_0= \emptyset$.  
By applying the constructions outlined in Section \ref{WCSsection} we can determine the dimension of this relative homology group and construct an explicit basis for it.  In doing so, we recover the same geometric objects that arise in Morse theory, which analyzes the topology of $X$ by studying the properties of the differential functions defined on it. \\
In the present case, the function we will use to carry out the analysis is the height function 
\begin{equation}
    h\equiv \mathrm{Re}(\gamma f(\mathbf{z})).
\end{equation}
\nn
The set $\Sigma$ of critical points of this function $h$ coincides with the one of $f$ because of the Cauchy-Riemann equations. A critical point is said to be non-degenerate if the Hessian matrix associated to $h$ in that point is invertible. If all critical points are non-degenerate, the height function is a well-defined Morse function.
 The number of negative eigenvalues of the Hessian equips critical points of an index, called Morse index. For non-degenerate holomorphic functions on complex manifolds of complex dimension $n$, the Morse indices are all equal to $n$. Consequently, the Betti inequalities, which provide lower bounds on the dimensions of the homology groups, are saturated
\begin{equation}
       rank[H_k (X,D_N,\R)]= \begin{cases} 0,\quad k<n,\\ \texttt{\#}\Sigma,\quad k=n.
   \end{cases}
   \label{BettiRelation}
\end{equation}
 This provides a direct way to compute the dimension of $H_n(X,D_N, \mathbb{Z})$. Let us now determine a basis for this group.
 A Morse function $h$ such that \eqref{BettiRelation} holds true, is called \textit{perfect}. It can be proven that a sufficient and easily checkable condition for $h$ to be perfect is that the differences between the indices of distinct critical points never equal $\pm 1$.\\
If $h$ is perfect, Morse theory provides a way to construct a relative $n-$cycle $\Gamma_i$ for each critical point in $\Sigma$. Let us make the simplifying assumption that all the critical points $\sigma_i \in \Sigma$ are isolated points in $X$, and the corresponding distinct critical values form the set
\begin{equation}
 S=\left\lbrace t_i \in \mathbb{C} \vert \, f(\sigma_i)=t_i\right\rbrace.
\end{equation}

\begin{figure}[h!]
    \centering
    \includegraphics[width=0.8\linewidth]{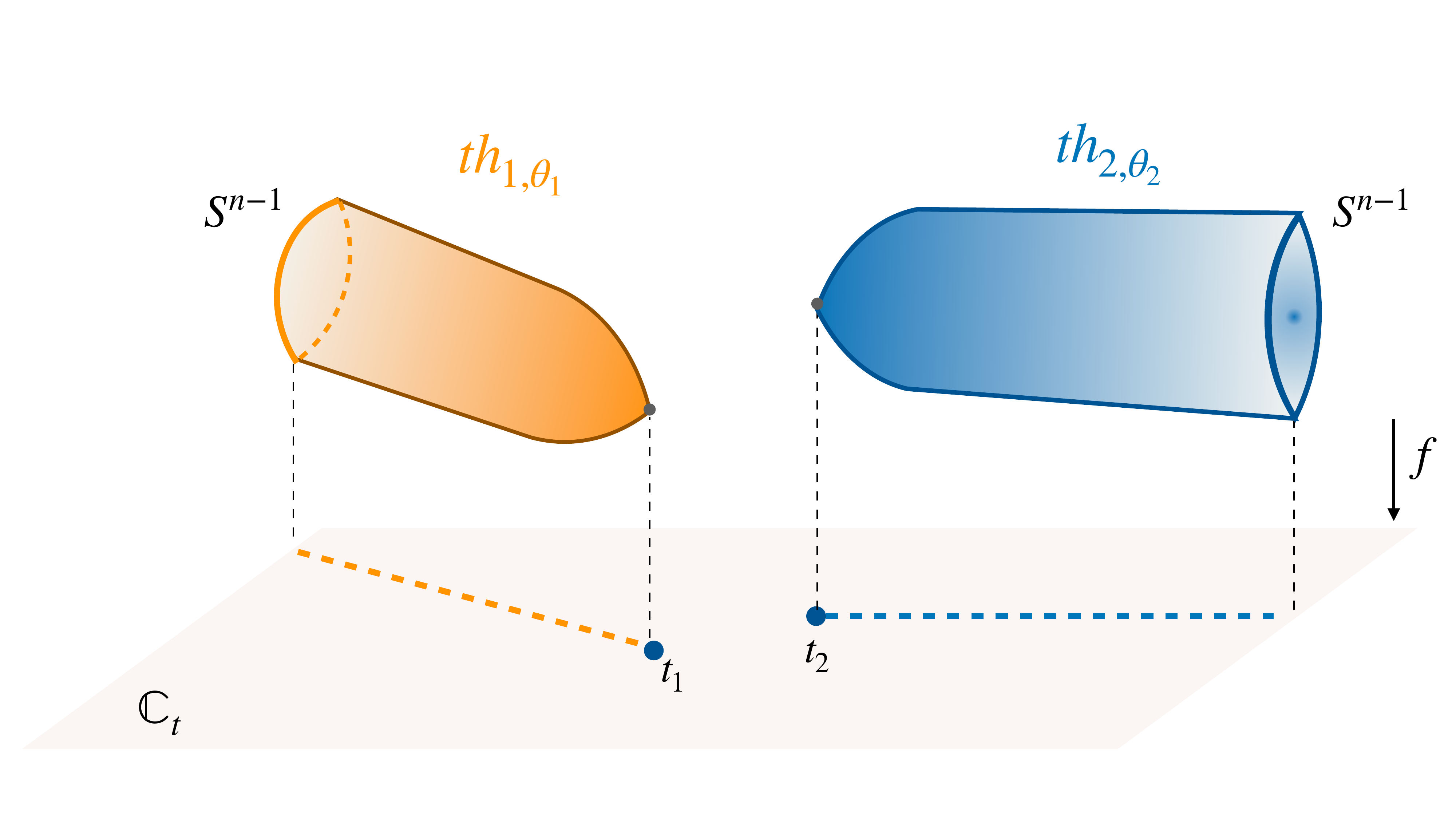}
    \caption{\small{Pictorial representation of two thimbles: $th_{1,\theta_1}$ and $th_{2, \theta_2}$ constructed as continuations of the vanishing cycles $\Delta_i$ and $\Delta_j$, diffeomorphic to $S^{n-1}$, along the paths $t_1 + e^{i \theta_1} \mathbb{R}_{\geq 0}$ and $t_2 + e^{i \theta_2} \mathbb{R}_{\geq 0}$ in $\mathbb{C}_t$.}}
    \label{fig_th2}
\end{figure}

For each critical point $\sigma_i \in X$, there is a unique vanishing cycle $\Delta_i \subset f^{-1}(t_i)$ diffeomorphic to $S^{n-1}$. The explicit method to construct these cycles is reported in example \ref{A3}.  Moreover, for each critical point $\sigma_i$ and a generic direction $\theta \in \mathbb{R} / 2 \pi \mathbb{Z}$, we can construct the Lefschetz thimble $th_{i, \theta} \sim \mathbb{R}^n \subset X$ as the continuation of the vanishing cycle along the path $t_i + e^{i \theta} \mathbb{R}_{\geq 0} \subset \mathbb{C}_t$ (see Figure \ref{fig_th2} for a conceptual visualization), ill-defined only if $\theta = \arg (t_j-t_i)$ for some $t_j \neq t_i$.\\
Among these thimbles, we aim to select a basis for the relative homology. This is achieved by considering the continuations of vanishing cycles along special paths of the form $t_i + e^{i \theta_i} \mathbb{R}_{\geq 0}$ in $ \mathbb{C}_t$ which start from the critical points $t_i$ and reach $t= \infty$ while maintaining a constant phase $\theta_i = \rm{Im}(\gamma f(t_i))$. These paths are solutions of the gradient flow equations:
\begin{equation}
\frac{du^i}{dt} =+ g^{ij} \frac{\partial h}{\partial u^j},
\label{eq:gradient_flow_step-up}
\end{equation}
where $u^i$ are real coordinates on $X$ and $g_{ij}$ is a Riemannian metric on $X$. These paths define the steepest ascent Lefschetz thimbles $\Gamma_i^+$, which have the key properties that the function $h$  increases monotonically along them, and, if $h=\mathrm{Re}(\gamma f)$ for a holomorphic function $f$, then the imaginary part $\rm{Im}(\gamma f)$ remains constant along the thimble. If $\Gamma_i^+$ contains exactly one critical point, it corresponds to a good Lefschetz thimble. Otherwise, it is referred to as a \textbf{Stokes line}\footnote{Note that we use the term "Stokes rays" to refer to the semi-infinite lines in the complex plane $\mathbb{C}_{\gamma}$  as established in the definition from Section \ref{sec:betti}. Instead, we use the term "Stokes lines" to denote Lefschetz thimbles in $\mathbb{C}^n$ that contain more than one critical point. Each time $\gamma \in \mathbb{C}_{\gamma}$ lies on a Stokes ray, Stokes lines appear in $\mathbb{C}^n$ as a manifestation of the associated Stokes phenomenon.}. Assuming no Stokes lines are present in our set of thimbles, the number of thimbles exactly matches the rank of the relative homology group. To prove that they indeed generate the homology group, namely that they are independent cycles, we need to establish a method for uniquely decomposing any element $\Gamma \in H_n (X, D_{N}, \mathbb{Z})$  as a linear combination of the form
\begin{equation}
    \Gamma = \sum_{\sigma \in \Sigma} n_{\sigma} \Gamma^+_{\sigma}.
    \label{path_decomposition}
\end{equation}
Such a decomposition exists, and the coefficients $n_{\sigma}$ that appear on it are integer numbers representing the intersection between the cycle $\Gamma$ and the basis of steepest descent Lefschetz thimbles $\Gamma_i^-$ which span to the dual homology group $H_n(X,D_{-N}, \mathbb{Z})$, where  $D_{-N}= \left\lbrace  \mathbf{z} \in X \vert \mathrm{Re}(\gamma f(z)) \leq -N\right\rbrace$ with $N \in \mathbb{R}$ taken to be sufficiently large.
The new thimbles $\Gamma_i^-$ are solutions of the gradient flow equations with opposite sign:

\begin{equation}
\frac{du^i}{dt} =- g^{ij} \frac{\partial h}{\partial u^j}.
\end{equation}

They represent the downward-flowing cycles associated with each critical point $\sigma_i \in \Sigma$. These cycles retain the property that  $ \rm{Im}(\gamma f(z))$ remains constant along them, while the function $h$ decreases monotonically.  In the absence of Stokes rays, the intersection pairing is given by:

\begin{equation}
    \langle \Gamma^+_i , \Gamma^-_j \rangle = \delta_{ij}.
    \label{pairing:up_down_thimbles}
\end{equation}

It is straightforward to evaluate this formula for perfect Morse functions with no flows between distinct critical points. Indeed, if there are no flows between two distinct points $ \sigma_i, \sigma_j \in \Sigma$, the corresponding Lefschetz thimbles do not intersect. This is because they are associated with different constant values of the phase $\rm{Im}(\gamma f)$, which remains constant along each thimble. Conversely, the thimbles $\Gamma^{\pm}_{\sigma}$ follow paths along which the function $h$ is monotonically increasing or decreasing. As a result, they intersect exactly once, at the critical point  $\sigma$ itself. 

Therefore, a generic integration contour $\Gamma \in H_n(X,D_{N}, \mathbb{Z})$  can be decomposed in terms of the paths $\Gamma^+_i$ for generic $\gamma \in \mathbb{C}^{\ast}$, away from a Stokes ray, as in \eqref{path_decomposition}, with coefficients uniquely determined by

\begin{equation}
    n_{\sigma} = \langle \Gamma, \Gamma^-_{\sigma} \rangle.
\end{equation}

They count, with appropriate orientation, the number of downward flows from each critical point $\sigma$ to $\Gamma$. \\
Let us consider the example where $\gamma$ is purely imaginary, and the function $f$ is given by the quotient $P_1 (\mathbf{z})/P_2(\mathbf{z})$, where $P_1$ and $P_2$ are polynomials with real coefficients. We take the integration cycle $\Gamma=\Gamma_{\mathbb{R}}$ to be the product of $n$ real lines in $\mathbb{C}^n$. Let us partition the set of critical points $\Sigma$ into three subsets:

\begin{equation}
    \Sigma = \Sigma_{\mathbb{R}} + \Sigma_{\leq 0} + \Sigma_{> 0},
\end{equation}

where $\Sigma_{\mathbb{R}}$ denotes the set of critical points lying on the real axis, $\Sigma_{\leq 0}$ consists of critical points off the real axis for which the associated value of $h$ satisfies $h \leq0$, and $\Sigma_{> 0}$ includes those off the real axis for which $h > 0$. \\
For critical points lying on the real line, since $\gamma \in \mathrm{Im}$, the function $h = \mathrm{Re}(\gamma f)$, vanishes identically. In particular, we have $h_{\sigma}=0$ for all $\sigma \in \Sigma_{\mathbb{R}}$. Because $h$ strictly decreases along downward gradient flows, there can be no such flows starting at 
$\sigma \in \Sigma_{\mathbb{R}}$ that remain on the real line. Consequently, the only intersection between the downward Lefschetz thimble $\Gamma^-_{\sigma}$ and the real cycle $\Gamma_{\mathbb{R}}$ is the point $\sigma$ itself:

\begin{equation}
    \sigma \, \in \, \Sigma_{\mathbb{R}} \quad \quad \Longrightarrow \quad \quad n_{\sigma}= \langle \Gamma_{\mathbb{R}}, \Gamma^-_{\sigma} \rangle = 1.
\end{equation}

If $\sigma \in \Sigma_{\leq 0}$,  no downward flows originating from $\sigma$ intersect $\Gamma_{\mathbb{R}}$. This follows from the fact that $h$ is strictly decreasing along downward flows, and by definition, $h \leq 0$ for points in $\Sigma_{\leq}$. Thus, we have:

\begin{equation}
    \sigma \, \in \, \Sigma_{\leq 0} \quad \quad \Longrightarrow \quad \quad n_{\sigma}=\langle \Gamma_{\mathbb{R}}, \Gamma^-_{\sigma} \rangle =0. 
\end{equation}

Finally, if $\sigma \in \Sigma_{>0}$,  it is in principle possible for downward flows originating from $\sigma$ to intersect the real section $\Gamma_{\mathbb{R}}$. The precise number of such intersections depends on the specific geometry of the function and must be determined case by case. Altogether, we obtain the decomposition:

\begin{equation}
    \Gamma_{\mathbb{R}} = \sum_{\sigma \in \Sigma_{\mathbb{R}}} \Gamma^+_{\sigma} + \sum_{\sigma\in \Sigma_{>0}} n_{\sigma} \Gamma^+_{\sigma}.
\end{equation}

\subsection{Stokes rays}
In the previous paragraph, we described the relative homology $H_n (X,D_{N}, \mathbb{Z})$, constructed a basis of thimbles for it, and defined an intersection pairing with the dual homology $H_n (X, D_{-N}, \mathbb{Z})$ to express a generic cycle $\mathcal{C} \in H_n (X,D_{N}, \mathbb{Z})$ in terms of this basis. However, as we explained in Section \ref{sec:betti}, the presence of Stokes rays affects the well-definedness of certain Lefschetz thimbles, making the previous construction insufficient. In this section, we explain why some Lefschetz thimbles become ill-defined in the presence of Stokes phenomena and how the framework introduced earlier can be adapted to restore consistency.\\
The key perspective we adopt is to construct a description of the homology group $H_n \left( X, D_{N}, \mathbb{Z}\right)$ that ensures a well-defined pairing \eqref{pairing:up_down_thimbles} for any value of $\gamma \in \mathbb{C}_{\ast}$. This is achieved by first establishing the structure for a specific $\gamma$ where no Stokes rays appear, as we did in the previous section, and then extending it across the entire $\mathbb{C}_{\ast}$. 

\begin{figure}
    \centering
    \includegraphics[width=1\linewidth]{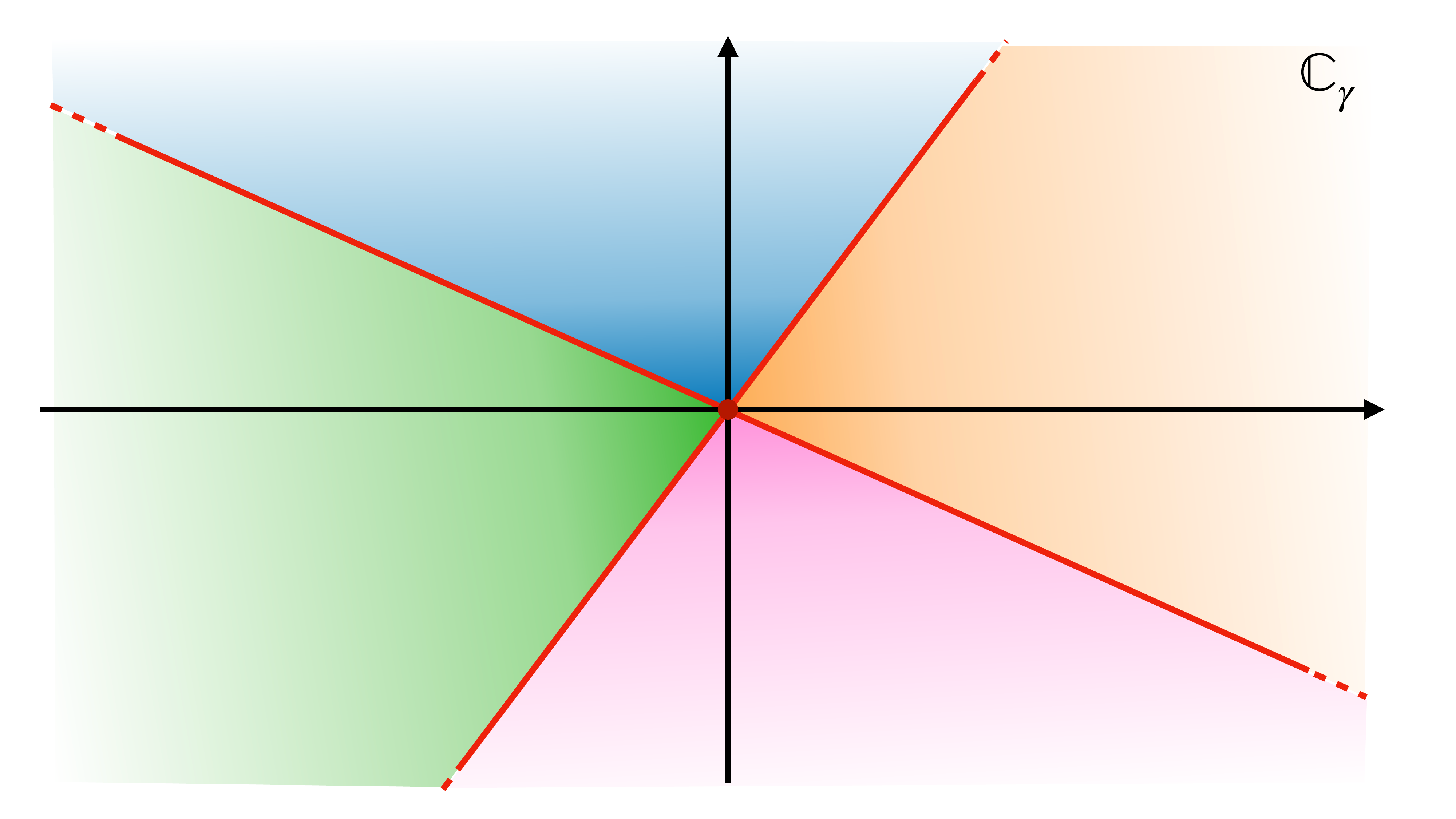}
    \caption{\small{ The $\mathbb{C}_{\gamma}$ plane: The red lines indicate the Stokes lines, across which discontinuities arise in the definition of the Lefschetz thimble structure for the corresponding exponential integrals. The differently colored regions represent the distinct petals of the fan.}}
    \label{fig_petals}
\end{figure}

Stokes lines are solutions of the gradient flow equation \eqref{eq:gradient_flow_step-up} that cross at least two critical points of the function $\gamma f(\mathbf{z})$. Since the imaginary part of this function is preserved along the flows, we have
\begin{equation}
    \rm{Im}(\gamma f(\mathbf{z})) \, \vert_{\sigma_i} = \rm{Im}(\gamma f(\mathbf{z})) \, \vert_{\sigma_j} 
    \label{StokesRaysFormula}
\end{equation}
for any point in the thimbles $\Gamma^+_{i}$ and  $\Gamma^+_{j}$. Moreover, since the number of critical points is finite, there can only be a finite number of Stokes lines. By assumption, $\gamma f(\mathbf{z})$ takes distinct values at different critical points, meaning that Stokes lines appear only for specific values of  $\gamma$. For  $\sigma_i \neq \sigma_j \, \in \, \Sigma$, the loci 
\begin{equation}
    l= \left\lbrace  \gamma \,  \in \, \mathbb{C}_{\ast} \,\ \vert\ \, \rm{Im}(\gamma f(\mathbf{z})) \, \vert_{\sigma_i} = \rm{Im}(\gamma f(\mathbf{z})) \, \vert_{\sigma_j} \right\rbrace
\end{equation}
in the complex plane $\mathbb{C}_{\gamma}$, define regions where the Lefschetz thimble structure undergoes discontinuities: they are the Stokes rays discussed in Section \ref{sec:betti}. These rays always pass through the origin; however, since $\left\lbrace 0 \right\rbrace \notin \mathbb{C}_{\gamma}$, they remain disconnected and form straight lines radiating outward from the center. As a result, the complex $\gamma$-plane  is divided into a fan-like structure composed of distinct sectors—referred to as petals in Figure \ref{fig_petals}. 
\nn
At this point, the procedure is to fix $\gamma$ within a specific petal of the fan, say the zeroth region $(0)$, away from any Stokes rays, and define the Lefschetz thimble structure for the corresponding integral. We can then vary $\gamma$ along the complex plane. As we cross a Stokes ray $s_{\theta}$, associated with a Stokes line between the critical points $\sigma_i $ and $\sigma_j$, for which $h_{\sigma_i} < h_{\sigma_j}$, the corresponding thimbles $\Gamma^+_{i}$ and $\Gamma^+_{j}$ undergo a discontinuous jump to the adjacent region $(1)$ of the form: 
\begin{equation}
    \left( \begin{matrix}   \Gamma^{+(1)}_{i} \\ \Gamma^{+(1)}_{j} \end{matrix} \right) = \left( \begin{matrix} 1 & \Delta_{ij} \\ 0 & 1 \end{matrix} \right) \, \left( \begin{matrix}   \Gamma^{+(0)}_{i} \\ \Gamma^{+(0)}_{j} \end{matrix} \right) \quad ,\quad \text{for } \, \, h_{\sigma_i} < h_{\sigma_j}
    \label{equation:jump}
\end{equation}
\nn
where the integers $\Delta_{ij}$ receive a contribution $\pm 1$ for each upflow line from $\sigma_i$ to $\sigma_j$, with the sign depending on cycles orientation and on the direction from which $\gamma$ crosses the Stokes line.  This is nothing but the intersection number of the corresponding vanishing cycles expressed by the Picard-Lefschetz formula \eqref{PicardLefschetzFormula}, up to a sign depending on the relative orientation of the cycles, when we cross the cut line starting from $t_j=f(\sigma_j)$ in the plane $\mathbb{C}_t$:
\begin{equation}
    \Delta_{ij} = (\pm 1) \Delta_i \circ \Delta_j.
     \label{TotalIntersection}
\end{equation}
This means that the new thimble $\Gamma_i^{+(1)}$ in the region $(1)$ is associated to the new vanishing cycle
\begin{equation}
    \Delta_i^{(1)}= \Delta_i^{(0)} \pm \left( \Delta_i \circ \Delta_j \right) \Delta_j^{(0)}.
\end{equation}
\nn
In order for the decomposition \eqref{path_decomposition} to be continuous, the coefficients $n_{\sigma_i}$ and $n_{\sigma_j}$ transform across the ray by 

\begin{equation}
    \left( \begin{matrix}   n_{\sigma_i} \\ n_{\sigma_j} \end{matrix} \right) \quad \longmapsto \quad  \, \left( \begin{matrix} 1 & 0 \\ - \Delta_{ij} & 1 \end{matrix} \right) \, \left( \begin{matrix}   n_{\sigma_i}  \\ n_{\sigma_j} \end{matrix} \right) \, .
\end{equation}

\begin{figure}[h!]
    \centering
    \includegraphics[width=1 \linewidth]{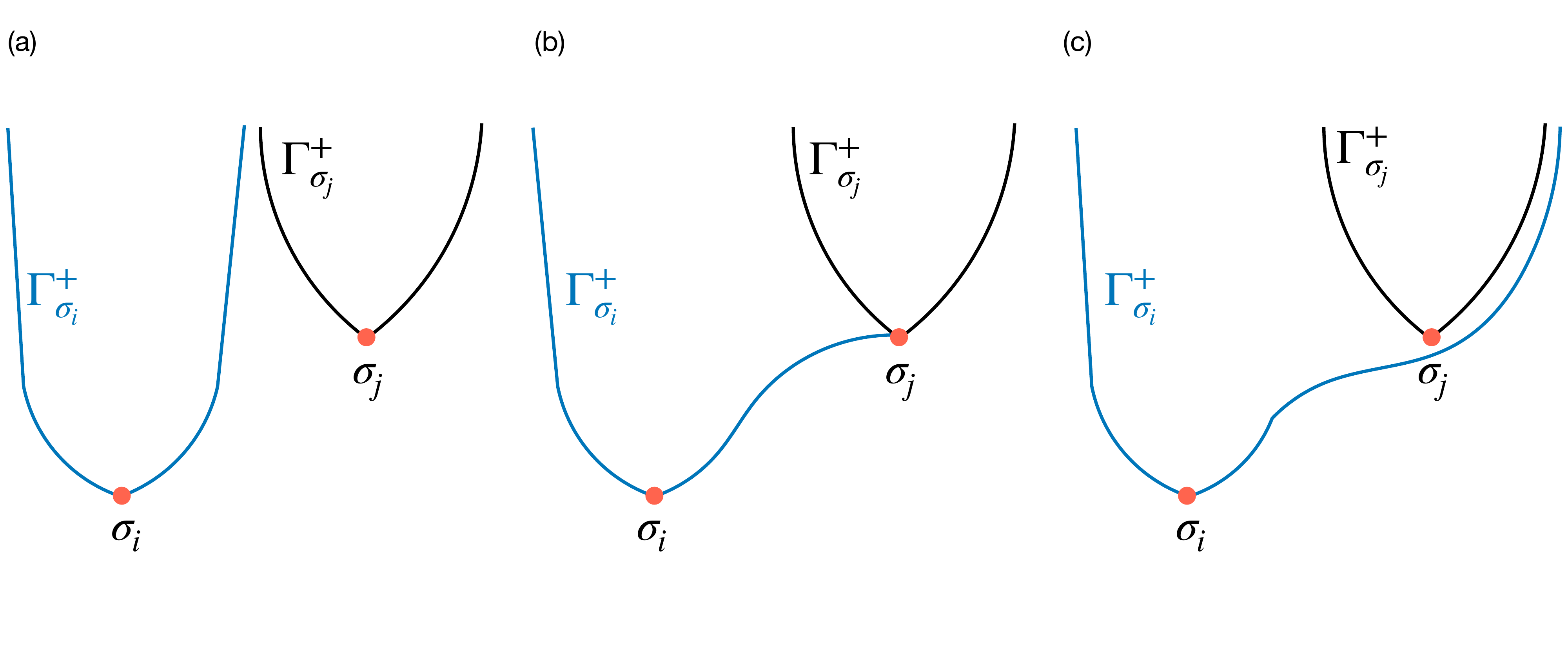}
    \caption{\small{Jump of the thimble $\Gamma^+_{i}$ across a Stokes ray. From left to right, $\gamma$ crosses a Stokes ray: (b), (c).}}
    \label{fig_jump}
\end{figure}

To understand the reason for these jumps and the meaning of the integer coefficients 
$\Delta_{ij}$ appearing in the jump matrix in \eqref{equation:jump} let us consider a simple one-dimensional example. Suppose that for a suitable 
$\gamma$, away from any Stokes line, we have two critical points $\sigma_i$ and $\sigma_j$ with distinct values of $h_{\sigma_i} < h_{\sigma_j}$ and distinct imaginary parts $\rm{Im}(\gamma f(\mathbf{z}))$, for which we can define two distinct thimbles without any intersection (see Figure \ref{fig_jump} (a)). As we move $\gamma$ towards a Stokes line, the thimble $\Gamma^+_{i}$ is continuously deformed until it crosses the thimble 
 $\Gamma^+_{j}$ at the critical point $\sigma_j$ (see Figure \ref{fig_jump} (b)). At this point, the first thimble is no longer well defined. As we continue moving 
$\gamma$ across the Stokes line, the support of the thimble $\Gamma^+_{i}$ continues to be deformed on the other side of the thimble $\Gamma^+_{j}$, as shown in Figure \ref{fig_jump} (c). The comparison between the representations (a) and (c) in Figure \ref{fig_jump} illustrates the jump.

The number of upward flows from $\sigma_i$ that intersect the point $\sigma_j$ along a Stokes ray, counted with appropriate sign based on the orientation, gives the number $\Delta_{ij}$ in the matrix \eqref{equation:jump}.

{\bf Remark:} An interesting relation between the total monodromy acting on the thimbles after a transformation $\gamma \rightarrow e^{2 \pi i } \gamma$ and the transformation of a basis of $(n-1)-$forms dual to the vanishing cycles has been pointed out in \cite{Cecotti:1992ccv}. Let us start from a regular point $\gamma$ and let us transport it along a circle in a clockwise direction with $\vert \gamma \vert$ fixed. Each time that $\gamma$ crosses a Stokes ray $l_{ij}$, corresponding to the crossing of the Stokes line connecting the critical points $\sigma_i$ and $\sigma_j$, we have a change on the thimble basis given by \eqref{equation:jump}. Let us indicate the matrix giving the jump as
\begin{equation}
    M_{ij} = \mathbb{1} + A_{ij} ,
\end{equation}
where the only non-zero entry in the matrix $A_{ij}$ is $(ij)=\Delta_{ij}$ counting the intersection number among the vanishing cycles $\Delta_i$ and $\Delta_j$. After a tour of $\pi$ around the origin, the total change on the basis of thimbles is
\begin{equation}
    S= \prod_{l_{ij}} M_{ij}.
\end{equation}
In the second half sector, beyond $\pi$, each time $\lambda$ crosses a Stokes line we have a jump given by:
\begin{equation}
    M_{ji}^{-t} = \mathbb{1} - A_{ji}.
\end{equation}
The total jump along this second half circle is represented by the matrix:
\begin{equation}
    S^{-t}=\prod_{l_{ij}} M_{ji}^{-t}.
\end{equation}
The full monodromy is defined via
\begin{equation}
    H=S^{-t}S.
\end{equation}
This matrix is invariant under deformations of the function $\gamma f({z})$ and it is quasi-unipotent.\footnote{This means that some power of it is unipotent: the sum of the identity plus a nilpotent matrix} Then, its eigenvalues are always roots of the unity
\begin{equation*}
    \text{Eigenvalues of H}\subseteq \left\lbrace e^{2 \pi i q_k}, \qquad q_k\in\Q \right\rbrace.
\end{equation*}

\subsection{Twisted de Rham Cohomology}\label{TDC}
Let us now move to the cohomological side of the exponential pairing, explicitly showing its construction in the one variable case $X=\C$, considering as holomorphic function the polynomial $f=\mathcal{P}_\ell\in \C[z]$ of degree $\ell$. Adding to $\C$ the divisor $D_v=p=\{\infty\} $, we end up with the good normalization $\overline{X}=\Pro^1$, on which $\mathcal{P}_\ell$ naturally extends to $\overline{\mathcal{P}}_\ell:\Pro^1\rightarrow\Pro^1$. Denoting the twisted differential as $\nabla\equiv (\gamma^ {-1} d+d\mathcal{P}_\ell\wedge)$, let consider the complex of sheaves: 

\begin{equation}
    (\Omega^\bullet_{\Pro^1,p},\nabla): 0\rightarrow \Ol_{\Pro^1}(*p) \xrightarrow{\nabla}\Omega^1_{\Pro^1}(*p)\rightarrow 0.
\end{equation}
\nn
On $\Pro^1$, meromorphic functions with only allowed pole at infinity are in fact polynomials on $\C$, thus we have:
\begin{equation}
\begin{split}
    &\Ol_{\Pro^1}(*p)\cong \C[z],\\
    &\Omega^1_{\Pro^1}(*p)\cong \{P(z)dz|P(z)\in \C[z]\}.
    \end{split}
\end{equation}
\nn
By definition \eqref{GlobalDerham}, the global twisted de Rham cohomology is
\begin{equation}
    H^\bullet_{dR,t}(\C,d\mathcal{P}_n)=\mathbb{H}(\Pro^1,(\Omega^\bullet_{\Pro^1,p},\nabla))=H^\bullet(R\Gamma(\Pro^1,(\Omega^\bullet_{\Pro^1,p},\nabla))).
\end{equation}
\nn
We can compute the hypercohomology by means of the Serre spectral sequence \eqref{SerreSQ} with second page

 \begin{equation}
E_2^{p,q}=R^p\Gamma(\Pro^1,H^q(\Omega^\bullet_{\Pro^1,p},\nabla)).
 \end{equation}
 \nn
 Let us firstly determine the cohomology of $(\Omega^\bullet_{\Pro^1,p},\nabla)$:
 \begin{equation}
     H^\bullet(\Omega^\bullet_{\Pro^1,p},\nabla)=\ker\nabla\oplus {\rm coker}\nabla.
 \end{equation}
 \nn
 Thus, the computation reduces to the calculation of the kernel and the cokernel of the twisted differential. For $g(z)\in \C[z]$, the stalk of $\ker\nabla$ is 

\begin{equation}
    \ker\nabla=\{g(z)\in\C[z]\ |\ \gamma^{-1} g'(z)+\mathcal{P}'_\ell(z)g(z)=0\}.
\end{equation}
\nn
For $\gamma \in\C^*$, the constrain on $g(z)$ becomes:

\begin{equation}
    g(z)=e^{-\gamma\mathcal{P}_\ell(z)},
\end{equation}
\nn
that cannot be a polynomial unless $\ell=0$. Therefore:

\begin{equation}
    \ker\nabla=\emptyset.
\end{equation}
\nn
The cokernel of $\nabla$ measures the failure of $\nabla$ to be surjective: away from the critical points of $\mathcal{P}_\ell$, $\nabla$ is locally surjective and its cokernel vanishes. On the other hand, near each critical point the equation $\nabla g=\eta$ for $\eta\in \Omega^1_{\Pro^1}(*p)$ may not have a polynomial solution for $g$. This generates a one dimensional obstruction to surjectivity, thus the stalk ${\rm coker}(\nabla)_\sigma\cong \C$. The cokernel of $\nabla$ is a direct sum of skyscraper sheaves with support on critical points

\begin{equation}
    {\rm coker}\nabla=\bigoplus_{\sigma\in\Sigma}\C.
\end{equation}
\nn
An alternatively and, for our purposes, more interesting way to determine it consists to notice that the cokernel of $\nabla$ is isomorphic to the Jacobian ring associated to $\mathcal{P}_\ell$ 
\begin{equation}
    {\rm coker}\nabla \cong \frac{\C[z]}{\rm{Im}(\nabla)}\cong\frac{\C[z]}{(\mathcal{P}'_\ell(z))}=J_{\mathcal{P}_\ell}.
\end{equation}
\nn
One can indeed prove the image of $\nabla$ is isomorphic to the ideal generate by $\mathcal{P}'_\ell$. 
The Jacobian ring, as a $\C$ vector space, is:
\begin{equation}
    J_f=span\{1,z,\dots,z^{\ell-2}\}\cong \C^{\mu},
    \label{CokerJacobian}
\end{equation}
\nn
with $\mu=\ell-1$ the total Milnor number.\\
Because of the vanishing of the higher cohomology groups of $(\Omega^\bullet_{P^1}(*p),\nabla)$, the terms $E_2^{p,q}$ vanish for $q>1$ and because of the acyclicality of skyscraper sheaves $E_2^{p,q}=0$ for $p>0$. Thus, the only possibly non zero ``turning page'' differential could be $d_2:E_2^{0,1}\rightarrow E_2^{2,0}$, but $E_2^{2,0}$ is also zero. Hence, the spectral sequence degenerates at page $E_2$. Therefore,
\begin{equation}
\begin{split}
    H^n(R\Gamma(\Pro^1,(\Omega^\bullet_{\Pro^1,p},\nabla)))&=\bigoplus_{p+q=n}E_\infty^{p,q}=\bigoplus_{p+q=n}E_2^{p,q}\\&=\bigoplus_{p+q=n}R^p\Gamma(\Pro^1,H^q(\Omega^\bullet_{\Pro^1,p},\nabla)).
    \end{split}
\end{equation}
\nn
That is:
\begin{equation}
\begin{alignedat}{2} 
    &\mathbb{H}^0(\Pro^1,(\Omega^\bullet_{\Pro^1,p},\nabla))\cong H^0(\Pro^1,\ker\nabla)\cong 0,\\ &\mathbb{H}^1(\Pro^1,(\Omega^\bullet_{\Pro^1,p},\nabla))\cong H^1(\Pro^1,\ker\nabla)\oplus H^0(\Pro^1,{\rm coker}\nabla)\\& \hspace{2.6cm} \cong H^0(\Pro^1,\C^{\ell-1})\cong \C^{\ell-1}\otimes H^0(\Pro^1,\Z)\cong \C^{\ell-1},\\
    \end{alignedat}
\end{equation}
\nn
where the first isomorphism in the last line is given by the universal coefficient theorem.
Finally, the global de Rham cohomology is
\begin{equation}
\begin{split}
    &H^0_{dR,t}(\C,d\mathcal{P}_\ell)\cong 0,\\
    &H^1_{dR,t}(\C,d\mathcal{P}_\ell)\cong \C^{\ell-1}.
    \end{split}
    \label{1DTwistedPoly}
\end{equation}
\nn
As we can see, the global twisted de Rham cohomology is independent on the possible coalescence of critical points. The point is that the global Jacobian ring has dimension equal to the total Milnor number \eqref{CokerJacobian}, which takes into account any possible multiplicity $m_i$. So, although the support of ${\rm coker}(\nabla)$ and its local structure changes

\begin{equation}
    {\rm coker}(\nabla)=\bigoplus_{\Sigma}\C_{\sigma_i}^{m_i},
\end{equation}
\nn
its global sections remain always $\Gamma({\rm coker}(\nabla))\cong \C^{\ell-1}$.
The sheaf ${\rm coker}\nabla$ is not sensitive to the coalescence of critical points, due to its naively construction as a direct sum of skyscraper shaves, which loses information about the local structure. In order to recover such information, we need to turn it into a perverse sheaf and to consider a suitable extension of the twisted de Rham complex, i.e. of $\nabla$.\\
Given a connection defined on an open dense subset $U\subset X$ (smooth locus), its extension along a divisor $D=X\backslash U$ comes with a prescription about its behaviour near $D$. The ones we are possibly interested in are the so called Middle and Logarithmic extensions: the first one, arising in the context of perverse sheaves, avoids the addition of unnecessary singularities while preserving key invariants; on the contrary, the second one allows for the connection to have logarithmic singularities near the divisors. Although $\nabla$ is defined on the whole complex plane $\C$, it does not define a local system $\La$ on it, because locally constancy fails on $\Sigma$, due to the obstructions arising in solving $\nabla s=0$: flat sections do not freely generate the cohomology. In fact, $\nabla$ defines a local system on $\C\backslash (D\cup\Sigma)$. Such obstructions arise as a consequence of the non trivial monodromy around critical points (and around branch points for a multivalued function).
Indeed, near a critical point $\sigma_i$, the expansion $\mathcal{P}_\ell(z)\sim \mathcal{P}_\ell(\sigma_i)+c(z-\sigma_i)^{m_i+1}$ shows that the monodromy has a Jordan block of size $m_i$, and it becomes unipotent in full degenerate case. Thus, the number of critical points influences the rank of $\rm coker\nabla$ by reducing it by the size of the Jordan blocks of the monodromy matrices.
Explicitly, if $\mathcal{P}_\ell$ has $\ell-1$ distinct critical points, the monodromy acts on flat sections via distinct eigenvalues, thus $\La$ ha no invariant subspaces and ${\rm coker}\nabla\cong\C^{\ell-1}$: each critical point contributes with independent obstructions. Instead, if $\mathcal{P}_\ell$ has only one critical point with multiplicity $\ell-1$, the monodromy matrix becomes unipotent ($\ell-1$ equal eigenvalues), introducing $\ell-2$ relations among obstructions and thus ${\rm coker}\nabla=\C$. The solutions to $\nabla s=0$ is $s(z)=e^{-\gamma\mathcal{P}_\ell(z)}(\sum_{i=0}^{\ell-1} (c_i\log^i(z-\sigma))$. 
Thus:

\begin{equation}
    ({\rm coker}\nabla)_{mid} = \bigoplus_{\sigma\in\Sigma} \C=\C^{n_c},
\end{equation}
\nn
with $n_c$, the number of distinct critical points. Taking into account the monodromy in this way, equivalently, means to restrict on sections with moderate growth, that is to consider the middle (mid) extension $(\Omega^\bullet_{\C,D})_{mid}$. Notice that no prescription along the divisor $D$ is added. In particular, the mid extension is independent on $D$.\\
We want just to add a comment on perversity, without dwelling too much on the subject;\footnote{Readers interested in a deeper understanding of perversity are referred to the lecture notes \cite{goresky2021lecturenotessheavesperverse}.} in this context, we could forget about it, since its consideration is necessary only for categorical reasons, but totally irrelevant for the purposes of the present calculations. Consider, for instance, a skyscraper sheaf $\delta$. It fails to be perverse because it does not satisfy co-support conditions, however we can easily make it into a perverse sheaf by just shifting it by $[-1]$, meaning now $H^{i-1}(\delta[-1])=H^i(\delta)$. The ``perversification'' of ${\rm coker}\nabla$ then just imply a unit shift to the left of the spectral sequence, leaving, in fact, hypercohomology unchanged.\\
Finally, supposing only one critical point has multiplicity $m$ 

\begin{equation}
\begin{split}
    &H^0_{LdR,t}(\C,d\mathcal{P}_\ell)_{(m)}\cong 0,\\
    &H^1_{LdR,t}(\C,d\mathcal{P}_\ell)_{(m)}\cong \C^{\ell-m},
    \end{split}
    \label{1DDeRham}
\end{equation}
We will need this refinement in the next section when considering the case of degenerate points.

\section{Exponential integral of holomoprhic functions}


\subsection{The Pearcey’s integral}\label{PearceyIntegrals}
As a first application of the Lefschetz thimble decomposition discussed above, we examine a Pearcey's integral \cite{Olver}, appearing in \cite{Cacciatori:2024ccm} as the grand-canonical partition function of the gauged Skyrme model, describing baryonic layers living at finite baryon density within a constant magnetic field.
We want to study the integral
\begin{equation}
    \mathcal{P}(a)= \int_{- \infty}^{+\infty} dz \,e^{-a \left( z^4 +bz^2+cz+d\right)},
    \label{Integral:Pearcey_real}
\end{equation}
for generic values of the real parameters $a,b,c,d$. This generalizes the case studied in \cite{Cacciatori:2022mbi}.\\ 
Following the prescription of the previous section, let proceed extending the polynomial argument of the exponential to a holomorphic function over $\C_z$, by complexifying both the variable $z$ and the parameters. In particular, the real parameter $a$ is promoted to the complex parameter $\gamma$, over which we will build the wall crossing structure. We then analyze the integral over a generic contour $\Gamma$ in $\C_z$
\begin{equation}
    \mathcal{P}(\gamma) = \int_{\Gamma} dz\,e^{-\gamma \left( z^4 +bz^2+cz+d\right)},
\end{equation}
and seek a basis of integration cycles along which the integral remains convergent. Once a basis and a intersection product (in homology) are identified, the real integration contour can be decomposed, with integral coefficients, in terms of such basis. As a result, the integral in \eqref{Integral:Pearcey_real} becomes a linear combination of integrals evaluated over the basis. For large values of the parameters, these basis integrals admit an asymptotic expansion, which is then transferred to the initial integral. The expectation is that for different values of the parameters, both the basis for the integration contours and the decomposition of the real line in terms of them will be modified.\\
The set of critical points of the holomorphic function
\begin{equation}
    f(z)\equiv z^4+bz^2+cz+d,
\end{equation}
 i.e. the solutions of the cubic equation $f'(z)=4z^3+2bz+c=0$, can be compactly written as

\begin{equation}
    \Sigma = \left \{-\frac{b}{\sqrt[3]{3\tilde{\Delta}}}+\frac{\sqrt[3]{\tilde{\Delta}}}{2\sqrt[3]{9}},\frac{(1\pm i\sqrt{3})b}{2\sqrt[3]{3\tilde{\Delta}}}-\frac{(1\pm i\sqrt{3})\sqrt[3]{\tilde{\Delta}}}{4\sqrt[3]{9}}\right\},
\end{equation}
\nn
where $\tilde{\Delta}\equiv -9c+\sqrt{3\Delta}$, with $\Delta$ the discriminant. According to the sign of $\Delta$, three different situation arise: 
\begin{equation}
    \Delta\equiv 8b^3+27c^2 \quad \quad \quad \begin{cases}
     > 0  \quad \quad 1 \, \text{ real and } 2 \text{ complex conjugate solutions,}\\
    < 0  \quad \quad 3 \, \text{ real different solutions,}\\
      =0  \quad \quad 3 \,\text{ real solutions with at least a multiple root.}\\
    
    \end{cases}
\end{equation}
We will analyze these three cases separately, since each one defines a different connected region in the parameter space $U=\left\lbrace (b,c) \in \mathbb{C}^2 \right\rbrace$, and on each region we can define a distinct local system of Betti homologies $H^{\bullet}_{GB,\gamma}(\mathbb{C}, f_{(a,b)}, \mathbb{Z})$ equipped with its own wall-crossing structure.
\paragraph{Positive discriminant.}\mbox{}\\
Let us firstly consider the case where $\Delta >0$. \\
For concreteness, we fix the parameters to $(b,c,d)=(3/2,7,-1)$ and carry out the explicit computations for this choice. The critical points are computed to be

\begin{equation}
    \sigma_i\in \Sigma=\left \{-1,\frac{1}{2}(1+i\sqrt{6}),\frac{1}{2}(1-i\sqrt{6})\right \},
\end{equation}

\begin{figure}[!htb]
\centering
\hspace{-1cm}
\begin{minipage}{0.5\textwidth}
\centering
  \includegraphics[width=0.98\linewidth]{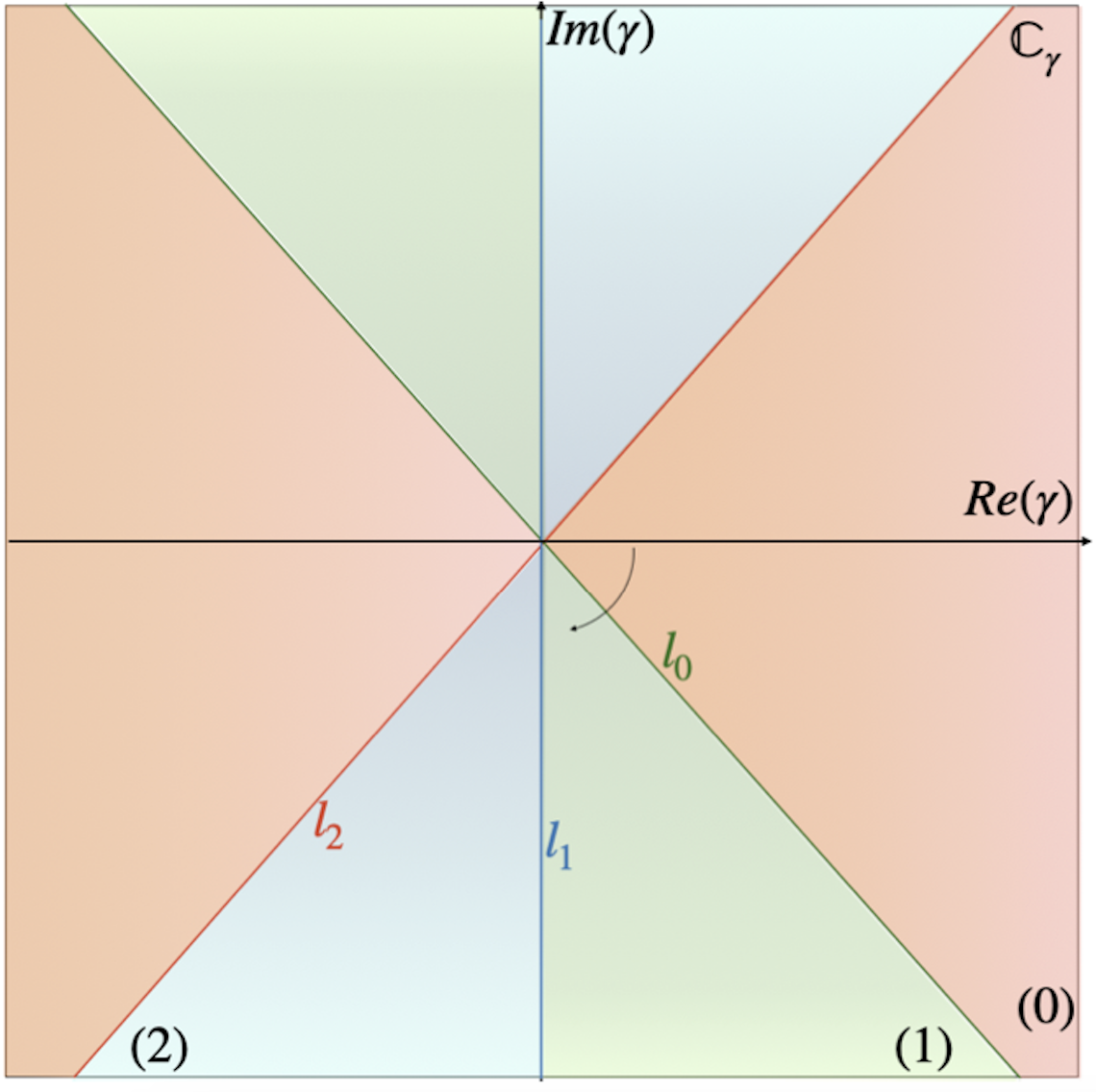}
\end{minipage}
\begin{minipage}{0.5\textwidth}
\centering
  \includegraphics[width=1.05\linewidth]{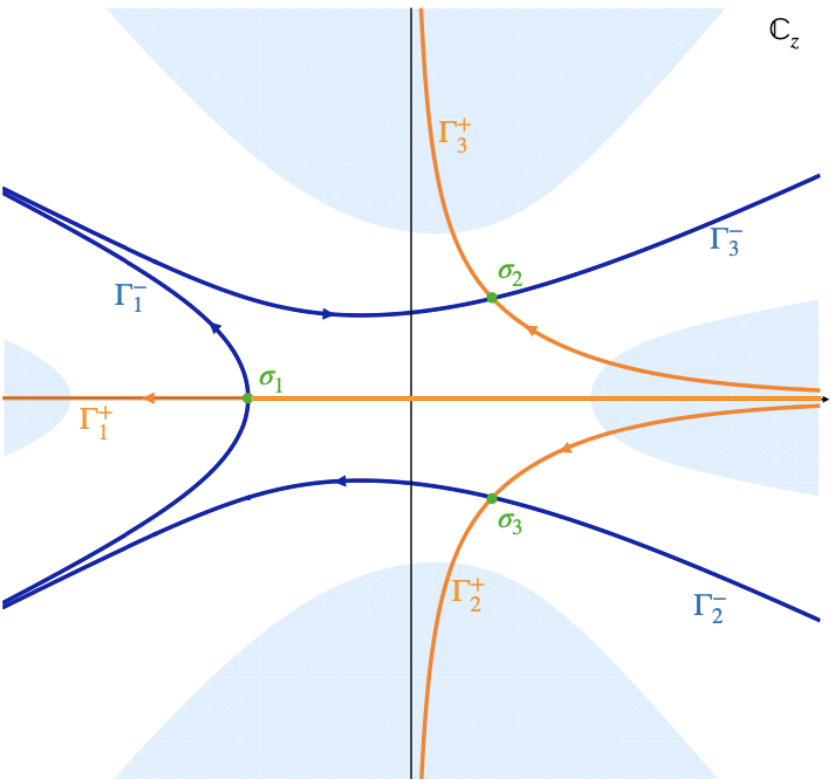}
\end{minipage}
\caption{\small{Positive discriminant case $(\gamma,b,c,d)=(1,3/2,7,-1)$. (Left) Stokes' lines on the $\C_\gamma$ plane. (Right) Ascendant paths $\Gamma_i^+$ spanning $H_1(\C_z,D_N,\Z)$ (in orange), descendant paths $\Gamma_i^-$ spanning $H_1(\C_z,X_{-N},\Z)$ (in blue).}}
\label{PearcyPositiveDiscriminant}
\end{figure}

\nn
where $f(z)$ takes (respectively) the critical values 

\begin{equation}
    t_i \equiv f(\sigma_i)\in \mathcal{S}=\left \{-\frac{11}{2},\left (\frac{11}{16}+3i\sqrt{6}\right ),\left (\frac{11}{16}-3i\sqrt{6}\right)\right\}.
\end{equation}
\nn
The non-degeneracy of the Hessian at each critical point, together with the fact that all Morse indices equal one guarantee the saturation of \eqref{BettiRelation}, thus $dim H_1(X,D_N,\Z)=3$. Here $D_N\subset \C$ denotes the union of the four connected regions in the complex $z=(u,v)$ (shaded blue in Figure \ref{PearcyPositiveDiscriminant}) where the Morse function $h(u,v)=\mathrm{Re}( \gamma f(u,v))>N$.\\
Using \eqref{StokesRaysFormula}, we find that the Stokes' rays are the three lines

\begin{equation}
\begin{alignedat}{3}
    &l_0:\mathrm{Re}(\gamma)=-\frac{11}{16}\sqrt{\frac{3}{2}}\rm{Im}(\gamma), \quad \quad &&\mbox{where}\quad \rm{Im}(\gamma f(z))|_{\sigma_1}=\rm{Im}(\gamma f(z))|_{\sigma_2},\\
    &l_1:\mathrm{Re}(\gamma)=0, \quad \quad &&\mbox{where}\quad \rm{Im}(\gamma f(z))|_{\sigma_2}=\rm{Im}(\gamma f(z))|_{\sigma_3},\\
    &l_2:\mathrm{Re}(\gamma)=\frac{11}{16}\sqrt{\frac{3}{2}}\rm{Im}(\gamma),\quad \quad &&\mbox{where}\quad \rm{Im}(\gamma f(z))|_{\sigma_1}=\rm{Im}(\gamma f(z))|_{\sigma_3},
\end{alignedat}
\end{equation}
\nn
resulting in a splitting of the $\C_{\gamma}$ plane in three regions with different thimbles structures. Let us fix $\gamma=1$, lying in the first petal of the fan (orange region labeled with $(0)$ on the right side of Figure \ref{PearcyPositiveDiscriminant}).
We identify the three thimbles $\Gamma_i\equiv \Gamma_{\sigma_i}$ as the paths passing through a critical point and keeping constant the imaginary part of the Morse function (Figure \ref{PearcyPositiveDiscriminant}):

\begin{equation}
\begin{split}
    &H_1(X,D_N,\Z)= \mathrm{span}\{\Gamma_1^+,\Gamma_2^+,\Gamma_3^+\}\cong \Z^3,\\
    &H_1(X,D_N,\Z)^\vee= \mathrm{span}\{\Gamma_1^-,\Gamma_2^-,\Gamma_3^-\}\cong \Z^3.
    \end{split}
\end{equation}
\nn
Let us set $f(z)=t$ and look at the preimage 

\begin{equation}
    f^{-1}(t)=\begin{pmatrix} z_1(t)\\z_2(t)\\z_3(t)\\z_4(t)\end{pmatrix}.
\end{equation}
\nn
When approaching a critical point $\sigma_i$, the four point fiber degenerates to a three point set, identifying a vanishing cycle $\Delta_i$. We have

\begin{equation}
    \Delta_1=\{z_3\}-\{z_4\} ,\quad \quad \Delta_2=\{z_1\}-\{z_4\}\quad \mbox{and}\quad \Delta_3=\{z_1\}-\{z_4\}.
    \label{VanishingPearcey}
\end{equation}
\nn
The monodromy matrices acting on this base of vanishing cycles are computed to be (see example  \ref{A3} for details):

\begin{equation}
    M_1=\begin{pmatrix}-1&0&0\\-1&1&0\\-1&0&1\end{pmatrix} \quad \mbox{and}\quad M_2=M_3=\begin{pmatrix}1&-1&0\\0&-1&0\\0&0&-1\end{pmatrix}.
    \label{MonodromyMatricesPositiveDet}
\end{equation}
\nn
By using \eqref{PicardLefschetzFormula} and \eqref{TotalIntersection} we compute the intersection numbers

\begin{equation}
    \Delta_{12}= 1 \quad, \quad \Delta_{13}=1 \quad\mbox{and}\quad \Delta_{23}= -2.
\end{equation}
\nn
Note that the intersection numbers $\Delta_{ij}$ are defined in \eqref{TotalIntersection} up to a sign depending on the orientation of the cycles. 
When crossing a Stokes' line, the base of thimbles undergoes a change of the type \eqref{equation:jump}. Let $\Gamma_i^{+(k)}$ be the vector of thimbles in the $(k)$-th sector of the fan. With the clockwise ordering showed in Figure \ref{PearcyPositiveDiscriminant} and $T^{\theta_{(k)}}$\footnote{Matrix representation of the operator dual to $T_{\theta}$ in \eqref{Tij}.} the jump matrix associated to the Stokes line $l_{(k)}\to s_{\theta_k}$ connecting the $(k)$-th and the $(k+1)$-th sectors of the fan, 

\begin{equation}
     \Gamma_i^{+(k+1)}=T_{ij}^{\theta_{(k)}}\Gamma_j^{+(k)},
\end{equation}
\nn
we have\\
\begin{equation}
\begin{alignedat}{3}
    &\mathrm{Re}(\gamma f(z))|_{\sigma_1}<\mathrm{Re}(\gamma f(z))_{\sigma_2} \quad \quad\quad \quad &&\mbox{along}\,\, l_0,\\
    &\mathrm{Re}(\gamma f(z))|_{\sigma_2}>\mathrm{Re}(\gamma f(z))_{\sigma_3} \quad \quad\quad \quad &&\mbox{along}\,\, l_1,\\
    &\mathrm{Re}(\gamma f(z))|_{\sigma_1}>\mathrm{Re}(\gamma f(z))_{\sigma_3} \quad \quad\quad \quad &&\mbox{along}\,\, l_2,\\ 
\end{alignedat}
\end{equation}
\nn
so that

\begin{equation}
    T^{\theta_{(0)}}=\begin{pmatrix}1&1&0\\0&1&0\\0&0&1\end{pmatrix} \quad,\quad T^{\theta_{(1)}}=\begin{pmatrix}1&0&0\\0&1&0\\0&-2&1\end{pmatrix}\quad,\quad T^{\theta_{(2)}}=\begin{pmatrix}1&0&0\\0&1&0\\1&0&1\end{pmatrix}.
\end{equation}
\noindent
These matrices define the wall crossing structure in $\mathbb{C}_{\gamma} \times U_{b,c}^{\Delta > 0}$, in which the walls are defined on subregions  of this space where exactly two critical values are aligned
\begin{equation}
    W_{ij}= \left\lbrace (\gamma, u) \in \mathbb{C}^{\ast} \times U_{b,c}^{\Delta > 0} \vert \rm{Im} \left( \gamma (t_i(u)-t_j(u))\right)=0 \right\rbrace.
\end{equation}
They correspond to walls of the second type in the sense of \cite{kontsevich:2008kos}.\\
After a complete round of $2\pi$, we get the full monodromy matrix
\begin{equation}
    S=\begin{pmatrix}
        0&2&-1\\1&-2&2\\1&-1&1
    \end{pmatrix},
\end{equation}
\nn
with eigenvalues $-1,-1,1$.\\
Let us now consider the case of non positive discriminant. Then, singular points lie on the real axes of $\C_z$, this meaning that a Stokes' line appears along $\rm{Im}(\gamma)=0$, splitting the $\C_\gamma$ plane into two regions, corresponding to the upper and lower half planes. It is worthwhile to emphasize that the a priori naive choice of a real $\gamma$, in this case, would give rise to a wrong description, since we would have set precisely on the Stokes' line. In order to proceed, let us thus set $\gamma=i$.

\begin{figure}[!htb]
\centering
\hspace{-1cm}
\begin{minipage}{0.5\textwidth}
\centering
  \includegraphics[width=1\linewidth]{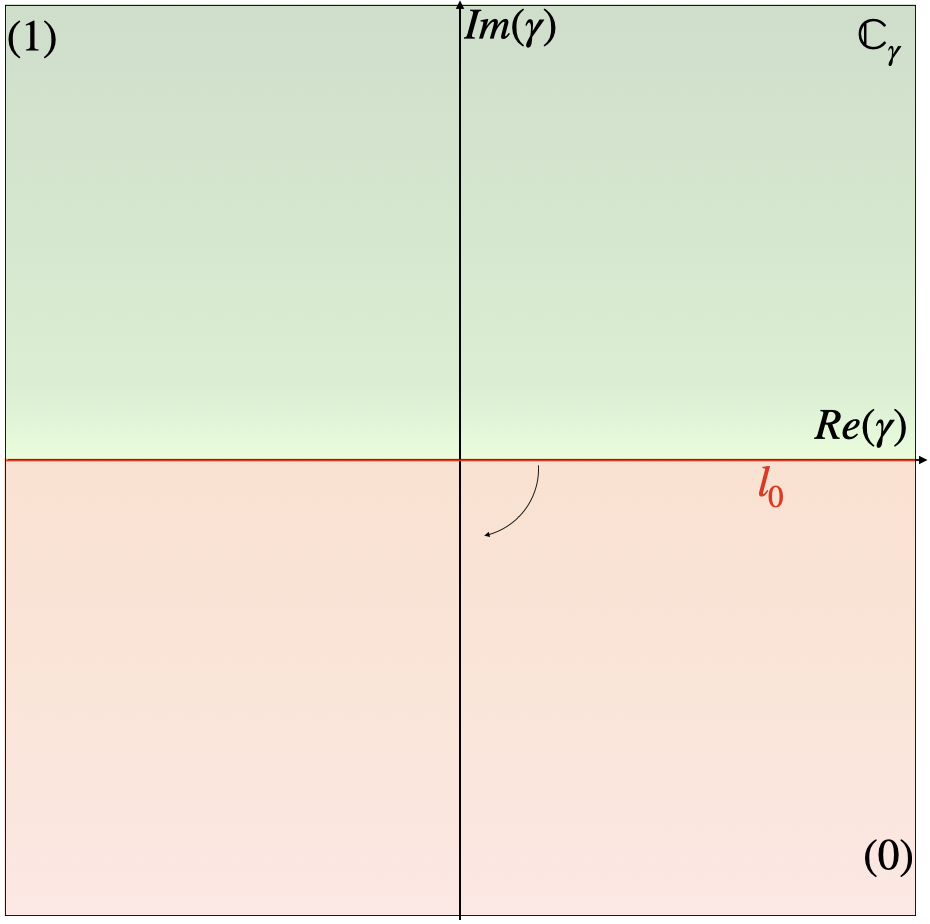}
\end{minipage}
\begin{minipage}{0.5\textwidth}
\centering
  \includegraphics[width=0.99\linewidth]{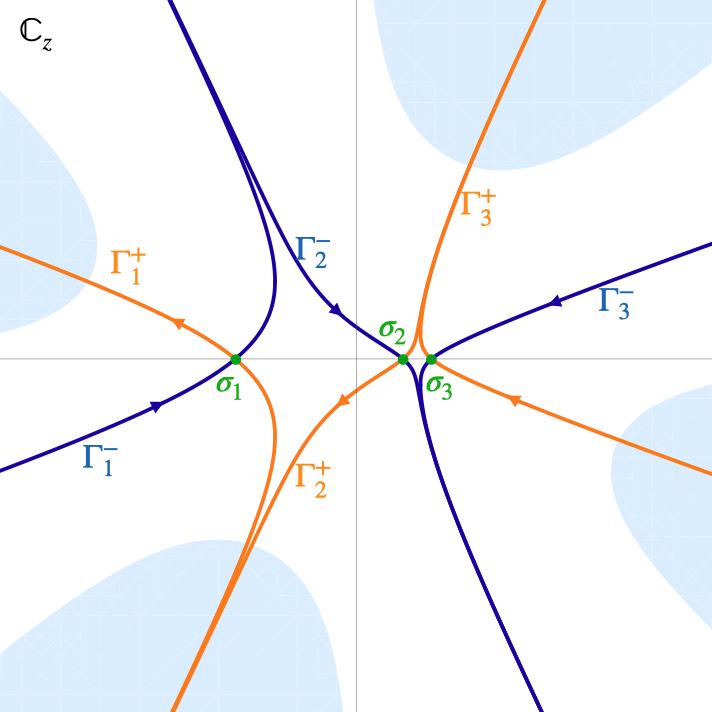}
\end{minipage}
\caption{\small{Negative discriminant case $(\gamma,b,c,d)=(i,-1,1/2,-1)$. (Left) Stokes' lines on the $\C_\gamma$ plane. (Right) Ascendant paths $\Gamma_i^+$ spanning $H_1(\C_z,D_N,\Z)$ (in orange), descendant paths $\Gamma_i^-$ spanning $H_1(\C_z,X_{-N},\Z)$ (in blue).}}
\label{PearcyNegativeDiscriminant}
\end{figure}

\paragraph{Negative discriminant.}\mbox{}\\
Firstly, let us consider the case $\Delta<0$, characterized by three different real critical points. The local Betti homology generated by local thimbles, shown in Figure \ref{PearcyNegativeDiscriminant} (right), is pretty much the same as in the positive discriminant case, as it is of course unaffected by the reality of critical points. However, as emphasized earlier, the relevant peculiarity appears in the thimbles structure, due to Stokes' line $l_0$ on the real axis of the $\C_\gamma$ plane, (see Figure \ref{PearcyNegativeDiscriminant}(left)). Setting $(b,c,d)=(-1,1/2,-1)$, we get

\begin{equation}
    \Sigma=\left \{\frac{1}{2},\frac{1}{4}(1\pm \sqrt{5})\right \} \quad \mbox{and} \quad \mathcal{S}=\left \{-\frac{15}{16},\frac{1}{32}(-41\pm 5 \sqrt{5})\right \}.
\end{equation}
\nn
Proceeding as above, we identify three vanishing cycles

\begin{equation}
    \Delta_1=\{z_1\}-\{z_2\}, \quad \Delta_2=\{z_3\}-\{z_4\}, \quad \Delta_3=\{z_1\}-\{z_4\}
\end{equation}
\nn
and the corresponding monodromy matrices

\begin{equation}
    M_1=\begin{pmatrix}-1&0&0\\0&1&0\\-1&0&1\end{pmatrix},
    \,M_2=\begin{pmatrix}1&0&0\\0&-1&0\\0&-1&1\end{pmatrix} \quad \mbox{and}\quad M_3=\begin{pmatrix}1&0&-1\\0&1&-1\\0&0&1\end{pmatrix}.
\end{equation}
\nn
In order to determine the jump matrices, we have to compute the intersection numbers among thimbles. However, the Morse function vanishes in all critical points. In order to avoid it, we slightly move away from the imaginary axis setting $\gamma=i+\delta$. We get 

\begin{equation}
    \mathrm{Re}(\gamma f(z))|_{\sigma_3}>\mathrm{Re}(\gamma f(z))|_{\sigma_1}>\mathrm{Re}(\gamma f(z))_{\sigma_2}, \quad \quad\quad \quad \mbox{along}\,\, l_0.
\end{equation}
\nn
Note that in this case the ray $l_0$ corresponds to a Stokes line intersecting three distinct critical values. The natural generalization of the jump matrix \eqref{equation:jump}  in this case accounts for the double jump of $\Gamma^{+(0)}_2$:  the first caused by crossing the branch cut emanating from $t_1$ in the $\mathbb{C}_t$ plane, and the second by crossing the cut associated with $t_3 \in \mathbb{C}$. Then, when we cross the line $l_0$ in the clockwise direction, the corresponding jump matrix is given by the following upper triangular matrix:

\begin{equation}
      \begin{pmatrix}\Gamma^{+(1)}_2\\\Gamma^{+(1)}_1\\\Gamma^{+(1)}_3\end{pmatrix}=\begin{pmatrix}1&\Delta_{21}&\Delta_{23}-\Delta_{21}\Delta_{13}\\0&1&\Delta_{13}\\0&0&1\end{pmatrix}\begin{pmatrix}\Gamma^{+(0)}_2\\\Gamma^{+(0)}_1\\\Gamma^{+(0)}_3\end{pmatrix},  
\end{equation}

\nn
where intersection numbers among vanishing cycles $\Delta_{ij}$ are computed with \eqref{PicardLefschetzFormula}. Reordering the base of thimbles, we get

\begin{equation}
   T^{(0)}=\begin{pmatrix}1&0&-1\\0&1&-1\\0&0&1\end{pmatrix},
   \label{jump:Negative_discriminant}
\end{equation}
\nn
and 
\begin{equation}
    S=\begin{pmatrix}1&0&-1\\0&1&-1\\1&1&-1\end{pmatrix},
\end{equation}
\nn
with eigenvalues $i,-i,1$.
The matrix \eqref{jump:Negative_discriminant} define the wall crossing structure in $\mathbb{C}_{\gamma}^{\ast} \times U_{(b,c)}^{\Delta<0}$, in which the walls are defined by subregions of this space where three critical values are aligned. They correspond to walls of the first type in the sense of \cite{kontsevich:2008kos}.

\begin{figure}[!h]
\centering
\hspace{-1cm}
\begin{minipage}{0.5\textwidth}
\centering
  \includegraphics[width=1\linewidth]{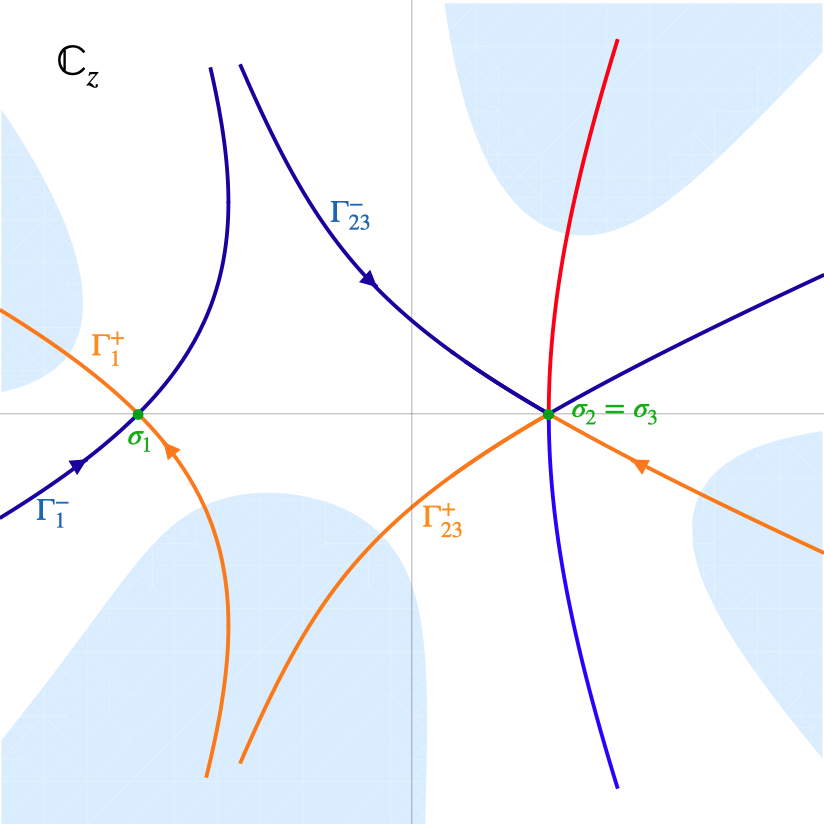}
\end{minipage}
\begin{minipage}{0.5\textwidth}
\centering
  \includegraphics[width=0.99\linewidth]{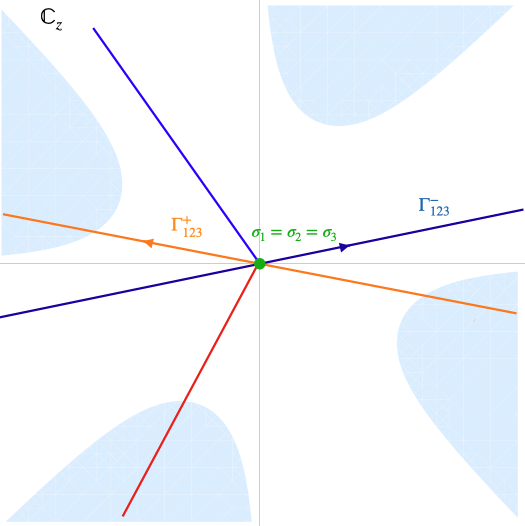}
\end{minipage}
\caption{\small{Vanishing discriminant case. (Left)Double Degeneration, $(\gamma,b,c,d)=(i,-6,8,-1)$. Ascendant paths $\Gamma_i^+$ spanning $H_1(\C_z,D_N,\Z)$ (in orange), descendant paths $\Gamma_i^-$ spanning $H_1(\C_z,X_{-N},\Z)$ (in blue). (Right)Triple Degeneration, $(\gamma,b,c,d)=(i,0,0,-1)$. Ascendant paths $\Gamma_i^+$ spanning $H_1(\C_z,D_N,\Z)$ (in orange), descendant paths $\Gamma_i^-$ spanning $H_1(\C_z,X_{-N},\Z)$ (in blue). (Both) Coalescing ascendant (red) and descendant (light blue) paths.}}
\label{PearcyVanishingDiscriminant}
\end{figure}

\paragraph{Vanishing discriminant.}\mbox{}\\
The last case we want to discuss is $\Delta=0$, where two or all three critical points coalesce. Comparing Figures \ref{PearcyNegativeDiscriminant}(right) and \ref{PearcyVanishingDiscriminant}(left) we observe that as $\sigma_2$ and $\sigma_3$ coalesce, the upward branches of $\Gamma_2^+$ and $\Gamma_3^+$, as well as the downward branches of $\Gamma_2^-$ and $\Gamma_3^-$,  begin to overlap with opposite orientations. As a result, the combination of these four paths yields only two independent thimbles:

\begin{equation}
    \Gamma_2^++\Gamma_3^+=\Gamma_{23}^+ \quad \mbox{and}\quad  \Gamma_2^-+\Gamma_3^-=\Gamma_{23}^-.
\end{equation}
\nn
This is a consequence of the fact that the vanishing cycles $\Delta_2$ and $\Delta_3$, appearing in \eqref{VanishingPearcey}, identify the same homology class when the associated critical points coalesce.\\
Similarly, when the triple degeneration occurs\footnote{For $b=c=0$.}, shown in Figure \ref{PearcyVanishingDiscriminant}(right), four branches overlap, leading to a further reduction in the number of independent thimbles. We are left with:

\begin{equation}
    \Gamma_1^++\Gamma_{23}^+=\Gamma_{123}^+ \quad \mbox{and}\quad  \Gamma_1^-+\Gamma_{23}^-=\Gamma_{123}^-.
\end{equation}
\nn
Therefore, the Betti homology turn out to be

\begin{equation}
\begin{alignedat}{2}
    &H_1(X,D_N,\Z)_{(2)}=\mathrm{span}\{\Gamma_1^+,\Gamma_{23}^+\} \quad \quad &&\mbox{and}\quad H_1(X,D_N,\Z)_{(2)}^\vee=\mathrm{span}\{\Gamma_1^-,\Gamma_{23}^-\},\\
    &H_1(X,D_N,\Z)_{(3)}=\mathrm{span}\{\Gamma_{123}^+\}, \quad \quad &&\mbox{and}\quad H_1(X,D_N,\Z)_{(3)}^\vee=\mathrm{span}\{\Gamma_{123}^-\},\\
\end{alignedat}
\end{equation}
\nn
where bracket subscripts denote the multiplicity of coalescent critical points. 
In the cohomology side, we can compute the relative cohomology using (\ref{1DDeRham}). We obtain

\begin{equation}
\begin{alignedat}{2}
    &H^1(X,D_N,\C)_{(2)}=\mathrm{span}\{1,z\} \cong \C^2,\\
    &H^1(X,D_N,\C)_{(3)}=\mathrm{span}\{1\}\cong \C.
\end{alignedat}
\end{equation}
\nn
Notice that the universal coefficient theorem for cohomology explicitly shows the duality \footnote{$\mathrm{Ext}(H_0(X,D_N,\Z)_{(i)},\C)=0.$}
\begin{equation}
    H^1(X,D_N,\C)_{(i)}\cong \mathrm{Hom}(H_1(X,D_N,\Z),\C)_{(i)}.
\end{equation}

\section{Exponential integrals for closed $1-$forms}\label{EICFFI}

In the previous section, we associated to each triple $\left( X, D_0,f \right)$ four local systems over $\mathbb{C}_{\gamma}^{\ast}$ corresponding to the global and local de Rham and Betti cohomologies. Moreover, we showed that, using this language, exponential integrals  can be naturally interpreted as periods pairing these two types of (co)homologies. 
Here, still following \cite{Kontsevich:2024mks}, we extend the discussion to the more general setting of a triple $(X, D_0, \alpha)$ where $X$ is a complex smooth algebraic variety, $D_0 \subset X$ is a normal crossing divisor, and $\alpha$ is a closed algebraic 1-form. This generalized setup extended the above framework to exponential integrals defined by multivalued functions, which is precisely what we need to treat Feynman integrals in this fashion. \\

\paragraph{Twisted de Rham Cohomology}\mbox{}\\
The first important consequence of dealing with a multivalued function in the exponential is that the twisted de Rham cohomology side of the pairing is defined with respect to the differential
\begin{equation}
    \nabla_{\alpha} \, = \, d \, - \,  \alpha \,  \wedge,
\end{equation}
where $\alpha$ is a closed $1$-form on the complex algebraic variety $X$ that is not necessarily exact. This means that, in general, $\alpha$ cannot be written globally as $\alpha = d f$ for some function $f$, but additional contributions may enter into its definition. To identify the various potential contributions, one needs to choose a suitable compactification $\overline{X}$ of $X$. However, the final result should ultimately be independent of the specific choice (see \cite{Kontsevich:2024mks}). \\
This ``good'' compactification for $X$ must be constructed using a set of normal crossing divisors $D_h, D_v$ and $D_{\log}$
\begin{equation}
    \overline{X} - X = D_h \cup D_v \cup D_{\log},
\end{equation}
such that
\begin{itemize}
    \item[\textit{(i)}] Among all the normal crossing divisors $\overline{D}_0, D_h, D_v$ and $D_{\log}$ only $D_v$ and $D_{\log}$ can have common irreducible components;
    \item[\textit{(ii)}] For any point $x \in \overline{X}$ there exists a small analytic neighborhood $U$ and a closed meromorphic $1-$form $\alpha$ locally given by the expression
\begin{equation}
    \alpha = \alpha_{reg} + \alpha_{log} + \alpha_{\infty}.
\end{equation}
The first contribution represent a regular form on $U$. The second contribution, $\alpha_{\log}$, can be expressed in local coordinates near a divisor $D_{log}$ as
\begin{equation}
    \alpha_{log} = \sum_i c_i d \log z_i,
\end{equation}
where $z_i$ are local coordinates in which $D_{log}$ is given by $\prod_i z_i=0$, and $c_i \in \mathbb{C} \setminus \left\lbrace 0 \right\rbrace$ represent the periods of $\alpha$ around the loci $z_i=0$, computed as
\begin{equation}
    c_i = \frac{1}{2 \pi i} \oint_{S^1_i} \alpha,
\end{equation}
the circle $S^1_i$ encircling $D_{log}$ in a smooth point.
The final contribution admits the form $\alpha_{\infty}=dF$ in local coordinates near $D_v$, where $F$ is an analytic function of the form
\begin{equation}
    F = \frac{c}{\prod_j z_j^{k_j}} \left( 1 + o (1) \right) \, , \quad k_j \geq 1
\end{equation}
and $D_v$ is locally defined by $\prod_j z_j=0$ in those coordinates.
\end{itemize}
\nn
Once we have such a compactification, we can construct the sheaf $\Omega^{\bullet}_{\overline{X},D_{\log}}$ of differential forms on $\overline{X}$ with possible logarithmic poles along the divisors $D_{\log}$ and use this sheaf to construct the global de Rham cohomology at any $\gamma \in \mathbb{C}^{\ast}$ using the following definition.\\

\defn[Global twisted de Rham]{
Let $\gamma \in \mathbb{C}_{\gamma}^{\ast}$. The (twisted) de Rham cohomology is the graded abelian group constructed by the hypercohomology
\begin{equation}
   H^{\bullet}_{GdR,\gamma} \left( X, D_0, \alpha \right) \, = \, \mathbb{H}^{\bullet} \left( \overline{X}, (\Omega^{\bullet}_{\overline{X},D}, \gamma^{-1}d + \alpha \wedge \, ) \right).
   \label{GlobaldR:forms}
\end{equation}
}

Also in the case of closed forms, one can define a local version of the cohomology and establish a global-to-local isomorphism, analogous to equation \eqref{dR:global_to_local} for exact differentials. To provide this definition we need to introduce the subsheaf $\Omega^{\bullet}_{\overline{X}, \alpha}$ of $\Omega^{\bullet}_{\overline{X}}(\log (\overline{D}_0 +D_h+D_v+D_{\log}))(- \overline{D}_0)$ consisting of forms $\eta \in \Omega^{\bullet}_{\overline{X}}(\log (\overline{D}_0 +D_h+D_v+D_{\log}))(- \overline{D})$ such that $\alpha \wedge \eta \in \Omega^{\bullet}_{\overline{X}}(\log (\overline{D}_0 +D_h+D_v+D_{\log}))(- \overline{D}_0)$, which means that the $1$-form $\alpha$ can have poles of order one along the divisors $D_v \cup D_h \cup D_{\log}$.\\
\defn[Local twisted de Rham]{
Let us indicate with $\mathcal{Z} (\alpha)$ the set of zeros of $\alpha$ on $X$ and the zeros of the restriction of $\alpha$ on $D_0 \cup D_h$. We define the local (twisted) de Rham cohomology as
\begin{equation}
   H^{\bullet}_{LdR,\gamma} \left( X, D_0, \alpha \right) \, = \bigoplus_{i \in I} \, \mathbb{H}^{\bullet} \left( U_{form } (z_i), (\Omega^{\bullet}_{U_{form(z_i),\alpha}} \left[ \left[ \gamma^{-1} \right] \right], \gamma^{-1}d + \alpha \wedge \, ) \right),
   \label{LocaldR:forms}
\end{equation}
where $U_{form}(z_i)$ is the formal neighborhood of the component $z_i$ of $\mathcal{Z} (\alpha)$.}

The following isomorphism holds:
\begin{equation}
    H_{GdR, \gamma}^{\bullet} \left( X, \alpha \right) \, \simeq \, H_{LdR, \gamma}^{\bullet} \left( X, \alpha \right).
\end{equation}

\subsection{Betti Cohomology}\label{SubsectionBetti}
The direct construction of the global Betti cohomology is extremely technical, we just briefly recall it here for fixing notations, referring to \cite{Kontsevich:2024mks} for details. 
Let $\widetilde{X}$ be the real oriented blow-up of $\overline{X}$ along $D=D_h\cup D_v\cup D_{log}$, that is the manifold with boundary, and possibly corners, obtained from $\overline{X}$ replacing $D$ with the $S^1-$ bundle of its normal bundle. Let $\pi:\widetilde{X}\rightarrow\overline{X}$ and $\Pi:X\rightarrow\widetilde{X}$ be the natural projection and embedding, respectively. Let $\La_{\alpha,\gamma}$ be the local system on $X$ of flat sections of the trivial vector bundle on $X$ with respect to $\nabla=(d+\gamma\alpha)$, i.e.

\begin{equation}
    \La_{\alpha,\gamma}(U)=\ker \nabla|_{U}, \quad \quad U\subset X.
\end{equation}
\nn
Let $D^{\R}_k=\{y \in \de \widetilde{X}|\pi(y)\in D_k\}$, $k=h, v, log$, be the portion of the boundary of $\widetilde{X}$ whose points are projected onto $D_k$, and let us split each of them according to the growth behavior of $F$,

\begin{equation}
\begin{split}
    &D^{\R\,\pm}_v=\{\tilde{z}\in D^{\R}_v| \pm \mathrm{Re}(\gamma F(\pi(\tilde{z}))\geq 0)\},\\
    &D^{\R\,\pm}_{log,i}=\{\tilde{z}\in D^{\R}_{log}| \mp \mathrm{Re}( \gamma c_i)> 0)\},
    \end{split}
    \label{PlusMinusDivisors}
\end{equation}
\nn
where $c_i$ is the residue of $\alpha$ in the $i-th$ irreducible component of $D^\R_{log}$.
Next, let us define $\widetilde{X}^-$ as the subset of $\widetilde{X}$ obtained by removing the normal directions where $F$ has a ``bad'' behavior: 
\begin{equation}
    \widetilde{X}^-=\widetilde{X}\backslash \left ( (D^\R_{log}-D^{\R\,+}_{log})\cup(D^\R_{v}-D^{\R\,-}_{v})\cup D_h \right ),
\end{equation}
\nn
with $D^{\R\,\pm}_{log}=\overline{\cup_i D^{\R\,\pm}_{log,i}}$.
Let $i:\widetilde{X}^-\rightarrow \widetilde{X}$ its open inclusion. \\[0.5cm]
\defn[Global Betti cohomology]{
The global Betti cohomology is defined as
\begin{equation}
\begin{split}
    H^{\bullet}_{GB, \gamma} \left( X, \alpha \right)& \equiv H^{\bullet} \left( \widetilde{X}, \Pi_\ast (\La_{\alpha,\gamma})\otimes i_!(\underline{\Z}_{\widetilde{X}^-})\right)\\ &\cong H^{\bullet} \left( \widetilde{X}, i_!(\underline{\Z}_{\widetilde{X}^-}\otimes i^\ast \Pi_\ast (\La_{\alpha,\gamma})\right)\cong H^{\bullet} \left( \widetilde{X}, i_!i^\ast \Pi_\ast (\La_{\alpha,\gamma})\right) \\&\cong H^{\bullet} \left( \widetilde{X}, D^{\R\,+}_v\cup D^{\R\,-}_{log},\Pi_\ast (\La_{\alpha,\gamma})\right).
    \label{GlobalBetti1form}
    \end{split}
\end{equation}}

Let us describe the construction of the \textit{local} Betti cohomology in a way similar to that presented in section  \ref{sec:betti} for the case of holomorphic functions. 
Fixing a Riemannian metric on $X$, for each $z_i \in \mathcal{Z}(\alpha)$, we can always choose a sufficiently small $\varepsilon-$neighborhood $U_{\varepsilon,i} \left( z_i  \right) \subset X$ and a holomorphic function $f_i$ defined on it such that, locally in this neighborhood,
\begin{equation}
    \alpha = d f_i.
\end{equation}
For each $\theta \in \mathbb{R} / 2 \pi \mathbb{Z}$, with $\theta = \arg \left( \gamma \right)$, we can define the graded $\mathbb{Z}-$module $H^{\bullet}_{LB,z_i,\gamma}$, analogous of \eqref{Betti_local:ti}, via the relative cohomology with respect to preimage of the point $t_i +\varepsilon e^{i \theta}$ in the boundary of $\overline{U}_{\varepsilon,i} \left( z_i  \right)$:
\begin{equation}
    H^{\bullet}_{LB,z_i, \gamma} \left( X, \alpha \right) = H^{\bullet} \left(U_{\varepsilon,i}(z_i), \overline U_{\varepsilon,i}(z_i) \cap f_i^{-1} \left( t_i+\varepsilon e^{i \theta} \right) \right),
\end{equation}
with $t_i=f_i \left( z_i \right)$.\\\\
\defn[local Betti cohomology]{
For fixed $\gamma$, the direct sum
\begin{equation}
    H^{\bullet}_{LB, \gamma} \left( X, \alpha \right) = \bigoplus_{z_i \in \mathcal{Z}(\alpha)} H^{\bullet}_{LB,z_i,\gamma} \left( X, \alpha \right)
\end{equation}
is called the local Betti cohomology.\\\\}
 Since the divisor $D_h$ is empty, as we will see below, the only zeros of $\alpha$ contributing to $\mathcal{Z}(\alpha)$ arise from the domain $X= \mathbb{C}^n \setminus \left\lbrace \mathcal{B}=0 \right\rbrace$. 
If the roots of $d \mathcal{B}$ do not lie on the hypersurface $\mathcal{B}=0$, the set $\mathcal{Z}(\alpha)$ coincides with the set $\Sigma$ of critical points of the polynomial $\mathcal{B}$ within $X$.  Since these are the only points contributing to the construction of the local Betti (co)homology, we can proceed in these cases analogously to the approach described for holomorphic functions.\\
Let us now discuss the global-to-local isomorphism for these Betti cohomologies. First, let us fix a new definition of Stokes rays in terms of zeros of the $1$-form $\alpha$ rather than in terms of critical points as done in the case of holomorphic functions.\\[0.5cm]
\defn[Stokes ray]{
We call the ray $s_{\theta} = \left\lbrace \gamma \, \vert \, \arg{\gamma}= \pi - \theta_{ij} = \theta \right\rbrace = \mathbb{R}_{\geq 0} \cdot e^{i \theta} \subset \mathbb{C}_{\gamma}$ with $\theta= \arg \left( \int_{\Gamma_{ij} } \alpha \right)$, where $\Gamma_{ij}$ is the homotopy class of paths in $X$ joining the two points $z_i$ and $z_j$ in $\mathcal Z(\alpha)$, a Stokes ray. \\ Rays with vertex at the origin that are not Stokes rays are called generic rays.}

If $\gamma$ does not lie on a Stokes ray, given the local system $\mathcal{L}_{\alpha, \gamma}$ associated with the holomorphic $1$-form $\gamma \alpha$, we always have a well defined isomorphism
\begin{equation}
    \varphi_{\arg{\gamma}} \, : \, H^{\bullet}_{GB, \gamma} (X, \alpha) \, \simeq  \, H^{\bullet}_{LB, \gamma} (X, \alpha).
    \label{isomorphism:local_to_global_forms}
\end{equation}
Close to a Stokes ray $s_{\theta}$, there exist two isomorphisms, $\varphi_{\theta^+}$ and $\varphi_{\theta^-}$,  corresponding to angles immediately adjacent to the ray.  The discrepancy between these isomorphisms is captured by the Stokes automorphism
\begin{equation}
    \varphi_{\theta^-}^{-1} \circ \varphi_{\theta^+} \, : \, H^{\bullet}_{LB, \gamma} (X, \alpha) \, \rightarrow \, H^{\bullet}_{LB, \gamma} (X, \alpha). 
    \label{Stokes_automorphism:closed_forms}
\end{equation}
Just as for holomorphic functions, we can also associate a wall-crossing structure to the pair $\left( X, \alpha \right)$ by using the maps \eqref{Stokes_automorphism:closed_forms} as $\gamma$ varies along $S^1_{\theta}$ in $\mathbb{C}^{\ast}_{\gamma}$. For a continuous family of pairs $\left( X, \alpha \right)$, as far as $\pi_0 \left( \mathcal{Z}(\alpha)\right)$ is locally constant, the corresponding wall crossing structures form a continuous family of WCS. 

Now, let us assume that the zeros of $\alpha$ are isolated and simple. Then, for each $z_i \in \mathcal{Z}(\alpha)$, and its associated holomorphic function $\gamma f_i$, we can construct the thimble $th_{i , \theta_{\gamma}}$ emanating from $z_i \in X$ by tracing the vanishing cycles $\Delta_i (s)$ on the level sets $f_i - f_i(z_i)=s$ for $s \in \left[0; + \infty\right)$ along the direction $\theta_{\gamma}$. We say that $th_{i , \theta_{\gamma}}$ is compatible with $\alpha$ in the direction $\theta_{\gamma}$ if, for any point $x \in \Delta_i(s)$, as $s \rightarrow + \infty$, we have $\gamma f_i(x) \rightarrow + \infty$. In this case, we can define the pairing between the de Rham cohomology and the Betti homology class represented by $th_{i , \theta_{\gamma}}$ via the exponential integral
\begin{equation}
    I_i (\gamma)= \int_{th_{i , \theta_{\gamma}}} e^{- \gamma f_i} \, \mu,
    \label{Pairing:dR_Betti_forms}
\end{equation}
which is well defined only when the integral is convergent. Following a construction analogous to the one described in Section \ref{WCSsection}, the integral \eqref{Pairing:dR_Betti_forms} can rewritten as

\begin{equation}
    I_i (\gamma) = e^{- \gamma f_i (z_i)} \int_{0}^{+ \infty} ds\ e^{- \gamma s} vol_{\Delta_i} (s).
\end{equation}

where $vol_{\Delta_i} (s)$ denotes the volume of the vanishing cycle $\Delta_i (s)$ on the level set $f_i - f_i(z_i)=s$. Since $vol_{\Delta_i} (s)$ typically increases as $s \rightarrow + \infty$, the converge of the integral \eqref{Pairing:dR_Betti_forms} is ensured if and only if this volume growth is at most exponential.  However, this condition may fail when $\dim_{\mathbb{C}} X \geq 3$. 

 \section{Exponential integral for Feynman integrals}
Feynman integrals in the Baikov representation \eqref{Baikov}, can be seen as exponential integrals involving the multivalued logarithmic functions 
\begin{equation}
    f(z_1,z_2,\dots,z_n)= \log \mathcal{B} \left( z_1, z_2,\dots,z_n \right).
\end{equation}
Despite the logarithmic function does not define a global holomorphic function, its derivative is a well defined closed holomorphic one form. Thus, we want to specialize the above discussion to the case where $\alpha$ precisely includes such a contribution and the induced pairing describes a exponential integral with a logarithmic exponent.\\
The domain of definition of $f$ is the complex manifold  $X=\mathbb{C}^n \setminus \left\lbrace \mathcal{B}=0 \right\rbrace$, which excludes the zero locus of $\mathcal{B}$. A natural compactification of this space is the complex projective space $\overline{X}=\mathbb{P}^n= \mathbb{C}^n  \cup \mathbb{P}^{n-1}$, obtained introducing the additional divisors corresponding to the hyperplane at infinity and the zeros of the homogenized Baikov polynomial:

\begin{equation}
    D_{\overline{\mathcal{B}}}\equiv\{[z_1,\dots,z_n,\zeta]\in \Pro^n\ |\ \overline{\mathcal{B}}(z_1,\dots,z_n,\zeta)=0\},
\end{equation}
where $\overline {\mathcal B}$ is the extension of $\mathcal B$ to the compactification, with values in $\mathbb P^1$.  Notice that $f$ does not extend, but this is totally irrelevant.
The two added divisors intersect non trivially in $D_{\overline{\mathcal{B}}}\cap \Pro^{n-1} = \{[z_1,\dots,z_n,0]\in \Pro^n\ |\ \overline{\mathcal{B}}(z_1,\dots,z_n,0)=0\}$, i.e. the points at infinity of the compactification in $\Pro^n$ of the variety defined by the Baikov polynomial. 
Let us call the hyperplane at infinity
\begin{equation}
    D_\infty \equiv \{[z_1,\dots,z_n,0]\in \Pro^n\ \},
\end{equation}
\nn
and finally write
\begin{equation}
    \overline{X}- X= D_{\overline{\mathcal{B}}}\cup D_\infty.
\end{equation}
\nn
Consider now the globally-defined 1-form 

 \begin{equation}
    \alpha = d \log \overline{\mathcal{B}} = \frac{d \overline{\mathcal{B}}}{\overline{\mathcal{B}}},
\end{equation}
\nn
which clearly shows logarithmic poles along $D_{\overline{\mathcal{B}}}$ with residue $c_{\overline{\mathcal{B}}}=+1$, thus $D_{\overline{\mathcal{B}}}\subset D_{log}$. In order to study the behavior at infinity, let us choose the coordinate $\eta^{-1} = \overline{\mathcal{B}}$: as $\eta$ approaches zero, $\alpha=-d\eta/\eta$ exhibits again a logarithmic singularity, but with opposite residue: $c_\infty=-1$. We then finally conclude:

\begin{equation}
    D_{log}=D_{\overline{\mathcal{B}}}\cup D_\infty \quad \quad \mbox{and}\quad\quad  D_h=D_v = \emptyset.
\end{equation}
\nn
The $S^1-$bundle with Euler class $1$ over $D_\infty\cong S^{n}$ is given by the Hopf fibration: $D^\R_\infty\cong S^{2n-1}$. On the other hand $D^\R_{\overline{\mathcal{B}}}$ has Euler class $e=d c_1$, with $d$ the degree of $\mathcal{B}$ and $c_1=c_1(\Ol_{D_{\overline{\mathcal{B}}}})$ the first Chern class of the normal bundle of $D_{\overline{\mathcal{B}}}$ in $\C\Pro^n$.
Applying definitions \eqref{PlusMinusDivisors} we get 

\begin{equation}
    D^{\R+}_{log} =\begin{cases}D^\R_\infty, \quad\quad \mbox{if}\quad \mathrm{Re}(\gamma)>0,\\ D^\R_{\overline{\mathcal{B}}},\quad\quad\, \mbox{if}\quad \mathrm{Re}(\gamma)<0 ,\end{cases}
\end{equation}
\nn
and 

\begin{equation}
    D^{\R-}_{log} =\begin{cases}D^\R_\infty, \quad\quad \mbox{if}\quad \mathrm{Re}(\gamma)<0,\\ D^\R_{\overline{\mathcal{B}}},\quad\quad\, \mbox{if}\quad \mathrm{Re}(\gamma)>0,\end{cases}
\end{equation}
thus the global Betti cohomology \eqref{GlobalBetti1form} becomes
\begin{equation}
    H^{\bullet}_{GB, \gamma} \left( X, \alpha \right)\cong \begin{cases}H^{\bullet} \left( \widetilde{X}, D^\R_\infty ,\Pi_\ast (\La_{\alpha,\gamma})\right), \quad\quad \mbox{if}\quad \mathrm{Re}(\gamma)>0,\\
    H^{\bullet} \left( \widetilde{X}, D^\R_{\overline{\mathcal{B}}}\,, \Pi_\ast (\La_{\alpha,\gamma})\right),\quad\quad\, \mbox{if}\quad \mathrm{Re}(\gamma)<0 .\end{cases}
    \label{GlobalBettiLog}
\end{equation}
\nn
For better readability, we now simplify the notation by setting $\La \equiv \La_{\alpha,\gamma}$.\\
The key of the whole discussion lies in the behavior of the direct image $\Pi_\ast (\La)$ that, by definition, is the extension of $\La$ to $\widetilde{X}$ by zero, 
i.e. the sheaf associating to $U\subset \widetilde{X}$ the group of sections of $\La$ on $U\cap X$. Notice that no nonzero sections are entirely supported on the boundary, and since $X$ is a deformation retract of $\widetilde{X}$, we get 

\begin{equation}
    H^k(\widetilde{X},\Pi_\ast \La)\cong   H^k(X,\La).
\end{equation}
\nn
Therefore, the long exact sequence for the pair $(\widetilde{X},D^\R_k)$ reduces to 

\begin{equation}
\begin{split}
    \cdots\longrightarrow H^k(\widetilde{X},D^\R_k,\Pi_\ast \La) &\longrightarrow H^k(X,\La)\longrightarrow H^k(D^\R_k, \Pi_\ast \La)\longrightarrow \\&\longrightarrow H^{k+1}(\widetilde{X},D^\R_k,\Pi_\ast \La) \longrightarrow \cdots .
    \end{split}
    \label{LES6}
\end{equation}
\nn
In order to compute \eqref{GlobalBettiLog}, we have thus to determine $H^k(D^\R_k, \Pi_\ast \La)$ and $H^k(X,\La)$.\\ 
Now, let $M_k\in GL(r,\C)$ be the monodromy matrix around $D_k$, with $r$ the rank of $\La$. In order for a global section on $X$ to be flatly extendable to the boundary $D^\R_k$, it must belong to $\ker(M_k-I)$. Indeed, subspaces that are invariant under the action of monodromies determine directions towards the boundary, along which global sections can be analytically continued. If $M_k$ is semisimple  with no eigenvalues $1$ and no product relations (like, e.g., $\det\,(M_1 M_2)=1$) occur, no nonzero section of $\La$ extends along the boundary $\de\widetilde{X}$. In the opposite situation, when all $M_k$ are the identity matrix, all global sections flatly extend along the boundary. In the general case, whenever $M_k$ has some unitary eigeinvalue and/or relations among monodromies appear, some global sections extend, whereas others do not. 
Intuitively, the cohomology relative to a portion of the boundary kills all global sections that can be flatly extended to that boundary. Thus it depends on the possibility to extend global sections, encoded in  $H^k(X,\La)$, and on possible obstructions due to the topology of the boundary, encoded in  $H^k(D^\R_k, \Pi_\ast \La)$.\\ 
Moreover, unless the local system is trivial (e.g. a constant sheaf), the isomorphism $H^k(-,\La)\cong H^k(-,\C)\otimes\La$ fails to be true, due to the torsion\footnote{Measured by the $Ext(\C,\La)$.} of $\La$ on $\C$, and the cohomologies with coefficients in $\La$, significantly depends on the behaviour of the local system, encoded in the monodromy matrices defining it. This contribution must be computed case by case, however, we can compute, in fully generality, the contribution coming from the cohomologies with constant coefficients.  \\
We assume $D_{\overline{\mathcal{B}}}$ to be smooth. In particular, $D_{\mathcal B}\equiv\{z\in\mathbb C^n|\mathcal B(z)=0 \}$ is a smooth affine variety. Therefore, we can proceed as follows. First, the Alexander duality theorem ensures that\footnote{Here smoothness is irrelevant}
\begin{align}
    H^{k}(\mathbb C^n\backslash D_{\mathcal B}, \mathbb C)\simeq \tilde H_{2n-k-1}(D_{\mathcal B},\mathbb C),
\end{align}
where $\tilde H_*$ is the reduced homology. Next, since $D_{\mathcal B}$ is smooth and has complex codimension $1$, despite being noncompact, we can use the Poincar\'e duality to get (for $k\geq 1$)
\begin{align}
    H^{k}(\mathbb C^n\backslash D_{\mathcal B}, \mathbb C)\simeq \tilde H^{k-1}(D_{\mathcal B},\mathbb C).
\end{align}
Finally, because $\tilde H^0=H^0/\mathbb C$ and $\tilde H^q=H^q$ for $q\geq 1$, we get
\begin{equation}
    H^k(X,\C) \cong \begin{cases}\C,\hspace{5cm} k=0\\ 0,\hspace{5.05cm} k=1  \\ H^{k-1}(D_{\mathcal B},\mathbb C)\cong \C^{m_k} , \hspace{1.65cm} k=2,\ldots,n
    \end{cases}
    \label{Primi}
\end{equation}
\nn
with $m_k\equiv \dim H^{k-1}(D_\mathcal{B},\mathbb C)$. The cohomologies of the boundaries components are computed via the Serre spectral sequence \eqref{SerreSQ} applied to the circle bundles $(S^1\rightarrow D^\R_j\rightarrow D_j)$. The second page $E_2^{p,q}=H^p(D_j,H^q(S^1))$ is 

\begin{equation}
    E_2^{p,q}=H^p(D_j,\C)\quad \quad \mbox{for}\quad\quad q=0,1
\end{equation}

 and it vanishes otherwise.\\
 Both spectral sequences degenerate at page $E_2$. The case of $D_{\overline {\mathcal B}}$ clearly depends on $\mathcal B$ and must be computed case by case.\footnote{Notice $D^\R_{D_\mathcal{B}}$ is twisted by $\Ol(degree(\mathcal{B}))$.} In particular, in the Serre spectral sequence the obstruction to lift the classes of $S^1$ to that of the total space of the bundle is $d_2^{p,1}: E^{p,1}_2\rightarrow E^{p+2,0}_2$. It acts as the cup product by the Chern class of the $U(1)$ bundle.\\ 
For $D_\infty$ we have that 

\begin{equation}
    H^{2j}(\mathbb P^{n-1})\cong \begin{cases} \C,\hspace{2cm} k=0,\dots,n-1\\0,\hspace{2cm} \mbox{otherwise.}
    \end{cases}
\end{equation}

Notice that if we call ${{\rm H}}\in\mathbb P^{n-1}$ the hyperplane at infinity, then the generator of $H^{2j}$ is ${\rm H}^j$. On the other hand, ${\rm H}$ is exactly the first Chern class of the $U(1)$ bundle. Thus, $d_2^{p,1}=\cdot \cup H$ is the cup product by $H$. It maps $H^{2j}(\mathbb P^{n-1},\mathbb C)\rightarrow H^{2j+2}(\mathbb P^{n-1},\mathbb C)$ injectively for $j=0,1,\ldots,n-2$ and $H^{2n-2}(\mathbb P^{n-1},\mathbb C)$ to  $0$. Therefore, the only surviving terms are 

\begin{equation}
    E^{0,0}_2\cong \mathbb C\cong H^0(D_\infty^{\mathbb R},\mathbb C)
\end{equation}

and 
\begin{equation}
    E^{2n-2,1}_2\cong \mathbb C\cong H^{2n-1}(D_\infty^{\mathbb R},\mathbb C).
\end{equation}
Thus 
\begin{equation}
    H^k(D^\R_\infty,\C)\cong \begin{cases} \C \quad \mbox{for}\, k=0,2n-1\\ 0 \quad \mbox{otherwise}, \end{cases}
\end{equation}
\nn
and we finally obtain:

\begin{equation}
        H^\bullet (\widetilde{X},D^\R_\infty,\C)\cong \C^{m_2}\oplus..\oplus\C^{m_n},
\end{equation}
concentrated in degrees $2\leq k\leq n$.\\
If $p=D_\infty\cap D_{\overline{\mathcal{B}}}$ is a critical point for $\overline{\mathcal{B}}$, $\La$ contains the monodromy $M_\infty$ around $p$, and the twisted cohomology could be affected by its action. However, even in that case, because $\pi_1(S^{2n-1})=0$, any action is impossible. Therefore: 

\begin{equation}
    H^k(D^\R_\infty,\La)\cong  H^k(S^{2n-1},\C)\otimes\La.
    \label{InfinityCohology}
\end{equation}

\nn
In the following subsection we will use these results to explicitly compute \eqref{GlobalBettiLog} in a concrete example.

\subsection{Elliptic fibers} \label{SI}
We will now apply the generalized framework for multivalued functions described above, to a concrete example closely related to the family of $l-$loops banana integrals we saw in section \ref{SecBanana}. Remind the two-loops banana diagram, the sunrise graph

\begin{figure}[h!]
    \centering
\begin{tikzpicture}
\draw[black, thick] (0,0) -- (1,0);
\draw[black, thick] (5,0) -- (6,0);
\draw[black] plot [smooth,tension=1.5] coordinates {(1,0) (3,1) (5,0)};
\draw[black] plot [smooth,tension=1.5] coordinates {(1,0) (3,-1) (5,0)};
\draw[black](1,0) -- (5,0);
\draw node [above] at (3,1) {$m_1$};
\draw node [above] at (3,0) {$m_2$};
\draw node [below] at (3,-1) {$m_{3}$};
\draw node [above] at (0.5,0) {$p$};
\draw node [above] at (5.5,0) {$p$};
\draw node [above] at (6.5,-0.3) {,};
\end{tikzpicture}
    \label{Sunrise2}
\end{figure}

is associated to a Feynman Integral whose  Symanzik representation \eqref{2DSunriseSymanzik}

\begin{equation}
    I=\int_\Delta \frac{\mu}{(\mu_1x+\mu_2y+z)(xz+yz+xy)-\xi xyz},
    \label{2DSunriseSymanzik2}
\end{equation}

is related to the periods of the family of elliptic curves defined by the second Symanzik polynomial, that, in the general case, is controlled by four parameters: the external momentum and the masses of the three internal propagators.\\
In order to make the details of the formalism above as clear as possible, we choose to apply it to the simplest case of the well-known Legendre family of elliptic curves, thus avoiding technical complications.\\
Consider then the family of integrals

\begin{equation}
\begin{split}
\mathcal{I}&= \int_{\Gamma} \frac{dx \wedge dy}{\left[ y^2+x (x-1)(x- \lambda) \right]^{ \gamma}} = \int_{\Gamma} e^{- \gamma \log \left[ y^2+x (x-1)(x- \lambda) \right]} dx \wedge dy \\&=\int_{\Gamma} e^{- \gamma \log \mathcal{B}(x,y;\lambda)} dx \wedge dy,\end{split}
\end{equation}
\nn
associated to the Legendre family of elliptic curves $\mathcal{E}_\lambda\equiv\{(x,y)\in \C^2| \mathcal{B}(x,y,\lambda)=0\}$, where $\lambda \in \mathcal{M}_{cs}=\mathbb{C} \setminus \left\lbrace 0,1,\infty \right\rbrace$ is a complex structure parameter. In the physical interpretation, $\lambda$ can be thought of as parameterizing the masses of internal particles or the external momenta.\\
In this example we have $X=\C^2\backslash \mathcal{E}_\lambda$, $\gamma \in \mathbb{C}^{\ast}_{\gamma}$ and the integration contour $\Gamma \in H_2 \left( \C^2, D_0 \right)$ is a singular Borel-Moore $2-$chain with boundaries on the divisor $D_0$ defined as
\begin{equation}
    D_0 = \lim_{N \rightarrow \infty} X_N = \lim_{N \rightarrow \infty} \left\lbrace (x,y) \in X\ \vert\ Re \left( \gamma \log \mathcal{B}(x,y;\lambda) \right) \geq N \right\rbrace.
\end{equation}
\nn
The natural choice for the compactification is
\begin{equation}
  \overline{X} \, = \, \Pro^2 \,  = \, \mathbb{C}^2 \setminus \left\lbrace \mathcal{E}_{\lambda}=0 \right\rbrace \, \cup \, \left\lbrace \overline{\mathcal{E}}_{\lambda}=0 \right\rbrace  \, \cup \, \mathbb{P}^1,
\end{equation}
 where $\mathcal{B}(x,y;\lambda)$ extends to
\begin{equation}
    \overline{\mathcal{B}}(x,y,\eta;\lambda)=y^2\eta-x(x-\eta)(x-\eta\lambda), 
\end{equation}
\nn
with 
\begin{equation}
    d\,\log \overline{\mathcal{B}} =\frac{2\eta y dy+[y^2+x^2+x\lambda(x-2\eta)]d\eta+[-3x^2-\eta^2\lambda+2x\eta(1+\lambda)]dx}{y^2\eta-x(x-\eta)(x-\eta\lambda)}.
\end{equation}
\nn
By analyzing the behavior of $\overline{\mathcal{B}}$ on $\Pro^2$ we can identify the types of divisors introduced in our compactification. In particular, we find

\begin{equation}
    D_h=D_v=\emptyset \quad\mbox{and}\quad D_{log}=D_{\overline{\mathcal{B}}}\cup D_\infty,
\end{equation}
\nn
with 
\begin{equation}
\begin{split}
    &D_{\overline{\mathcal{B}}}=  \overline{\mathcal{E}}_{\lambda}  = \{[x:y:\eta]\in \Pro^2| \overline{\mathcal{B}}=0\},\\
    &D_\infty = \mathbb{P}^1 = \{[x:y:0]\in \Pro^2\}, 
    \end{split}
    \label{intersectionpoint}
\end{equation}
\nn
intersecting at $D_{\overline{\mathcal{B}}}\cap D_\infty =[0:1:0]$.\\
Since the divisors $D_v$ and $D_h$ are empty, the $1$-form $\alpha$ receives a singular contributions only from the divisor $D_{\log}$. Therefore, we can write
\begin{equation}
    \alpha = \alpha_{\log} + \alpha_{reg}.
\end{equation}
Around $\mathcal{E}_\lambda\subset D_{log}$, the closed $1$-form $\alpha$ is the holomorphic form
\begin{equation}
    \alpha_{\log} = \frac{d \mathcal{B}}{\mathcal{B}},
\end{equation}
whose zeros are
\begin{equation}
\begin{split}
     \mathcal{Z}(\alpha)&= \left\lbrace z_i=(x_i,y_i) \in X \vert d \mathcal{B} (x,y;\lambda)/ \mathcal{B}(x,y,\lambda)=0 \right\rbrace \\
    &  = \left\lbrace \left( \frac{1}{3} \left[ 1+ \lambda \mp \sqrt{1 + \lambda (\lambda-1)}\right] ,0 \right) \right\rbrace, \\
\end{split}
\end{equation}
with corresponding critical values
\begin{equation}
 S= \left\lbrace \log t_i \in \mathbb{C} \vert \log \mathcal{B} \left( z_i; \lambda\right) \right\rbrace,
\end{equation}
with
\begin{equation}
     \begin{cases} & t_1= \frac{1}{27} \left(\sqrt{\lambda^2-\lambda+1}-\lambda-1\right) \left(\lambda^2-4 \lambda-(\lambda+1)\sqrt{\lambda^2-\lambda+1}+1\right)\\
    & t_2=  -\frac{1}{27} \left(\sqrt{\lambda^2-\lambda+1}+\lambda+1\right) \left(\lambda^2-4\lambda+(\lambda+1)\sqrt{\lambda^2-\lambda+1}+1\right).\\ \end{cases}
\end{equation}
The map 
\begin{equation}
    \mathcal{B} \, : \quad X \quad \longmapsto \quad \mathbb{C}_t^{\ast},
\end{equation}
defines a non-trivial Lefschetz fibration over $\mathbb{C}_t^{\ast}= \mathbb{C} \setminus \left\lbrace t_1, t_2 \right\rbrace$ for each $\lambda \in \mathcal{M}_{cs}=\C_\lambda\setminus \{0,1,\infty\}$. \\
At fixed $\lambda$, the generic fiber $\mathcal{B}^{-1}(t)$ is the elliptic curve:
\begin{equation}
   \mathcal{F}_t \, : \quad  y^2+x (x-1)(x- \lambda)=t.
\end{equation}
The badness of the fibration at the critical values in $\left\lbrace t_1, t_2 \right\rbrace$ is measured in terms of the local monodromies acting on the homology group $H_{1}(\mathcal{F}_t)$ through the matrices:
\begin{equation}
    M_{i} \, \, : \, \, H_1 (\mathcal{F}_t) \quad \longmapsto \quad H_1 (\mathcal{F}_t) \, \, \quad , \,\,\, i=1,2.
\end{equation}
To determine a basis for this homology and the therein representation of the monodromies we follow the description revised in section \ref{PLT}. We fix a non-critical point $t_0$ in $\mathbb{C}_t$ and construct two paths
\begin{equation}
    u_i \, \, : \, \, \left[0;1 \right] \quad \longmapsto \quad \mathbb{C}_t,
\end{equation}
each connecting the non-critical value $t_0=u_i(0)$  to a critical value $t_i =u_i(1)$ without crossing any other critical point. For each path $u_i(t)$ we can define a family of $1$-dimensional spheres in the level manifolds $\mathcal{F}_{u_i}$:
\begin{equation}
    S_i(s)= \sqrt{u_i(s)-t_i} S^1.
\end{equation}
They shrink to radius zero as we approach the critical point $t_i$. The homology classes $\Delta_i \in H_1 (\mathcal{F}_{t_0})$ represented by these spheres are the Picard-Lefschetz vanishing cycles along the paths $u_i$ and they form a basis for the homology $H_1 (\mathcal{F}_{t_0})$. \\
To provide a clear visualization of the construction, we fix the parameter $\lambda=3$ and carry out the explicit computations for this case. As long as $\pi_0 \left( \mathcal{Z} (\alpha) \right)$ remains invariant under variations of  $\lambda \in \mathcal{M}_{cs}$ the wall crossing structures associated with the pairs $(X_{\lambda}, \alpha_{\lambda})$ remain continuously connected to that of $\left( X_3, \alpha_3 \right)$.  The sets $\mathcal{Z}(\alpha)$ and $S$ are:
\begin{equation}
\begin{split}
    &\mathcal{Z}(\alpha) = \left\lbrace z_1=\left( \frac{1}{3} \left(\sqrt{7}+4\right), 0 \right) , z_2 = \left(\frac{1}{3} \left(-\sqrt{7}+4\right), 0 \right)\right\rbrace \, , \\
    & S=\left\lbrace t_1= - \frac{2}{27} \left(7 \sqrt{7}+10\right), t_2= \frac{2}{27} \left(7 \sqrt{7}-10\right) \right\rbrace. \\
\end{split}
\end{equation}
The level manifold $\mathcal{F}_{t_0}$ at the regular point $t_0$ is the graph of the two-valued function
\begin{equation}
    y= \pm \sqrt{t_0-(x-3) (x-1) x},
\end{equation}
namely, the double-covering of the $x$ plane, branched at the points:

\begin{equation}
\begin{split}
 & x_1= \frac{1}{6} \left(-2^{2/3} c_1-\frac{14 \sqrt[3]{2}}{c_1}+8\right), \\
   & x_2 = \frac{1}{12} \left(2^{2/3} \left(1-i \sqrt{3}\right) c_1+\frac{14 \sqrt[3]{2} \left(1+i \sqrt{3}\right)}{c_1}+16\right), \\
   & x_3= \frac{1}{12} \left(2^{2/3} \left(1+i \sqrt{3}\right) c_1+\frac{14 \sqrt[3]{2} \left(1-i \sqrt{3}\right)}{c_1}+16\right),
\end{split}
\end{equation}
with 

\begin{equation}
    c_1=\sqrt[3]{3 \sqrt{3} \sqrt{27 t^2+40 t-36}-27 t-20}.
\end{equation}
Let us choose the first cut from $x_1$ to $x_3$ and the second cut from $x_2$ to infinity.\\
As we move the value of $t$ from $t_0$ to one of the critical values, the level manifold $\mathcal{F}_t$ is deformed and it becomes singular. In particular, when we approach $t_1$ we have that the branch point $x_2$ moves until it overlaps with $x_3$, while when we approach $t_2$ the point $x_1$ moves towards the point $x_3$. From this construction we can draw the vanishing cycles $\Delta_1$ and $\Delta_2$ in $\mathcal{F}_{t_0}$ associated to the paths $u_1$ and $u_2$, respectively. The cycle $\Delta_1$ encircles the points $x_2$ and $x_3$, while $\Delta_2$ encircles $x_1$ and $x_3$.

Tracing the change in the positions of the three points $x_j$ as we follow the counterclockwise-oriented closed loop $\tau_i \in \pi_1 \left( \mathbb{C}_t \setminus \left\lbrace t_1, t_2\right\rbrace, t_0 \right)$ encircling the critical point $t_i$ we can deduce the corresponding monodromy action on the ordered basis $\left\lbrace \Delta_1, \Delta_2 \right\rbrace$ of vanishing cycles. In particular, we obtain
\begin{equation}
    M_1 = \left( \begin{matrix}  1 & 0 \\ 1 & 1  \end{matrix} \right) \quad \quad , \quad \quad M_2 = \left( \begin{matrix}  1 & -1 \\ 0 & 1  \end{matrix} \right).
    \label{monodromies:cubic_points}
\end{equation}
 Using the Picard-Lefschetz formula \eqref{ExplicitPicardLef}, we can derive the intersection form on $H_1 \left( \mathcal{F}_t, \mathbb{Z}  \right)$ from these monodromies, expressed with respect to the chosen basis of vanishing cycles:
\begin{equation}
     \Delta_i \circ \Delta_j = \left( \begin{matrix} 0 & 1 \\ -1 & 0 \end{matrix}\right).
    \label{intersection:cubic_vanishing_cycles}
\end{equation}
These local monodromies characterize the type of singularity occurring at the critical points $t_1$ and $t_2$. In this particular case, where the fiber is an elliptic curve, we can refer to the singularity in  $t_1$ as a MUM singularity, while the singularity in $t_2$ as a conifold point. \\

\paragraph{Betti cohomology.}\mbox{}\\
The divisors \eqref{intersectionpoint} intersect with normal crossing, thus the real oriented blow-up of $\Pro^2$ along $D_{log}$ is given by union of blow-ups along the two divisors. The normal bundles of $D_\infty\cong S^2$ and $D_{\overline{\mathcal{B}}}\cong T^2$ in $\Pro^2$ are respectively the complex line bundles $\Ol(1)$ and $\Ol(3)$. The corresponding circle bundles respectively are the Hopf fibration $S^3\rightarrow S^2$ and the 3-dimensional Heisenberg nilmanifold ${\rm{Nil}}^3\cong H_3(\R)/H_3(\Z) $, so $\widetilde{X}$ has boundary

\begin{equation}
    \de \widetilde{X}=S^3 \cup {\rm{Nil}}^3,
\end{equation}
\nn
where $\de D^\R_\infty$ and $ \de D^\R_{\overline{\mathcal{B}}}$ are glued along the corner $S^1\times S^1$, preimage of the intersection point \footnote{Notice that it has multiplicity 3.}.\\
We want to compute 
\begin{equation}
\begin{split}
    &H^\bullet_{GB,\gamma}(X,\alpha)(\widetilde{X},S^3,\Pi_\ast(\La_{\alpha,\gamma})), \quad\quad \mathrm{Re}(\gamma)>0,\\  &H^\bullet_{GB,\gamma}(X,\alpha)(\widetilde{X},D^R_{\overline{\mathcal{B}}},\Pi_\ast(\La_{\alpha,\gamma})), \quad \,\,\,\mathrm{Re}(\gamma)<0.
    \end{split}
\end{equation}
\nn
using the results discussed in section \ref{SubsectionBetti} and the monodromy matrices obtained in \eqref{monodromies:cubic_points}. 
The contributions coming from $D^{\mathbb R}_\infty \simeq S^3$, can be easily computed by the straight application of \eqref{InfinityCohology}, yielding: 
\begin{align}
    H^{\bullet}(D^\R_{\infty},\mathcal L)\simeq H^{\bullet}(S^3,\mathbb C)\otimes \mathcal L\simeq \mathbb C^2\oplus 0\oplus 0\oplus C^2.
    \label{HS}
\end{align}
\nn
The situation is much more involved in the case of $D^\R_{\overline{\mathcal{B}}}$, defined by the fibration 

\begin{align}
    S^1\hookrightarrow D^\R_{\overline{\mathcal{B}}} \rightarrow D_{\overline{\mathcal{B}}},
\end{align}
which is nontrivial (since is generated by a nontrivial normal bundle of degree 9). The monodromies are nontrivial around the elliptic curve, so we can choose to assign them to a basis of generators of its homotopy $\pi_1$, say $\rho(a)=M_1$, $\rho(b)=M_2$. The Heisenberg structure gives a central extension such that the commutator (in the group theoretical sense) $[a,b]=\iota$ generates the homotopy of the fibre. This means that the representation must respect this relation and we must have
\begin{align}
    \rho(\iota)=M_1 M_2 M_1^{-1}M_2^{-1} =\begin{pmatrix} 2 & 1 \\ 1 & 1
    \end{pmatrix}.
\end{align}
To compute the twisted cohomology we now use the isomorphism
\begin{align}
    H^{\bullet}(D^\R_{\overline{\mathcal{B}}},\mathcal L)=H^{\bullet}(\pi_1(D^\R_{\overline{\mathcal{B}}}),V_\rho),
\end{align}
where $V_\rho$ is the representation space of $\rho$ seen as left $\rho$-module. The Nilmanifold can be represented by a CW-complex obtained by gluing a 3-cell to three 2-cells, next to three 1-cells and finally to a 0-cell. So we have that the j-chains $C^j$ satisfy
\begin{equation}
\begin{split}
    &C^0(D^\R_{\overline{\mathcal{B}}},\mathcal L)\simeq V_\rho, \qquad C^1(D^\R_{\overline{\mathcal{B}}},\mathcal L)\simeq V_\rho^{3},\\&C^2(D^\R_{\overline{\mathcal{B}}},\mathcal L)\simeq V_\rho^{3}, \qquad C^3(D^\R_{\overline{\mathcal{B}}},\mathcal L)\simeq V_\rho,
\end{split}
\end{equation}
with $V_\rho\simeq \mathbb C^2$. Also, we can use that $H^j(\pi_1(D^\R_{\overline{\mathcal{B}}}),V_\rho)\simeq H^{3-j}(\pi_1(D^\R_{\overline{\mathcal{B}}}),V_{\rho^*})$, where $\rho^*$ is is the dual representation. These representations are irreducible so one finds that 
\begin{align}
   H^j(\pi_1(D^\R_{\overline{\mathcal{B}}}),V_{\rho^*})\simeq H^j(\pi_1(D^\R_{\overline{\mathcal{B}}}),V_\rho), \qquad j=0,1, 
\end{align}
so that we can reduce ourselves to the computations for $j=0,1$. The differential in the group cohomology is the standard one. We have to consider explicitly the differentials. If 
\begin{align}
    f=\begin{pmatrix}
        f_1\\ f_2
    \end{pmatrix} \in C^0\equiv \mathbb C^2
\end{align}
is a 0-chain, one has that $d^0f$ has to be a 1-chain so, for $g=a,b,\iota$, is defined by
\begin{align}
    d^0f(g)=\rho(g)v-v.
\end{align}
This means that the elements of $H^0$ are the invariant vectors.\\
Thus
\begin{align}
    0=d^0f(a)=\begin{pmatrix}
        0\\ f_1
    \end{pmatrix}, \qquad 0=d^0f(b)=\begin{pmatrix}
        -f_2\\ 0
    \end{pmatrix}, \qquad 0=d^0f(\iota)=\begin{pmatrix}
        f_1+f_2\\ f_1
    \end{pmatrix},
\end{align}
which gives $H^0(\pi_1(D^\R_{\overline{\mathcal{B}}}),V_\rho)=0$.\\
Similarly, for $f\in C^1\simeq \mathbb C^2\oplus \mathbb C^2\oplus \mathbb C^2$, one has (2-chains acts on pairs $(g,g')$ of elements of $\pi_1$)
\begin{align}
    d^1f((g,g'))=\rho(g)f(g')-f(gg')+f(g).
\end{align}
The explicit calculation of the kernel of $d^1$ is direct but tedious, so we skip details to the computation. After quotienting by the image of $d^0$ one gets:

\begin{align}
    H^{\bullet}(D^\R_{\overline{\mathcal{B}}},\La)\cong 0\oplus \C^2\oplus \C^2\oplus 0.
    \label{HDB}
\end{align}
\nn
The last piece we have to compute is  $H^{\bullet}(X,\mathcal L)$. Here, we can use again the Alexander duality:
\begin{align}
    H^k(X,\mathcal L)\simeq H^{4-k-1}_c(\mathcal{E}_\lambda,\mathcal L^*)^*. 
\end{align}
\nn
Next we use Poincar\'e duality to get \footnote{Notice in finite dimensions biduals cancel out naturally.}
\begin{align}
    H^k(X,\mathcal L)\simeq H^{k-1}(\mathcal{E}_\lambda,\mathcal L). 
\end{align}
\nn
Therefore, we have to compute $H^k(\mathcal{E}_\lambda,\mathcal L)$. The elliptic curve $\mathcal{E}_\lambda$ is affine so $H^2(\mathcal{E}_\lambda,\mathcal L)=0$. The group $H^0(\mathcal{E}_\lambda,\mathcal L)$ is determined by the vectors of $V_\rho$ invariant under the action of $M_1-I$ and $M_2-I$ which, like before, give $H^0(\mathcal{E}_\lambda,\mathcal L)=0$. The cellular decomposition of $\mathcal{E}_\lambda$ consists in a 2-cell, two 1-cells (the 0-cell is missing; in cohomology this corresponds to $H^2=0$). Thus the chains are $C^0\equiv V_\rho$, $C^1=V_\rho^2$, $C^2=0$. It follows that $\ker d^1=V_\rho^2\simeq \mathbb C^4$, while ${\rm Im}\, d^0\simeq \mathbb C^2$. We conclude that
\begin{align}
   H^{\bullet}(X,\mathcal L)=0\oplus 0\oplus \mathbb C^2\oplus 0\oplus 0. 
   \label{Hx}
\end{align}
Finally, we can use \eqref{HS},\eqref{HDB} and \eqref{Hx} in the long exact sequence \eqref{LES6}, getting:
\begin{equation}
\begin{split}
     &H^\bullet_{GB,\gamma}(X,\alpha)(\widetilde{X},S^3,\Pi_\ast(\La_{\alpha,\gamma}))\cong 0\oplus \C^2\oplus \C^2\oplus 0\oplus \C^2,  \\
     &H^\bullet_{GB,\gamma}(X,\alpha)(\widetilde{X},D^\R_{\overline{\mathcal{B}}},\Pi_\ast(\La_{\alpha,\gamma}))\cong 0 \oplus 0 \oplus \C^2\oplus \C^4\oplus 0.
     \end{split}
     \label{EllipcticGlobalBetti}
\end{equation}

\begin{figure}[h!]
    \centering
    \includegraphics[width=1\linewidth]{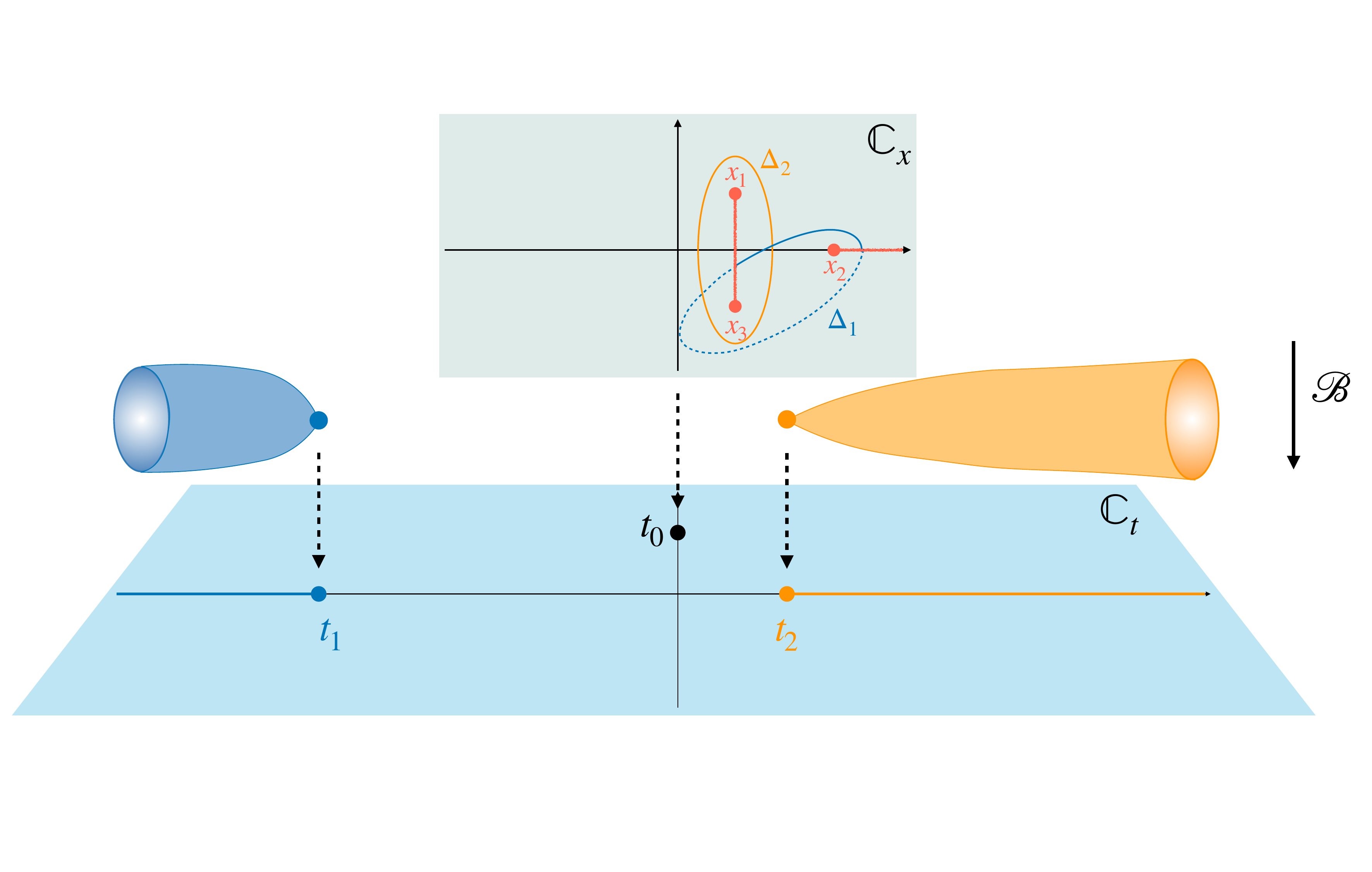}
    \caption{\small{Representation of the construction of the thimbles in the double fibration $\log: (P:X \rightarrow \mathbb{C}_t) \rightarrow \mathbb{C}_T$.}}
    \label{fig:thimbles_elliptics}
\end{figure}

\paragraph{Thimbles construction.}\mbox{}\\
At this stage, we have all the necessary ingredients to construct the thimbles associated with the vanishing cycles $\Delta_1$ and $\Delta_2$, which form a basis for the local Betti homology groups $H_2^{LB,z_i, \gamma} \left( X, \alpha \right)$. As in the case of holomorphic functions, we begin by studying the homology for a fixed $\gamma \in \mathbb{C}^{\ast}_{\gamma}$, and then we analyze its analytic continuation, equipped with a wall-crossing structure.\\
Let us fix $\gamma =1$. The steepest ascend thimble associated with the vanishing cycle $\Delta_i$ is defined as the trace over a path in $\mathbb{C}_t^{\ast}$, starting from $t_i$, along which the imaginary part $\rm{Im}(T)=\rm{Im}(\gamma \log t)$ remains constant to the value $\rm{Im}(\gamma \log t_i)=\rm{Im}(\log t_i)=arg(t_i)$, while the real part $\mathrm{Re}(\gamma \log t)=\mathrm{Re}(\log t)$ increases monotonically from $\log \vert t_i \vert$ to $+ \infty$. A graphical illustration of the results is provided in Figure \ref{fig:thimbles_elliptics}.
\nn
The Stokes rays in the plane $\mathbb{C}_{\gamma}$ are
\begin{equation}
\begin{split}
    & s_{\theta_1} = \left\lbrace \gamma \, \vert \, \arg{\gamma}= \arctan{\frac{\pi}{\log \vert t_1 \vert - \log \vert t_2 \vert}} \right\rbrace, \\
    & s_{\theta_2} = \left\lbrace \gamma \, \vert \, \arg{\gamma}= \arctan{\frac{\pi}{\log \vert t_1 \vert - \log \vert t_2 \vert}} + \pi \right\rbrace, \\
\end{split}
\end{equation}
that never stay along the real axis.\\
Now, if $\gamma$ does not belong to a Stokes ray we can consider the collection of integrals evaluated along the thimbles:
\begin{equation}
    I_i (\gamma)= \int_{th_{i, \theta_{\gamma}}} e^{-\gamma \log \mathcal{B}(x,y)} dx \wedge dy,
\end{equation}
where $\log \mathcal{B}$ is exactly the function $f$ such that $df= \alpha_{\log}$.\\
Using the parameterization $th_{i, \theta_{\gamma}} = \Delta_i (s) \times \mathbb{R}_{s \geq 0} $, we have 
\begin{equation}
    I_i (\gamma) = e^{- \gamma \log t_i} \int_0^{+ \infty} e^{- \gamma s} vol_{\Delta_i}(s) ds,
\end{equation}
where $vol_{\Delta_i} (s)$ is the volume of the vanishing cycle in the fiber $\mathcal{B}^{-1}(s)$ defined with respect to the Gelfand-Leray form $\frac{dx \wedge dy}{d\mathcal{B}/\mathcal{B}}$, namely:
\begin{equation}
    vol_{\Delta_i} (s)= \int_{\Delta_i (s)} \frac{dx \wedge dy}{d \mathcal{B}/ \mathcal{B}} = \int_{\Delta_i(s)} (t_i+s) \frac{dx}{2y(s)}.
\end{equation}
In the integrand we can recognize the holomorphic form $\omega^{1,0}= dx/y$,  so the resulting integrals are precisely the periods of this form with respect to the basis of vanishing cycles for the family of varieties $\mathcal{F}_t$:
\begin{equation}
    \mathbf{\Pi}(t) = \left( \begin{matrix} \int_{\Delta_1} \frac{dx}{y} \\ \int_{\Delta_2} \frac{dx}{y} \end{matrix} \right).
\end{equation}
Since we know the monodromies \eqref{monodromies:cubic_points} around the critical values $t_i$, we can compute the expansions for $0 \leq s \leq \epsilon$ using the Nilpotent Orbit Theorem \cite{Schmid73}:
\begin{equation}
    \mathbf{\Pi} (\tilde{s}) = e^{\tilde{s} N_i} \left( \mathbf{a}_0 + \mathbf{a}_1 e^{2 \pi i \tilde{s}}+\dots \right),
\end{equation}
where $N_i$ is the Nilpotent matrix encoding the unipotent part of the monodromy $M_i$:
\begin{equation}
    N_i = \log \left( M_i^{(u)}\right).
\end{equation}
The variable $\tilde{s}$ is related to the coordinate $s$ by the following transformation
\begin{equation}
    \tilde{s}= \frac{1}{2 \pi i} \log s,
\end{equation}
and we have
\begin{equation}
    \mathbf{a}_0 = \left( \begin{matrix} 1 \\ 0
    \end{matrix} \right) \, \, , \quad \text{or} \quad \, \, \mathbf{a}_0 = \left( \begin{matrix} 0 \\ 1 \end{matrix} \right)
\end{equation}
for $th_{1, \theta_{\gamma}}$ and $th_{2, \theta_{\gamma}}$ respectively.\\
This explicit construction shows the existence of two Lefschetz thimbles that generate the local Betti homology, which is isomorphic to the global version \eqref{EllipcticGlobalBetti}, via the isomorphism \eqref{isomorphism:local_to_global_forms}.\\
In order to compare the two Master Integrals $I_i(\gamma)$ for $\gamma \rightarrow \infty$ within the same sector of the $\mathbb{C}_{\gamma}$ plane, we associate to each of them the corresponding power series expansion in $\gamma$, as in \eqref{series:Exp_integral_gamma}. Following the discussion in Section \ref{Sec:WCS}, the coefficients $c_{i, \lambda}$ governing the series can be computed from the closed formula \eqref{coefficients:c_series_MIs}, which is entirely determined by the local monodromies \eqref{monodromies:cubic_points} around the critical points. In the present case, both monodromy matrices have a single Jordan block of dimension $2 \times 2$ associated with the eigenvalue $+1$. Consequently, for both cases we obtain the following non-trivial coefficient:
\begin{equation}
\begin{split} c_{i,0} = \frac{1}{\gamma^{-1}}  \big[ a_{0,0} \gamma^{-1} +a_{0,1} &\big( \gamma ^{-\lambda -1} \Gamma (\lambda +1) \psi ^{(0)}(\lambda +1)\\&-\gamma ^{-\lambda -1} \log (\gamma ) \Gamma (\lambda +1) \big)_{\lambda=0} \big].\end{split}
\end{equation}
This shows that the two Master Integrals associated with the different homology classes represented by thimbles are of the same order in the same sector of the $\mathbb{C}_{\gamma}$ plane.

%% file: Conclusions.tex
\chapter{Conclusions and Outlooks}
\thispagestyle{empty}                     

\chaptermark{Conclusions and Outlooks}

The analysis of physical systems across various domains, from quantum mechanics to statistical physics and from quantum field theory to string theory, often necessitates the computation of increasingly complex integrals. Developing systematic methods to address their computation remains a central challenge in theoretical physics.
One of the most powerful techniques for simplifying certain classes of integrals, those expressible as periods of de Rham cocycles over closed cycles on smooth manifolds, is provided by Stokes' theorem. By fixing appropriate bases in the relevant cohomology and homology spaces, the integration of basis cocycles over basis cycles yields a set of simpler integrals encoded in the period matrix. Stokes' theorem then allows any integral within the family to be reduced to a linear combination of these fundamental integrals, whose coefficients are interpretable as intersection numbers in either cohomology or homology.
A natural and compelling extension of this framework would be to encompass broader classes of integrals, particularly those encountered in physics. However, a significant obstacle arises in cases involving multivalued or otherwise intricate integrals, where geometric intuition is lost, and the appropriate cohomology/homology needed to define the pairing required to interpret the integral as a period is no longer evident.\\
In this thesis, based on \cite{Angiusnostro}, we proposed a systematic approach to identify the appropriate (co)-homological structures to apply in a large classes of physical integrals. Leveraging recent mathematical developments \cite{Kontsevich:2024mks}, we employ twisted de Rham cohomology and Betti homology over complex manifolds to rigorously treat exponential-type integrals as periods. This framework accommodates a wide range of physically relevant integrals, including quantum mechanical partition functions, conformal correlators, and, importantly, Feynman integrals. In the latter case this interpretation becomes viable through a generalization of established techniques for exponential integrals involving holomorphic functions in the exponent, extended to accommodate multivalued functions. Indeed, Feynman integrals expressed in the Baikov representation, as in any parametric representation as well, naturally admit such a reformulation, where the role of the multivalued function is played by the logarithm of a polynomial, the Baikov polynomial $\B$.\\
A key ingredient in this framework is the study of the complex analytic continuation of a real parameter $\gamma$, which appears as a prefactor in the exponent, related to the regularizator parameter that appear in dimensional regularization. This continuation induces a wall crossing structure, known in physics as the Cecotti-Vafa wall crossing structure \cite{Cecotti:1992ccv}, on the complex $\gamma$-plane, $\mathbb{C}_{\gamma}^{\ast}$, where four distinct local systems can be defined: local and global versions of twisted de Rham and Betti (co)homologies. The complex plane $\mathbb{C}_{\gamma}$ is partitioned into sectors by Stokes rays, and within each sector, one can define a canonical basis for each of the four (co)homologies. As $\gamma$ crosses a Stokes ray in this fan, these bases undergo discontinuous transformations encoded by Stokes automorphisms.
Correctly taking into account the presence of such jumps and the careful avoiding of stokes lines is a mandatory requirement to obtain a right basis of master integral and the right projections onto it.\\
After introducing in chapter \ref{C3} the necessary main mathematical concepts and providing in chapter \ref{C2} a overview of intersection theory applied to perturbative integrals in QFT and ST; we propose a reformulation of the formalism suited to the physical contexts of interest, we presented our main ideas for applying these techniques in physics and outline the objectives we aim to achieve.\\\\
 \textbf{Main results.}
 We explicitly determined the Twisted de Rham cohomology and Betti homology for the exponential integral of a one variable polynomial, including the cases of possible degeneracy due to the coalesce of some critical point. We perform an explicit analysis of the wall-crossing structure and associated Stokes phenomena for the Lefschetz thimble decomposition of a class of exponential Pearcey integrals arising in the grand-canonical partition function of gauged Skyrme models, which describe nuclear matter in various pasta phases. We studied how the thimbles and the Stokes lines structure change with respect to the double and triple degeneracy of the critical points.\\
We introduced, using a language accessible to physicists, the recent mathematical framework developed by Kontsevich and Soibelman \cite{Kontsevich:2024mks} to study wall-crossing structures in exponential integrals involving multivalued functions arguing that this is the appropriate framework for interpreting Feynman integrals in the Baikov representation, as well as any other parameter representation, as periods. 
We studied and described the case where the multivalued function is the logarithm of a polynomial $\B$. 
    After choosing a suitable compactification, we determined the relevant divisors and their real oriented blowups, whose identification is necessary for the correctly construction of the local system inducing the desiderate exponential pairing. 
    Although the final result clearly depends on $\B$ and it must be computed case by case, we managed to express Betti cohomology with constant coefficient in terms of the de Rham cohomology of the affine variety $\V[\B]$, defined by $\B$; in fact, we reduced the evaluation of the relevant Betti cohomology to the much easier cohomology of $\V[\B]$ and the “torsion” of the local system. \\
We used the result above to explicitly compute  the Betti cohomology of the Legendre family of elliptic curves, that is closely related to the FI associated to the two loops Banana diagram, namely the sunrise graph. The associated thimble decomposition yields a basis of simplified integrals, expressible in terms of standard elliptic integrals of the first and second kind.\\
We propose that the decomposition of Feynman integrals into simpler components via this formalism matches the standard notion of Master Integral decomposition. This correspondence offers a more geometric and algebraic perspective on the structure of these integrals. Moreover, we argue that this setup allows for a sharp counting of MIs, that could otherwise be distorted by the improper control of the behavior at infinity, the arising of trivial monodromy matrices that forbid non middle cohomology groups to vanish, and the avoiding of ambiguities linked to Stokes phenomena due to the analytic continuation in the parameter $\gamma$.\\
We analyzed the large-parameter asymptotic expansion of exponential integrals expressed over a basis of Lefschetz thimbles, where the expansion coefficients correspond to periods of standard (co)homology classes associated with families of algebraic varieties. The existence of a well-defined pairing between the global de Rham and Betti (co)homology imposes a constraint on these periods: the volume growth of these standard cycles must not exceed an exponential rate.  \\

\textbf{Outlooks}\\
The present work supplies a promising starting analysis and a robust conceptual and technical foundation for a systematic program of further investigations applying exponential period methods to a wide range of physical interest problems.\\
The immediate continuations of the analysis presented here, currently part of the in progress project \cite{Angius25}, consists in the detailed study of the implementation of these methods to: the Master Integral decompositions for families of Feynman integrals with free external kinematic parameters (such as masses of the internal particles and external momenta), and the determination of their canonical differential equation through the study of the Gauss-Manin connection over the moduli space of the corresponding geometries; the application to integrals involving a larger number of variables, as, for instance, higher loops Banana integrals, associated to families of higher dimensional Calabi–Yau manifolds. A further natural extension of this method relies on integrals with singular geometries, whose treatment is still an open problem.
Moreover, we plan to extend the exponential period viewpoint to string theory amplitudes, where vertex operator insertions and Koba-Nielsen factors originate multivalued exponents; being confident that such new perspective could provide significant progresses in understanding KLT relations at higher genus. \\
These directions aim both to test the practicality of the framework on challenging, phenomenologically relevant examples and to stimulate the development of new tools for intersection numbers evaluation, Stokes phenomena comprehension, and thimble integration.

%% file: bibliography.bib
@article{Adams18,
      author         = "Adams, L. and Weinzierl, S.",
      title          = "{The $\varepsilon$-form of the differential equations for
                        Feynman integrals in the elliptic case}",
      journal        = "Phys. Lett.",
      volume         = "B781",
      year           = "2018",
      pages          = "270-278",
      doi            = "10.1016/j.physletb.2018.04.002",
      eprint         = "1802.05020",
      archivePrefix  = "arXiv",
      primaryClass   = "hep-ph",
      reportNumber   = "MITP-18-011",
      SLACcitation   = "%%CITATION = ARXIV:1802.05020;%%"
}

@article{adolphson_sperber_1997, title={On twisted de Rham cohomology}, volume={146}, DOI={10.1017/S0027763000006218}, journal={Nagoya Mathematical Journal}, publisher={Cambridge University Press}, author={Adolphson, A. and Sperber, S.}, year={1997}, pages={55-81}}

@article{Ahmed:2023htp,
    author = "Ahmed, T. and Crisanti, G. and Gasparotto, F. and Hasan, S. M. and Mastrolia, P.",
    title = "{Two-loop vertices with vacuum polarization insertion}",
    eprint = "2308.05028",
    archivePrefix = "arXiv",
    primaryClass = "hep-ph",
    reportNumber = "MITP-23-041",
    doi = "10.1007/JHEP01(2024)010",
    journal = "JHEP",
    volume = "01",
    pages = "010",
    year = "2024"
}

@article{Alvarez:2020zui,
    author = "Alvarez, P. D. and Cacciatori, S. L. and Canfora, F. and Cerchiai, B. L.",
    title = "{Analytic SU(N) Skyrmions at finite Baryon density}",
    doi = "10.1103/PhysRevD.101.125011",
    journal = "Phys. Rev. D",
    volume = "101",
    number = "12",
    pages = "125011",
    year = "2020"
}

@article{Anastasiou06,
    author = "Anastasiou, C. and Britto, R. and Feng, B. and Kunszt, Z. and Mastrolia, P.",
    title = "{D-dimensional unitarity cut method}",
    eprint = "hep-ph/0609191",
    archivePrefix = "arXiv",
    reportNumber = "ITFA-2006-35",
    doi = "10.1016/j.physletb.2006.12.022",
    journal = "Phys. Lett. B",
    volume = "645",
    pages = "213--216",
    year = "2007"
}

@article{Angiusnostro,
    author = "Angius, R. and Cacciatori, S. L. and Massidda, A.",
    title = "{Wall crossing structure from quantum phenomena to Feynman Integrals}",
    eprint = "2506.03252",
    archivePrefix = "arXiv",
    primaryClass = "hep-th",
    month = "6",
    year = "2025"
}

@article{Angius25,
      title="{Exponential integrals for Feynman diagrams}", 
      author="{R. Angius, S. L. Cacciatori, A. Massidda, S. Noja and P. Mastrolia}",
      year={In progress},

}

@article{Aomoto75vanishing,
author = "Aomoto, K.",
doi = "10.2969/jmsj/02720248",
fjournal = "Journal of the Mathematical Society of Japan",
journal = "J. Math. Soc. Japan",
month = "04",
number = "2",
pages = "248--255",
publisher = "Mathematical Society of Japan",
title = "{On vanishing of cohomology attached to certain many valued meromorphic functions}",
volume = "27",
year = "1975"
}

@article{Aomoto77_structure,
  title={On the structure of integrals of power product of linear functions},
  author={Aomoto, K.},
  journal={Sci. Papers College Gen. Ed. Univ. Tokyo},
  volume={27},
  number={2},
  pages={49--61},
  year={1977}
}

@book{Aomoto11,
  title="{Theory of Hypergeometric Functions}",
  author={Aomoto, K. and Kita, M.},
  series={Springer Monographs in Mathematics},
  doi={10.1007/978-4-431-53938-4},
  year={2011},
  publisher={Springer Japan}
}

@article{Aomoto2,
  title={Un théorème du type de Matsushima-Murakami concernant l'intégrale des fonctions multiformes},
  author={K. Aomoto},
  journal={J. Math. pures et appl.},
  year={1973},
  volume={52},
  pages={1-11}
}

@unpublished{Ashoke,
      title={String Field Theory: A Review}, 
      author={A. Sen and B. Zwiebach},
      year={2024},
      eprint={2405.19421}, 
}

@article{Argeri07,
    author = "Argeri, M. and Mastrolia, P.",
    title = "{Feynman Diagrams and Differential Equations}",
    eprint = "0707.4037",
    archivePrefix = "arXiv",
    primaryClass = "hep-ph",
    reportNumber = "ZU-TH-19-07",
    doi = "10.1142/S0217751X07037147",
    journal = "Int. J. Mod. Phys. A",
    volume = "22",
    pages = "4375--4436",
    year = "2007"
}

@book{Arkani-Hamed:2012zlh,
    author = "Arkani-Hamed, N. and Bourjaily, J. L. and Cachazo, F. and Goncharov, A. B. and Postnikov, A. and Trnka, J.",
    title = "{Grassmannian Geometry of Scattering Amplitudes}",
    doi = "10.1017/CBO9781316091548",
    publisher = "Cambridge University Press",
    month = "4",
    year = "2016"
}

@article{Arkani17,
      author         = "Arkani-Hamed, N. and Bai, Y. and Lam, T.",
      title          = "{Positive Geometries and Canonical Forms}",
      journal        = "JHEP",
      volume         = "11",
      year           = "2017",
      pages          = "039",
      doi            = "10.1007/JHEP11(2017)039",
      eprint         = "1703.04541",
      archivePrefix  = "arXiv",
      primaryClass   = "hep-th",
      SLACcitation   = "%%CITATION = ARXIV:1703.04541;%%"
}

@article{Arkani-Hamed:2024jbp,
    author = "Arkani-Hamed, N. and Figueiredo, C. and Vaz{\~a}o, F.",
    title = "{Cosmohedra}",
    eprint = "2412.19881",
    archivePrefix = "arXiv",
    primaryClass = "hep-th",
    month = "12",
    year = "2024"
}

@article{Damgaard:2019ztj,
    author = "Damgaard, D. and Ferro, L. and Lukowski, T. and Parisi, M.",
    title = "{The Momentum Amplituhedron}",
    eprint = "1905.04216",
    archivePrefix = "arXiv",
    primaryClass = "hep-th",
    reportNumber = "LMU-ASC 21/19",
    doi = "10.1007/JHEP08(2019)042",
    journal = "JHEP",
    volume = "08",
    pages = "042",
    year = "2019"
}

@article{Arkanijha,
    author = "Arkani-Hamed, N. and Trnka, J.",
    title = "{The Amplituhedron}",
    eprint = "1312.2007",
    archivePrefix = "arXiv",
    primaryClass = "hep-th",
    doi = "10.1007/JHEP10(2014)030",
    journal = "JHEP",
    volume = "10",
    pages = "030",
    year = "2014"
}

@book{Arkanizlh,
    author = "Arkani-Hamed, N. and Bourjaily, J. L. and Cachazo, F. and Goncharov, A. B. and Postnikov, A. and Trnka, J.",
    title = "{Grassmannian Geometry of Scattering Amplitudes}",
    eprint = "1212.5605",
    archivePrefix = "arXiv",
    primaryClass = "hep-th",
    reportNumber = "PUPT-2435",
    doi = "10.1017/CBO9781316091548",
    publisher = "Cambridge University Press",
    month = "4",
    year = "2016"
}

@book{Arnold88, 
title="{Singularities of Differentiable Maps, Vol.II}", 
publisher ="{Birkh\"{a}usser, Boston}", 
author="{V.I. Arnold, S.M. Gusein–Zade and A.N. Var\v{c}enko,}", 
year={1988}}

@book{hartshorne66,
  author    = {Hartshorne, R.},
  title     = {Residues and Duality},
  series    = {Lecture Notes in Mathematics},
  volume    = {20},
  publisher = {Springer-Verlag},
  year      = {1966},
}

@book{hartshorne1977algebraic,
  title={Algebraic Geometry},
  author={Hartshorne, R.},
  series={Graduate Texts in Mathematics},
  year={1977},
  publisher={Springer}
}

@article{Baikov96-1,
    author = "Baikov, P. A.",
    title = "{Explicit solutions of n loop vacuum integral recurrence relations}",
    eprint = "hep-ph/9604254",
    archivePrefix = "arXiv",
    reportNumber = "INP-96-10-418",
    month = "4",
    year = "1996"
}

@article{Baikov96-2,
      author         = "Baikov, P. A.",
      title          = "{Explicit solutions of the multiloop integral recurrence
                        relations and its application}",
      booktitle      = "{New computing techniques in physics research V.
                        Proceedings, 5th International Workshop, AIHENP '96,
                        Lausanne, Switzerland, September 2-6, 1996}",
      journal        = "Nucl. Instrum. Meth.",
      volume         = "A389",
      year           = "1997",
      pages          = "347-349",
      doi            = "10.1016/S0168-9002(97)00126-5",
      reportNumber   = "INP-96-42-449",
}

@article{Baikov05,
    author = "Baikov, P. A.",
    title = "{A Practical criterion of irreducibility of multi-loop Feynman integrals}",
    doi = "10.1016/j.physletb.2006.01.052",
    journal = "Phys. Lett. B",
    volume = "634",
    pages = "325--329",
    year = "2006"
}

@book{Balser94,
  author       = {Balser, W.},
  title        = {From Divergent Power Series to Analytic Functions: Theory and Applications of Multisummable Power Series},
  publisher    = {Springer},
  year         = {1994}
}

@book{barth2012,
  title={Compact Complex Surfaces},
  author={Barth, W. and Peters, C. and van de Ven, A.},
  series={Ergebnisse der Mathematik und ihrer Grenzgebiete. 3. Folge / A Series of Modern Surveys in Mathematics},
  year={2012},
  publisher={Springer Berlin Heidelberg}
}

@article{Beauville,
title = {Les familles stables de courbes elliptiques sur P1 admettant quatre fibres singulieres},
journal = {C. R. Acad. Sc. Paris },
volume = {294 Série I},
pages = {657},
year = {1982},
author = {A. Beauville}
}

@article{Bern:1994cg,
      author         = "Bern, Z. and Dixon, L. J. and Dunbar, D. C. and
                        Kosower, D. A.",
      title          = "{Fusing gauge theory tree amplitudes into loop
                        amplitudes}",
      journal        = "Nucl. Phys.",
      volume         = "B435",
      year           = "1995",
      pages          = "59-101",
      doi            = "10.1016/0550-3213(94)00488-Z",
      eprint         = "hep-ph/9409265",
      archivePrefix  = "arXiv",
      primaryClass   = "hep-ph",
      reportNumber   = "SLAC-PUB-6563, SACLAY-SPH-T-94-95, UCLA-TEP-94-29,
                        SWAT-94-36",
      SLACcitation   = "%%CITATION = HEP-PH/9409265;%%"
}

@article{Bern96-1,
   title={Massive loop amplitudes from unitarity},
   volume={467},
   DOI={10.1016/0550-3213(96)00078-8},
   number={3},
   journal={Nuclear Physics B},
   publisher={Elsevier BV},
   author={Bern, Z. and Morgan, A.G.},
   year={1996},
   month=may, pages={479–509} }

@article{Bern96-2,
   title={Progress in one-loop QCD computations},
   volume={46},
   DOI={10.1146/annurev.nucl.46.1.109},
   number={1},
   journal={Annual Review of Nuclear and Particle Science},
   publisher={Annual Reviews},
   author={Bern, Z. and Dixon, L. and Kosower, D. A.},
   year={1996},
   month=dec, pages={109–148} }

@article{Bern00-3,
   title={A two-loop four-gluon helicity amplitude in QCD},
   volume={2000},
   DOI={10.1088/1126-6708/2000/01/027},
   number={01},
   journal={Journal of High Energy Physics},
   publisher={Springer Science and Business Media LLC},
   author={Bern, Z. and Dixon, L. and Kosower, D. A},
   year={2000},
   month=jan, pages={027–027} }

@article{Bern04-4,
   title={Two-Loop g->gg Splitting Amplitudes in QCD},
   volume={2004},
   DOI={10.1088/1126-6708/2004/08/012},
   number={08},
   journal={Journal of High Energy Physics},
   publisher={Springer Science and Business Media LLC},
   author={Bern, Z. and Dixon, L. and Kosower, D.},
   year={2004},
   month=aug, pages={012–012} }

@article{Bern:2011qt,
    author = "Bern, Z. and Huang, Y.",
    title = "{Basics of Generalized Unitarity}",
    eprint = "1103.1869",
    archivePrefix = "arXiv",
    primaryClass = "hep-th",
    reportNumber = "UCLA-11-TEP-103",
    doi = "10.1088/1751-8113/44/45/454003",
    journal = "J. Phys. A",
    volume = "44",
    pages = "454003",
    year = "2011"
}

@article{BCJ,
author = "{Z. Bern, J.J. Carrasco and H. Johansson}",
year = {2010},
title = {Perturbative Quantum Gravity as a Double Copy of Gauge Theory},
volume = {105},
journal = {Physical review letters}
}

@article{Bhardwaj:2023vvm,
    author = "R. Bhardwaj, A. Pokraka, L. Ren and C.Rodriguez",
    title = "{A double copy from twisted (co)homology at genus one}",
    journal = "JHEP",
    volume = "07",
    pages = "040",
    year = "2024"
}

@article{BLOCH2015328,
title = {The elliptic dilogarithm for the sunset graph},
journal = {Journal of Number Theory},
volume = {148},
pages = {328-364},
year = {2015},
doi = {https://doi.org/10.1016/j.jnt.2014.09.032},
author = {S. Bloch and P. Vanhove}
}

@article{Bogner:2007mn,
    author = "Bogner, C. and Weinzierl, S.",
    title = "{Periods and Feynman integrals}",
    eprint = "0711.4863",
    archivePrefix = "arXiv",
    primaryClass = "hep-th",
    reportNumber = "MZ-TH-07-19",
    doi = "10.1063/1.3106041",
    journal = "J. Math. Phys.",
    volume = "50",
    pages = "042302",
    year = "2009"
}

@article{Bosma:2017ens,
      author         = "Bosma, J. and Sogaard, M. and Zhang, Y.",
      title          = "{Maximal Cuts in Arbitrary Dimension}",
      journal        = "JHEP",
      volume         = "08",
      year           = "2017",
      pages          = "051",
      doi            = "10.1007/JHEP08(2017)051",
      eprint         = "1704.04255",
      archivePrefix  = "arXiv",
      primaryClass   = "hep-th",
      SLACcitation   = "%%CITATION = ARXIV:1704.04255;%%"
}

@book{Bredon,
  title="{Sheaf Theory}",
  author="{G. E. Bredon}",
  year={1967},
  publisher="{McGraw-Hill, New York}"
}

@misc{bree25,
      title={The geometric bookkeeping guide to Feynman integral reduction and $\varepsilon$-factorised differential equations}, 
      author={I. Bree and F. Gasparotto and A. Matijašić and P. Mazloumi and D. Melnichenko and S. Pögel and T. Teschke and X. Wang and S. Weinzierl and K. Wu and X. Xu},
      year={2025},
      eprint={2506.09124},
      archivePrefix={arXiv},
      primaryClass={hep-th},
      url={https://arxiv.org/abs/2506.09124}, 
}

@article{Brieskorn1970,
author = {Brieskorn, Egbert},
journal = {Manuscripta mathematica},
pages = {103-162},
title = {Die Monodromie der isolierten Singularitäten von Hyperflächen.},
url = {http://eudml.org/doc/153994},
volume = {2},
year = {1970},
}

@article{Britto:2004ap,
    author = "Britto, R. and Cachazo, F. and Feng, B.",
    title = "{New recursion relations for tree amplitudes of gluons}",
    eprint = "hep-th/0412308",
    archivePrefix = "arXiv",
    doi = "10.1016/j.nuclphysb.2005.02.030",
    journal = "Nucl. Phys. B",
    volume = "715",
    pages = "499--522",
    year = "2005"
}

@article{Britto:KLT,
    author = "R. Britto, S. Mizera, C. Rodriguez and O. Schlotterer",
    title = "{Coaction and double-copy properties of configuration-space integrals at genus zero}",
    journal = "JHEP",
    volume = "05",
    pages = "053",
    year = "2021"
}

@article{Britto08,
    author = "Britto, R. and Feng, B. and Mastrolia, P.",
    title = "{Closed-Form Decomposition of One-Loop Massive Amplitudes}",
    eprint = "0803.1989",
    archivePrefix = "arXiv",
    primaryClass = "hep-ph",
    reportNumber = "CERN-PH-TH-2008-050, ITFA-2008-06",
    doi = "10.1103/PhysRevD.78.025031",
    journal = "Phys. Rev. D",
    volume = "78",
    pages = "025031",
    year = "2008"
}

@article{Britto05,
      author = "Britto, R. and Cachazo, Freddy and Feng, Bo",
    title = "{Generalized unitarity and one-loop amplitudes in $N=4$ super-Yang-Mills}",
    doi = "10.1016/j.nuclphysb.2005.07.014",
    journal = "Nucl. Phys. B",
    volume = "725",
    pages = "275--305",
    year = "2005"}

@article{Broedel:2018qkq,
      author         = "Broedel, Johannes and Duhr, Claude and Dulat, Falko and
                        Penante, Brenda and Tancredi, Lorenzo",
      title          = "{Elliptic Feynman integrals and pure functions}",
      journal        = "JHEP",
      volume         = "01",
      year           = "2019",
      pages          = "023",
      doi            = "10.1007/JHEP01(2019)023",
      eprint         = "1809.10698",
      archivePrefix  = "arXiv",
      primaryClass   = "hep-th",
      reportNumber   = "CP3-18-58, CERN-TH-2018-211, HU-Mathematik-2018-09,
                        HU-EP-18/29, SLAC-PUB-17336",
      SLACcitation   = "%%CITATION = ARXIV:1809.10698;%%"
}

@article{Brunello2,
  title = {Fourier calculus from intersection theory},
  author = {Brunello, G. and Crisanti, G. and Giroux, M. and Mastrolia, P. and Smith, S.},
  journal = {Phys. Rev. D},
  volume = {109},
  issue = {9},
  pages = {094047},
  numpages = {16},
  year = {2024},
  publisher = {American Physical Society},
  doi = {10.1103/PhysRevD.109.094047},
}

@misc{brunello2024,
      title={Intersection Numbers from Companion Tensor Algebra}, 
      author={G. Brunello and V. Chestnov and P. Mastrolia},
      year={2024},
      eprint={2408.16668},
      archivePrefix={arXiv},
      primaryClass={hep-th},
      url={https://arxiv.org/abs/2408.16668}, 
}

@article{Brunello:2023rpq,
    author = "Brunello, G. and Chestnov, V. and Crisanti, G. and Frellesvig, H. and Mandal, M. K. and Mastrolia, P.",
    title = "{Intersection numbers, polynomial division and relative cohomology}",
    eprint = "2401.01897",
    archivePrefix = "arXiv",
    primaryClass = "hep-th",
    doi = "10.1007/JHEP09(2024)015",
    journal = "JHEP",
    volume = "09",
    pages = "015",
    year = "2024"
}

@article{Bonisch:2022bdf,
    author = " Bönisch, K. and Duhr, Claude and Fischbach, Fabian and Klemm, Albrecht and Nega, Christoph",
    title = "{Feynman integrals in Dimensional Regularization and extensions of Calabi-Yau Motives}",
    eprint = "2108.05310",
    archivePrefix = "arXiv",
    primaryClass = "hep-th",
    doi = "10.1007/JHEP09(2022)156",
    journal = "JHEP",
    volume = "09",
    number = "156",
    year ={2022},
}

@article{Bonisch:2021bfk,
    author = " Bönisch, K. and Fischbach, F. and Klemm, A. and Nega, C. and Safari, R.",
    title = "{Analytic structure of all Loop Banana Amplitudes}",
    eprint = "2008.10574",
    archivePrefix = "arXiv",
    primaryClass = "hep-th",
    doi = "10.1007/JHEP05%282021%29066",
    journal = "JHEP",
    volume = "2021",
    number = "05",
year={},
}

@article{Cacciatori:banana,
    author = "Cacciatori, S. L. and Epstein, H. and Moschella, U.",
    title = "{Banana integrals in configuration space}",
    journal = "Nucl. Phys. B",
    volume = "995",
    pages = "116343",
    year = "2023"
}

@article{Cacciatori:2021neu,
    author = "Cacciatori, S. L. and Canfora, F. and Lagos, M. and Muscolino, F. and Vera, A.",
    title = "{Analytic multi-Baryonic solutions in the SU(N)-Skyrme model at finite density}",
    doi = "10.1007/JHEP12(2021)150",
    journal = "JHEP",
    volume = "12",
    pages = "150",
    year = "2021"
}

@article{Cacciatori:2021nli,
    author = "Cacciatori, S. L. and Conti, M. and Trevisan, S.",
    title = "{Co-Homology of Differential Forms and Feynman Diagrams}",
    eprint = "2107.14721",
    archivePrefix = "arXiv",
    primaryClass = "hep-th",
    doi = "10.3390/universe7090328",
    journal = "Universe",
    volume = "7",
    number = "9",
    pages = "328",
    year = "2021"
}

@article{Cacciatori:2022kag,
    author = "Cacciatori, S. L. and Canfora, F. and Lagos, M. and Muscolino, F. and Vera, A.",
    title = "{Cooking pasta with Lie groups}",
    doi = "10.1016/j.nuclphysb.2022.115693",
    journal = "Nucl. Phys. B",
    volume = "976",
    pages = "115693",
    year = "2022"
}

@misc{Cacciatori:2022mbi,
    author = "Cacciatori, S. L. and Mastrolia, P.",
    title = "{Intersection Numbers in Quantum Mechanics and Field Theory, arXiv:2211.03729}",
    eprint = "2211.03729",
    archivePrefix = "{arXiv}",
    primaryClass = "hep-th",
    year = "2022"
}

@article{Cacciatori:2024zbe,
    author = "Cacciatori, S. L. and Epstein, H. and Moschella, U.",
    title = "{Loops in anti de Sitter space}",
    doi = "10.1007/JHEP08(2024)109",
    journal = "JHEP",
    volume = "08",
    pages = "109",
    year = "2024"
}

@article{Cacciatori:2024zrv,
    author = "Cacciatori, S. L. and Epstein, H. and Moschella, U.",
    title = "{Loops in de Sitter space}",
    doi = "10.1007/JHEP07(2024)182",
    journal = "JHEP",
    volume = "07",
    pages = "182",
    year = "2024"
}

@article{Cacciatori:2024ccm,
    author = "Cacciatori, S. L. and Canfora, F. and Muscolino, F.",
    title = "{Pearcey integrals, Stokes lines and exact baryonic layers in the low energy limit of QCD}",
    doi = "10.1016/j.nuclphysb.2024.116477",
    journal = "Nucl. Phys. B",
    volume = "1000",
    pages = "116477",
    year = "2024"
}

@book{Calabi1957,
  author    = {E. Calabi},
  title     = {On Kähler manifolds with vanishing canonical class},
  booktitle = {Algebraic Geometry and Topology: A Symposium in Honor of S. Lefschetz},
  publisher = {Princeton University Press},
  year      = {1957},
  pages     = {78--89},
}

@article{CANDELAS198546,
title = {Vacuum configurations for superstrings},
journal = {Nuclear Physics B},
volume = {258},
pages = {46-74},
year = {1985},
author = "{P. Candelas, G. T. Horowitz, A. Strominger and E. Witten}",
}

@article{Caron-Huot:2021xqj,
    author = "Caron-Huot, S. and Pokraka, A.",
    title = "{Duals of Feynman integrals. Part I. Differential equations}",
    eprint = "2104.06898",
    archivePrefix = "arXiv",
    primaryClass = "hep-th",
    doi = "10.1007/JHEP12(2021)045",
    journal = "JHEP",
    volume = "12",
    pages = "045",
    year = "2021"
}

@article{Caron-Huot:2021iev,
    author = "Caron-Huot, S. and Pokraka, A.",
    title = "{Duals of Feynman Integrals. Part II. Generalized unitarity}",
    eprint = "2112.00055",
    archivePrefix = "arXiv",
    primaryClass = "hep-th",
    doi = "10.1007/JHEP04(2022)078",
    journal = "JHEP",
    volume = "04",
    pages = "078",
    year = "2022"
}

@article{Capuano:2025ehm,
    author = "Capuano, M. and Ferro, L. and Lukowski, T. and Palazio, A.",
    title = "{Canonical Differential Equations for Cosmology from Positive Geometries}",
    eprint = "2505.14609",
    archivePrefix = "arXiv",
    primaryClass = "hep-th",
    month = "5",
    year = "2025"
}

@article{Cattani:2008ec,
    author = {Cattani, E.},
    title = "{Mixed Lefschetz Theorems and Hodge-Riemann Bilinear Relations}",
    journal = {International Mathematics Research Notices},
    volume = {2008},
    pages = {rnn025},
    year = {2008},
    month = {01},
    doi = {10.1093/imrn/rnn025},
}

@article{Cecotti:1992ccv,
    author = "Cecotti, S. and Vafa, C.",
    title = "{On classification of N=2 supersymmetric theories}",
    eprint = "hep-th/9211097",
    archivePrefix = "arXiv",
    reportNumber = "HUTP-92-A064, SISSA-203-92-EP",
    doi = "10.1007/BF02096804",
    journal = "Commun. Math. Phys.",
    volume = "158",
    pages = "569--644",
    year = "1993"
}

@article{Chen:2022lzr,
    author = "Chen, J. and Jiang, X. and Ma, C. and Xu, X. and Yang, L. L.",
    title = "{Baikov representations, intersection theory, and canonical Feynman integrals}",
    eprint = "2202.08127",
    archivePrefix = "arXiv",
    primaryClass = "hep-th",
    doi = "10.1007/JHEP07(2022)066",
    journal = "JHEP",
    volume = "07",
    pages = "066",
    year = "2022"
}

@misc{chen25,
      title={On an approach to canonicalizing elliptic Feynman integrals}, 
      author={J. Chen and L. L. Yang and Y. Zhang},
      year={2025},
      eprint={2503.23720},
      archivePrefix={arXiv},
      primaryClass={hep-th},
      url={https://arxiv.org/abs/2503.23720}, 
}

@article{Chestnov:2022alh,
    author = "Chestnov, V. and Gasparotto, F. and Mandal, M. K. and Mastrolia, P. and Matsubara-Heo, S. J. and Munch, H. J. and Takayama, N.",
    title = "{Macaulay matrix for Feynman integrals: linear relations and intersection numbers}",
    eprint = "2204.12983",
    archivePrefix = "arXiv",
    primaryClass = "hep-th",
    doi = "10.1007/JHEP09(2022)187",
    journal = "JHEP",
    volume = "09",
    pages = "187",
    year = "2022"
}

@article{Chestnov:2022xsy,
    author = "Chestnov, V. and Frellesvig, H. and Gasparotto, F. and Mandal, M. K. and Mastrolia, P.",
    title = "{Intersection numbers from higher-order partial differential equations}",
    eprint = "2209.01997",
    archivePrefix = "arXiv",
    primaryClass = "hep-th",
    doi = "10.1007/JHEP06(2023)131",
    journal = "JHEP",
    volume = "06",
    pages = "131",
    year = "2023"
}

@article{Chestnov:2024mnw,
    author = "Chestnov, V. and Fontana, G. and Peraro, T.",
    title = "{Reduction to master integrals and transverse integration identities}",
    eprint = "2409.04783",
    archivePrefix = "arXiv",
    primaryClass = "hep-ph",
    reportNumber = "ZU-TH 42/24",
    doi = "10.1007/JHEP03(2025)113",
    journal = "JHEP",
    volume = "03",
    pages = "113",
    year = "2025"
}

@article{Chetyrkin:1981qh,
      author         = "Chetyrkin, K. G. and Tkachov, F. V.",
      title          = "{Integration by Parts: The Algorithm to Calculate beta
                        Functions in 4 Loops}",
      journal        = "Nucl. Phys.",
      volume         = "B192",
      year           = "1981",
      pages          = "159-204",
      doi            = "10.1016/0550-3213(81)90199-1",
      SLACcitation   = "%%CITATION = NUPHA,B192,159;%%"
}

@article{cho1995,
author = "K. Cho and K. Matsumoto",
fjournal = "Nagoya Mathematical Journal",
journal = "Nagoya Math. J.",
pages = "67--86",
publisher = "Duke University Press",
title = "{Intersection theory for twisted cohomologies and twisted Riemann's period relations I}",
url = "http://projecteuclid.org/euclid.nmj/1118775097",
doi = {10.1017/S0027763000005304},
volume = "139",
year = "1995"
}

@article{Choudary1994KoszulCA,
  title={Koszul complexes and hypersurface singularities},
  author={A. D. R. Choudary and A. Dimca},
  journal={American Mathematical Society},
  year={1994},
  url={https://api.semanticscholar.org/CorpusID:120170776}
}

@article{Crisanti:2024onv,
    author = "Crisanti, G. and Smith, S.",
    title = "{Feynman integral reductions by intersection theory with orthogonal bases and closed formulae}",
    eprint = "2405.18178",
    archivePrefix = "arXiv",
    primaryClass = "hep-th",
    reportNumber = "MPP-2024-104",
    doi = "10.1007/JHEP09(2024)018",
    journal = "JHEP",
    volume = "09",
    pages = "018",
    year = "2024"
}

@article{Cutkosky60,
    author = "Cutkosky, R. E.",
    title = "{Singularities and discontinuities of Feynman amplitudes}",
    doi = "10.1063/1.1703676",
    journal = "J. Math. Phys.",
    volume = "1",
    pages = "429--433",
    year = "1960"
}

@article{Decataldo09,
  title={The decomposition theorem, perverse sheaves and the topology of algebraic maps},
  author={de Cataldo, M. and Migliorini, L.},
  journal={Bulletin of the American Mathematical Society},
  volume={46},
  number={4},
  pages={535--633},
  year={2009}
}

@article{Decataldo02,
title = {The Hard Lefschetz Theorem and the topology of semismall maps},
author={},
journal = {Annales Scientifiques de l’École Normale Supérieure},
volume = {35},
number = {5},
pages = {759-772},
year = {2002},
doi = {https://doi.org/10.1016/S0012-9593(02)01108-4},
}

@article{DeRham,
    author = {G. De Rham},
    title = "{Sur l’analysis situs des variétés à n dimensions}",
    journal = {J. Maths. Pures Appl},
    year = {1931}
}

@article{dimca2008,
      title={A generalization of Griffiths theorem on rational integrals, II}, 
      author={Alexandru Dimca and Morihiko Saito and Lorenz Wotzlaw},
      year={2008},
      eprint={math/0702105},
      archivePrefix={arXiv},
      primaryClass={math.AG},
}

@article{dimca2024,
      title={Koszul complexes and spectra of projective hypersurfaces with isolated singularities}, 
      author={Alexandru Dimca and Morihiko Saito},
       journal={},
      year={2024},
      eprint={1212.1081},
      archivePrefix={arXiv},
      primaryClass={math.AG},
}

@article{Dlapa:2020cwj,
    author = "Dlapa, C. and Henn, J. and Yan, K.",
    title = "{Deriving canonical differential equations for Feynman integrals from a single uniform weight integral}",
    eprint = "2002.02340",
    archivePrefix = "arXiv",
    primaryClass = "hep-ph",
    reportNumber = "MPP-2020-10",
    doi = "10.1007/JHEP05(2020)025",
    journal = "JHEP",
    volume = "05",
    pages = "025",
    year = "2020"
}

@article{Doran:2023yzu,
    author = "Doran, C. F. and Harder, A. and Vanhove, P. and Pichon-Pharabod, E.",
    title = "{Motivic Geometry of two-Loop Feynman Integrals}",
    journal = "Quart. J. Math. Oxford Ser.",
    volume = "75",
    number = "3",
    pages = "901--967",
    year = "2023",
}

@misc{duhr2025-CY,
      title={Aspects of canonical differential equations for Calabi-Yau geometries and beyond}, 
      author={C. Duhr and S. Maggio and C. Nega and B. Sauer and L. Tancredi and F. J. Wagner},
      year={2025},
      eprint={2503.20655},
      archivePrefix={arXiv},
      primaryClass={hep-th},
      url={https://arxiv.org/abs/2503.20655}, 
}

@article{Duhr:2025xyy,
    author = "Duhr, C. and Maggio, S. and Porkert, F. and Semper, C. and Sohnle, Y. and Stawinski, S. F.",
    title = "{Canonical differential equations and intersection matrices}",
    eprint = "2509.17787",
    archivePrefix = "arXiv",
    primaryClass = "hep-th",
    reportNumber = "BONN-TH/2025-30, UUITP--27/25",
    month = "9",
    year = "2025"
}

@article{Duhr:2023dkn,
    author = " Duhr, C.  and Klemm, A. and Nega, C. and Tancredi, L.",
    title = "{The ice cone family and integrals for Calabi-Yau varieties}",
    eprint = "2212.09550",
    archivePrefix = "arXiv",
    primaryClass = "hep-th",
    doi = "10.1007/JHEP02(2023228)",
    journal = "JHEP",
    volume = "02",
    number = "228",
    year ={2023},
}

@article{Duhr:2024bzt,
    author = "Duhr, C. and Gasparotto, F. and Nega, C. and Tancredi, L. and Weinzierl, S.",
    title = "{On the electron self-energy to three loops in QED}",
    eprint = "2408.05154",
    archivePrefix = "arXiv",
    primaryClass = "hep-th",
    reportNumber = "BONN-TH-2024-12, MITP/24-065, TUM-HEP-1518/24",
    doi = "10.1007/JHEP11(2024)020",
    journal = "JHEP",
    volume = "11",
    pages = "020",
    year = "2024"
}

@book{Dundas, 
title="{A Short Course In Differential Topology.}", 
publisher ="{Cambridge University Press}", 
author={Bj{\o}rn Ian Dundas}, 
year={2018}}

@book{Ebeling1,
title = {The monodromy groups of isoleted singularities of complete intersections},
publisher = {Springer Berlin, Heidelberg},
series ={Lectures Notes in Mathematics},
year = {1980},
author = {W. Ebeling}
}

@article{Ebeling,
    author = {Ebeling, W.},
    title = "{The Milnor Lattices of the Elliptic Hypersurface Singularities}",
    journal = {Proceedings of the London Mathematical Society},
    volume = {s3-53},
    number = {1},
    pages = {85-111},
    year = {1986},
    month = {07},
    doi = {10.1112/plms/s3-53.1.85},
}

@misc{ebeling2005monodromy,
      title={Monodromy}, 
      author={W. Ebeling},
      year={2005},
      eprint={math/0507171},
      archivePrefix={arXiv},
      primaryClass={math.AG},
      url={https://arxiv.org/abs/math/0507171}, 
}

@article{Eilen45,
    author = "{S.Eilenberg and S.MacLane}",
    title ={General Theory of Natural Equivalences} ,
    journal ={American Mathematical Society} ,
    volume = "58",
    pages = "231-294",
    year = {1945}
}

@article{Esnault1992,
author = "H. Esnault, V. Schechtmann, E. Viehweg",
journal = {Inventiones mathematicae},
number = {3},
pages = {557-562},
title = {Cohomology of local systems on the complement of hyperplanes.},
url = {http://eudml.org/doc/144034},
volume = {109},
year = {1992}
}

@article{Fontana_2023,
   title={Reduction to master integrals via intersection numbers and polynomial expansions},
   volume={2023},
   DOI={10.1007/jhep08(2023)175},
   number={8},
   journal={Journal of High Energy Physics},
   publisher={Springer Science and Business Media LLC},
   author={Fontana, G. and Peraro, T.},
   year={2023},
   month=aug }

@book{Fre,
    author = "P. Fre  and P. Soriani",
    title = "{The N=2 wonderland: From Calabi-Yau manifolds to topological field theories}",
    year = "1995",
    publisher = "World Scientific"

}

@article{Frellesvig:2021vem,
    author = "Frellesvig, H. and Mattiazzi, L.",
    title = "{On the Application of Intersection Theory to Feynman Integrals: the univariate case}",
    eprint = "2102.01576",
    archivePrefix = "arXiv",
    primaryClass = "hep-ph",
    doi = "10.22323/1.383.0017",
    journal = "PoS",
    volume = "MA2019",
    pages = "017",
    year = "2022"
}

@article{Frellesvig:2024ymq,
    author = "Frellesvig, H.",
    title = "{The loop-by-loop Baikov representation {\textemdash} Strategies and implementation}",
    eprint = "2412.01804",
    archivePrefix = "arXiv",
    primaryClass = "hep-th",
    doi = "10.1007/JHEP04(2025)111",
    journal = "JHEP",
    volume = "04",
    pages = "111",
    year = "2025"
}

@article{Frellesvig:2017aai,
      author         = "Frellesvig, H. and Papadopoulos, C. G.",
      title          = "{Cuts of Feynman Integrals in Baikov representation}",
      journal        = "JHEP",
      volume         = "04",
      year           = "2017",
      pages          = "083",
      doi            = "10.1007/JHEP04(2017)083",
      SLACcitation   = "%%CITATION = ARXIV:1701.07356;%%"
}

@article{Frellesvig:2019kgj,
      author         = "Frellesvig, H. and Gasparotto, F. and Laporta,
                        S. and Mandal, M. K. and Mastrolia, P. and
                        Mattiazzi, L. and Mizera, S.",
      title          = "{Decomposition of Feynman Integrals on the Maximal Cut by
                        Intersection Numbers}",
      journal        = "JHEP",
      volume         = "05",
      year           = "2019",
      pages          = "153",
      doi            = "10.1007/JHEP05(2019)153",
      eprint         = "1901.11510",
      archivePrefix  = "arXiv",
      primaryClass   = "hep-ph",
      SLACcitation   = "%%CITATION = ARXIV:1901.11510;%%"
}

@article{Frellesvig:2019uqt,
      author         = "Frellesvig, H. and Gasparotto, F. and Mandal,
                        M. K. and Mastrolia, P. and Mattiazzi, L. and
                        Mizera, S.",
      title          = "{Vector Space of Feynman Integrals and Multivariate
                        Intersection Numbers}",
      journal        = "Phys. Rev. Lett.",
      volume         = "123",
      year           = "2019",
      number         = "20",
      pages          = "201602",
      doi            = "10.1103/PhysRevLett.123.201602",
      eprint         = "1907.02000",
      archivePrefix  = "arXiv",
      primaryClass   = "hep-th",
      SLACcitation   = "%%CITATION = ARXIV:1907.02000;%%"
}

@article{Frellesvig:2020qot,
    author = "Frellesvig, H. and Gasparotto, F. and Laporta, S. and Mandal, Manoj K. and Mastrolia, P. and Mattiazzi, L. and Mizera, S.",
    title = "{Decomposition of Feynman Integrals by Multivariate Intersection Numbers}",
    eprint = "2008.04823",
    archivePrefix = "arXiv",
    primaryClass = "hep-th",
    doi = "10.1007/JHEP03(2021)027",
    journal = "JHEP",
    volume = "03",
    pages = "027",
    year = "2021"
}

@article{Frellesvig:2021hk,
    author = "Frellesvig, H.",
    title = "{On Epsilon Factorized Differential Equations for Elliptic Feynman Integrals}",
    eprint = "2110.07968",
    archivePrefix = "arXiv",
    primaryClass = "hep-th",
    month = "10",
    year = "2021"
}

@book{fulton2013riemann,
  title={Riemann-Roch Algebra},
  author={Fulton, W. and Lang, S.},
  series={Grundlehren der mathematischen Wissenschaften},
  year={1985},
  publisher={Springer}
}

@article{Gabrielov1979,
author = {Gabrielov, A.M.},
title = {Polar Curves and Intersection Matrices of Singularities.},
journal = {Inventiones mathematicae},
volume = {54},
pages = {15-22},
year = {1979},
url = {http://eudml.org/doc/142670},
}

@article{Gasparotto:2022mmp,
    author = "Gasparotto, F. and Rapakoulias, A. and Weinzierl, S.",
    title = "{Nonperturbative computation of lattice correlation functions by differential equations}",
    eprint = "2210.16052",
    archivePrefix = "arXiv",
    primaryClass = "hep-th",
    reportNumber = "MITP/22-090",
    doi = "10.1103/PhysRevD.107.014502",
    journal = "Phys. Rev. D",
    volume = "107",
    number = "1",
    pages = "014502",
    year = "2023"
}

@article{Giroux:2022wav,
    author = "Giroux, Mathieu and Pokraka, Andrzej",
    title = "{Loop-by-loop differential equations for dual (elliptic) Feynman integrals}",
    eprint = "2210.09898",
    archivePrefix = "arXiv",
    primaryClass = "hep-th",
    doi = "10.1007/JHEP03(2023)155",
    journal = "JHEP",
    volume = "03",
    pages = "155",
    year = "2023"
}

@misc{goresky2021lecturenotessheavesperverse,
      title={Lecture notes on sheaves and perverse sheaves}, 
      author={M. Goresky},
      year={2021},
      eprint={2105.12045},
      archivePrefix={arXiv},
      primaryClass={math.AG},
      url={https://arxiv.org/abs/2105.12045}, 
}

@article{Gorges23,
   title={On a procedure to derive $\varepsilon$-factorised differential equations beyond polylogarithms},
   volume={2023},
   DOI={10.1007/jhep07(2023)206},
   number={7},
   journal={Journal of High Energy Physics},
   publisher={Springer Science and Business Media LLC},
   author={Görges, L. and Nega, C. and Tancredi, L. and Wagner, F. J.},
   year={2023},
   month=jul }

@article{Goto:2013Laur,
author = {Goto, Y.},
title = "{Twisted Cycles and Twisted Period Relations for Lauricella's Hypergeometric Function $F_C$}",
journal = {International Journal of Mathematics},
volume = {24},
number = {12},
pages = {1350094},
year = {2013},
doi = {10.1142/S0129167X13500948},
eprint = "1308.5535",
archivePrefix  = "arXiv",
primaryClass   = "math.AG"
}

@article{Yoshiaki_GOTO2015203,
  title="{Intersection Numbers and Twisted Period Relations for the Generalized Hypergeometric Function $_{m+1}F_{m}$}",
  author={Y. Goto},
  journal={Kyushu Journal of Mathematics},
  volume={69},
  number={1},
  pages={203-217},
  year={2015},
  doi={10.2206/kyushujm.69.203}
}

@article{goto2015,
author = "Goto, Y. and Matsumoto, K.",
doi = "10.1215/00277630-2873714",
fjournal = "Nagoya Mathematical Journal",
journal = "Nagoya Math. J.",
month = "03",
pages = "61--94",
publisher = "Duke University Press",
title = "{The monodromy representation and twisted period relations for Appell's hypergeometric function $F_{4}$}",
volume = "217",
year = "2015"
}

@article{Goto2022homology,
	author = {Y. Goto and S.J. Matsubara-Heo},
	doi = {https://doi.org/10.1016/j.indag.2021.12.002},
	journal = {Indagationes Mathematicae},
	number = {3},
	pages = {546-580},
	title = {Homology and cohomology intersection numbers of GKZ systems},
	volume = {33},
	year = {2022}
}

@article{griffiths1,
 URL = {http://www.jstor.org/stable/1970746},
 author = {P. A. Griffiths},
 journal = {Annals of Mathematics},
 number = {3},
 pages = {460--541},
 publisher = {[Annals of Mathematics, Trustees of Princeton University on Behalf of the Annals of Mathematics, Mathematics Department, Princeton University]},
 title = {On the Periods of Certain Rational Integrals: I,II},
 urldate = {2025-02-24},
 volume = {90},
 year = {1969}
}

@article{Grimm:2025zhv,
    author = "Grimm, T. W. and Hoefnagels, A. and van Vliet, M.",
    title = "{A Reduction Algorithm for Cosmological Correlators: Cuts, Contractions, and Complexity}",
    eprint = "2503.05866",
    archivePrefix = "arXiv",
    primaryClass = "hep-th",
    month = "3",
    year = "2025"
}

@article{Grimm:2024tbg,
    author = "Grimm, T. W. and Hoefnagels, A.",
    title = "{Reductions of GKZ systems and applications to cosmological correlators}",
    eprint = "2409.13815",
    archivePrefix = "arXiv",
    primaryClass = "hep-th",
    doi = "10.1007/JHEP04(2025)196",
    journal = "JHEP",
    volume = "04",
    pages = "196",
    year = "2025"
}

@incollection{grothendieck62,
  author    = {Grothendieck, A.},
  title     = {Th{\'e}or{\`e}mes de dualit{\'e} pour les faisceaux alg{\'e}briques coh{\'e}rents},
  booktitle = {Fondements de la g{\'e}om{\'e}trie alg{\'e}brique},
  publisher = {Secr{\'e}tariat math{\'e}matique},
  year      = {1962},
}

@article{Grozin:2011mt,
      author         = "Grozin, A. G.",
      title          = "{Integration by parts: An Introduction}",
      booktitle      = "{Computer Algebra and Particle Physics: CAPP 2011 DESY,
                        Zeuthen, Germany, March 21-25, 2011}",
      journal        = "Int. J. Mod. Phys.",
      volume         = "A26",
      year           = "2011",
      pages          = "2807-2854",
      doi            = "10.1142/S0217751X11053687",
      reportNumber   = "TTP11-09",
      SLACcitation   = "%%CITATION = ARXIV:1104.3993;%%"
}

@article{Henn:2013pwa,
      author         = "Henn, J. M.",
      title          = "{Multiloop integrals in dimensional regularization made
                        simple}",
      journal        = "Phys. Rev. Lett.",
      volume         = "110",
      year           = "2013",
      pages          = "251601",
      doi            = "10.1103/PhysRevLett.110.251601",
      eprint         = "1304.1806",
      archivePrefix  = "arXiv",
      primaryClass   = "hep-th",
      SLACcitation   = "%%CITATION = ARXIV:1304.1806;%%"
}

@article{HuseinSade, 
author = {S M Husein-Zade},
title = {The monodromy groups of isoleted singularities of hypersurfaces},
journal = {Russian Mathematical Surveys},
volume = {32},
number = {2},
pages = {23},
year = {1977},
month = {apr},
doi = {10.1070/RM1977v032n02ABEH001615},
}

@article{Kaderli:2019dny,
    author = "Kaderli, A.",
    title = "{A note on the Drinfeld associator for genus-zero superstring amplitudes in twisted de Rham theory}",
    eprint = "1912.09406",
    archivePrefix = "arXiv",
    primaryClass = "hep-th",
    reportNumber = "HU-EP-19/40, HU-Mathematik-2019-09",
    doi = "10.1088/1751-8121/ab9462",
    journal = "J. Phys. A",
    volume = "53",
    number = "41",
    pages = "415401",
    year = "2020"
}

@article{Kalyanapuram:2020vil,
    author = "Kalyanapuram, N. and Jha, R. G.",
    title = "{Positive Geometries for all Scalar Theories from Twisted Intersection Theory}",
    eprint = "2006.15359",
    archivePrefix = "arXiv",
    primaryClass = "hep-th",
    doi = "10.1103/PhysRevResearch.2.033119",
    journal = "Phys. Rev. Res.",
    volume = "2",
    number = "3",
    pages = "033119",
    year = "2020"
}

@article{Kawai:1985xq,
      author         = "Kawai, H. and Lewellen, D. C. and Tye, S. H. H.",
      title          = "{A Relation Between Tree Amplitudes of Closed and Open
                        Strings}",
      journal        = "Nucl. Phys.",
      volume         = "B269",
      year           = "1986",
      pages          = "1-23",
      doi            = "10.1016/0550-3213(86)90362-7",
      reportNumber   = "CLNS-85/667",
      SLACcitation   = "%%CITATION = NUPHA,B269,1;%%"
}

@article{kita2,
author = {Kita, M. and Yoshida, M.},
title = "{Intersection Theory for Twisted Cycles II - Degenerate Arrangements}",
journal = {Mathematische Nachrichten},
volume = {168},
number = {1},
publisher = {WILEY-VCH Verlag},
doi = {10.1002/mana.19941680111},
pages = {171--190},
year = {1994},
}

@article{kita1,
author = {Kita, M. and Yoshida, M.},
title = "{Intersection Theory for Twisted Cycles}",
journal = {Mathematische Nachrichten},
volume = {166},
number = {1},
publisher = {WILEY-VCH Verlag},
doi = {10.1002/mana.19941660122},
pages = {287--304},
year = {1994},
}

@article{Klemm:2020knr,
    author = " Klemm, A. and Nega, C. and Safari, R.",
    title = "{The l-loop Banana Amplitude from GKZ Systems and relative Calabi-Yau Periods}",
    eprint = "1912.06201",
    archivePrefix = "arXiv",
    primaryClass = "hep-th",
    doi = "10.1007/JHEP04(2020)088",
    journal = "JHEP",
    volume = "2020",
    number = "4",
    year={},
}

@article{Kontsevich:2001kza,
  title={Periods},
  author={ Kontsevich, M. and  Zagier, D.},
journal = {Mathematics Unlimited — 2001 and Beyond},
pages = {771–808},
publisher={Springer Berlin Heidelberg},
year={2001},
}

@article{Kontsevich:2013rda,
    author = "Kontsevich, M. and Soibelman, Y.",
    title = "{Wall-crossing structures in Donaldson-Thomas invariants, integrable systems and Mirror Symmetry}",
    doi = "10.1007/978-3-319-06514-4_6",
    journal = "Lect. Notes Union. Mat. Ital.",
    volume = "15",
    pages = "197--308",
    year = "2014"
}

@misc{kontsevich:2008kos,
      title={Stability structures, motivic Donaldson-Thomas invariants and cluster transformations}, 
      author={M. Kontsevich and Y. Soibelman},
      year={2008},
      eprint={0811.2435},
      archivePrefix={arXiv},
      primaryClass={math.AG},
      url={https://arxiv.org/abs/0811.2435}, 
}

@article{Kontsevich:2022ana,
      title={Analyticity and resurgence in wall-crossing formulas}, 
      author={M. Kontsevich and Y. Soibelman},
      year={2022},
      eprint={2005.10651},
      archivePrefix={arXiv},
      primaryClass={math.AG},
      url={https://arxiv.org/abs/2005.10651}, 
}

@misc{Kontsevich:2024mks,
      title="{Holomorphic Floer theory I: exponential integrals in finite and infinite dimensions, arXiv:2402.07343}", 
      author={M. Kontsevich and Y. Soibelman},
      year={2024},
      eprint="2402.07343",
      archivePrefix={arXiv},
      primaryClass={math.SG},
}

@article{Kotikov1991123,
title = {Differential equation method. The calculation of N-point Feynman diagrams},
journal = {Physics Letters B},
volume = {267},
number = {1},
pages = {123-127},
year = {1991},
doi = {https://doi.org/10.1016/0370-2693(91)90536-Y},
author = {A.V. Kotikov},
}

@article{KOTIKOV1991158,
title = "{Differential equations method. New technique for massive Feynman diagram calculation}",
journal = "Physics Letters B",
volume = "254",
number = "1",
pages = "158 - 164",
year = "1991",
doi = "https://doi.org/10.1016/0370-2693(91)90413-K",
author = "A.V. Kotikov"
}

@book{Kulikov_1998, place={Cambridge}, series={Cambridge Tracts in Mathematics}, title={Mixed Hodge Structures and Singularities}, publisher={Cambridge University Press}, author={Kulikov, Valentine S.}, year={1998}, collection={Cambridge Tracts in Mathematics}}

@article{Lairez:2023nih,
    author = "Lairez, P. and Pichon-Pharabod, E. and Vanhove, P.",
    title = "{Effective homology and periods of complex projective hypersurfaces}",
    eprint = "2306.05263",
    archivePrefix = "arXiv",
    primaryClass = "math.AG",
    doi = "10.1090/mcom/3947",
    journal = "Math. Comput.",
    volume = "93",
    number = "350",
    pages = "2985--3025",
    year = "2024"
}

@article{Laporta:2001dd,
    author = "Laporta, S.",
    title = "High precision calculation of multiloop Feynman integrals by difference equations",
    eprint = "hep-ph/0102033",
    archivePrefix = "arXiv",
    doi = "10.1142/S0217751X00002159",
    journal ="Int. J. Mod. Phys. A",
    year = "2000",
    volume = "15",
    pages = "5087--5159"
}

@article{Lee:2009dh,
      author         = "Lee, R. N.",
      title          = "{Space-time dimensionality D as complex variable:
                        Calculating loop integrals using dimensional recurrence
                        relation and analytical properties with respect to D}",
      journal        = "Nucl. Phys.",
      volume         = "B830",
      year           = "2010",
      pages          = "474-492",
      doi            = "10.1016/j.nuclphysb.2009.12.025",
      eprint         = "0911.0252",
      archivePrefix  = "arXiv",
      primaryClass   = "hep-ph",
      SLACcitation   = "%%CITATION = ARXIV:0911.0252;%%"
}

@article{Lee:2013hzt,
      author         = "Lee, R. N. and Pomeransky, A. A.",
      title          = "{Critical points and number of master integrals}",
      journal        = "JHEP",
      volume         = "11",
      year           = "2013",
      pages          = "165",
      doi            = "10.1007/JHEP11(2013)165",
}

@article{Lefschetz:1924,
  title={L'analysis Situs et la Géométrie Algébrique},
  author={Lefschetz, S.},
journal={Gauthier-Villars},
year={1924},
}

@article{Levelt61,
  author       = {Levelt, A. H. M.},
  title        = {Jordan decomposition for a class of singular differential equations},
  journal      = {Arkiv f{\"o}r Matematik},
  year         = {1961}
}

@book{Lipman,
author = {Lipman, J.},
year = {2009},
month = {01},
pages = {},
publisher ={},
title = {Notes on Derived Functors and Grothendieck Duality.},
volume = {1960},
journal = {Lecture Notes in Math.}
}

@article{Lunev_1994,
   title={Differential equations for definition and evaluation of Feynman integrals},
   volume={50},
   DOI={10.1103/physrevd.50.6589},
   number={10},
   journal={Physical Review D},
   publisher={American Physical Society (APS)},
   author={Lunev, F. A.},
   year={1994},
   month=nov, pages={6589–6593} }

@article{Ma:2021cxg,
    author = "Ma, C. and Wang, Y. and Xu, X. and Yang, L. L. and Zhou, B.",
    title = "{Mixed QCD-EW corrections for Higgs leptonic decay via $H W^+ W^-$ vertex}",
    eprint = "2105.06316",
    archivePrefix = "arXiv",
    primaryClass = "hep-ph",
    doi = "10.1007/JHEP09(2021)114",
    journal = "JHEP",
    volume = "09",
    pages = "114",
    year = "2021"
}

@article{Malgrange1994,
    author = "B. Malgrange",
    title = "{Chap. IV: Regular connexions after Deligne}",
    journal = {Algebraic D-Modules} ,
    year = {1994},
    pages = "151-172",
}

@mastersthesis{Massidda,
    author = "Massidda, A.",
    title = "{A modern approach to String Amplitudes and Intersection Theory}",
    eprint = "2403.09741",
    archivePrefix = "arXiv",
    primaryClass = "hep-th",
    type = "Other thesis",
    month = "3",
    year = "2024"
}

@article{Mastrolia:2018uzb,
      author         = "Mastrolia, P. and Mizera, S.",
      title          = "{Feynman Integrals and Intersection Theory}",
      journal        = "JHEP",
      volume         = "02",
      year           = "2019",
      pages          = "139",
      doi            = "10.1007/JHEP02(2019)139",
      eprint         = "1810.03818",
      archivePrefix  = "arXiv",
      primaryClass   = "hep-th",
      SLACcitation   = "%%CITATION = ARXIV:1810.03818;%%"
}

@article{matsubaraheo2019algorithm,
    title={An algorithm of computing cohomology intersection number of hypergeometric integrals},
    author={S. J. Matsubara-Heo and N. Takayama},
    year={2019},
    eprint={1904.01253},
    archivePrefix={arXiv},
    primaryClass={math.AG},
    journal = "Nagoya Mathematical Journal",
    pages = "1--17",
    volume = "",
    doi = "10.1017/nmj.2021.2"
}

@article{Matsubara-Heo:2020lzo,
    author = "Matsubara-Heo, S. J.",
    title = "{Computing cohomology intersection numbers of GKZ hypergeometric systems}",
    eprint = "2008.03176",
    archivePrefix = "arXiv",
    primaryClass = "math.AG",
    doi = "10.22323/1.383.0013",
    journal = "PoS",
    volume = "MA2019",
    pages = "013",
    year = "2022"
}

@article{Matsubara-Heo:2021dtm,
	author = {Matsubara-Heo, S. J.},
	doi = {10.2969/jmsj/87738773},
	journal = {Journal of the Mathematical Society of Japan},
	pages = {1 -- 32},
	publisher = {Mathematical Society of Japan},
	title = {{Localization formulas of cohomology intersection numbers}},
	year = {2022}
}

@article{matsumoto1994,
  title="{Quadratic Identities for Hypergeometric Series of Type $(k,l)$}",
  author={K. Matsumoto},
  journal={Kyushu Journal of Mathematics},
  volume={48},
  number={2},
  pages={335-345},
  year={1994},
  doi={10.2206/kyushujm.48.335}
}

@article{matsumoto1998,
author = "Matsumoto, K.",
fjournal = "Osaka Journal of Mathematics",
journal = "Osaka J. Math.",
number = "4",
pages = "873--893",
publisher = "Osaka University and Osaka City University, Departments of Mathematics",
title = "Intersection numbers for logarithmic $k$-forms",
url = "https://projecteuclid.org:443/euclid.ojm/1200788347",
volume = "35",
year = "1998"
}

@article{Mazloumi:2024wys,
    author = "Mazloumi, P. and Stieberger, S.",
    title = "{One-loop double copy relation from twisted (co)homology}",
    journal = "JHEP",
    volume = "10",
    pages = "148",
    year = "2024"
}

@book{Milnor:1963jmi,  
title={Morse theory},
publisher={Princeton University Press},
author={Milnor, J.}, 
year={1963}, 
}

@book{Porteous_1971, 
title="{Singular Points of Complex Hypersurfaces.}", 
publisher ="{Princeton University Press}", 
series ={Annals of Mathematics Studies},
author={J. Milnor}, 
year={1969}}

@Article{Mimachi2003,
author="Mimachi, K. and Yoshida, M.",
title="{Intersection Numbers of Twisted Cycles and the Correlation Functions of the Conformal Field Theory}",
journal="Communications in Mathematical Physics",
year="2003",
volume="234",
number="2",
pages="339--358",
doi="10.1007/s00220-002-0766-4"
}

@Article{Mimachi2004,
author="Mimachi, K.
and Yoshida, Masaaki",
title="Intersection Numbers of Twisted Cycles Associated with the Selberg Integral and an Application to the Conformal Field Theory",
journal="Communications in Mathematical Physics",
year="2004",
month="Aug",
day="01",
volume="250",
number="1",
pages="23--45",
doi="10.1007/s00220-004-1138-z"
}

@article{Mizera:2016jhj,
      author         = "Mizera, S.",
      title          = "{Inverse of the String Theory KLT Kernel}",
      journal        = "JHEP",
      volume         = "06",
      year           = "2017",
      pages          = "084",
      doi            = "10.1007/JHEP06(2017)084",
      eprint         = "1610.04230",
      archivePrefix  = "arXiv",
      primaryClass   = "hep-th",
      SLACcitation   = "%%CITATION = ARXIV:1610.04230;%%"
}

@article{Mizera:2017rqa,
      author         = "Mizera, S.",
      title          = "{Scattering Amplitudes from Intersection Theory}",
      journal        = "Phys. Rev. Lett.",
      volume         = "120",
      year           = "2018",
      number         = "14",
      pages          = "141602",
      doi            = "10.1103/PhysRevLett.120.141602",
      eprint         = "1711.00469",
      archivePrefix  = "arXiv",
      primaryClass   = "hep-th",
      SLACcitation   = "%%CITATION = ARXIV:1711.00469;%%"
}

@article{Mizera:2019gea,
      author         = "Mizera, S.",
      title          = "{Aspects of Scattering Amplitudes and Moduli Space
                        Localization}",
      journal = {arxiv: hep-th},
      year           = "2019",
      eprint         = "1906.02099",
      archivePrefix  = "arXiv",
      primaryClass   = "hep-th",
      SLACcitation   = "%%CITATION = ARXIV:1906.02099;%%"
}

@article{Mizera:2019vvs,
    author = "Mizera, S. and Pokraka, A.",
    title = "{From Infinity to Four Dimensions: Higher Residue Pairings and Feynman Integrals}",
    eprint = "1910.11852",
    archivePrefix = "arXiv",
    primaryClass = "hep-th",
    doi = "10.1007/JHEP02(2020)159",
    journal = "JHEP",
    volume = "02",
    pages = "159",
    year = "2020"
}

@Inbook{Nicolaescu2007,
author="L. Nicolaescu",
title="An Invitation to Morse Theory",
year="2007",
publisher="Springer New York",
address="New York, NY",
pages="151--191",
doi="10.1007/978-0-387-49510-1_4"
}

@misc{Ohara98,
    author = {K. Ohara},
    title = "{Intersection numbers of twisted cohomology groups associated with Selberg-type integrals}",
    year = {1998},
    url={http://www.math.kobe-u.ac.jp/HOME/ohara/Math/980523.ps}
}

@article{OST2003,
  title="{Quadratic Relations for Generalized Hypergeometric Functions $_p F_{p-1}$}",
  author={K. Ohara and Y. Sugiki and N. Takayama},
  journal={Funkcialaj Ekvacioj},
  volume={46},
  number={2},
  pages={213-251},
  year={2003},
  doi={10.1619/fesi.46.213}
}

@book{Olver,
author = "F. W.I. Olver",
title = "{Asymptotics and Special Functions}",
publisher= {CRC Press},
place = {New York},
year={1997},
}

@article{Pham:1967,
   title="Introduction \'a l'\'etude topologique des singularit\'es de Landau",
   number={164},
   journal={M\'emorial des sciences math\'ematiques},
   publisher={Gauthier-Villars},
   author={Pham, Fr\'ed\'eric},
   year={1967},
   month=sep }

@misc{pichon2025,
      title={Periods of fibre products of elliptic surfaces and the Gamma conjecture}, 
      author={E. Pichon-Pharabod},
      year={2025},
      eprint={2505.07685},
      archivePrefix={arXiv},
      primaryClass={math.AG},
      url={https://arxiv.org/abs/2505.07685}, 
}

@article{Pogel23,
   title={Bananas of equal mass: any loop, any order in the dimensional regularisation parameter},
   volume={2023},
   DOI={10.1007/jhep04(2023)117},
   number={4},
   journal={Journal of High Energy Physics},
   publisher={Springer Science and Business Media LLC},
   author={Pögel, S. and Wang, X. and Weinzierl, S.},
   year={2023},
   month=apr }

@article{Pogel23-2,
   title={Taming Calabi-Yau Feynman Integrals: The Four-Loop Equal-Mass Banana Integral},
   volume={130},
   DOI={10.1103/physrevlett.130.101601},
   number={10},
   journal={Physical Review Letters},
   publisher={American Physical Society (APS)},
   author={Pögel, S. and Wang, X. and Weinzierl, S.},
   year={2023},
   month=mar }

@article{Pogel22,
   title={The three-loop equal-mass banana integral in $\varepsilon$-factorised form with meromorphic modular forms},
   volume={2022},
   DOI={10.1007/jhep09(2022)062},
   number={9},
   journal={Journal of High Energy Physics},
   publisher={Springer Science and Business Media LLC},
   author={Pögel, S. and Wang, X. and Weinzierl, S.},
   year={2022},
   month=sep }

@article{Pokraka:2025zlh,
    author = "A. Pokraka, L. Ren and C. Rodriguez",
    title = "{A double copy from twisted (co)homology at genus g}",
    eprint = "2509.01598",
    year = "2025"
}

@article{Primo17,
      author         = "Primo, A. and Tancredi, L.",
      title          = "{Maximal cuts and differential equations for Feynman
                        integrals. An application to the three-loop massive banana
                        graph}",
      journal        = "Nucl. Phys.",
      volume         = "B921",
      year           = "2017",
      pages          = "316-356",
      doi            = "10.1016/j.nuclphysb.2017.05.018",
      eprint         = "1704.05465",
      archivePrefix  = "arXiv",
      primaryClass   = "hep-ph",
      reportNumber   = "TTP17-021",
      SLACcitation   = "%%CITATION = ARXIV:1704.05465;%%"
}

@article{Remiddi:1997ny,
      author         = "Remiddi, E.",
      title          = "{Differential equations for Feynman graph amplitudes}",
      journal        = "Nuovo Cim.",
      volume         = "A110",
      year           = "1997",
      pages          = "1435-1452",
      eprint         = "hep-th/9711188",
      archivePrefix  = "arXiv",
      primaryClass   = "hep-th",
      reportNumber   = "DFUB-97-15",
      SLACcitation   = "%%CITATION = HEP-TH/9711188;%%"
}

@book{scheidemann,
  title={Introduction to Complex Analysis in Several Variables},
  author={Scheidemann, V.},
  series={Introduction to Complex Analysis in Several Variables},
  url={https://books.google.it/books?id=p3k-jW1USlAC},
  year={2005},
  publisher={Birkh{\"a}user Basel}
}

@article{Schmid73,
    author = "W. Schmid",
    title = "{Variation of hodge structure: The singularities of the period mapping}",
    journal = {Invent Math} ,
    doi = "10.1007/BF01389674",
    volume = "22",
    year = {1973},
    pages = "211-319",
}

@article{Serre51,
  title={Homologie Singuliere Des Espaces Fibres},
  author={J. P. Serre},
  journal={Annals of Mathematics},
  year={1951},
  volume={54},
  pages={425},
  url={https://api.semanticscholar.org/CorpusID:123635490}
}

@article{Serre55,
author = {Serre, J. P.},
journal = {Commentarii mathematici Helvetici},
pages = {9-26},
title = {Un théorème de dualité.},
url = {http://eudml.org/doc/139092},
volume = {29},
year = {1955},
}

@book{shafarevich1,
  title={Basic Algebraic Geometry 1:Varieties in Projective space},
  author={Shafarevich, I.R.},
  series={Basic Algebraic Geometry},
  year={1994},
  publisher={Springer-Verlag}
}

@book{shafarevich2,
  title={Basic Algebraic Geometry 2:Schemes and complex manifolds},
  author={Shafarevich, I.R.},
  series={Basic Algebraic Geometry},
  year={1994},
  publisher={Springer-Verlag}
}

@article{Stieberger:KLT,
    author = "S. Stieberger",
    title = "{One-Loop Double Copy Relation in String Theory}",
    journal = "Phys. Rev. Lett.",
    volume = "132",
    number = "19",
    pages = "191602",
    year = "2024"
}

@article{Tarasov96,
      author         = "Tarasov, O. V.",
      title          = "{Connection between Feynman integrals having different
                        values of the space-time dimension}",
      journal        = "Phys. Rev.",
      volume         = "D54",
      year           = "1996",
      pages          = "6479-6490",
      doi            = "10.1103/PhysRevD.54.6479",
      eprint         = "hep-th/9606018",
      archivePrefix  = "arXiv",
      primaryClass   = "hep-th",
      reportNumber   = "DESY-96-068, JINR-E2-96-62",
      SLACcitation   = "%%CITATION = HEP-TH/9606018;%%"
}

@article {Tjurina,
    author = {Tjurina, G. N.},
     title = {Locally semi-universal flat deformations of isolated
              singularities of complex spaces},
   journal = {Izv. Akad. Nauk SSSR Ser. Mat.},
  fjournal = {Izvestiya Akademii Nauk SSSR. Seriya Matematicheskaya},
    volume = {33},
      year = {1969},
     pages = {1026--1058},
}

@article{Turrittin55,
  author       = {Turrittin, H. L.},
  title        = {Convergent solutions of ordinary linear homogeneous differential equations in the neighbourhood of an irregular singular point},
  journal      = {Proc. Amer. Math. Soc. / Trans.},
  year         = {1955},
}

@inproceedings{vanhove18,
    author = "Vanhove, Pierre",
    title = "{Feynman integrals, toric geometry and mirror symmetry}",
    booktitle = "{KMPB Conference}: {Elliptic Integrals, Elliptic Functions and Modular  Forms in Quantum Field Theory}",
    doi = "10.1007/978-3-030-04480-0_17",
    pages = "415--458",
    year = "2019"
}

@inproceedings{verdier69,
  author    = {Verdier, J. L.},
  title     = {Base change for twisted inverse images of coherent sheaves},
  booktitle = {Algebraic Geometry ({B}ombay {C}olloquium)},
  pages     = {393-408},
  publisher = {Oxford University Press},
  year      = {1969},
}

@book{Voisin_2002, place={Cambridge}, series={Cambridge Studies in Advanced Mathematics}, title={Hodge Theory and Complex Algebraic Geometry I}, publisher={Cambridge University Press}, author={Voisin, C.}, year={2002}, collection={Cambridge Studies in Advanced Mathematics}}

@book{Wasow65,
  author       = {Wasow, W.},
  title        = {Asymptotic Expansions for Ordinary Differential Equations},
  publisher    = {Interscience / Dover},
  year         = {1965},
}

@article{Weinzierl:2020nhw,
    author = "Weinzierl, S.",
    title = "{Correlation functions on the lattice and twisted cocycles}",
    eprint = "2003.05839",
    archivePrefix = "arXiv",
    primaryClass = "hep-th",
    reportNumber = "MITP/20-013",
    doi = "10.1016/j.physletb.2020.135449",
    journal = "Phys. Lett. B",
    volume = "805",
    pages = "135449",
    year = "2020"
}

@article{Weinzierl:2022eaz,
    author = "Weinzierl, S.",
    title = "{Feynman Integrals}",
    eprint = "2201.03593",
    archivePrefix = "arXiv",
    primaryClass = "hep-th",
    month = "1",
    journal = {arxiv: hep-th},
    year = "2022"
}

@article{Witten:2010cx,
      author         = "Witten, E.",
      title          = "{Analytic Continuation Of Chern-Simons Theory}",
      booktitle      = "{Chern-Simons gauge theory: 20 years after. Proceedings,
                        Workshop, Bonn, Germany, August 3-7, 2009}",
      journal        = "AMS/IP Stud. Adv. Math.",
      volume         = "50",
      year           = "2011",
      pages          = "347-446",
      eprint         = "1001.2933",
      archivePrefix  = "arXiv",
      primaryClass   = "hep-th",
      SLACcitation   = "%%CITATION = ARXIV:1001.2933;%%"
}

@article{Yau1978,
  author  = {S. T. Yau},
  title   = {On the Ricci curvature of a compact Kähler manifold and the complex Monge–Ampère equation. I},
  journal = {Communications on Pure and Applied Mathematics},
  volume  = {31},
  number  = {3},
  pages   = {339--411},
  year    = {1978},
}

@book{yoshida2013hypergeometric,
  title="{Hypergeometric Functions, My Love: Modular Interpretations of Configuration Spaces}",
  author={Yoshida, M.},
  series={Aspects of Mathematics},
  doi={10.1007/978-3-322-90166-8},
  year={2013},
  publisher={Vieweg+Teubner Verlag}
}
